\documentclass[a4paper,12pt]{book}

\usepackage[utf8]{inputenc}
 \usepackage[left=2.5cm,right=2.5cm,top=3.25cm,bottom=3cm]{geometry}
\usepackage{mathtools}
\mathtoolsset{showonlyrefs}
\usepackage{amsfonts, amsthm}
\usepackage{braket}
\usepackage{dsfont}
\usepackage{bbm}
\usepackage{bm}
\usepackage{graphicx}
\usepackage{psfrag}
\usepackage{hyperref}

\usepackage{indentfirst}
\usepackage[nottoc,notlot,notlof]{tocbibind}

\newcommand{\cA}{\mathcal{A}}
\newcommand{\cB}{\mathcal{B}}
\newcommand{\cC}{\mathcal{C}}
\newcommand{\cD}{\mathcal{D}}
\newcommand{\cE}{\mathcal{E}}
\newcommand{\cF}{\mathcal{F}}
\newcommand{\cG}{\mathcal{G}}
\newcommand{\cH}{\mathcal{H}}

\newcommand{\cJ}{\mathcal{J}}
\newcommand{\cK}{\mathcal{K}}
\newcommand{\cL}{\mathcal{L}}
\newcommand{\cM}{\mathcal{M}}

\newcommand{\cP}{\mathcal{P}}
\newcommand{\cQ}{\mathcal{Q}}

\newcommand{\cS}{\mathcal{S}}
\newcommand{\cT}{\mathcal{T}}

\newcommand{\cV}{\mathcal{V}}

\newcommand{\cZ}{\mathcal{Z}}

\newcommand{\bA}{\mathbf{A}}

\newcommand{\bC}{\mathbf{C}}

\newcommand{\bI}{\mathbf{I}}
\newcommand{\bJ}{\mathbf{J}}

\newcommand{\bM}{\mathbf{M}}

\newcommand{\bQ}{\mathbf{Q}}

\newcommand{\bT}{\mathbf{T}}

\newcommand{\bV}{\mathbf{V}}

\newcommand{\bX}{\mathbf{X}}
\newcommand{\bY}{\mathbf{Y}}

\newcommand{\kC}{\mathfrak{C}}
\newcommand{\kK}{\mathfrak{K}}
\newcommand{\kM}{\mathfrak{M}}

\newcommand{\csig}{\bm{\sigma}}
\newcommand{\csigma}{\bm{\sigma}}
\newcommand{\slambda}{\sqrt{\lambda}}

\newcommand{\psit}{{\bf\Psi}}

\newcommand{\setZ}{\mathbb{Z}}
\newcommand{\setN}{\mathbb{N}}
\newcommand{\setC}{\mathbb{C}}
\newcommand{\setR}{\mathbb{R}}

\newcommand{\e}{\mathrm{e}}
\newcommand{\cc}{\mathrm{c}}

\newcommand{\be}{\begin{equation}}
\newcommand{\ee}{\end{equation}}

\newcommand{\bea}{\begin{eqnarray}}
\newcommand{\eea}{\end{eqnarray}}
\newcommand{\bal}{\begin{align}}
\newcommand{\eal}{\end{align}}
\newcommand{\beq}{\begin{equation}}
\newcommand{\eeq}{\end{equation}}
\newcommand{\dr}{\partial}
\newcommand{\tr}{\mathrm{Tr}}
\newcommand{\trd}{{\tr_\cD}}
\newcommand{\Tr}{\mathrm{Tr}}
\newcommand{\lnz}{\mathrm{log} Z}

\newcommand{\id}{\mathds{1}}
\newcommand{\Det}{\mathrm{Det}}
\newcommand{\bbone}{{\bf 1}}

\newtheorem{theorem}{Theorem}
\newtheorem{definition}{Definition}
\newtheorem{proposition}{Proposition}
\newtheorem{corollary}{Corollary}
\newtheorem{lemma}{Lemma}

\numberwithin{theorem}{chapter}
\numberwithin{definition}{chapter}
\numberwithin{corollary}{chapter}
\numberwithin{lemma}{chapter}
\numberwithin{proposition}{chapter}

\newcommand{\prf}{{\noindent \bf Proof\; \; }}

\usepackage[usenames,dvipsnames,svgnames,table]{xcolor}
\usepackage{calc} 
\usepackage{tikz} 
\usepackage[absolute]{textpos} 

\newcommand{\PhDTitleEN}{Quartic Tensor Models} 

 \label{Personnal}
\newcommand{\PhDname}{Thibault Delepouve} 
\newcommand{\NNT}{2017SACLS085} 
\newcommand{\PhDworkingplace}{\`a l'Universit\'e Paris-Sud et \`a l'\'Ecole Polytechnique} 
\newcommand{\defenseplace}{Orsay} 
\newcommand{\defensedate}{15 Mai 2017} 

\label{Jury}
\newcommand{\jurynameA}{Christoph KOPPER}
\newcommand{\jurygenderA}{Pr.} 
\newcommand{\juryadressA}{\'Ecole Polytechnique}

\newcommand{\juryroleA}{Président du jury} %
\newcommand{\jurynameB}{Manfred SALMHOFER}
\newcommand{\jurygenderB}{Pr.} 
\newcommand{\juryadressB}{Universit\"at Heidelberg}

\newcommand{\juryroleB}{Rapporteur}
\newcommand{\jurynameC}{Adrian TANASA}
\newcommand{\jurygenderC}{Pr.} 
\newcommand{\juryadressC}{Universit\'e de Bordeaux}

\newcommand{\juryroleC}{Rapporteur}
\newcommand{\jurynameD}{Nicolas CURIEN}
\newcommand{\jurygenderD}{Pr.} 
\newcommand{\juryadressD}{Universit\'e Paris-Sud}

\newcommand{\juryroleD}{Examinateur}
\newcommand{\jurynameE}{Razvan GURAU}
\newcommand{\jurygenderE}{Dr.} 
\newcommand{\juryadressE}{\'Ecole Polytechnique}

\newcommand{\juryroleE}{Co-directeur de th\`ese}
\newcommand{\jurynameF}{Vincent RIVASSEAU }
\newcommand{\jurygenderF}{Pr.} 
\newcommand{\juryadressF}{Universit\'e Paris-Sud}

\newcommand{\juryroleF}{Directeur de th\`ese}

 	\definecolor{upsac}{HTML}{62003C}

\begin{document}
\pagestyle{empty}
\newgeometry{textheight=150ex,textwidth=40em,top=30pt,headheight=30pt,headsep=30pt,inner=80pt}

\begin{tikzpicture}[remember picture,overlay,color=upsac]
	\draw[very thick]
		([yshift=-135pt,xshift=50pt]current page.north west)--     
		([yshift=-135pt,xshift=-30pt]current page.north east)--    
		([yshift=35pt,xshift=-30pt]current page.south east)--      
		([yshift=35pt,xshift=50pt]current page.south west)--cycle; 
\end{tikzpicture}

\begin{textblock}{13}(1.15,3)
  NNT : \NNT
\end{textblock}

\begin{textblock}{1}(1.15,1)
\includegraphics[height=1.8cm]{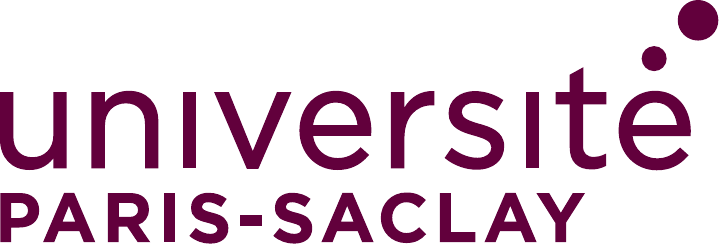} 
\label{Logo Paris Saclay}
\end{textblock}


\vspace{5cm}
\color{upsac} 
  \begin{center}    
   { \LARGE\textsc{Th\`ese de doctorat\\ de l'Universit\'e Paris-Saclay}} \\
    \large{ \textsc{pr\'epar\'ee \PhDworkingplace}
    } 
   \color{black} 
%
     \vspace{1.5cm} 
   \Large{par}
   
   \vspace{1cm}
   
   \Large{\textsc{\PhDname}} 
    \vfill
    \huge{\textbf{\PhDTitleEN}} 
    \vfill
    \bigskip
\end{center}
\color{black}
\begin{flushleft}
Th\`ese pr\'esent\'ee et soutenue \`a \defenseplace, le \defensedate. \\
\bigskip
Composition du Jury :
\end{flushleft}

\begin{center}
\begin{tabular}{llll}

    \jurygenderA & \textsc{\jurynameA}  & \juryadressA & (\juryroleA) \\

    \jurygenderB & \textsc{\jurynameB}  & \juryadressB & (\juryroleB) \\

    \jurygenderC & \textsc{\jurynameC}  & \juryadressC & (\juryroleC) \\

    \jurygenderD & \textsc{\jurynameD}  & \juryadressD & (\juryroleD) \\

    \jurygenderE & \textsc{\jurynameE}  & \juryadressE & (\juryroleE) \\

    \jurygenderF & \textsc{\jurynameF}  & \juryadressF & (\juryroleF) \\

  \end{tabular}    
\end{center}

\restoregeometry
\chapter*{Remerciements}

Je tiens \`a remercier Razvan Gurau et Vincent Rivasseau. Au del\`a de leur encadrement, cette th\`ese n'aurait \'et\'e 
possible sans leur engagement, leurs contributions et leurs conseils tout au long de ces trois ans. Leur passion pour le sujet, mais aussi pour les math\'ematiques et la 
physique en g\'en\'eral, furent une grande source de motivation. 

Merci \`a Manfred Salmhofer et Adrian Tanasa pour leur relecture minutieuse et leurs rapports d\'etaill\'es. 
Merci aussi \`a Nicolas Curien et Christoph Kopper qui me font l'honneur de participer au jury.

Merci \`a l'Universit\'e Paris-Sud et \`a l'\'Ecole Polytechnique. Merci \`a l'\'Ecole Normale Sup\'erieure Paris-Saclay pour le financement de mon projet.
Merci au Laboratoire de physique th\'eorique d'Orsay, au Centre de physique th\'eorique de l'\'Ecole Polytechnique et tout particuli\`erement au personnel et \`a 
la direction de ces deux instituts, pour leur efficacit\'e, leur disponibilit\'e et leur cordialit\'e.

Merci \`a tous les collaborateurs, proches ou lointains, dont l'aide fut déterminante pour le succès de mes travaux. 
Valentin Bonzom, pour nos conversations très instructives, et son importante implication dans nos travaux communs. 
Fabien Vignes-Tourneret, pour ses remarques cruciales ayant permis l'aboutissement des \'etudes constructives.

Merci à tous les collègues de laboratoires et de conférences, Damir Becirevic, Dario Benedetti, Joseph Ben Geloun, Thomas Krajewski et bien d'autres, 
dont les conversations amicales ont égayé des années de recherche qui, sinon, auraient été trop aust\`eres.
Merci \`a mes camarades de th\`eses, désormais amis, St\'ephane, Vincent et Luca, et aux comp\`eres de conf\'erences, Eduardo et Tajron,
pour d'innombrables raisons qui pourraient faire l'objet d'un ouvrage \`a elles seules.

Merci enfin \`a ma famille pour son soutien et ses encouragements, ainsi qu' \`a mes amis de Paris,
de Lille et d'ailleurs sur lesquels je peux toujours compter, m\^eme parfois apr\`es des ann\'ees d'\'eloignement.

Merci enfin \`a Indira, qui m'a accompagn\'e, soutenu et aimé pendant ces ann\'ees parfois difficiles.

\pagestyle{plain}
\tableofcontents

\chapter*{Introduction}
\addcontentsline{toc}{chapter}{Introduction}
%
%
%
\subsubsection*{A brief History of Tensor models}

Tensor models provide a generalisation of matrix models, and were first developed to study random geometries in any dimension.

Matrix models are probability distributions for square matrices, that are invariant under unitary transformation.
They where first introduced by Wishart for the statistical analysis of large samples \cite{Wishart1928}. However, they later 
acquired another use when a relation was discovered with random surfaces.
Indeed, the perturbative expansion of matrix models evaluates their moments and partition functions as a sum over maps. 
Maps correspond to graphs drawn on a surface, and thus procure triangulations of those surfaces. To each map corresponds a weight given by the Feynman rules, 
forming a probability distribution over maps. 
Therefore, from a matrix model arises a probability distribution over random discretised surfaces. 
Furthermore, the size $N$ of the matrix $\bM\in M_N(\setC)$ offers a new small parameter,
$1/N$, according to which the perturbative series can be re-arranged \cite{tHooft1974}.
This expansion in powers of $1/N$ happens to be indexed by the genus of the maps, and maps of a given genus grows at a manageable rate with the number of vertices,
can be enumerated explicitly and form a summable family \cite{BIPZ1978}. 
Moreover, the $1/N$ expansions sorts the discretised surfaces according to their topology, and at large $N$, the contribution of spheres dominates as the ones of 
higher-genus surfaces vanishes. The corresponding series has a critical point at the boundary of its convergence domain, where the theory reaches a phase which favours
larger maps, leading to a continuum theory of random planar surfaces, the Brownian map \cite{LeGall2007,LeGall1105}, 
and allowing for the quantification of two dimensional gravity\cite{DFGZJ1993}.

Early work on tensor models started in the 1990's, looking for a generalisation of random matrices with
tensors of rank higher than 2 \cite{ABJ1991,Gross1992,Sasakura1991}. 
Unfortunately, the progress soon stalled while a $1/N$ expansion remained to be found. The interest for tensor models was renewed two decades later, when invariant
{\it coloured} tensor models were derived from Group Field Theories.

Group Field Theories are quantum field theories based on a Lie-group manifold, first introduced by Boulatov in 1992 \cite{Bou92},
that were developed to provide a satisfactory, background independent,
field theoretical formulation for Loop Quantum Gravity. The path integral formulation of Loop Quantum Gravity writes as a sum over spin-foams, 
which are graphs bearing some additional data that ties to the geometry of space-time \cite{Oriti2001,Perez2003}. 
Group field theories aim at building a theory that generates these spin-foam as the Feynman graphs of a quantum field. 
As space-time geometry should arise from the field and not precede it, the field must be \emph{background independent} : 
it must not be based on a space manifold like usual quantum fields encountered, for example, in electrodynamics. In order to sow the necessary clues for the emergence of
space-time geometry, the field is based on the manifold of the Lorentz group, 
the rotation group or multiple copies of them \cite{PFKR99,Freidel2005,Oriti0912,Oriti1110}. By Fourier transformation, the field variables of a group field theory
over $D$ copies of a compact Lie group have a tensor structure, and corresponding group field theories are tensor models with additional data.

In 2009, R. Gurau introduced a coloured group field theory that tackles an issue with the geometries generated by group field theories \cite{Gurau0907,Gurau0911}. 
Ordinary group field theories generate graphs that triangulates manifolds and pseudo-manifolds but also highly singular spaces \cite{Gurau1006}. 
Each $D$-valent vertex corresponds to a $D$-simplex, with half edges corresponding to their boundary $(D-1)$-simplices. 
Connecting two vertices by an edge thus corresponds  to the gluing of the $D$-simplices by identifying the corresponding $(D-1)$-simplices. Unfortunately,
usual group field theories do not contain enough structure to forbid highly singular gluings. The coloured group field theories generates 
bipartite edge-coloured graphs which are subjects to much stricter rules that forbid such unwanted singularities. 

The $1/N$ expansion for the coloured models was soon discovered \cite{Gurau1011,GuRi1101,Gurau1102} and models without the geometric data of 
group field theories came into prominence under the name of {\it coloured tensor models} \cite{GuRy1109}. 
The leading order of the $1/N$ expansion corresponds to a family of graphs called {\it melons} \cite{BGRR1105}, which triangulates the sphere $S^D$ while displaying 
a tree-like structure \cite{GuRy1302}. This family of melonic graphs being summable, the nice properties of the $1/N$ expansion of matrix models where recovered for 
coloured tensor models. 

The last step toward modern invariant tensor models was to extend those results to general, non-coloured, invariant tensor models \cite{Gurau1111,BGR1202}. 

\subsubsection{Recent Developments}

The rise of invariant tensor models did not diminish the interest for Group Field Theory, and the joint work on both subjects gave birth to 
the broad topic of {\it Tensorial Group Field Theory} \cite{Oriti1211}, which gather any field theory on a Lie group with non-local {\it tensor invariant} interactions, 
regardless  of the nature of the group, the presence or absence of geometrical data, or the link to quantum gravity. In a sense, tensor models and the simpler
tensorial group field theories can be viewed as toy models, simplified versions of quantum gravity theories which allow a deep and careful study of the
graph expansion and renormalisation properties. A broad survey of perturbative renormalisability has been conducted \cite{Carrozza2014} along with the non-perturbative 
study of renormalisation \cite{BeBeOr1411,CarLah1612}.

A key property of tensor models, the domination of (tree-like) melonic graphs, has long been regarded as their main weakness. From the random geometry standpoint, 
they seemed bound to generate branched polymers \cite{GuRy1109}, and an effort to find tensor model generating new, non melonic triangulations was started, 
with the first glimpse of results arising from an enhanced quartic model \cite{BonDelRiv1502} which mixes melonic and planar (matrix-like) behaviours.
Most recently, however, the melonic behaviour of tensor models caught the attention of the Quantum Holography community, when  A. Kitaev presented a model of
 $1+1$ dimensional black hole based on the Sachdev-Ye model of a spin-fluid state \cite{Kit2015,SacYe1992}, which also generates melonic graphs \cite{PolRos1601}. 
 It was suggested in \cite{Witten1610} that a tensor model could replicate the asymptotic behaviour of the Sachdev-Ye-Kitaev model without its quenched disorder,
 resulting in the creation of the Gurau-Witten tensor model \cite{Gurau1611}.

In the present thesis, we focus on {\it quartic} tensor models, which are models restricted to quartic invariant interaction terms $\bT^4$, and their map expansions.
The particularity of the quartic models lies in the use of the Hubbard-Stratonovich intermediate field transformation. The intermediate field transformation is a well
established trick which physicists uses to decompose a 4-fields interaction into two 3-fields interactions, introducing a virtual {\it intermediate} field  to connect 
the new cubic interactions. 
For tensor models, it allows to write the Feynman expansion as an expansion over maps, which display some advantages over regular graphs. 
For example, in the map expansion, the {\it melonic} graphs, which dominates the $1/N$ expansion, correspond to plane trees, which are easier to characterise.
Planarity is also easier to handle. 

Furthermore, the intermediate field representation allowed to study quartic tensor models constructively. As usual with field theories, the Feynman expansions
of tensor models diverge and most of the results on moments, cumulants and partition functions are established term-by-term in the perturbation series but do not 
stand for the whole, re-summed, function. The intermediate field formalism allows to write {\it constructive}, convergent map expansions and establish results and 
bounds on the moments and partition functions \cite{Gurau1304,DGR1403,DG1504}.

\subsubsection{Plan of the thesis}

The present thesis is divided into three parts. 

The first part lays the foundations of quartic tensor models. The first chapter introduces the basic notions of tensor and invariants, 
and defines tensor models as measures for random tensors. This is followed by a quick overview of the basic properties of quartic matrix 
models, invariant tensor models and tensor field theories. Chapter 2 introduces the intermediate field representation and map expansions,
the main tool we use for the study of quartic tensor models. After presenting a bijection between bipartite coloured graphs and multi-coloured maps, 
we introduce a re-writing of the tensor model as a multi-matrix model which Feynman expansion writes in terms of maps. 
We use this map expansion to establish the main perturbative results on the cumulants of the tensor model. Finally, we briefly study the properties of
the intermediate field and its non-trivial vacuum, as presented in \cite{DG1504}.

The second part presents two constructive map expansions for tensors. In chapter 3, we introduce the main constructive notions which will be used in this thesis, 
along with the two forest expansion formulas used in the following chapters. In chapter 4, we perform a constructive expansion of the cumulants of the 
most general standard invariant tensor model. This allows us to prove rigorously the perturbative results of chapter 2, furthermore, we establish the Borel summability
of the Feynman expansion. In chapter 5, we introduce a constructive expansion for the simplest renormalisable tensor field theory : the rank  3 quartic model with invert 
Laplacian covariance, using multiscale analysis and an improved version of the Loop Vertex Expansion. This part summarises the work published in \cite{DGR1403} 
and \cite{DR1412}.

The third part is an introduction to enhanced models. In chapter 6, we introduce the enhanced invariant tensor model at any rank and discuss the behaviour of 
its map expansion. In Chapter 7, we briefly introduce the field theory counterpart for the enhanced model at rank 4 and discuss its renormalisation.  This last part is
based on the work introduced in \cite{BonDelRiv1502} and some yet unpublished collaboration with Vincent Lahoche.

\part{Foundations}
\chapter{Quartic tensor models}\label{chap:QTM}

\section{Introduction to random tensors}

In this first section we introduce the basic definitions and properties of general tensors. We define the notion of tensor invariants and 
introduce the notion of quartic tensor models \cite{BGR1202,DGR1403} in all generality.

\subsection{Tensors}

Let us consider a Hermitian inner product space $V$ of dimension $N$ and $\{e_n| n=1,\dots N\}$ an orthonormal basis in $V$. 
The dual of $V$, $V^{\vee}$ is identified with the complex conjugate $\bar V$ via the conjugate linear isomorphism, 
\[
z \to  z^{\vee}(\cdot)=\langle z, \cdot\rangle  \; . 
\]
We denote $e^n \equiv e_n^{\vee} = \langle e_n , \cdot \rangle$ the basis dual to $e_n $. Then,
\[
 \Bigl(\sum_n z^n e_n  \Bigr)^{\vee}(\cdot) =  \sum_n \overline{ z^n }\langle e_n, \cdot\rangle = \sum_{n} (z^{\vee })_n e^n (\cdot)
 \Rightarrow  (z^{\vee })_n = \overline{ z^n } \; .
\]

\subsubsection*{Covariant tensor}
A covariant tensor of rank $D$ is a multilinear form ${\bT}:  V^{\otimes D} \to \mathbb{C}$. 
We denote its components in the tensor product basis by 
\[ 
\bT_{n^1\dots n^D} \equiv {\bf T} (e_{n^1},\dots, e_{n^D}) \; , \qquad {\bf T} = \sum_{n^1,\dots n^D} \bT_{n^1\dots n^D} \; \;e^{n^1} 
\otimes \dots \otimes e^{n^D} \; .
\]
A priori $T_{n^1\dots n^D}$ has no symmetry properties, hence its indices have a well defined position. 
We call the position of an index its \emph{colour}, and we denote $\cD$ the set of colours $\{1,\dots D\}$. 

A tensor can be seen as a multilinear map between vector spaces. There are in fact as many choices as there are subsets $\cC\subset \cD$:
for any such subset the tensor is a multilinear map ${\bf T}: V^{\otimes \cC} \to \bar V^{ \otimes \cD \setminus \cC}$:
\[
 {\bf T}(z^{(c)}, c \in  \cC ) = \sum_{n^c, c\in  \cC} \bT_{n^1\dots n^D} \prod_{c\in   \cC} [z^{(c)}]^{n^c} \; .
\]
We denote $n^{\cC}= (n^c,c\in \cC)$ the indices with colours in $\cC$.
The complementary indices are then denoted $n^{\cD\setminus \cC} =( n^c,c\notin \cC)$. In this notation the set of all the indices 
of the tensor should be denoted $n^{\cD}=(n^1,\dots n^D)$. We will use whenever possible the shorthand notation $n\equiv n^\cD$.
The matrix elements of the linear map (in the appropriate tensor product basis) are
\[
  \bT_{ n^{\cD\setminus \cC} n^{\cC} } \equiv \bT_n \equiv \bT_{n^1\dots n^D} \; 
  \text{ with } n^c \in n^{\cC} \cup n^{\cD \setminus \cC} \;,\;\; \forall c \; .
\]

\subsubsection*{Dual tensor}
As we deal with complex inner product spaces, the dual tensor $ {\bf T}^{\vee}$ is defined by 
\[ {\bf T}^{\vee} \left( (z^{(1)})^{\vee}, \dots  (z^{(D)})^{\vee} \right) \equiv
  \overline{ {\bf T} \left( z^{(1)}, \dots  z^{(D)} \right) } \; .
\]
Taking into account that 
\[
   \sum_{n^{\cD}} \overline{\bT_{n^1\dots n^D}} \;  \overline{ (z^{(1)} )^{n^1} } \dots \overline{ (z^{(D)} )^{n^D} }
= \sum_{n^{\cD}} \overline{\bT_{n^1\dots n^D}} \;   ( z^{(1)\vee})_{n^1}  \dots  (z^{(D)\vee})_{n^D} \;,
\]
we obtain the following expressions for the  dual tensor and its components
\[  {\bf T}^{\vee} = \sum_{n^1, \dots n^D} \overline{\bT_{n_1\dots n_D}}\; \; e_{n^1} \otimes \dots e_{n^D} \; , \;\;
 ({\bf T}^{\vee})^{n^1\dots n^D} =  \overline{\bT_{n_1\dots n_D}} \; .
\]
The dual tensor is a conjugated multilinear map 
${\bf T}^{\vee} :  \bar V^{\otimes \cD \setminus\cC} \to   V^{\otimes \cC}  $
with matrix elements 
\[
 ( \bT^{\vee})^{ n^{\cC}  n^{\cD \setminus \cC} } \equiv  \overline{ \bT_{ n^1\dots  n^D} } \; \text{ with } 
  n^c \in n^{\cC} \cup  n^{\cD \setminus \cC} \; ,\;\; \forall c \; .
\]
From now on we denote $ \bar \bT_{n_1\dots n_D} \equiv \overline{\bT_{n_1\dots n_D}} $, and we identify $\bT^{\vee}=\bar \bT$. 
We write all the indices in subscript,
and we denote the contravariant indices with a bar.
Indices are always understood to be listed in increasing order of their colours.
We denote $ \delta_{n^{\cC} \bar n^{\cC}} = \prod_{c\in \cC} \delta_{n^c \bar n^c}  $ and 
$\tr_{\cC}$ the partial trace over the indices $n^c, c\in \cC$.

\subsection{Tensor invariants}

\subsubsection*{Definition}
Under unitary base change, covariant tensors transform under the tensor product of $D$ fundamental 
representations of $U(N)$: the group acts independently on each index of the tensor. 
For $U^{(1)}...U^{(D)}\in U(N)$,
\[
 {\bf T} \to  \left( U^{(1)} \otimes ... \otimes U^{(D)} \right){\bf T} , 
 \qquad  \bar {\bf T} \to \bar{\bT} \left( U^{(1)*} \otimes ... \otimes U^{(D)*} \right). 
\]
In components, it writes,
\[
 \bT_{a^{\cD}}\to \sum_{m^{\cD}} U^{(1)}_{a^1 m^1}...U^{(D)}_{a^D m^D}\ \bT_{m^{\cD}},
 \qquad  \bar{\bT}_{\bar a^{\cD}}\to \sum_{m^{\cD}}  \bar U^{(1)}_{\bar a^1 \bar m^1}... 
 \bar U^{(D)}_{\bar a^D \bar m^D}\ \bar \bT_{\bar m^{\cD}}.
\]

 A \emph{trace invariant} monomial is an invariant quantity under the action of the external tensor product of $D$ 
 independent copies of the unitary group $U(N)$  which is built by contracting indices of a product of tensor entries.
 
 \subsubsection*{Bipartite graphs}
 Trace invariant monomials can be graphically represented by bipartite edge-coloured graphs.
 
 \begin{definition}
  A bipartite $D$-coloured graph is a graph $\cG=(\cV^h,\cV^s,\{\cE^c\}_{c\in\cD})$ with vertices $v\in\cV$ an edges $e\in\cE$ such that
  \begin{itemize}
   \item $\cV$ is bipartite, it is the disjoint union of the set of hollow vertices $\cV^h$ and the set solid vertices $\cV^s$.  
   \item $\cE$ is $D$-coloured, it is the disjoint union of $D$ subsets $\cE^c,\ c\in\cD$ of edges of colour $c$.
   \item For each edge $e\in\cE$, $e=(v,\bar v)$ with $v\in\cV^h $ and $\bar v\in\cV^s$.
   \item Each vertex is D-valent with all incident edges having distinct colours.
  \end{itemize}
 \end{definition}
 \begin{figure}[h]
 \centering
  \includegraphics{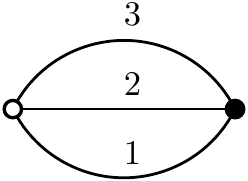}
\caption{The only bipartite 3-coloured graph with two vertices.}
\label{Simplegraph}
  \end{figure}
 For such graphs, we have
  \[
   |\cV^h|=|\cV^s|=|\cE^c|\ \forall\ c\in\cD\ .
  \]

  The graphical representation of an invariant consist of representing each term $\bT$ by a hollow vertex, each $\bar \bT$ by a solid vertex. 
  Then whenever an index $n^c$ on a tensor term $\bT$ is identified with an index $\bar n^c$ on a dual tensor term $\bar \bT$, 
  the two corresponding vertices are connected by an edge of colour $c$.
  
\subsubsection*{Quadratic and quartic invariants}

The tensor and its dual can be composed as linear maps to yield a map from $V^{\otimes \cC}$ to $V^{\otimes \cC}$
\[
 [ \bar\bT \cdot_{\cD\setminus \cC }  {\bf T} ]_{\bar n^{  \cC} n^{  \cC}  }
 = \sum_{n^{\cD\setminus\cC} , \bar n^{\cD \setminus \cC} }
  \bar \bT_{ \bar n^1\dots \bar n^D}
  \delta_{\bar n^{\cD \setminus \cC}  n^{\cD \setminus \cC}   }  \bT_{ n^1\dots n^D}   \; .
\]

The unique quadratic trace invariant is the (scalar) Hermitian pairing of $\bar{\bf T}$ and ${\bf T}$ which 
writes:
\[
 \bar\bT \cdot_{\cD  }  {\bf T} = \sum_{n^{\cD} \bar n^{\cD}}
  \bar \bT_{ \bar n^1\dots \bar n^D}  \delta_{ \bar n^{\cD} n^{\cD}}  \bT_{ n^1\dots n^D}   \; ,
\]
for rank $D=3$, this invariant is represented by the two-vertices graph of Fig.\ref{Simplegraph}.

A connected quartic trace invariant $V_{\mathcal{C}}$ for ${\bf T}$ is
specified by a subset of indices $\cC \subset \cD $:
\begin{equation}\label{eq:quarticinvariant}
 V_{\cC}(\bar\bT,{\bf T} ) = \tr_{\cC} \Big[ \left[ \bar{\bf T} \cdot_{\cD\setminus \cC }  {\bf T} \right] \cdot_{\cC}
 \left[\bar{\bf T} \cdot_{\cD\setminus \cC }  {\bf T} \right]  \Big] \;,
\end{equation}
where we denoted $\cdot_{\cC}$ the product of operators from $V^{\otimes \cC}$ to $V^{\otimes \cC}$.
In components this invariant writes:
\[
 \sum_{n, \bar n, m, \bar m} 
  \left(  \bar{\bT}_{\bar{n}}\  \delta_{\bar n^{\cD\setminus \cC} n^{\cD\setminus \cC}  }  \  \bT_{n}    \right) \ 
   \delta_{n^{\cC}\bar m^{\cC}} \delta_{   \bar n^{\cC}  m^{\cC}}\ 
   \left( \bar{\bT}_{\bar{m}}\  \delta_{\bar m^{\cD\setminus \cC}  m^{\cD\setminus \cC}  }  \ \bT_{m}    \right)   \; .
\]
Note that for any $\cC\subset\cD$, $V_\cC = V_{\cD\setminus\cC}$.
 \begin{figure}[h]
 \centering
  \includegraphics{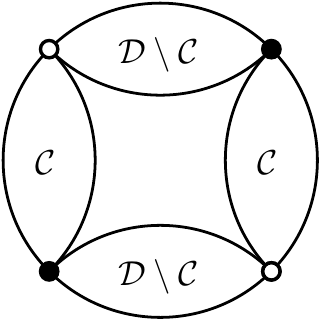}
\caption{General structure of the connected quartic invariants.}
\label{Quarticinvariantgraph}
  \end{figure}
 \subsection{Gaussian measure and tensor models}
\subsubsection*{Gaussian measure}
 A random tensor model is a measure on the space of tensors. The models are built from Gaussian measures, 
 by adding extra perturbation terms. In the case of quartic models, those terms are exponentials of quartic invariant polynomials.
 
 A Gaussian measure $\mu_\bC $ is characterised by its {\emph covariance} $\bC$, a rank $2D$ operator, as
 \begin{equation}
  \braket{\bT_n\bar{\bT}_{\bar n}}_{\mu_\bC}= \bC_{n\bar n} \ .
 \end{equation}
Whenever the inverse operator $\bC^{-1}$ exists, the Gaussian measure can be expressed as
\begin{equation}\label{eq:Gaussian}
 d\mu_\bC(\bT) = \Det\,\bC^{-1} \,
 e^{ - \sum_{n,\bar n} {\bT}_n \bC^{-1}_{n\bar n} \bar{\bT}_{\bar n}   }\, \Bigl( \prod_n \frac{ d \bar{\bT}_{{n}}  d\bT_n }{2 \imath \pi} \Bigr) \;,
\end{equation}
Standard (invariant) tensor models are built from the standard Gaussian measure with identity covariance $\bC=\id$, 
reducing the Gaussian measure to the exponential of the quadratic tensor invariant $\bT\cdot_\cD\bar{\bT}$. Non invariant tensor models,
such as those arising from group field theories, use more complicated covariances such as inverse Laplacians 
($\frac1{|n|^2}\delta_{n\bar n}$) or projectors (e.g. $\delta(\sum_c n^c = 0)\,\delta_{\bar n n}$).

 \subsubsection*{Invariant quartic models}
An invariant quartic tensor model is then the (invariant) perturbed Gaussian measure for a random tensor:
\begin{align}\label{eq:model}
d\nu &=  e^{-N^{1-D}\sum_{\mathcal{C} \in \mathcal{Q}} \frac12 \lambda_\cC V_{\mathcal{C}}(\bar{\bf T},{ \bf T} ) }\ d\mu_\id(\bar\bT,\bT)
\\
&= 
 e^{ - \big(  \bar{\bf T} \cdot_{\cD  }  {\bf T} 
  +N^{1-D}\sum_{\mathcal{C} \in \mathcal{Q}} \frac12 \lambda_\cC V_{\mathcal{C}}(\bar{\bf T},{ \bf T} ) \big) } \;
  \Bigl( \prod_n \frac{ d \bar{\bT}_{{n}}  d\bT_n }{2 \imath \pi} \Bigr)\ ,
\end{align}
where  $\mathcal{Q}$ is some set of $\cC$s and the coupling constants $\lambda_\cC$ are complex numbers. 
The factor $\frac12$ accounts for the symmetry of the quartic invariance, 
and the factor $N^{1-D}$ will ensure that the free energy is bounded by a polynomial bound $N^K$ at large $N$ \cite{Gurau1111}, 
as we will see in Chapter \ref{chap:IF}.
The argument of the exponential is called the \emph{action} of the model.
\begin{equation}\label{eq:action}
 S(\bar\bT,\bT)=  \bar{\bf T} \cdot_{\cD  }  {\bf T} 
  +N^{1-D}\sum_{\mathcal{C} \in \mathcal{Q}} \frac12 \lambda_\cC V_{\mathcal{C}}(\bar{\bf T},{ \bf T} ) \ .
\end{equation}
As $V_\cC = V_{\cD\setminus\cC}$, a same model can be represented with several choices of $\cQ$. 
This choice makes no difference as for the definition of the model but will 
have important consequences on the intermediate field representation.

The moment-generating function of the measure $d\nu$ is defined as :
\begin{align}\label{eq:momentgeneratingZ}
& Z(\bJ, \bar{\bJ})=\int e^{ \sum_{n} \bT_{n} \bar{\bJ}_{n} + \sum_{\bar n}\bar{\bT}_{\bar{n} }\bJ_{\bar{n} }} \; d\nu \;,
\end{align}
and its cumulants are thus written :
\begin{align}\label{eq:culumants}
& \kappa(\bT_{n_1}\bar{\bT}_{\bar{n}_1}...\bT_{n_k}\bar{\bT}_{\bar{n}_k})
=\frac{\partial^{(2k)} \Bigl( \log Z(\bJ,\bar \bJ) \Bigr) }{\partial \bar{\bJ}_{n_1}
\partial \bJ_{\bar{n}_1}...\partial \bar{\bJ}_{n_k}\partial \bJ_{\bar{n}_k}} \Bigg{\vert}_{\bJ =\bar \bJ =0}.
\end{align}

\section{Perturbative expansion}
\subsection{The Feynman expansion}
As for any field theory, the moments of a tensor model can be perturbatively evaluated using a Feynman graph expansion. 

The measure $\nu$ can be expanded in a power series as
\begin{align}
 d\nu =  d\mu_\bC(\bar\bT,\bT)\
 \sum_{b=0}^{\infty}\frac1{b!}
 \left[\sum_{\cC\in\cQ}
  \frac{-\lambda_\cC}{2N^{D-1}} V_{\mathcal{C}}(\bar{\bf T},{ \bf T} )
 \right]^{b} \ .
\end{align}
Therefore, the expected value of a polynomial $P(\bar\bT,\bT)$ can be evaluated as 
\begin{align}\label{eq:expandedexpected}
 \braket{P(\bar\bT,\bT)}_\nu \approx 
 \sum_{b=0}^{\infty}\frac1{b!}
 \left\langle
 P(\bar\bT,\bT) \left[\sum_{\cC\in\cQ}
  \frac{-\lambda_\cC}{2N^{D-1}} V_{\mathcal{C}}(\bar{\bf T},{ \bf T} )
 \right]^{b}
 \right\rangle_{\mu_\bC}\ .
\end{align}
Remark that an illegitimate operation has been made while extracting the infinite summation from the expectation bracket. 
\subsubsection*{Wick contractions}
The Wick theorem states that for a Gaussian measure of covariance $\bC$, the expectation of a product of $p$ covariant and $p$ contravariant tensor entries
writes
\begin{align}
 \left\langle \prod_{k=1}^p\bT_{n^k}\bar\bT_{\bar n^k}\right\rangle_{\mu_\bC} = \sum_{\sigma\in S_p} \prod_{k=1}^p \bC_{n^k \bar{n}^{\sigma(k)}}\ ,
\end{align}
where $S_p$ is the set of permutations over $p$ elements. Graphically, if tensor entries are represented by hollow and solid vertices, 
the sum runs over every ways to pair each hollow vertex with a solid one.

\begin{figure}[h]
 \centering
  \includegraphics{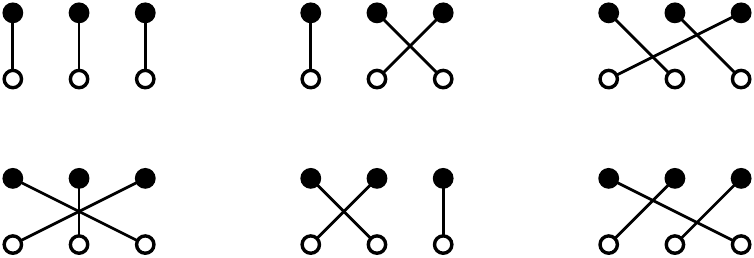}
\caption{The 6 possible pairings for 3 solid and 3 hollow vertices.}
\label{fig:wickpairs}
  \end{figure}
  
  Furthermore, this allows us to re-write the Gaussian measure as a differential operator \cite{GurauHDR}:
  \begin{corollary}\label{corollary:diffop}
   Let $\mu_\bC$ be the normalised Gaussian measure of covariance $\bC$, then for any function $A(\bar\bT,\bT)$,
   \[
    \langle A(\bar\bT,\bT) \rangle_{\mu_\bC} =\left[e^{\frac{\partial}{\partial \bT} \bC \frac{\partial}{\partial \bar \bT}}  A(\bar\bT,\bT) \right]_{\bar\bT=\bT=0} \ .
   \]
  \end{corollary}

\subsubsection*{Feynman graphs}
The expansion can be represented graphically as a Feynman graph expansion.
We label each interaction term and assign colour sets using,
\begin{align}
  \left[\sum_{\cC\in\cQ}
  \frac{-\lambda_\cC}{2N^{D-1}} V_{\mathcal{C}}(\bar{\bf T},{ \bf T} )
 \right]^{b}
 = \sum_{\substack{\cC(i)\in\cQ\\ \forall i\in\{1\cdots b\}}}\prod_{i=1}^b\  \frac{-\lambda_{\cC(i)}}{2N^{D-1}} V_{\mathcal{C}(i)}(\bar{\bf T},{ \bf T} ) \ .
\end{align}
%
%
%
 We represent each interaction term by the corresponding bipartite $D$-coloured graph, which we will call {\it bubble}, each bubble carries a label $i\in\{1\dots b\}$. 
A polynomial $P(\bar\bT,\bT)$ of degree $2k$ is represented as $k$ isolated hollow vertices and $k$ solid ones, labelled  by $j\in\{1\dots k\}$. 
The contraction of tensor entries into covariance operation by 
the Wick theorem is then represented with colour 0 edges connecting the corresponding hollow and solid vertices.
The resulting graph is a $D+1$-coloured bipartite graph with $2k$ external legs, with two sets of labels : each $D$-coloured bubble carries a label $i$ and each 
external leg carries a label $j$.
\begin{definition}
  An external leg is a monovalent vertex which only incident edge is of colour 0.
\end{definition}
Each bubble being symmetric, exchanging simultaneously the covariant tensor entries and the dual tensor entries of a bubble leaves \eqref{eq:expandedexpected} unchanged.
Therefore, for a given order $b$, there are $2^b$ Wick pairings  corresponding to
a same given graph $\cG$, and which contributions to \eqref{eq:expandedexpected} are equal.

The perturbative expansion of the expectation value of a polynomial of order $2k$ writes as
\begin{align}\label{eq:feynexp}
  \braket{P(\bar\bT,\bT)}_\nu \approx \sum_{b\in\setN}\frac1{b!} \sum_{\cG(b)} A(\cG)
\end{align}
where the sum runs over all $D+1$-coloured bipartite graphs with $2k$ external legs and whith $b$ connected 
$D$-coloured components.

The amplitude $A(\cG)$ associated with a Feynman graph $\cG$ is
\begin{align}\label{eq:feynmanamplitude}
 A(\cG)&=\prod_{\cC\in\cQ}\prod_{\cB^\cC\subset\cG} \left(-\frac{\lambda_\cC}{N^{D-1}}
 \sum_{n_\cB, \bar n_\cB, m_\cB, \bar m_\cB}   \delta_{\bar n_\cB^{\cD\setminus \cC} n_\cB^{\cD\setminus \cC}  }  \ 
   \delta_{n_\cB^{\cC}\bar m_\cB^{\cC}} \delta_{   \bar n_\cB^{\cC}  m_\cB^{\cC}}\ 
     \delta_{\bar m_\cB^{\cD\setminus \cC}  m_\cB^{\cD\setminus \cC}  }  \   \;
 \right)\crcr
 &\times \prod_{\e\in E^0} \bC_{n_\e\bar n_\e}
 \ .
\end{align}
where the $\cB$ are the bubbles of $\cG$. 
The amplitude of a graph writes as a product of covariances $\bC$, which indices are identified according to the structure of the quartic bubbles.
%

%
%
%
\begin{figure}[h]
 \centering
  \includegraphics{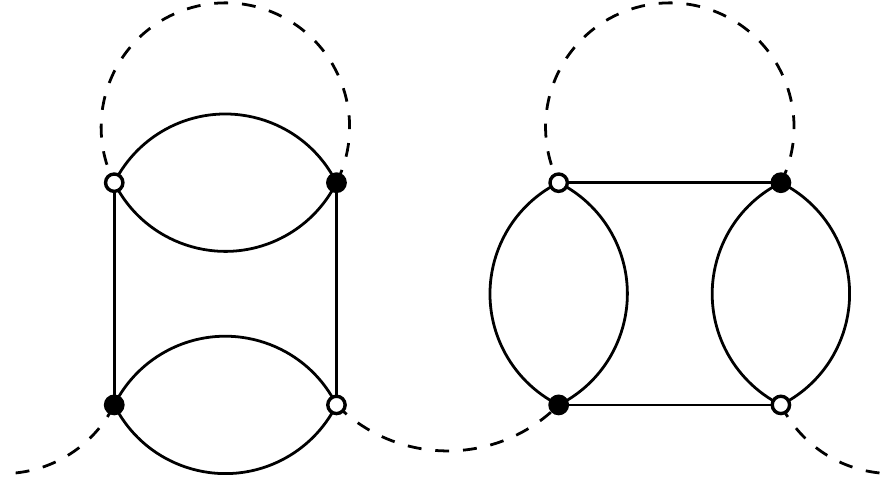}
\caption{A connected 2-point Feynman graph for the rank-3 quartic tensor model. Colour 0 edges are represented as dashed lines.}
\label{Feynmangraph}
  \end{figure}

%

Denoting $s(\hat\cG)$ the number of different ways to label the bubbles of a $D+1$-coloured graph with unlabelled bubbles $\hat\cG$, 
the Feynamnn expansion \eqref{eq:feynexp}
writes as a sum over $D+1$-coloured graphs with un-labelled bubbles and labelled external legs,
\begin{align}
 \braket{P(\bar\bT,\bT)}_\nu &\approx \sum_{\cG} \cA(\hat\cG) \ ,\\
 \cA(\hat\cG)&= \frac{1}{b(\hat\cG)!}
 \sum_{\rm labellings} A(\hat\cG) = \frac{s(\hat\cG)}{b(\hat\cG)!}A(\hat\cG)
\end{align}
Note that a connected graph $\hat\cG$ with $2k>0$ external legs bears no symmetry, therefore $s(\hat\cG) = b(\hat\cG) !$ and the above expression simplifies,
\[
  \braket{P(\bar\bT,\bT)}_\nu \approx \sum_{\hat\cG} A(\hat\cG)\ .
\]

\subsubsection{Partition function and vacuum graphs}

The partition function of the tensor model writes
\begin{align}
 Z = \int d\nu = \int  e^{-N^{1-D}\sum_{\mathcal{C} \in \mathcal{Q}} \frac12 \lambda_\cC V_{\mathcal{C}}(\bar{\bf T},{ \bf T} ) }\ d\mu_\id(\bar\bT,\bT) \ .
\end{align}
It can be evaluated perturbatively using the same approximation as in \eqref{eq:expandedexpected},
\begin{align}
 Z \approx \sum_{b\in\setN}\frac1{b!}
  \left\langle
 P(\bar\bT,\bT) \left[\sum_{\cC\in\cQ}
  \frac{-\lambda_\cC}{2N^{D-1}} V_{\mathcal{C}}(\bar{\bf T},{ \bf T} )
 \right]^{b}
 \right\rangle_{\mu_\bC}
 \ .
\end{align} 
Its Feynman expansion writes as a sum over \emph{vacuum} bipartite $D+1$-coloured graphs with labelled bubbles.
\begin{definition}
 A vacuum graph is a graph without external legs.
\end{definition}

\subsection{Connected graphs}

\subsubsection*{The free energy}
The free energy is defined as the logarithm of the partition function
\[
 F = \log Z \ ,
\]
its Feynman expansion writes as a sum over all \emph{connected} vacuum graphs.

\begin{theorem}\label{thm:log}
 For a quantity X written as a weighted sum over all bipartite $D+1$ coloured graphs
 \[
  X=\sum_{ \cG} A(\cG)\ ,
 \]
 and if the weight of any graph $\cG$ factorises as the product of the weight of its connected components $\cK$,
 \[
  A(\cG) = \prod_{\cK\subset\cG } A(\cK)\ .
 \]
 Then the logarithm of X writes as the sum over connected graphs,
 \[
  \log X = \sum_{\rm connected\ \cG^\cc} A(\cG^\cc)\ .
 \]
\end{theorem}
\prf
The exponential of a sum over connected graphs writes,
 \begin{align}
  {\rm exp}\left( \sum_{\rm connected\ \cG^\cc } A(\cG^\cc)\right) &= \sum_{n=0}^\infty \frac1{n!} \left( \sum_{\cG^\cc} A(\cG^\cc)\right)^n\\
  &= \sum_{n=0}^\infty \frac1{n!} \left[ \prod_{i=1}^n \left(\sum_{\rm connected\ \cG^\cc(i) } A(\cG^\cc(i))\right)\right] \\
  &= \sum_{n=0}^\infty \frac1{n!} \left[ \sum_{ \{ \cG^\cc(i),\ 1\leq i \leq n \} } \prod_{i=1}^n A(\cG^\cc (i)) \right]
  \ .
 \end{align}
 The last line writes as a sum over graphs with $n(\cG)=n$ connected components labelled from $1$ to $n$. As there are $n!$ permutations of such labellings, 
 \[
  \sum_{ \{ \cG^\cc(i),\ 1\leq i \leq n \} } A(\cG^\cc (i)) = n ! \sum_{ \cG,\ n(\cG)=n }\ \prod_{\cK\subset\cG} A(\cK) \ ,
 \]
which leads to,
\[
 X= {\rm exp}\left( \sum_{\rm connected\ \cG^\cc } A(\cG^\cc)\right)
\]
\qed
\subsubsection*{Cumulants}
The cumulants of a tensor model are defined in \eqref{eq:culumants} as derivatives of the generating function
\begin{align}
& \kappa(\bT_{n_1}\bar{\bT}_{\bar{n}_1}...\bT_{n_k}\bar{\bT}_{\bar{n}_k})
=\frac{\partial^{(2k)} \Bigl( \log Z(\bJ,\bar \bJ) \Bigr) }{\partial \bar{\bJ}_{n_1}
\partial \bJ_{\bar{n}_1}...\partial \bar{\bJ}_{n_k}\partial \bJ_{\bar{n}_k}} \Bigg{\vert}_{\bJ =\bar \bJ =0}.
\end{align}
The cumulant generating function, being the logarithm of the moment generating function, can be expanded over connected graphs as
\begin{align}\label{eq:expandlog}
  \log Z(\bJ,\bar \bJ) & = \sum_{k,k'\geq0}\ \sum_{ \cG^\cc_{k,k'}}
  \ \prod_{\e\in E^0} \bC_{n_\e\bar n_\e}\ \prod_{p=1}^k \bar \bJ_{n_p} \prod_{p=1}^{k'}\bJ_{\bar n_p}
\crcr
 &\times\prod_{\cB\subset\cG^\cc} \left(-\frac{\lambda_\cC}{N^{D-1}}
 \sum_{n_\cB, \bar n_\cB, m_\cB, \bar m_\cB}  
\delta_{\bar n_\cB^{\cD\setminus \cC} n_\cB^{\cD\setminus \cC}  }  \delta_{\bar m_\cB^{\cD\setminus \cC}  m_\cB^{\cD\setminus \cC}  } \ 
   \delta_{n_\cB^{\cC}\bar m_\cB^{\cC}} \delta_{   \bar n_\cB^{\cC}  m_\cB^{\cC}}
         \;
 \right),
\end{align}
where $\cG^\cc_{k,k'}$ are connected graphs with $k$ external legs of the covariant type, 
labelled from $1$ to $k$, and $k'$ of the dual type, labelled from $1$ to $k'$.

     The cumulants can then be approximated by evaluating the derivatives of  \eqref{eq:expandlog}, using
 \begin{align}
 \frac{\partial^{(2k)}  }{\partial \bar{\bJ}_{m_1}
\partial \bJ_{\bar{m}_1}...\partial \bar{\bJ}_{m_k}\partial \bJ_{\bar{m}_k}} \ 
  \prod_{d=1}^k\ \bar \bJ_{n_p}\bJ_{\bar n_p} = 
  \sum_{\pi \bar\pi}  \prod_{d=1}^k\ \delta_{m_{\pi(d)} n_p}
  \delta_{\bar m_{\bar \pi(d)} \bar n_p} \ ,
 \end{align}
 where $\pi$ and $\bar \pi$ runs over permutations of $k$ elements,
 the cumulant of order $2k$ writes as
 \begin{align}\label{eq:expandedcumulants}
  \kappa(\bT_{m_1}\bar{\bT}_{\bar{m}_1}...\bT_{m_k}\bar{\bT}_{\bar{m}_k})  =\sum_{\pi \bar\pi}\  \sum_{ \cG^c_{k,k}}\
  \prod_{\e\in E^0} \bC_{n_\e\bar n_\e}\ \prod_{d=1}^k\ \delta_{m_{\pi(d)} n_p}
  \delta_{\bar m_{\bar \pi(d)} \bar n_p}
\crcr
 \times\prod_{\cB\subset\cG^\cc} \left(-\frac{\lambda_\cC}{N^{D-1}}
 \sum_{n_\cB, \bar n_\cB, m_\cB, \bar m_\cB}  
\delta_{\bar n_\cB^{\cD\setminus \cC} n_\cB^{\cD\setminus \cC}  }  \delta_{\bar m_\cB^{\cD\setminus \cC}  m_\cB^{\cD\setminus \cC}  } \ 
   \delta_{n_\cB^{\cC}\bar m_\cB^{\cC}} \delta_{   \bar n_\cB^{\cC}  m_\cB^{\cC}}
         \;
 \right) .
\end{align}
$\pi$ and $\bar\pi$ represents the possible permutations between the labelled tensor entries $\bT_{m_i},\ \bar{\bT}_{\bar{m}_i}$ with $i\in\{1\dots k\}$
and the external legs (also labelled) of the graph $\cG^c_{k,k}$.

 \section{Matrix Models}\label{sec:MatrixModels}
 
 The simplest case of tensor models is for a rank $D=2$, where tensors are simply matrices.
 Tensor models where originally developed as a generalisation of random matrix models, 
which had many successes in physics, notably to understand 2 dimensional quantum gravity.

\subsubsection*{The quartic matrix model}
For $D=2$, the dual matrices correspond to Hermitian conjugates
\[
\bar \bM_{\bar n^1\bar n^2}=\bM_{\bar n^2\bar n^1}^{*}
 \,
\]
and the invariant polynomials are merely traces of a product of matrices. The quadratic invariant becomes
\[
 \bar \bM\cdot_{\cD}\bM = \tr(\bM^*\bM) \ ,
\]
and there is a unique connected quartic invariant,
\[
V(\bar{\bf M},{\bf M} )=
 \tr_{1} \Big[ \left[ \bar{\bf M} \cdot_{2 }  {\bf M} \right] \cdot_{1}
 \left[\bar{\bf M} \cdot_{2 }  {\bf M} \right]  \Big] = \tr(\bM^*\bM\bM^*\bM) \ .
\]
The (invariant) quartic matrix model is the measure :
\begin{align}\label{eq:matrixmodel}
 d\nu = \left(\prod_{a,b}\frac{d\bM_{ab}d\bM^*_{ba}}{2i\pi}\right) 
 e^{ -\tr(\bM^*\bM) - \frac{\lambda}{2N} \tr\ (\bM^*\bM)^2
 }
 \ ,
\end{align}
and the moment-generating function is
\begin{align} 
& Z(\bJ, \bJ^*)=\int e^{ \tr\ \bM {\bJ}^* + \tr\ \bM^* \bJ} \ d\nu \ .
\end{align}


%

The observables of the matrix model are the invariants $\tr\ (\bM^*\bM)^n$, and their expectations are derivatives of the moment generating function.

\subsubsection*{Feynman maps}

The Feynman expansion of the quartic matrix model can be defined in terms of directed maps.
\begin{definition}\label{definition_map}
 A directed map is a quadruplet $\cM=(H, \sigma, \epsilon, H_{\epsilon})$ such that:
 \begin{itemize}
  \item $H$ is a finite set of half-edges.
  \item $\sigma$ is a permutation on $H$.
  \item $\epsilon$ is an involution on $H$ with no fixed point.
  \item $H_{\epsilon}$ is a subset of $H$ such that $\epsilon(H_{\epsilon})=H\setminus H_\epsilon$.
 \end{itemize}
\end{definition}
\begin{definition}\label{definition_vertexedgecorner}
For a directed map  $\cM=(H, \sigma, \epsilon, H_{\epsilon})$,
\begin{enumerate}
 \item A vertex is a cycles of the permutation $\sigma$.
A cycle of length $p$ is called $p$-valent.
\item A corner is a pair of half-edges $(h,\sigma(h))$. If $v$ is a vertex and $h\in v$, then $(h,\sigma(h))$ 
is called a corner of the vertex $v$. 
\item A (directed) edge is a pair of half-edges $(h,\epsilon(h))$, with $h\in H_\epsilon$.
\end{enumerate}
\end{definition}
Therefore, a directed map can be seen as a directed graph, which edges are composed of two half-edges, directed from the one in $H_\epsilon$
toward the one in  $H\setminus H_\epsilon$,
and with a cyclic ordering of the half-edges incident to each vertex.
Note that a $p$-valent vertex also has $p$ corners.

The Feynman expansion of the $k$ point function of the quartic tensor model is then 
the sum over directed maps $\cM=(H, \sigma, \epsilon, H_{\epsilon})$
with arbitrary many 4-valent vertices and $k$ external legs, and with alternating direction of incident half-edges, where:
\begin{itemize}
 \item $H$ is the set of matrix entries.
 \item $\sigma$ is composed of 4-valent vertices and $k$ 1-valent vertices. 
 The 4-valent vertices are the interaction trace terms $\tr (\bM^*\bM)^2$, the 1-valent are the external legs.
 \item $\epsilon$ pairs half-edges together according to the Wick contractions.
 \item $H_{\epsilon}$ is the set of matrice $\bM$ entries. $H\setminus H_{\epsilon}$ is the set of conjugate matrices $\bM^*$.
 \item For $h$ in a $4$-valent vertex, if $h\in H_\epsilon$ then $\sigma(h)\not\in H_\epsilon$ and vice-versa. 
 Namely, each corner is composed of a half edge $h\in H_{\epsilon}$ and a half edge in  $h\in H\setminus H_{\epsilon}$ .
\end{itemize}
%
%
%

The one to one correspondence between a directed map $\cM=(H, \sigma, \epsilon, H_{\epsilon})$ and a $2+1$-coloured bipartite graph
$\cG=(\cV^h,\cV^s,\{\cE^c\}_{c\in\cD})$ is built as follows.
\begin{itemize}
 \item The vertices and external legs of the coloured graph are half-edges of the map. $\cV^h \cup \cV^h_{ext}= H_\epsilon$, 
 and $\cV^s \cup \cV^s_{ext}= H\setminus H_\epsilon$.
 \item For $h\in\cV^s$, $\sigma(h)$ is the hollow vertex connected to $h$ by an edge of colour $1$. 
 For $h\in\cV^h$, $\sigma(h)$ is the solid vertex connected to $h$ by an edge of colour $2$.
\item For $h\in H$, $\epsilon(h)$ is the vertex connected to $h$ by an edge of colour $0$. 
\end{itemize}

The bubbles, being 2-coloured bipartite graphs, are cycles of vertices and coloured edges. The permutaion $\sigma$ is built such that its
 cycles are the bubbles and external legs of the coloured graph. This choice of $\sigma$ is equivalent to choosing an orientation
 for the bubbles, such that each edge of colour 1 is directed from a solid vertex (representing a dual tensor entry) toward a hollow one.
This orientation defines a cyclic ordering of the hollow and solid vertices of the interaction bubbles.

\begin{figure}[h]
 \centering
  \includegraphics{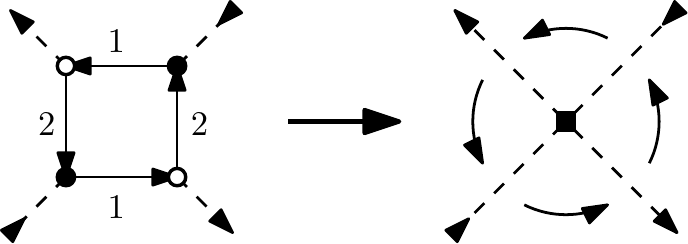}
\caption{Contraction of the oriented bubble. The arrows on the coloured edges represent the permutation $\sigma$. 
The resulting graph is not coloured and the cyclic ordering of half-edges is preserved.}
\label{directedgraphtomap}
  \end{figure}
  
  Quartic bubbles are therefore contracted into 4-valent map vertices, hollow and solid vertices of the bipartite coloured graph 
  being half-edges of the map and their cyclic ordering being preserved.  

Therefore, in the map representation, interactions are represented by mere vertices instead of bubbles. 
The edges of colour 0 becomes directed edges of the map, directed from hollow to solid vertices. 

The corners of the new map correspond to the edges of colour 1 and 2 of the $2+1$-coloured graph. 
Because of the bipartite nature of the coloured graphs, consecutive half edges are of opposite direction,  
corners are composed of an outward half-edge $h\in H_\epsilon = \cV^h$ and an inward  
$h'\in H\setminus H_\epsilon = \cV^s$. 

Note that a $2+1$ graph with labelled quartic bubbles corresponds to a map with labelled 4-valent map vertices. 

\subsubsection*{Ribbon graphs}
Ribbon graphs are a convenient way to represent directed maps with alternating direction of the incident half-edges, and 
therefore the Feynman expansion of the quartic matrix model. 

Each corner of the ribbon vertices are represented as a coloured strand, of colour 1 or 2 according to the coloured edge 
of the corresponding $2+1$-coloured graph.

\begin{figure}[h]
 \centering
  \includegraphics{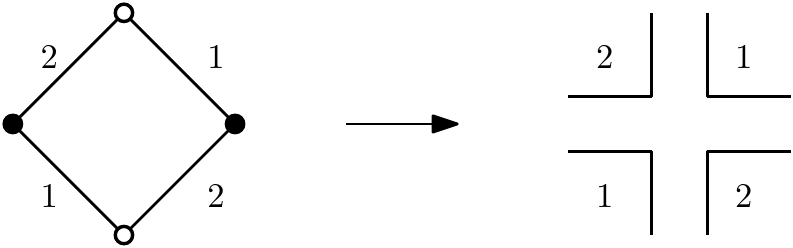}
\caption{Ribbon representation of a $D=2$ quartic bubble.}
\label{Bubbletoribbongraph}
  \end{figure}
  
 The Wick contraction are then represented by ribbon edges, made of a strand of each colour. The ribbon edges
 tie together the half-edges corresponding to matrix terms in the interactions, just like 0-coloured edges connect hollow 
 and full vertices in the $2+1$-coloured graphs. 
 
 \begin{figure}[h]
 \centering
  \includegraphics{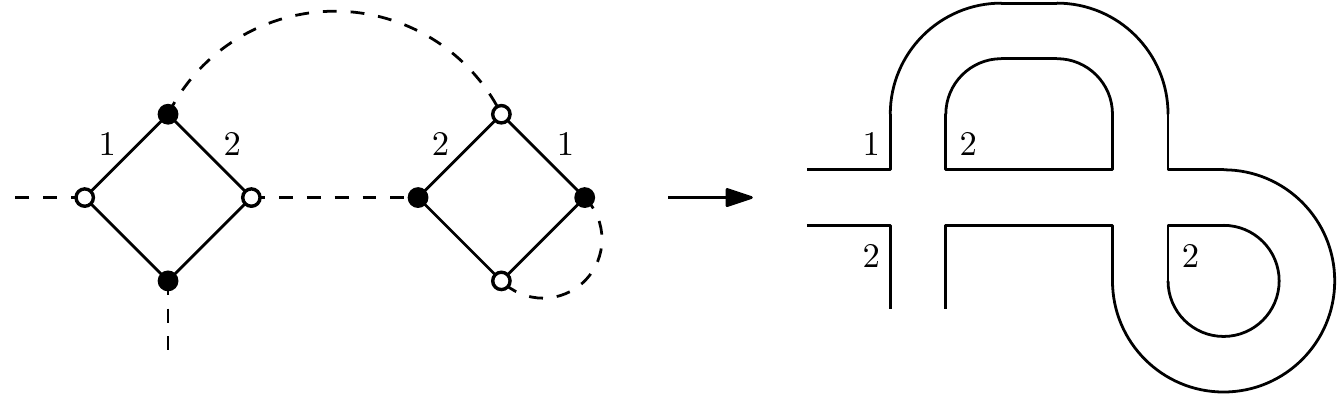}
\caption{Correspondence between a $2+1$-coloured graph and a ribbon graph.}
\label{Toribbongraph}
  \end{figure}
  
The ribbon representation keeps track of the structure of the vertices and of the index identifications. 
Open strands of ribbon graphs are called external faces and represent a identification of the indices of corresponding colour on both ends.
A closed strand, or \emph{internal face}, represents a free summation over an index, and contributes with a factor $N$ to the amplitude of the graph.

\subsubsection*{The 1/N expansion}

The $p$-th moments of the quartic matrix model can be evaluated via a Feynman expansion as a sum over ribbon graphs with $p$ external legs. 
The partition function is evaluated as a sum of ribbon graphs without external legs, of {\emph vacuum graphs} and 
its logarithm, the free energy, is then the sum over all connected vacuum graphs.
For simplicity we shall only consider the free energy, 
\begin{align}
 {\rm log}\ Z &= \sum_{\cG\ {\rm connected}}\frac{1}{|V(\cG)|\,!}\: A_\cG \crcr
 &=
 \sum_\cG\frac{1}{|V(\cG)|\,!}\prod_{v\in V(\cG)} \left(-\frac{\lambda}{N}
 \sum_{n_v, \bar n_v, m_v, \bar m_v}   \delta_{\bar m_v^1 m^1_v  }  \ 
   \delta_{\bar m^2_v n^2_v} \delta_{   \bar n^2_v  m^2_v}\ 
     \delta_{\bar n^1_v  n^1_v  }  \   \;
 \right)\crcr
 &\times \prod_{\e\in E^0} \delta_{n^1_\e\bar n^1_\e}\delta_{n^2_\e\bar n^2_\e} \ .
\end{align}
As there are no external legs, the ribbon graph $\cG$ only has \emph{internal faces}: closed strands of colour $1$ and $2$.
As the covariance is merely the identity for both colours, each face corresponds to the trace of a product of identity operator. 
Finally as $\tr_c\id_c=N$, the amplitude of a ribbon graph $\cG$ reduces to,
\begin{align}
 A_\cG=(-\lambda)^{|V(\cG)|}\ N^{|F(\cG)|-|V(\cG)|}\ ,
\end{align}
where $V(\cG)$ is the set of ribbon vertices and $F(\cG)$ is the set of faces of $\cG$.

%
 
As for quartic graphs $|E|=2|V|$, we have,
$ |F| - |V| = |F| + |E| - |V|$. This is Euler's characteristic
\[
 \omega=2-2g
\]
where $g$, the genus of the map, is defined as the genus of the lower genus surface on which the ribbon graph can be drawn
without crossings. We therefore have $g\ge0$,
and the Feynman expansion can be re-arranged as a power series in $1/N$, where to each order contribute the maps of a given genus,
\begin{equation}
 {\rm log}\ Z = N^2 \sum_{g\ge0} \frac{1}{N^{2g}}\sum_{\cG\slash g(\cG)=g} \frac{\lambda^{|V(\cG)|}}{|V(\cG)|\, !}\ .
\end{equation}
For large $N$, the contribution of non-planar maps is suppressed, and the free energy can be expressed as a sum over all planar ribbon graphs,
\begin{equation}
 N^{-2}\,{\rm log}\ Z = \sum_{\cG\slash g(\cG)=0} \frac{\lambda^{|V(\cG)|}}{|V(\cG)|\, !}\ +\ O(\frac1N)\ .
\end{equation}
This nice characteristic generates a lot of interest for matrix models,
which were notably used to quantise 2 dimensional (Liouville) gravity \cite{DFGZJ1993,tHooft1974}. 

%
%

\section{Invariant Tensor Models}\label{sec:tensmod}

\subsubsection*{The quartic invariant tensor model}

For rank $D\geq 3$ and a given set of interaction $\cQ$, the standard invariant tensor model is defined with a single coupling parameter $\lambda$ as,
\begin{align}\label{eq:standardmodel}
d\nu &=  e^{-\frac{\lambda}{2N^{D-1}}\sum_{\mathcal{C} \in \mathcal{Q}}  V_{\mathcal{C}}(\bar{\bf T},{ \bf T} ) }\ d\mu_\id(\bar\bT,\bT)
\end{align}
where $d\mu_\id$ is the normalised Gaussian measure of identity covariance $\id=\bigotimes_{c\in\cD}\id_c$.

\subsubsection*{Feynman graph expansion}

The free energy is evaluated perturbatively as a sum over connected $D+1$-coloured graphs.
The amplitude \eqref{eq:feynmanamplitude} of a Feynman graph writes as
\begin{align}
 A(\cG)&=\prod_{\cC\in\cQ}\prod_{\cB^\cC\subset\cG} \left(-\frac{\lambda}{N^{D-1}}
 \sum_{n_\cB, \bar n_\cB, m_\cB, \bar m_\cB}   \delta_{\bar n_\cB^{\cD\setminus \cC} n_\cB^{\cD\setminus \cC}  }  \ 
   \delta_{n_\cB^{\cC}\bar m_\cB^{\cC}} \delta_{   \bar n_\cB^{\cC}  m_\cB^{\cC}}\ 
     \delta_{\bar m_\cB^{\cD\setminus \cC}  m_\cB^{\cD\setminus \cC}  }  \   \;
 \right)\crcr
 &\times \prod_{\e\in E^0}\left(\prod_{c=1}^D \delta_{n^c_\e\bar n^c_\e}\right)
 \ ,
\end{align}
where, for an edge $\e=(v,\bar v),\ v\in\cV^h\ \bar v\in\cV^s$,  $n_\e=n_v$ is the index of tensor entry $\bT_{n_v}$ associated with the hollow vertex $v$
and $\bar n_\e = \bar{n}_{\bar{v}}$ is the index of tensor entry $\bar \bT_{\bar n_{\bar{v}}}$ associated with the hollow vertex $\bar v$.
Up to a prefactor, the amplitude writes as traces of products of identity operators, each contributing with a factor $N$ to the graph amplitude.
The notion of \emph{faces} used to compute the power of $N$ for ribbon graph amplitudes in matrix models can be extended to $D+1$ coloured graphs.
\begin{definition}\label{def:faces}
 Lets $\cG$ be a bipartite $D+1$-coloured graph with $2k$ external legs. The \emph{faces} of colour $c$ of $\cG$ are the connected $1+1$-coloured subgraphs of $\cG$
 with edges of colours $0$ and $c$.
 
 Let $\cV^{ext}$ be the set of external legs of $\cG$. 
 \begin{itemize}
  \item An \emph{internal} face $\cF$ is a face with no internal legs, $\cF\cap\cV^{ext}=\emptyset$.
  \item An \emph{external} face $\cF$ is a face with internal legs, $\cF\cap\cV^{ext}\not=\emptyset$.
 \end{itemize}
\end{definition}
Denoting $F_{int}$ the set of internal edges and $B$ the set of $D$-coloured bubbles of a graph $\cG$, the amplitudes writes 
\begin{align}
 A(\cG)&=\left(-\lambda\right)^{|B|} N^{|F_{int}|-(D-1)|B|} .
\end{align}

\subsubsection*{Cumulants}

According to \eqref{eq:expandedcumulants}, the Feynman expansion of the cumulants writes as a sum over connected $D+1$ coloured graphs with $2k$ external edges as
\begin{align}
  \kappa(\bT_{m_1}\bar{\bT}_{\bar{m}_1}...\bT_{m_k}\bar{\bT}_{\bar{m}_k})  =\sum_{\pi \bar\pi}\  \sum_{ \cG}\
  \prod_{\e\in E^0} \left(\prod_{c=1}^D \delta_{n^c_\e\bar n^c_\e}\right)\ \prod_{d=1}^k\ \delta_{m_{\pi(d)} n_d}
  \delta_{\bar m_{\bar \pi(d)} \bar n_d}
\crcr
 \times\prod_{\cB\subset\cG} \left(-\frac{\lambda}{N^{D-1}}
 \sum_{n_\cB, \bar n_\cB, m_\cB, \bar m_\cB}  
\delta_{\bar n_\cB^{\cD\setminus \cC} n_\cB^{\cD\setminus \cC}  }  \delta_{\bar m_\cB^{\cD\setminus \cC}  m_\cB^{\cD\setminus \cC}  } \ 
   \delta_{n_\cB^{\cC}\bar m_\cB^{\cC}} \delta_{   \bar n_\cB^{\cC}  m_\cB^{\cC}}
         \;
 \right) ,
\end{align}
which, up to a prefactor, writes as a product of identity operators. Therefore, each coloured index $n^c_p$ from a covariant external leg $p$ is identified
through a product of $\delta$'s with the index of same colour $\bar n^c_{\tau^c_\cG(p)}$ of a dual external leg $\tau^c_{\cG}(p)$. 
These pairings define a $D$-uplet of permutations $\tau^\cD_\cG$ over $k$ elements (the pairs external legs). 

Each internal face contributes with a factor $N$, and gathering the prefactors associated with each bubble,
\begin{align}\label{eq:feynampcumulants}
  \kappa(\bT_{m_1}\bar{\bT}_{\bar{m}_1}...\bT_{m_k}\bar{\bT}_{\bar{m}_k})  =\sum_{\pi \bar\pi}\  \sum_{ \cG}\
  \left(-\lambda\right)^{|B|} N^{|F_{int}|-(D-1)|B|}
  \prod_{d=1}^k\ 
  \prod_{c=1}^D \delta_{m^c_{\pi(d)} \bar m^c_{\tau^c_\cG(\bar \pi(d))}} .
\end{align}
Representing covariant external legs as hollow vertices and dual legs as solid vertices, 
and representing for each colour the different pairings by coloured edges between those vertices,
we define a $D$-coloured bipartite graph, with both hollow and solid vertices labelled from $1$ to $k$.
This graph is called the \emph{boundary graph} $\partial\cG$ of the Feynman graph $\cG$,
and has the structure of a (not necessarily connected) tensor invariant, with labelled vertices \cite{GuRy1109}.
Such graph can be canonically associated with a $D$-uple of permutations $\tau^\cD_{\partial\cG} = (\tau^c_{\partial\cG})_{c\in\cD}$. 
For each colour $c$,
$\tau^c_{\partial\cG}(d)$ is the label of the hollow vertex connected to the solid vertex $d$ by an edge of colour $c$.
In the later developments, the boundary graph and  the associated $D$-uple of permutations will often be identified, $\partial\cG=\tau^\cD_{\partial\cG}$.

The index contraction term $\delta_{m^c_{\pi(d)} \bar m^c_{\tau^c_\cG(\bar \pi(d))}}$ ensures that a graph can only
contribute to a cumulant that has the same tensor invariant structure as its boundary graph.
Therefore, only cumulants with the index structure of a tensor invariant can be non-zero and the cumulants can be rewritten as
\begin{align}
 \label{eq:structurecumulants}
\kappa(\bT_{m_1}\bar{\bT}_{\bar{m}_1}...\bT_{m_k}\bar{\bT}_{\bar{m}_k})
=\sum_{\pi, \bar\pi}\sum_{\tau^{\cD} }\mathfrak{K}(\tau^{\cD})\prod_{d=1}^k  
\prod_{c=1}^D\delta_{m_{\pi(d)}^c \bar{m}_{\tau_c\bar\pi(d)}^c} \;,
\end{align}
with 
\begin{align}
  \mathfrak{K}(\tau^{\cD})  = \sum_{\substack{\cG\, \slash \\ \tau^\cD_\cG = \tau^\cD}}\
  \left(-\lambda\right)^{|B|} N^{|F_{int}|-(D-1)|B|} .
\end{align}

Further studies of the invariant tensor models are made much easier in the intermediate field formalism, which will be introduced in Chapter \ref{chap:IF}. 

%
%
%
%
%
%

\section{Tensor Field Theories}\label{sec:TFT}

We call \emph{field theory} any tensor model built from a non-invariant Gaussian measure, with a non trivial Hermitian covariance $\bC\not=\id$. 
The action \eqref{eq:action} of such a model writes,
\begin{equation}
 S(\bar\bT,\bT)=  \sum_{n\bar n}\bar{\bf T}_{\bar n} \bC^{-1}_{\bar n n} {\bf T}_{ n} 
  +S_{\rm int}(\bar\bT,\bT) \ ,
\end{equation}
where the interaction part of the action $S_{\rm int}$ regroups the quartic terms. 
As $\bC$ is not invariant under the unitary group, the quadratic part for the action is not a tensor invariant. 
While we loose the nice properties of unitary invariance and the easy face-counting computation of Feynman amplitudes,
it allows the introduction of the notion of \emph{scale}. Scale is at the heart of all physics models and, through renormalisation, 
gives a new dynamics to statistical and quantum field theories.
The most common choice of covariance operator, inspired by quantum field theory, is
\[
 \bC_{\bar p p} = \frac{\delta_{\bar p p}}{\bar p\cdot p + m^2}
\]
and correspond to the propagator of a scalar field of mass $m$. The tensor indices act as momenta, and the tensor space abstractly becomes a discrete momentum space,  
which can be defined independently of any direct \emph{position} space.

Tensor field theories, defined directly with an abstract notion of momentum can appear unnatural within the tensorial framework, 
where all quantities are usually constructed from unitary invariance. 
However, they can also be related to scalar field theories on a compact group manifold $U(1)^D$ \cite{BengeRiv1111,CarOriRiv1207}.

\subsection{Tensor models as group field theories}
%
A scalar field theory over the Abelian group manifold $U(1)^D$ is defined as a generating functional,
\begin{equation}\label{genfunc}
\mathcal{Z}[J,\bar{J}]=\int  e^{-S[\psi,\bar{\psi}]+\int_{U(1)^D}\bar{J}\psi+\int_{U(1)^D}\bar{\psi}J}\ d\psi d\bar{\psi}\ ,
\end{equation}
where the field $\psi$ over $U(1)^4$ is a smooth map $\psi:U(1)^4\to\mathbb{C}$, 
and the source terms $J$ and $\bar{J}$ are both smooth maps from $U(1)^4$ to $\mathbb{C}$.

The action $S[\psi,\bar{\psi}]$ split into a kinetic part and an interaction part: $S[\psi,\bar{\psi}]=S_{\rm kin}[\psi,\bar{\psi}]+S_{\rm int}[\psi,\bar{\psi}]$.
We choose the kinetic part as in \cite{CarOriRiv1207},
\begin{equation}\label{skin}
S_{\rm kin}[\psi,\bar{\psi}]=\int_{U(1)^D}\prod_{i=1}^D dg_i\bar{\psi}(g_1,\dots g_D)\bigg(-\sum_{i=1}^D\Delta_i+m^{2}\bigg)\psi(g_1,\dots g_D)\ ,
\end{equation}
where $\Delta$ is the Laplace-Beltrami operator and $dg$ is the Haar measure over $U(1)$.
%
%
In Fourier space, denoting $\mathbf{\Psi}_{\vec p},\ \vec p=(p^1,\dots p^D)$ the Fourier components of $\psi$, 
the kinetic part of the action and the generating functional are 
\begin{align}
S_{\rm kin}[\psit,\bar{\psit}]&=\sum_{{\vec p} \in \mathbb{Z}^D}\left({\vec p}^{\:2}+m^{2}\right)\bar{\psit}_{\vec p} \psit_{\vec p} \ , 
\\
\mathcal{Z}[\mathbf{J},\bar{\bf J}]&=\int \prod_{\vec p \in \mathbb{Z}^D} d{\bf\Psi}_{\vec p}\;
d\bar{\bf\Psi}_{\vec p}\ e^{-\: S_{int}[\psit,\bar{\psit}]\: -\: S_{kin}[\psit,\bar{\psit}]\: +\: 
\sum_{\vec p}\:  \bar{\bf J}_{\vec p} \psit_{\vec p}\: +\: \bar{\psit}_{\vec p}{\bf J}_{\vec p}} \ . \label{genfuncfourier1}
\end{align}
The Fourier field $\psit$ is a map from $\setZ^D$ to $\setC$ and can be considered as the infinite size limit of a rank $D$ tensor.

In the framework of \emph{tensorial} group field theories \cite{Oriti1211}, the interaction part of the action is built from tensor invariants, 
which, in Fourier space, follows the same definiton as for invariant tensor models. The quartic invariants \eqref{eq:quarticinvariant} writes:
\begin{align}
 {V}_{\cC}(\psit,\bar\psit) &= {\rm Tr}_\cC \left[ \left(\psit\cdot_{\cD\backslash\cC}\bar\psit\right)
 \cdot_\cC\left(\psit\cdot_{\cD\backslash\cC}\bar\psit\right)\right]   \ ,     \label{quarticbubbles} 
\end{align}
In the direct space, these invariants can be written in terms of the group field $\psi$, 
\begin{align}
{V}_{\cC}(\psi,\bar\psi)&=\int_{U(1)^D} \prod_{i=1}^D dg_idg_i'dh_idh_i'\
\psi(g_1,\dots g_D)\\
&\times\bar{\psi}(h_1,\dots h_D)\ \psi(g_1',\dots g'_D)\ \bar{\psi}(h'_1,\dots h'_D)\label{directquartic}\\ 
&\times\prod_{c\not\in\cC}\delta (g^c-h^c)\delta (g'^c-h'^c)\prod_{c\in\cC}\delta (g^c-g'^c)\delta (h^c-h'^c).
\end{align}

A quartic tensorial group field action can then be built using a linear combination of quartic invariants as  the interaction part :
\[
S_{\rm int}[\psit,\bar\psit]=\sum_{\cC}\lambda_{\cC}{V}_\cC (\psit,\bar\psit)\ .
\]

\subsection{Perturbative expansion}


The Feynman expansion of tensor field theories is similar to the one of invariant models, and according to \eqref{eq:feynmanamplitude}, the amplitude of a graph writes
\begin{align}\label{eq:TFTfeynmanamplitude1}
 A(\cG)&=\prod_{\cC\in\cQ}\prod_{\cB^\cC\subset\cG} \left(-\lambda_\cC
 \sum_{n_\cB, \bar n_\cB, m_\cB, \bar m_\cB}   \delta_{\bar n_\cB^{\cD\setminus \cC} n_\cB^{\cD\setminus \cC}  }  \ 
   \delta_{n_\cB^{\cC}\bar m_\cB^{\cC}} \delta_{   \bar n_\cB^{\cC}  m_\cB^{\cC}}\ 
     \delta_{\bar m_\cB^{\cD\setminus \cC}  m_\cB^{\cD\setminus \cC}  }  \   \;
 \right)\crcr
 &\times \prod_{\e\in E^0} \bC_{n_\e\bar n_\e}
 \ .
\end{align}
However, as the covariance $\bC$ is not the identity, 
Feynman amplitudes are not products of traces of identity operator, and their amplitudes cannot be computed directly with a face-counting technique. 
However, as the covariance $\bC$ is diagonal, to each face of colour $c$ corresponds a single \emph{momentum}, 
or index $p^c_f$ that is summed over $\cZ$. For a graph $\cG$ with $F$ faces, the amplitude can be written as
\[
 A(\cG) = \left(\prod_{\cC} \lambda_\cC^{|B_c|}\right) \cA_{\cG}\ ,\qquad \cA_{\cG}=\sum_{\{p_f\}\in\setZ^F} \frac{1}{P(\{p_f\})}\ ,
\]
where $|B_{\cC}|$ is the number of bubbles of type $\cC$ in the graph. 
$P$, the product of all the inverse covariances in $\cG$, is a polynomial in the face-momenta $p_f$ which structure depends on the graph $\cG$.  
Depending on $P$, the sum over momenta can be divergent, in which case the amplitudes need some regularisation, 
and the theory will require renormalisation.

\subsection{Multi-scale analysis}
\subsubsection{Scale slices}

The multi-scale analysis \cite{Rivasseau1991}
consists in organising the momenta into \emph{scale slices}, which are analogous to the orders of magnitude for physical quantities. 
The goal is then to organise physical processes in quantum field theory, or merely summations over momentum indices and parts of Feynman graphs
for our tensor theories, according to the (momentum) scale on which they happen. This leads to a finer understanding of the renormalisation process, where 
local divergences arising from sub-graphs of higher scales are treated independently.   

Let $M$ be an integer with $M>1$. We define the $j$-th momentum slice $S^j$ as,
\begin{align}\label{eq:slices}
S^1 &= \left\{ p\in\setZ^D,\ \,m^2+p\cdot p \,\le\, M^{2} \right\}\crcr
 S^j &= \left\{ p\in\setZ^D,\ M^{2j-2}\,<\,m^2+p\cdot p \,\le\, M^{2j} \right\}\quad\forall j> 1.
\end{align}

Then, we decompose the covariance $\bC$ over slices.
Denoting $\bI_{x}$ the characteristic function of the set $x$, we define 
the following operators
: 
\begin{align}
(\bI_j)_{\bar n n} =\bI_{S^j}(n)\ \delta_{\bar nn}\ ,\qquad
\bI_{\le f} = \sum_{i=1}^j \bI_i \ . 
\label{propmombound1}
\end{align}
The covariance of scale $j$ is defined as,
\begin{equation}
 \bC_j = \bI_{j} \bC \ .
\end{equation}
Note that this operator is bounded by
\begin{equation}\label{eq:Cjbound}
 (\bC_j)_{n\bar n} \leq M^{2-2j} \prod_{c\in \cD} \delta_{n^c\bar n^c}\ \bI_{|n^c|\leq M^j} .
\end{equation}
where $\bI_x$ is the characteristic function of the event $x$.

An overall cut-off  can be chosen to renormalise the theory by choosing an upper bound $j_{\max}$ for the momentum scale 
and replacing the covariance of the model by the \emph{regularised} covariance $\bC_{\le j_{\max}}$, defined as
\begin{equation}
 \bC_{\le j_{\max}} = \bI_{\le j_{\max}} \bC = \sum_{j=1}^{j_{\max}} \bC_j
\end{equation}
such that the momenta run over a finite set $\cup_{j\le j_{\max}} S^j$\ .
%
%

\subsubsection*{Amplitude decomposition}

Using the shorthand notation
\[
 \delta^{\cB} = \delta_{\bar n_\cB^{\cD\setminus \cC} n_\cB^{\cD\setminus \cC}  }  \ 
   \delta_{n_\cB^{\cC}\bar m_\cB^{\cC}} \delta_{   \bar n_\cB^{\cC}  m_\cB^{\cC}}\ 
     \delta_{\bar m_\cB^{\cD\setminus \cC}  m_\cB^{\cD\setminus \cC}  },
\]
the regularised amplitude of a graph $\cG$ writes
\begin{align}
 \cA^{\rm reg}_{\cG}
 &=\sum_{\{n\}}\ \; \prod_{\cB\subset\cG} 
    \delta^{\cB}  \   \;
 \prod_{\e\in E^0} (\bC_{\le j_{\max}})_{n_\e\bar n_\e}
 \crcr
 &=\sum_{\{n\}}\ \;\prod_{\cB\subset\cG} 
    \delta^{\cB}  \   \;
 \prod_{\e\in E^0} \left[\sum_{j=1}^{j_{\max}}(\bC_{ j})_{n_\e\bar n_\e}\right]\crcr
 &=\sum_\mu {\cA}_{G}^\mu
 \ .
\end{align}
where $\mu=\{j_\e,\ \e\in E^0\}\in\{1\dots j_{\max}\}^{|E^0|}$ is composed of a discrete scale index for each colour 0 edge of $\cG$,
and ${\cA}_{G}^\mu$ is the corresponding amplitude,
\begin{align} \label{eq:amplitude3}
 {\cA}_{G}^\mu
 &=\sum_{\{n\}}\ \,\prod_{\cB\subset\cG}
    \delta^{\cB}  \   \;
 \prod_{\e\in E^0} \left[(\bC_{ j_\e(\mu)})_{n_\e\bar n_\e}\right]
\ .
\end{align}
The perturbative expansion thus writes as a sum over $D+1$-coloured graphs with scale attribution, i.e. 
to each edge of colour $0$ is attributed a scale parameter $j\le j_{\max}$. 
\subsubsection*{High subgraphs}
We define $\cG_i$ as the subgraph of $\cG$ obtained by deleting every color-$0$ edges of scale attribution
lower than $i$ (bearing $C_j,j<i$), then every isolated bubbles. The high subgraphs $\cG_i^k, k\in\{1...\kappa(i)\}$ 
are the $\kappa(i)$ connected components of $\cG_i$. 
Within a high subgraph, every solid and hollow vertices may not be connected to a colour-0 edge. This is analogous to the 
external legs of the full graph $\cG$ as vertices without colour-0 edges can be interpreted as bearing external edges 
(a colour-0 edge without covariance or scale and connected to an external leg).
This defines the \emph{internal} and the \emph{external} faces of a high subgraph $\cG_i^k$.

From \eqref{eq:Cjbound} and \eqref{eq:amplitude3}, one has the following theorem:

\begin{theorem}\label{thm:powercounting}
The amplitude $\mathcal{A}_{\mathcal{G}}^{\mu}$ with scales indices $\mu$ admit the following uniform bound:
\begin{equation}\label{eq:powercounting}
|\mathcal{A}_{\mathcal{G}}^{\mu}|\leq M^{2E^0(\cG)}\prod_{i}\prod_{k=1}^{\kappa(i)}M^{\omega(\mathcal{G}_i^k)},
\end{equation}
where, denoting $E^0 ,\ F_{\rm int}  ,\ F_{\rm ext}$ the sets of internal colour 0 edges, 
internal and external faces of a graph, the \textit{degree of divergence} $\omega$ is defined by:
\begin{equation}\label{eq:divdeg}
\omega(\mathcal{G}_i^k) = -2 |E^0(\mathcal{G}_i^k)|+|F_{\rm int}(\cG_i^k)| \ .
\end{equation}
\end{theorem}

\prf 
Using the bound \eqref{eq:Cjbound} on the covariances,
\begin{align}
 {\cA}_{\cG}^\mu&\leq 
 \sum_{\{n\}}\ \,\prod_{\cB\subset\cG}
    \delta^{\cB}  \   \;
 \prod_{\e\in E^0} \left[
 M^{2-2j_\e} \prod_{c\in \cD} \delta_{n^c_\e\bar n^c_\e}\ \bI_{|n^c_\e|\leq M^{j_\e}}
 \right]
 \ .
\end{align}
The contractions $\delta$ identify indices along the faces of the graph $\cG$ such that 
only one summation survives per face $f\in F_{\rm int}(\cG)$, therefore, 
\begin{align}
 \mathcal{A}_{\mathcal{G}}^{\mu}
 \leq 
 \prod_{\e\in E^0(\cG)} M^{2-2j_\e} \, \times
 \prod_{f\in F_{\rm int}(\cG)} M^{j_{\min} (f)}
  \ ,
\end{align}
where 
\[
j_{\min} (f) = \min_{\e\in f} j_\e \ .
\]
The  product over edges can be reorganised according to the high sub-graphs of $\cG$ as 
\begin{align}\label{eq:edgebound}
\prod_{\e\in {E}^0(\mathcal{G})}M^{-2j_\e}&=\prod_{\e\in E^0(\mathcal{G})}\prod_{i=1}^{j_\e}M^{-2}=
\prod_{i\geq 1}\prod_{l\in E^0(\cup_{k=1}^{\kappa(i)}\mathcal{G}_i^k)}M^{-2} \\
&=\prod_{i\geq1}\prod_{k=1}^{\kappa(i)} M^{-2 |E^0(\mathcal{G}_i^k)|} \ .
\end{align}
and the product over faces can then be reorganised as
\begin{align}\label{eq:facebound}
 \prod_{f\in F_{\rm int}(\cG)}M^{j_{\min}(f)} \ =\ \prod_{i\geq1}\prod_{f\in F_{\rm int}(\cup_{k=1}^{\kappa(i)}\mathcal{G}_i^k)} M 
 \ = \ \prod_{i\geq1}\prod_{k=1}^{\kappa(i)} M^{|F_{\rm int}(\cG_i^k)|} \ .
\end{align}
Finally, the contribution of the external faces has no sum and can be bounded independently of $M$. 

Gathering the bounds \eqref{eq:edgebound} and \eqref{eq:facebound} proves equation \eqref{eq:powercounting}.

\qed

 \subsection{Renormalisation}
 \subsubsection{Divergences}
 When the regularisation is lifted ($j_{\max}\to\infty$), the amplitude of a Feynman graph $\cG$ can diverge if it contains some sub-graphs with a 
 non-negative degree of divergence \eqref{eq:divdeg}. Therefore, a sub-graph with $\omega\ge0$ is called a {\it divergent} sub-graph, 
 while the {\it convergent} sub-graphs ($\omega<0$) cannot generate divergences.
 These divergences occur when the scale of a divergent sub-graph becomes large, then it contributes with an multiplicative factor $M^\omega>1$ to each scale $i$ in
 \eqref{eq:powercounting}. 
 
 \subsubsection{Counter-terms}
 Such divergent sub-graphs therefore need to be renormalised, by adding a counter-term to the initial action, of value equal to minus the amplitude of the sub-graph at
 a given value of the external momenta, and of the same index structure as the boundary graph of the divergent sub-graph. 
 For a divergent graph $\cG$ with $2k$ external edges and a boundary graph $\partial\cG$, the counter-term writes,
 \[
  \delta_{\cG}(\psit,\bar\psit)= -A^{reg}_{\cG}\big\vert_{\psit_1=\bar\psit_1=\dots\psit_k=\bar\psit_k=0}\ V_{\partial\cG}(\psit,\bar\psit)\ ,
 \]
 where $V_{\partial\cG}$ is the tensor invariant corresponding to the boundary graph ${\partial\cG}$. 
 Note that, as the amplitude depends on $j_{\max}$, the counter-term $\delta_{\cG}$ and the renormalised action
 \[
  S^{\rm ren} = S + \delta_{\cG}\ ,
 \]
both depend on the regularisation scale $j_{\max}$. 

\subsubsection{Renormalisability}
For the renormalisation procedure to be carried out, we must have some control over the number and nature of the divergence. 
A \emph{renormalisable} theory must obey the following conditions
\begin{itemize}
 \item the degree of divergence must be bounded : For all $k>0$, there is a finite integer $\omega_{\max}(k)$, such that, 
 for any sub-graph $\cG_k$ with $2k$ external edges, $\omega(\cG_k)\le \omega_{\max}(k) $.
 \item the number of external edges of a divergent sub-graph must be bounded : 
 There is a finite $k_{\max}$ such that, if $k>k_{\max}$, $\omega_{\max}(k)<0$.
\end{itemize}
The first condition forbids the appearance of arbitrarily strong divergences, which could not be properly renormalised by finite (at a given $j_{\max}$) counter-terms. 
The second condition ensures that the theory keeps a finite number of interaction terms. 

Tensor field theories can therefore be classified according to their renormalisation properties.
\begin{itemize}
 \item a theory that satisfy both conditions, and which generates only a finite number of different
 divergent (sub-)graphs is called {\it super-renormalisable}.
\item a theory that satisfy both conditions, and which generates infinitely many different
 divergent (sub-)graphs is called {\it just-renormalisable}.
\item a theory that does not satisfy both conditions is called \it{non-renormalisable}.
\end{itemize}

\subsubsection{Divergent degree}
The study of renormalisability thus requires a close inspection of the properties of the Feynman graphs and their divergent degree. 
Such a study will be conducted later, as it is made easier with the help of the intermediate field formalism of chapter \ref{chap:IF}. 
For now, let us use without proof the results of chapter \ref{chap:IF}. 
According to corollary  \ref{cor:omegagraph}, 
any bipartite $D+1$ coloured graph $\cG_k$ with $2k$ external edges, $|B|$ quartic $D$-coloured bubbles and $2|B|+k$ colour-0 edges follows :
  \begin{equation}\label{eq:bounddiv}
   (D-1)|B|-|F_{int}| \ge -D + (D-1) k +  C(\partial\cG_k) \ .
  \end{equation}
  where $C(\partial\cG_k)$ is the number of connected component of the boundary graph $\partial\cG_k$.
Note that, for a connected graph, the number of internal edges of colour 0 is related to the number of bubbles by $|E^0|=2|B|-k$.
Gathering \eqref{eq:divdeg} and \eqref{eq:bounddiv}, the divergence degree  satisfies,
\begin{equation}
 \omega(\mathcal{G}_k) = 2k -4 |B|+|F_{\rm int}| \le (D-5)|B| +D +(3-D)k - C(\partial\cG)\ ,
\end{equation}
which allows us to conclude that the quartic tensor field theory at rank $D$ is,
\begin{itemize}
 \item super-renormalisable for $D\le4$. As the divergence degree decreases with the number of bubbles, only a finite number of graphs can be divergent.
 \item just-renormalisable for $D=5$. The divergent degree is bounded by  $\omega_{\max}(k) =4-2k$, only graphs with at most 4 external legs can diverge. 
 However, divergent sub-graphs can be arbitrarily large, with  arbitrarily many quartic bubbles.
 \item non-renormalisable for $D\ge 6$. The divergent degree is not bounded, and larger graphs can generate higher order divergences.
\end{itemize}

\chapter{Intermediate field representation}\label{chap:IF}
The intermediate field representation is a convenient way to study the graph expansion of quartic tensor models. 
It is a bijection that represents the usual Feynman graphs as maps (graphs with a cyclic order of the edges incident to a vertex),
which often makes the study of their properties easier. If it can be considered as a purely graphical transformation, 
it also corresponds to a rewriting of the partition function with a new field, called the intermediate field.

\section{Graphical representation}\label{sec:graphIFrep}

Each Feynman graph of a quartic model can be represented by an intermediate field map using a graphical transformation where each bubble 
of the original graph becomes an edge, and the vertices of the intermediate field map are made of the Wick contraction edges 
of the original graph. For a given theory, there are several different intermediate field representation, 
each corresponding to a given choice of the set of interactions  $\cQ$.

\begin{definition}\label{definition_multicolouredmap}
 A ciliated $\cD$-multicoloured map is a quadruplet $\cM=(C, \{H_\cC\}_{\cC\subset\cD}, \sigma, \epsilon)$ such that:
 \begin{itemize}
 \item $C$ is a finite set of cilia.
  \item $H=\dot\cup_{\cC\subset\cD} H_{\cC}$ is a finite set of multicoloured half-edges, $H_\cC$ is the set of half-edges of colours $\cC$.  
  \item $\sigma$ is a permutation on $H\cup C$. 
  \item $\epsilon$ is an involution on $H$ with no fixed point, such that $\forall \cC$, $\epsilon(\cH_\cC)=\cH_\cC$.
 \end{itemize}
\end{definition}
Using the similar definition of graph objects as for directed maps (Definition \ref{definition_vertexedgecorner}), 
\begin{definition}\label{definition_colouredvertexedgecorner}
For a multicoloured map  $\cM=(C, \{H_\cC\}_{\cC\subset\cD}, \sigma, \epsilon)$,
\begin{enumerate}
 \item A vertex is a cycle of the permutation $\sigma$.
A cycle of length $p$ is called $p$-valent.
\item A corner is a pair of half-edges or cilia $(h,\sigma(h))$. If $v$ is a vertex and $h\in v$, then $(h,\sigma(h))$ 
is called a corner of the vertex $v$. 
\item An multicoloured edge of colours $\cC$ is a pair of half-edges $(h_\cC,\epsilon(h_\cC))$ with $h_\cC\in H_\cC$. 
\end{enumerate}
\end{definition}
 A ciliated multicoloured map can be seen as a graph where
 \begin{itemize}
  \item Every edge carries a set of colours $\cC\subset\cD$.
  \item Every vertex has a cyclic ordering of the incident edges.
  \item Vertices can bear cilia, that behave as half-edges.
 \end{itemize}

\subsection{Graph to map transformation}

\begin{figure}[h]
 \centering
  \includegraphics{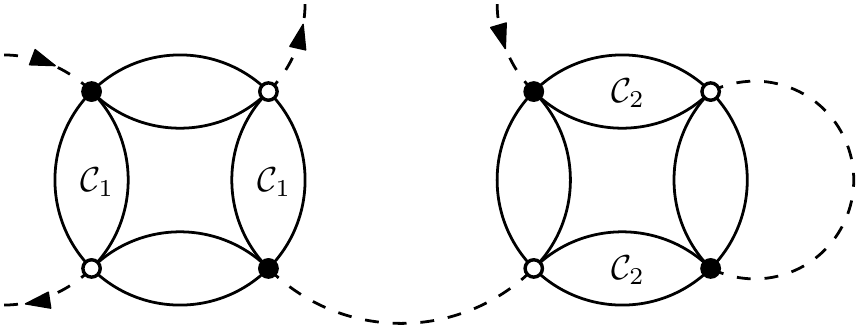}
\caption{This $D+1$-coloured graph will be used as a example to illustrate the transformation process.}
\label{FigIFstep0}
  \end{figure}
Let $G$ be a Feynman graph with interaction bubbles of type $V_\cC, \cC\in\cQ$, Wick contraction edges  of colour $0$ and external legs.
The colour 0 edges are directed from the hollow vertices toward the solid ones.
The intermediate field map $\cM(G)$ is the edge coloured map built from $G$ by the following three steps transformation.

\begin{enumerate}
 \item Each bubble of colours $\cC$ of the original graph is replaced by a multicoloured edge of colours $\cC$, 
 each pair of solid and hollow vertices connected by edges of colours in $\cD\setminus\cC$ being contracted into a single vertex. 
 The new (step 1)-vertices are trivalent, with two half-edges of colour $0$ and a multicoloured half-edge. 
 The direction of the colour 0 edges is preserved.
\begin{figure}[h]
 \centering
  \includegraphics{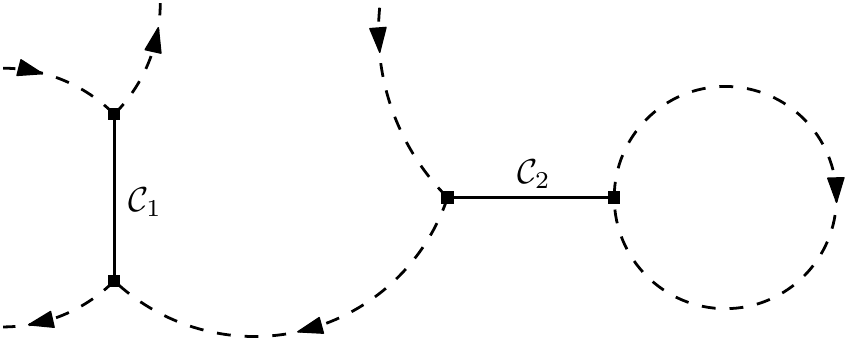}
\caption{The graph from Fig. \ref{FigIFstep0} after the first transformation step.}
\label{FigIFstep1}
  \end{figure}
 \item To each external leg of the covariant type is associated an external leg of the dual type by following the unique path 
 made of Wick contraction edges only (without going through multicoloured edges).
 Then each pair of external legs is replaced by a \emph{cilium}.
 At this stage of the transformation, each (intermediate stage)-vertex of the graph is connected to two colour-0 edges, 
 therefore there is a unique way to replace external legs by cilia. 
 The colour-0 edges now form closed cycles, or loops. By construction, 
 there can be an arbitrary number of vertices and 0-colour edges on each cycle, but only up to one cilium. 
 \begin{figure}[h]
 \centering
  \includegraphics{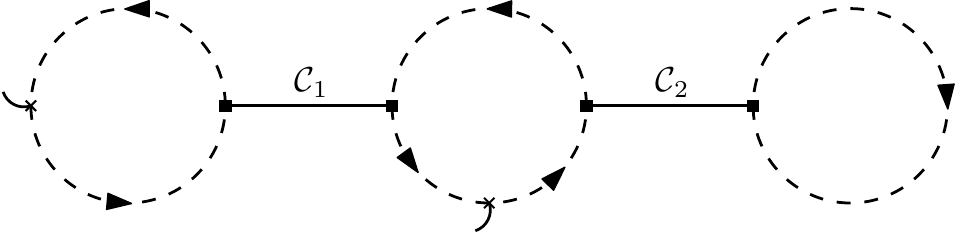}
\caption{The graph from Fig. \ref{FigIFstep1} after the second transformation step.}
\label{FigIFstep2}
  \end{figure}
 \item Each cycle of colour-0 edges is replaced by a vertex of the intermediate field representation map, also called a \emph{loop-vertex}. 
 The incident (multi)-coloured edges are ordered on the loop-vertex according to their former position on the colour-0 loop. 
 The vertex is oriented clockwise according to the direction of the former colour 0 edges.
 Cilia are treated as regular half edges during the process, and colour 0 edges become corners of the new map $\cM(G)$.  
 \begin{figure}[h]
 \centering
  \includegraphics{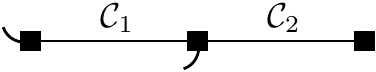}
\caption{The graph from Fig. \ref{FigIFstep2} after the third and final transformation step. 
This is the intermediate field map associated with the graph of Fig. \ref{FigIFstep0}}
\label{FigIFstep3}
  \end{figure}
\end{enumerate}

If the steps 2 and 3 are uniquely defined, the result of step 1 entirely depends on the choice of $\cQ$. 
Replacing $\cC \in\cQ$ by $\cD\setminus\cC$ does not change the model as the associated trace invariant is the same, but, 
as it changes the set of (multi)-coloured edges and the pairing of hollow and solid vertices into intermediate stage vertices,
it modifies the structure of the cycles of colour-0 edges and the pairing of external legs, 
leading to a completely different intermediate field map. 
\begin{figure}[h]
 \centering
  \includegraphics{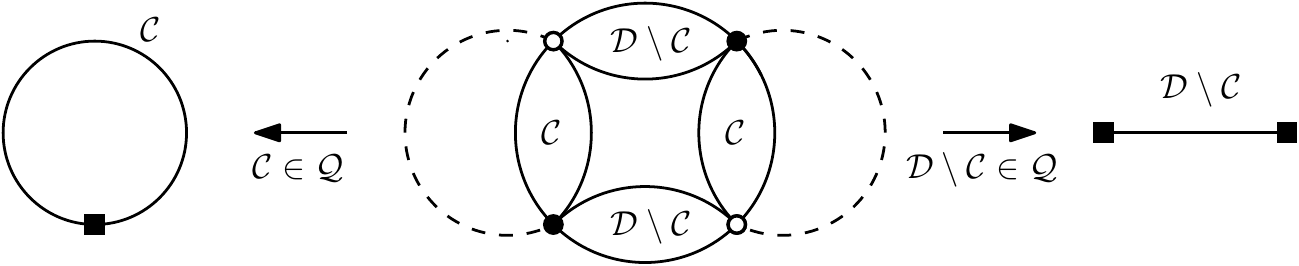}
\caption{Two different intermediate field maps arising from different choices of $\cQ$.}
\label{FigdifferentIF}
  \end{figure}

The two most common choices of $\cQ$ are
\begin{itemize}
 \item $\forall \cC \in \cQ,\ 1\in\cC$,
 \item $\forall \cC\in\cQ,\ |\cC|\leq|\cD|/2$, if $|\cC|=|\cD|/2$ then $1\in|\cC| $.
\end{itemize}
Only the latter will be used in the following developments.

Once the set of interactions $\cQ$ is chosen, the graph to map transformation defines a one to one correspondence between
\begin{enumerate}
 \item the bipartite $D+1$-coloured graphs with $2k$ external legs, and which $D$-coloured connected sub-graphs are quartic bubbles 
of type $\cQ$, 
\item the multicoloured maps with edges of colours $\cC\in\cQ$ and with $k$ ciliated vertices, each bearing a single cilium.
\end{enumerate}
As quartic bubbles become edges of the map, graphs with labelled bubbles correspond to maps with labelled edges. Cilia can be labelled according the label of the
corresponding dual external legs in the bipartite coloured graph. 

The inverse transformation can be performed easily by replacing corner by directed colour-0 edges, 
splitting the cilia into pairs of external edges, then replacing the multicoloured edges by quartic bubbles while being careful of 
placing hollow and solid vertices according to the direction of the colour 0 edges.

Note that the intermediate field map representation is very different from the map representation of matrix models developed
in section \ref{sec:MatrixModels}. Here, the interaction bubbles are represented as edges of the map, while they were represented as vertices
of the matrix map representation. The vertices of the intermediate field maps actually represent cycles of colour 0 edges 
(often called {\it loops} in quantum field theory). The colour 0 edges, that were still represented as edges of the matrix map, 
are therefore the corners of the intermediate field maps.

This above construction is a bijection between edge-multicoloured maps and bipartite $D+1$-coloured graphs with external legs, which $D$-coloured 
connected components are quartic bubbles. This restriction to quartic bubbles imposes a strict structure on the graphs and forbids the use of the transformation 
for more general $D+1$-coloured graphs. Notably, the Feynman graphs arising from tensor models with higher-order interactions cannot be treated with this transformation.
A generalised version of this transformation, for graphs with higher-order bubbles, was consequently introduced in \cite{BoLiRi1508}, using some more involved \emph{stuffed Walsh maps} instead of
edge-coloured maps. 

\subsubsection*{Multi-stranded graphs}

Multi-stranded graphs are a convenient way to represent multicoloured maps that keeps tracks of the tensor indices and their identifications.
The 0-valent vertex is represented as a multi-stranded vertex, made of $D$ concentric circular coloured strands, 
ordered from 1 to $D$, the innermost being of colour 1. Edges are represented as multicoloured ribbon edges. 
The addition of an edge of colours $\cC$ between two loop vertices opens the strands of colours $c\in\cC$ in both vertices 
and ties them together according the orientation of the strands. Each edge is thus composed of $2|\cC|$ strands.
\begin{figure}[h]
 \centering
  \includegraphics{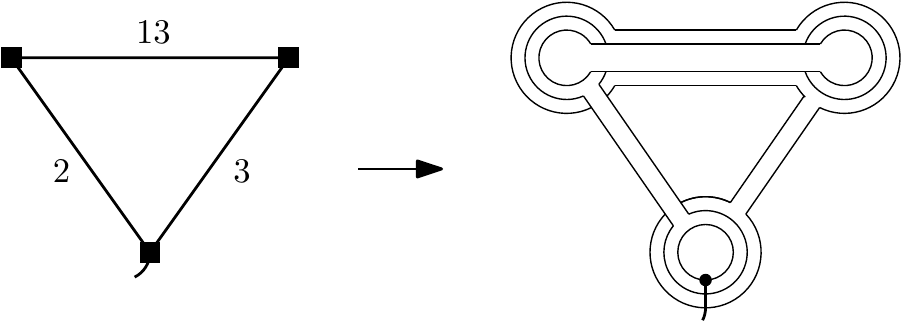}
\caption{The multi-stranded representation of a ciliated 3-multicoloured map. }
\label{FigMultistand}
  \end{figure}
  In this representation, the faces of a map are directly apparent. Internal faces are the closed lines of a multi-stranded graph, while external faces are lines
  connecting two cilia. Not that the edges of the multi-stranded graphs have the same structure as the corresponding $D$-coloured quartic bubbles,
  the hollow and solid vertices and the colour-0 edges connecting them, being replaced by $D$-stranded corners displaying explicitly their index contraction structure.



\section{The intermediate field formula}\label{sec:IFformula}

The intermediate field transformation is not only a graphical correspondence 
between bipartite coloured graphs and multicoloured maps.
The intermediate field maps can actually be obtained as a genuine Feynman expansion of some intermediate fields as
the moment-generating function of the quartic tensor models can be rewritten in terms of a multi-matrix model, 
with matrices of different sizes, and which Feynman expansion is directly written in terms of the multicoloured maps, 
allowing to perturbatively evaluate the moments of the quartic tensor model directly in terms of intermediate field maps. 

This re-writting of the moment-generating model is based on the Hubbard-Stratonovich transformation \cite{Hubbard,Stratonovich}, which has been 
a well-established tool in many-body physics and quantum field theory \cite{MatSal1954,Muh1962}, where it allows to decompose a 4 fields interaction into 3 fields interactions through
the introduction of a new field, the {\it intermediate field}, which gives its name to this chapter.

\subsection{Transformation}
Let us denote $\csig^{\cal{C}}_{a^{\cal C} b^{\cal C}}$ a $N^{|\cal{C}|} \times N^{|\cal{C}|}$ Hermitian matrix with  
line and column indices of colours in $\cal{C}$, $a^{\cal {C}} = (a^c | c\in \cal{C})$, and let us denote 
$\mathbf1^{\cD\setminus \cC}$ the identity matrix on the indices of colours $\cD\setminus \cC$. 
We define  
\begin{align}
& A(\csig)=\sum_{\mathcal{C}\in \cal{Q}}\ i\ \sqrt{\frac{\lambda_\cC}{ N^{D-1}}}\  \id^{\cD \setminus \cC }
\otimes\csig^{\mathcal{C}} \ ,\crcr
& \big[A(\csig) \big]_{nm} =   
\sum_{\mathcal{C}\in \cal{Q}} \ i\ \sqrt{\frac{\lambda_\cC}{ N^{D-1}}}\ \delta_{n^{\cD\setminus \cC} m^{\cD\setminus \cC}} \ 
\csig^{\mathcal{C}}_{n^{\cal{C}} m^{\cal{C}} }  \ .
\end{align}
 Note that, as $A(\csig)$ is anti Hermitian, $ [ \mathbf1^{\cD} + A(\csig) ]^{-1}$ is 
well defined for all $\lambda\in \mathbb{C} \setminus (-\infty,0)$.

The intermediate field representation of the quartic tensor model $\mu$ \eqref{eq:model} is:
\begin{theorem}\label{thm:IFrep}
The moment-generating function of $\mu$ \eqref{eq:momentgeneratingZ} with a set of interactions $\mathcal{Q}$ is:
\begin{align}\label{eq:ZLVE}
Z(\bJ,\bar \bJ)=\int 
e^{-
\ \tr_{\cD} \big[ \log\left(\id^{\cD} + A(\csig)\right)   \big] 
+  \sum_{nm } \bar \bJ_{n} 
\left[\frac{1}{ \id^{\cD} + A(\csig) }\right]_{nm}\bJ_{m} }  \; d\mu_\id(\csig) \ .
\end{align}
where $d\mu_\id(\csig)$ is the normalized Gaussian measure of identity covariance $\id$ over the Hermitian $\csig$ matrices.
\begin{equation}
 d\mu_\id (\csig) = \prod_{\cC\in\cQ}  d\mu_\id (\csig^\cC) = \prod_{\cC\in\cQ} e^{-\frac12\tr (\csig^\cC)^2}\left[d\csig^\cC\right] \ .
\end{equation}
\end{theorem}

\prf 

The Hubbard Stratonovich intermediate field representation relies on the observation that, 
for any Hermitian matrix $\cM$, 
\begin{equation} \label{eq:Hubbard}
  \int e^{ i \tr \bM \csig }  \; d\mu_\id (\csig)= e^{-\frac12 \tr \bM^2} \ ,
\end{equation}
where $d\mu_\id$ is the standard Gaussian measure of covariance $\id$ over Hermitian matrices.

We will now apply this formula for the quartic interaction terms. We have:
\begin{align}
& e^{-\frac{\lambda_\cC}{2N^{D-1}} \ \tr_{\cC} \left[  [ \bar{\bf T} \cdot_{\cD\setminus \cC} {\bf T} ] \cdot_{\cC}
  [ \bar{\bf T} \cdot_{\cD\setminus \cC} {\bf T} ] \right] } \crcr
  &= 
   \int
   e^{i\sqrt{\frac{\lambda_\cC}{ N^{D-1}} } \; \sum_{ n^{\cC} \bar n^{\cC}  }\; [ {\bf T}^{\vee} \cdot_{\cD\setminus \cC} {\bf T} ]_{ \bar n^{\cC}  n^{\cC}  } 
  \ \csig^{\cC}_{ \bar n^{\cC} n^{\cC} }       }\ d\mu_\id (\csig) \crcr
  \;\;
   &= \int  e^{  i \sqrt{\frac{\lambda_\cC}{ N^{D-1}}}\; 
    \sum_{n \bar n}\; \bar \bT_{\bar n}\; \left[  \mathbf1^{\cD\setminus \cC} 
   \otimes\csig^{\mathcal{C}} \right]_{\bar n n}\;  \bT_{ n } } \ d\mu_\id (\csig)\; .
\end{align}
The generating function is then:
\begin{align}
& Z(\bJ, \bar{\bJ}) =\int  e^{  -\sum_{n \bar n} \bar \bT_{\bar n }    \left[ \id^{\cD} \ +\  i
\left(  \sum_{\mathcal{C}} \sqrt{\frac{\lambda_\cC}{ N^{D-1}}} \id^{ \cD \setminus \cC }\otimes\csig^{\mathcal{C}} \right)\right]_{\bar n; n} 
   \bT_{n} } \crcr
& \;\; \times  e^{ \sum_{n} \bT_{n} \bar{\bJ}_{n} + \sum_{\bar n}\bar{\bT}_{\bar{n} }\bJ_{\bar{n} }}  \;
\left(\prod_{n} \frac{d\bT_{n} d \bar{\bT}_{n } } {2\imath \pi}\right) 
\prod_{\cC\in\cQ}d\mu_\id (\csig^\cC)
.
\end{align}
The integral over $\bT$ and $\bar \bT$ is now Gaussian with covariance
\[
 \left(\id^\cD + A(\csig)\right)^{-1}
 \ ,
\]
 and as
 \[
  \det\left(\id^\cD + A(\csig)\right)^{-1} = e^{- \tr_\cD \log\left(\id^\cD + A(\csig)\right)}
  \ ,
 \]
 a direct computation leads to eq. \eqref{eq:ZLVE}.

\qed

The intermediate field transformation rewrites the tensor model as a multi-matrix model with a non-polynomial interaction term,
\begin{equation} \label{eq:IFinteraction}
 V_{IF}(\csig) = \tr_{\cD} \left[ \log\left(\id^{\cD} + A(\csig)\right)   \right] \ .
\end{equation}
In order to perform the Feynman expansion of the intermediate field theory, one must perform a series expansion of the interaction.

\subsection{Feynman expansion} \label{subsec:IFFeyn}

The interaction term \eqref{eq:IFinteraction} for the intermediate field can be expanded as,
\begin{align}\label{eq:expandedIFinteraction}
 V_{IF}(\csig) &= \sum_{p\geq1} \;\frac{(-1)^p}p\; \tr_{\cD} \left[ A(\csig)^p  \right] \crcr
 &=\sum_{p\geq1} \;\frac1p\; \left(\frac{-i}{ \sqrt{N^{D-1}}}\right)^{p}\ \tr_{\cD} \left[ 
 \sum_{\mathcal{C}\in \cal{Q}}\ \sqrt{\lambda_\cC} \ \id^{\cD \setminus \cC }
\otimes\csig^{\mathcal{C}}\right]^p  
 \ .
\end{align}
According to the previous equation, the interaction part of the intermediate field action is a 
sum of traces of arbitrarily long cycles of $\csig$ fields, each of arbitrary colours.
Each term comes with a square-root coupling constant $\sqrt{\lambda^\cC}$ per corresponding field $\csig^\cC$ in the cycle.
In the Feynman expansion, this would translate in vertices of arbitrary valence and with coupling constants associated with the edges instead of the vertices.

Therefore, it can be easily checked that the Feynman expansion of the $2k$-point function computed with the intermediate 
field theory writes as a sum over all multicoloured maps with $k$ cilia ans colours in $\cQ$, as:
\begin{itemize}
 \item Edges are multicoloured, they are associated with sets of colour $\cC\in\cQ$ as they correspond 
 to Wick contractions of the matrix field $\csig^\cC$.
 \item Vertices has cyclic ordering of the half-edges. Each vertex is indeed a trace of a product of $A(\csig)$ operators. 
 \item Vertices are of arbitrary valence, and with arbitrary colours of the half edges.
 \item The cilia are the pairs $\bJ\bar \bJ$ of sources as they appear in \eqref{eq:ZLVE}. 
 As the source term in \eqref{eq:ZLVE} can be expanded in a similar way
 as the interaction term \eqref{eq:expandedIFinteraction}, ciliated vertices can also have arbitrary valence. 
\end{itemize}
Note that there are $p$ different rooted cycles of length $p$ of $A(\csig)$ terms corresponding
to a single cyclically-ordered vertex. This  contributes with a factor $p$ in the map amplitude for each non-ciliated vertex. 
Ciliated vertices are naturally rooted at the cilium and do not bring symmetry factors.
\begin{figure}[h]
 \centering
  \includegraphics{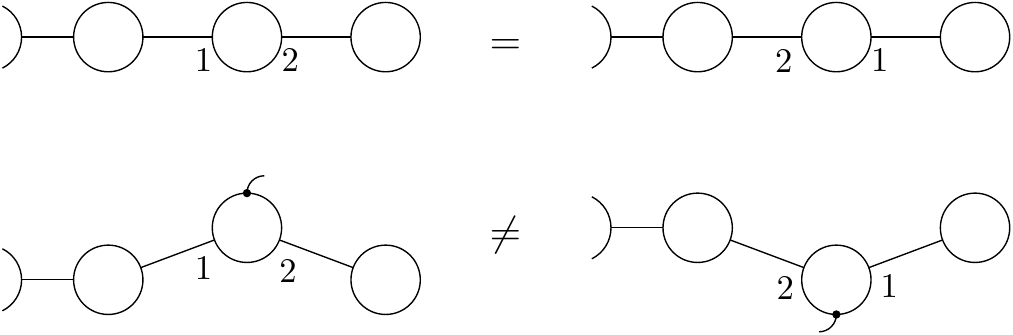}
\caption{The symmetry for a bivalent vertex. The rotated non-ciliated vertex contributes to the same map. This is no longer the case for a ciliated vertex.}
\label{graph:cycles}
  \end{figure}
  
  Thus, the perturbative expansion writes in term of multicoloured maps $\cM_{\rm v}$ with $|V(\cM_{\rm v})|$ labelled vertices and $k$ cilia as, 
\[
  \braket{\bT_{m_1}\bar{\bT}_{\bar{m}_1}...\bT_{m_k}\bar{\bT}_{\bar{m}_k}}_\nu \approx\sum_{\cM_{\rm v}}\frac{1}{|V(\cM_{\rm v})|\, !} A(\cM_{\rm v}) \ .
\]
Moreover, the structure of the interaction terms in \eqref{eq:expandedIFinteraction}, with tensor product of identity operators,
shows that the Feynman maps display the same multi-stranded structure as the maps obtained by the graphical intermediate-field transformation.

The amplitude associated with the Feynman maps of the intermediate field are computed by multiplying the coupling constants $\lambda$, 
and counting the internal faces of the map. Each edge of colours $\cC$ bears a coupling $\lambda_\cC$ and the faces are defined as 
for graphically transformed tensor field Feynman graphs.
Therefore, the amplitude associated with an intermediate field theory map is the same as the one associated to the same map when obtained 
 from a direct tensor field graph by the graphical transformation. 
\begin{align}\label{eq:IFfeynammoments}
  A(\cM_{\rm v}) =\sum_{\pi \bar\pi}\ \
  N^{|F_{int}(\cM_{\rm v})|}\prod_{\cC\in\cQ}\left(\frac{-\lambda_\cC}{N^{D-1}}\right)^{|E_\cC(\cM_{\rm v})|} 
  \prod_{d=1}^k\ 
  \prod_{c=1}^D \delta_{m^c_{\pi(d)} \bar m^c_{\tau^c_\cM(\bar \pi(d))}} ,
\end{align}
where $E_\cC(\cM)$ is the set of edges of $\cM$ of colours $\cC$ and the $D$-uple of permutation $\tau^c_\cM=\partial\cM$ is the boundary graph of $\cM$, 
which, for each colour, pairs the cilia (and corresponding source terms) together according to the external faces of $\cM$.

Finally, the sum can be rearranged as a sum over maps with labelled edges. Denoting respectively $s_{\rm v}(\cM)$ and $s_{\e}(\cM)$ the number of 
different vertex labellings and edge labellings of a map $\cM$ with unlabelled edges and vertices, we have $\frac{s_{\rm v}}{|V|!} = \frac{s_{\rm e}}{|E|!}$, therefore,
\[
 \sum_{\cM_{\rm v}}\frac{1}{|V(\cM_{\rm v})|\, !} A(\cM_{\rm v}) = \sum_{\cM}\frac{s_{\rm v}(\cM)}{|V(\cM_{\rm v})|\, !} A(\cM) 
 =\sum_{\cM_{\rm e}}\frac{1}{|E(\cM_{\rm v})|\, !} A(\cM_{\rm e})\ ,
\]
where $\cM_\e$ sums over maps with labelled edges. Therefore, the Feynman expansion of the intermediate field theory is truly the 
 intermediate field map representation of the expansion of the tensor model.
 
 For connected ciliated maps, ${s_{\rm v}}={|V|!}$ as the symmetry group of a map becomes trivial, 
 and the Feynman expansion of the cumulants can be written in terms of unlabelled connected maps as,
 \[
   \kappa(\bT_{m_1}\bar{\bT}_{\bar{m}_1}...\bT_{m_k}\bar{\bT}_{\bar{m}_k}) = \sum_{\cM {\rm connected}} A(\cM)\ .
 \]

Surprisingly, the intermediate field formalism was not originally introduced in the random tensor framework for the nice properties of its perturbative map expansion, 
which will be further discussed in the next section, but merely as a necessary step toward the constructive study of tensor models through the loop vertex expansion
\cite{Gurau1304}, which will be introduced in part two.

\section{The perturbative expansion}

Using Feynman expansion on the intermediate field theory, the partition function of a quartic tensor model can be perturbatively approximated as a sum
over all multicoloured maps with no cilium. For a model with a single coupling constant $\lambda$,
\begin{equation}
 \log Z \approx \sum_{\cM \rm \ connected} \frac{1}{|E| !}  \left(-\lambda\right)^{|E|} N^{|F_{int}|-(D-1)|E|}\ .
\end{equation}
The cumulants are evaluated as sums over ciliated connected (unlabelled) maps with the proper boundary graph.
\begin{align}\label{eq:IFcumu}
\kappa(\bT_{m_1}\bar{\bT}_{\bar{m}_1}...\bT_{m_k}\bar{\bT}_{\bar{m}_k})
\approx\sum_{\pi, \bar\pi}\sum_{\tau^{\cD} }\mathfrak{K}(\tau^{\cD})\prod_{d=1}^k  
\prod_{c=1}^D\delta_{m_{\pi(d)}^c \bar{m}_{\tau_c\bar\pi(d)}^c} \;,
\end{align}
with 
\begin{align}\label{eq:pertcumulant}
  \mathfrak{K}(\tau^{\cD})  = \sum_{\substack{\cM\, \slash  \partial\cM = \tau^\cD \\ \mathrm{connected}}}\
  \left(-\lambda\right)^{|E|} N^{|F_{int}|-(D-1)|E|} \ .
\end{align}
In any case, the amplitude of a multicoloured map $\cM$ can be written,
\[
 A(\cM) =  \left(-\lambda\right)^{|E(\cM)|} N^{-\Omega(\cM)}\ ,
\]
where the exponent of $1/N$ is,
\begin{equation}\label{eq:expon}
 \Omega(\cM) = (D-1)|E(\cM)|-|F_{int}(\cM)|\ .
\end{equation}
For any number of cilia $k$, this exponent is bounded from below. The logarithm of the partition function and the cumulants
can therefore be expressed as power series in $1/N$.

\subsection{A map glossary}
In this subsection, we define the useful terms associated with maps. 
\begin{itemize}
\item A leaf is a monovalent vertex.
\item A bridge is an edge whose deletion transform a connected map into two connected maps. 
\item A plane forest is a map which edges are all bridges.
\item A plane tree is a simply connected map. i.e. a connected plane forest.
\item Let $\cM$ be a connected map, a spanning tree $\cT$ of $\cM$ is a simply connected sub-map of $\cM$, that contains every vertices of $\cM$, $V(\cT)=V(\cM)$.  
\item If an edge $\e\in\cM$  is not a bridge, there is a spanning tree $\cT\subset\cM$ such that $\e\not\in\cT$. $\e$ is called a \emph{loop edge}.
\end{itemize}

\subsection{The $1/N$ expansion}\label{subsec:1/Nexp}

\subsubsection{Bound on the exponent}
We denote $C(\tau^\cD)$ the number of connected components of the $D$-coloured graph associated with the $D$-uple of permutations $\tau^\cD$, and
$\id_k^\cD$ the $D$-uple of identity permutations over $k$ elements.
\begin{theorem}\label{thm:1/Nexp}
 For any $D$-uple  $\tau^\cD$ of permutations over $k$ elements, we define the minimal exponent,
 \[
  \Omega_{\min}(\tau^\cD) = -D + (D-1) k +  C(\tau^\cD) \ .
 \]
 Then,
with the exponent $\Omega$ defined in \eqref{eq:expon},
 \begin{itemize}
 \item For $k>0$,
 for any connected multicoloured map $\cM$ with $\partial\cM=\tau^\cD$,  $\Omega(\cM)\geq\Omega_{\min}(\tau_\cD)\geq 0$.
 \item For a connected vacuum map $\cM$, $\Omega(\cM)\geq-D = \Omega_{\min}(\emptyset)$.
  \item For $k\geq0$, if $\Omega(\cM)=\Omega_{\min}(\partial\cM)$ then
 \begin{itemize}
  \item $\cM$ is a plane tree,
  \item For $\e\in E(\cM)$, either $\e$ is mono-coloured ($|\cC(\e)|=1$) or  all faces running through $\e$ are external 
  (each coloured strand in $\e$ belongs to an external face).
 \end{itemize}
 \item  For $k=0$ (and $\partial\cM=\emptyset$) or $k>0$ and $\partial\cM=\id_k^\cD$, then $\Omega(\cM)=\Omega_{\min}(\partial\cM)$ if and only if $\cM$ is a
 plane tree made solely of mono-coloured edges. 
\end{itemize}
\end{theorem}
Note that for $k>1$, and a given $\partial \cM$, their might not be any map such that $\Omega(\cM)=\Omega_{\min}(\partial\cM)$.

\prf
As for any connected map, $|E|\geq|V|-1$, in order to prove the first part of this theorem, it is sufficient to bound the number of internal faces of a given map by,
\begin{align}
     F_{\rm int} (\cM)  \le\, 1\ -\ (D-1) k\, -\, C(\partial\cM) 
     +\, (D-1)V(\cM)  \; ,
 \end{align}
 which is established in the following lemma.
\begin{lemma}\label{lem:facesbound}
 The number of internal faces of a connected multicoloured map $\cM$ with $V(\cM)$ vertices, $k(\partial\cM)$ cilia and $E(\cM)$ edges is bounded by:
\begin{align}
     F_{\rm int} (\cM)  \le& \ 1\ -\ (D-1) k(\partial\cM)\ -\ C(\partial\cM)  \crcr
     &+\ (D-1)V(\cM) \  +\ \left\lfloor\frac{D}{2}\right\rfloor\left[E(\cM)-V(\cM)+1\right] \; .
 \end{align}
 \end{lemma}
As $E(\cM)-V(\cM)+1=0$ only for trees, Lemma \ref{lem:facesbound} also proves that $\Omega(\cM)$ 
can be equal to $\Omega_{\min}(\partial\cM)$ only if $\cM$ is a tree. 

 Let $\cT$ be a plane tree and $\e\in E(\cT)$ an edge. For $c\in\cC(\e)$, either both strand of colour $c$ in $\e$ belong to external faces, 
 or both they belong to the same internal face. Let us assume that the strands of colour $c$ belong to an internal face and that $\cC\setminus\{c\}\neq\emptyset$,
 we define $\cT^*$ the multicoloured map obtained by replacing  $\e$ by an edge of colours $\cC\setminus\{c\}$. Then $F_{\rm int} (\cT^*)=F_{\rm int} (\cT)+1$, and 
 $\Omega(\cT)>\Omega(\cT^*)\geq\Omega_{\min}(\partial\cT)$. The prescription $\Omega(\cT)=\Omega_{\min}(\partial\cT)$ is  achievable only is all edges either are
 mono-coloured or do not contain strands belonging to internal faces.
 
 Finally, it can be easily shown that, for any tree $\cT$ made solely of mono-coloured edges and with $\partial\cT=\emptyset$ or $\partial\cT=\id^\cD_k$, 
 the number of internal faces is, 
 \[
  F_{\rm int}(\cT)= D + (D-1)(|V(\cT)|-1) -Dk \ .
 \]
Therefore, $\Omega(\cT)=-D+Dk=\Omega_{\min}(\partial\cT)$ . 
Moreover, a tree $\cT$ with a least one multi-coloured edge ($|\cC(\e)|\ge2$) would have a lower number of faces. 
Proving Lemma \ref{lem:facesbound} will achieve the proof of Theorem \ref{thm:1/Nexp} .
 
\subsubsection{Proof of lemma \ref{lem:facesbound}}
 In order to establish this bound, we first introduce a new coloured map $\cM^*$ associated to $\cM$ which will allow us to keep track
of the boundary $\partial\cM$ while modifying $\cM$. 
Starting from $\cM$, for each cilium $h$, 
we introduce $2D$ half edges on the corner bearing the cilium $h$, one at its left and one at its right for each colour $c\in \cD$ (that is, for each colour $c$,
one half edge precedes $h$ and the other succeeds $h$ when turning around the vertex). We then connect these new half edges into dashed edges
following the edges of $\partial\cM$: if the external strand of colour $c$ starting at the cilium $h$ ends at the cilium $h'$, 
we connect the half edge of colour $c$ following $h$ with the half edge of colour $c$
preceding $h'$. This construction is represented in Figure \ref{fig:tauedges}.

 \begin{figure}[htb]
\begin{center}
\includegraphics[height=1.8cm]{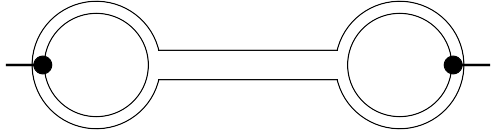}  
\caption{A map $\cM$ with two cilia. For simplicity, only the colours 1 and 2 have been represented.}\label{fig:simpleM}
\end{center}
\end{figure}

Each of the external faces of $\cM$ corresponds to an edge in the boundary graph $\partial \cM$, hence $\cM$ has exactly $Dk(\partial\cM)$ external faces. 
By construction $\cM^*$ has a new edge for each external face of $\cM$, which closes the external face of $\cM$ into an internal face of $\cM^*$.
It follows that:
\[
F_{\rm int} (\cM^*) = F_{\rm int} (\cM)+ Dk(\partial\cM)\ .
\]

On the other hand, as the new dashed edges exactly follow the external faces of $\cM$, it ensues that $\partial \cM^* = \partial \cM$.

\begin{figure}[htb]
\begin{center}
\includegraphics[height=3cm]{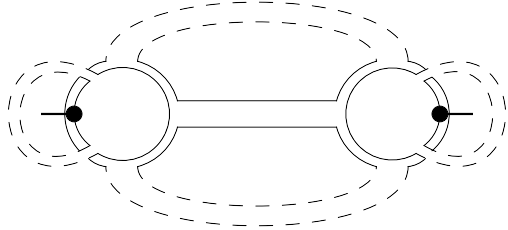}  
\caption{The map $\cM^*$ obtained from the map $\cM$ in Fig. \ref{fig:simpleM} by adding the dashed edges.}\label{fig:tauedges}
\end{center}
\end{figure}

The reason to introduce $\cM^*$ is the following. Below we will delete solid edges of $\cM^*$ (that is edges which belonged to $\cM$) 
and track the change in the number of internal faces of $\cM^*$ under these deletions. The crucial point is that these deletions 
modify the structure of the internal faces of $\cM^*$ but \emph{do not modify} its external faces, hence the boundary graph $\partial\cM^*$ will remain unchanged.
The introduction of $\cM^*$ allows us to modify a map while keeping its boundary graph unchanged.

Let $\cT$ be a spanning tree in $\cM$ (with all the vertices but no loop edges) and let us denote $\cT^*$ the 
map made of $\cT$ and all dashed edges of $\cM^*$. We have:
\[\partial\cT^*=\partial \cM^*=\partial\cM \;.\]

The map $\cT^*$ can be obtained from $\cM^*$ by deleting $E(\cM)-V(\cM)+1$ solid edges. As there are at most $\left\lfloor D/2\right\rfloor$ colours going through any 
of the solid edges, the deletion of such an edge in $\cM^*$ can either increase of decrease the number of faces by at most $\left\lfloor D/2\right\rfloor$, hence:
\begin{align*}
F_{\rm int} (\cM)+ Dk(\partial \cM)= F_{\rm int}(\cM^*)\ \leq\ F_{\rm int}(\cT^*) \,+\, \left\lfloor\frac{D}2\right\rfloor\left[E(\cM)-V(\cM)+1\right] \ .
\end{align*}

We call a \emph{leaf} of $\cT^*$ a vertex which becomes univalent when one erases all the dashed edges. 
The leaves of $\cT^*$ are the univalent vertices of $\cT$.
In order to complete the proof of Lemma \ref{lem:facesbound}, it is now enough to show that for any $\cT^*$ we have: 
\begin{align}\label{eq:sublemmbound}
 F_{\rm int}(\cT^*) + C(\partial\cT^*) \leq 1 + k(\partial\cT^*) + (D-1)V(\cT^*)  \ .
\end{align}

This bound is obtained by tracking the evolution of $F_{\rm int}(\cT^*) + C(\partial\cT^*)$ under the iterative deletion of the leaves. 
This deletion is somewhat involved, as it can change the boundary graph.

A few remarks are in order before proceeding further.
\begin{itemize}
 \item A cilium corresponds to a pair of black and white vertices of the boundary graph $\partial \cT^*$. 
 Those two vertices can belong to separate connected components of $\partial \cT^*$. 
 \item An external face can either connect two different cilia, 
 or start and end at the same cilium $h$. In the later case, the external face is said to be \emph{looped} at the cilium $h$. 
\end{itemize}

In Figure \ref{fig:tauedges} we can identify two looped external strands.

Let us denote $\ell$ a leaf of $\cT^*$ and let us denote $\hat\cT^*$ the tree obtained from $\cT^*$ after deleting $\ell$ as follows.

\paragraph{Deleting a non ciliated leaf.} 
If $\ell$ has no cilium, deleting it simply means deleting the vertex $\ell$ and the edge connecting it to the rest of $\cT^*$. 
If the edge connecting $\ell$ to the rest of $\cT^*$ has color $\cC$, the deletion of $\ell$ erases all the faces of colours 
$\cD\setminus\cC$ running trough $\ell$. The boundary graph is unaltered by this deletion, therefore:
\begin{align*}
 F_{\rm int} (\cT^*)  = F_{\rm int} (\hat\cT^*) + (D-|\cC|) \; , & \qquad 
  C(\partial\cT^*) =C(\partial\hat\cT^*) \Rightarrow \crcr
F_{\rm int} (\cT^*)+C(\partial\cT^*) &\leq F_{\rm int} (\hat\cT^*) + C(\partial\hat\cT^*) + (D-1)\ .
 \end{align*}

\paragraph{Deleting a ciliated leaf.} 
If $\ell$ has a cilium $h$, deleting it consists in :
\begin{itemize}
 \item {\it Step 1.} For all the external faces starting or ending at $h$ that are not looped at $h$, we cut the corresponding dashed edges into two half-edges. 
 We then reconnect the four dashed half edges into edges the other way around, respecting the colours. This is represented in Figure 
 \ref{fig:deletion1} below. Observe that after performing this step, all the external faces starting at $h$ 
 (and thus all external faces going trough $\ell$) are looped.

 \begin{figure}[htb]
\begin{center}
\includegraphics[height=3.5cm]{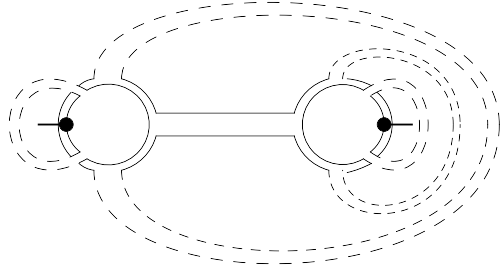}  
\caption{The first step of the deletion of a ciliated leaf : the dashed edges have been reconnected in such way that all the external strands are looped at the cilium $h$.}\label{fig:deletion1}
\end{center}
\end{figure}
 \item{\it Step 2.} The leaf now has only looped external strands and it is connected to the rest of the $\cT^*$ by a solid edge only.
 The cilium $h$ represents a connected component of the boundary graph consisting in a black and a white vertex connected by $D$ edges. 
 We erase the vertex $\ell$, its cilium, $h$, and the edge connecting $\ell$ to the rest of the map, as in Figure \ref{fig:deletion2}
\begin{figure}[htb]
\begin{center}
\includegraphics[height=3.2cm]{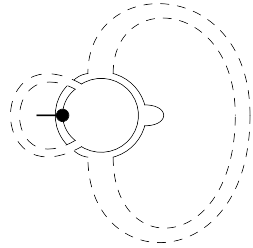}  
\caption{The second step of the deletion of a ciliated leaf.}\label{fig:deletion2}
\end{center}
\end{figure}
 \end{itemize}
 
{\it Boundary graph.} Upon deleting a ciliated leaf, the boundary graph changes: $\partial \hat \cT^*\neq \partial \cT^*$. 
There are two cases, each divided in two sub cases:
\begin{itemize}
\item None of the external faces of $\cT^*$ are looped at $h$. One needs to apply Step 1 for all the colours and then Step 2. 
There are two sub cases:
\begin{itemize}
 \item  the solid and hollow vertices associated to $h$ in $\partial \cT^*$ belong to the same connected component of $\partial\cT^*$. Then, in the boundary graph,
 Step 1 creates at least a new connected component, and Step 2 deletes exactly one connected component. Thus:
   \[  C(\partial\cT^*) \leq  C(\partial\hat\cT^*) \; .\]
 \item the solid and hollow vertices associated to $h$ in $\partial \cT^*$ belong to different connected components of $\partial\cT^*$. Then, in the boundary graph,
 Step 1 can not decrease the number of connected components, and Step 2 deletes exactly one connected component. Thus:
   \[  C(\partial\cT^*)  \leq C(\partial\hat\cT^*) +1 \; .\]
 \end{itemize}
\item At least one external face of $\cT^*$ is looped at $h$. There are two sub cases:
\begin{itemize}
 \item not all the external faces are looped at $h$. The solid and hollow vertices associated to $h$ in $\partial \cT^*$ 
 belong to the same connected component of $\partial\cT^*$.
  One must apply Step 1 at least once, and then step 2. As before, Step 1 creates at least a new connected component,
  and Step 2 deletes exactly one connected component,
  hence:
  \[
   C(\partial\cT^*) \leq  C(\partial\hat\cT^*) \; .
  \]
\item all the external strands are looped at $h$. The black and white vertices associated to $h$ in $\partial \cT^*$ 
belong to two different connected components of $\partial\cT^*$,
   and one must apply Step 2 directly. This decreases the number of connect components of the boundary graph by $1$:
  \[
   C(\partial\cT^*)  =  C(\partial\hat\cT^*) +1 \; .
  \]
\end{itemize}
\end{itemize}
 
 {\it Internal faces.} Let us denote $\e^\cC$ the solid edge of colours $\cC$ connecting $\ell$ to the rest of $\cT^*$. We have several cases:
 \begin{itemize}
  \item the external face of colour $c\in\cC$ is looped at $h$. Then there is only one internal face of colour $c$ in $\cT^*$ running through $\e^\cC$,
  which can not be erased by deleting $\ell$, hence:
 \[ F_{\rm int}^c(\cT^*)  =F_{\rm int}^c(\hat\cT^*) \; . \]
  \item the external face of colour $c\in\cC$ is not looped at $h$. Then there are either one or two internal faces of colour $c$ in $\cT^*$ running trough $\e^\cC$,
  hence the number of internal 
  faces of colour $c$ can not decrease by more than one:
      \[ F_{\rm int}^c(\cT^*)  \leq F_{\rm int}^c(\hat\cT^*) +1 \;. \]
    \item the external face of colour $c'\not\in\cC$ is looped. Then there is only one internal face of colour $c'$ through $\ell$, which is erased: 
       \[ F_{\rm int}^{c'}(\cT^*) = F_{\rm int}^{c'}(\hat\cT^*)+1 \; .\]
  \item the external face of colour $c'\not\in\cC$ is not looped. Then there is just one internal face of colour $c'$ through $\ell$, which is not erased: 
  \[F_{\rm int}^{c'}(\cT^*) = F_{\rm int}^{c'}(\hat\cT^*) \; .\] 
\end{itemize}

Combining the counting of the connected components of the boundary graphs with the counting of the internal faces we obtain three cases:
\begin{itemize}
 \item no external face is looped at $h$. Then:
     \[ 
      F_{\rm int} (\cT^*)+C(\partial\cT^*) \le \left[F_{\rm int} (\hat \cT^*) + \lfloor \frac{D}{2}\rfloor\right] + \left[ C(\partial \hat \cT^*) + 1\right]
      \le F_{\rm int} (\hat \cT^*)+C(\partial \hat \cT^*) + D \;. 
     \]
 \item all the external faces are looped at $h$. Then:
      \[
        F_{\rm int} (\cT^*)+C(\partial\cT^*) \le [F_{\rm int} (\hat \cT^*) + D-1] + [ C(\partial \hat \cT^*) + 1 ]
        \le F_{\rm int} (\hat \cT^*)+C(\partial \hat \cT^*) + D \;. 
      \]
+ D \;.
\item At least one external face is  looped at $h$, but not all. Then:
       \[
        F_{\rm int} (\cT^*)+C(\partial\cT^*) \le [ F_{\rm int} (\hat \cT^*)  + D ] + [C(\partial \hat \cT^*)] \le F_{\rm int} (\hat \cT^*)+C(\partial \hat \cT^*) + D \;.
      \]
\end{itemize}

In all cases, by deleting a ciliated leaf:
 \begin{equation}
  F_{\rm int}(\cT^*) + C(\partial\cT^*) \leq F_{\rm int}(\hat\cT^*) + C(\partial\hat\cT^*) + D \; .
 \end{equation}
Iterating up to the last vertex, we either end up with a vertex with no cilium or with a ciliated vertex with $D$ looped external strands.
Counting the number of internal faces and connected component of the two possible final maps (ciliated or not) gives us \eqref{eq:sublemmbound}, 
and we conclude.

 \qed
 \subsubsection{$1/N$ expansion}
 
 The size of the tensor gives us a new small parameter to re-arrange the graph expansion as a power series of $1/N$ instead of $\lambda$. 
 As the exponent \eqref{eq:expon} is bounded from below  by $\Omega$, the perturbative expansion of the cumulant \eqref{eq:pertcumulant} writes,
 \begin{align}
   \mathfrak{K}(\tau^{\cD})  = \sum_{\Omega\ge\Omega_{\min}(\tau^{cD})}\left(\frac1N\right)^{\Omega}
   \sum_{\substack{\cM\ \mathrm{connected} \slash \\ \partial\cM = \tau^\cD, \\ \Omega(\cM)=\omega }}\
  \left(-\lambda\right)^{|E(\cM)|}  \ ,
 \end{align}
which is the \emph{ $1/N$ expansion} of the cumulant $ \mathfrak{K}(\tau^{\cD})$.

 \subsubsection{$D+1$ coloured graphs}
 
 The internal faces of a multi-coloured map and of the corresponding $D+1$ coloured bipartite graph coincide, as stated in section \ref{sec:graphIFrep}. 
 Therefore, using the fact that the edges of a map are the bubbles of the corresponding $D+1$-coloured graph, one can immediately formulate the previous result 
 as a bound for the $D+1$ coloured graphs.
 \begin{corollary}\label{cor:omegagraph}
  Let $\cG$ be a bipartite $D+1$ coloured graph with $2k$ external edges, $|B(\cG)|$ quartic $D$-coloured bubbles and $2|B(\cG)|+k$ colour-0 edges. Then,
  \[
   (D-1)|B(\cG)|-|F_{int}(\cG)| \ge -D + (D-1) k +  C(\partial\cG) \ .
  \]
 \end{corollary}

\subsection{The two point cumulant}\label{sec:K2}
The map expansion of the two-point cumulant $\kK(\id_1^\cD)$ writes as a sum over (unlabelled) connected maps with a single cilium.
The minimal exponent, $\Omega_{\min}(\id_1^{\cD})=0$, corresponds to the family of mono-ciliated plane trees with mono-coloured edges,
and at large $N$, the cumulants writes,
\[
 \kK(\id_1^\cD) = \sum_{\cT}  \left(-\lambda\right)^{|E(\cT)|} + O\left(\frac1N\right) \ .
\]
A mono-ciliated plane tree can be canonically rooted at its cilium. The number of un-labelled rooted plane trees with $n$ edges is given by the Catalan number
$C_n= \frac1{n+1} \left(\genfrac{}{}{0pt}{}{2n}{n }  \right)$, and there are $D^n$ different edge-colouring of such a tree with mono-coloured edges. Therefore,
\[
 \kK(\id_1^\cD) \underset{ \infty}\to \sum_{n}  \left(-\lambda D\right)^{n} \frac1{n+1} \left(\genfrac{}{}{0pt}{}{2n}{n }  \right) \ .
\]
which, for $|4D\lambda|\le 1$ is absolutely convergent and sums to
\[
 \frac{-1+\sqrt{1+4D\lambda}}{2D\lambda} \ .
\]

\section{The intermediate field vacuum}

The intermediate field transformation of section \ref{sec:IFformula} defines a wholly new multi-matrix model $\nu(\csig)$. 
The partition function of this new model is, by construction, the same as the one of the original tensor model and the expectations and cumulants of 
the tensor model can be expressed using the intermediate field, which perturbative expansion is strongly related to the one of the original tensor model.  
The model itself, however, bears little resemblance with the original one, 
and its most basic field theoretical properties deserve a closer inspection \cite{DG1504,NguDarEyn1409}.   

 
 In the present section, we will study the intermediate field model for the standard melonic quartic model, which is defined with a single coupling parameter,
 \begin{align}
  &\cQ = \{ \{ c \}, c\in\cD\} \crcr
  &\forall c\in \cD, \lambda_{\{c\}} = \lambda \ .
 \end{align}

By Theorem \ref{thm:IFrep}, the intermediate field action is
\begin{equation}
 S(\csig)= \frac12 \tr_\cD\; \csig^2
 +\ \tr_{\cD} \big[ \log\left(\id^{\cD} + A(\csig)\right)   \big] 
\end{equation}
where $\csig=(\csig^c)_{c\in\cD}$ is a collection of $D$ Hermitian matrices of size $N$.

The interaction term can be expanded in powers of $\sqrt\lambda$, which at the first order gives,
\[
 \log\left(\id^{\cD} + A(\csig)\right) = i \sqrt{\frac{\lambda}{ N^{D-1}}}
\sum_{c\in \cal{D}}\  \id^{\cD \setminus \{c\} }
\otimes\csig^{c}\ +\ O(\sqrt\lambda^2) \ .
\]
According to the above equation, the intermediate field action shows a linear term in $\csig^c$ for each colour $c\in\cD$. 
Therefore, $\csig=0$ is not a solution of the equations of motion, and we must look for a non trivial intermediate field vacuum. 

%

\subsection{Equations of motion}

The equations of motion of the intermediate field writes,
\begin{align}\label{eq:eqmotion}
0=\frac{\partial S(\csig)}{\partial \csig^c_{ b^c a^c  }} =   \csig^c_{a^c b^c} +i \sqrt{\frac{\lambda}{N^{D-1}}} \; \sum_{ p^{\cD } q^{\cD } }
   R(\csig)_{ q^{\cD}   p^{\cD }   } \left( \delta_{  p^{\cD\setminus \{c\} } q^{\cD \setminus \{c\} } } \delta_{ p^{c} b^{c} } 
    \delta_{q^c a^c} \right)  \;,
\end{align}
where the resolvent $R$ is defined as
\begin{equation}
 R(\csig)\ =\ \frac{1}{\id^\cD+A(\csig)}\ = \ \frac{1}{\id^\cD+ i \sqrt{\frac{\lambda}{ N^{D-1}}}
\sum_{c\in \cal{D}}\  \id^{\cD \setminus \{c\} }
\otimes\csig^{c}} \ .
\end{equation}
In matrix form, the equation of motion \eqref{eq:eqmotion} writes,
\begin{align}\label{eq:matrixeqmotion}
0 =   \csig^c +i \sqrt{\frac{\lambda}{ N^{D-1}}} \; \tr_{\cD\setminus\{c\}}\left[\frac{1}{\id^\cD+ 
i \sqrt{\frac{\lambda}{ N^{D-1}}}\sum_{c\in \cD}\   \id^{\cD \setminus \{c\} }
\otimes\csig^{c}} \right]\ ,
\end{align}

A simple inspection of these equations reveals that, for $\lambda$ small enough, the stable vacuum of the theory is 
invariant under conjugation by the unitary group and invariant under colour permutation. The solution is of the form
$\csig^c = a\sqrt{N^{D-1}}\,\id\,\ \forall c$, where $a$ must be real as $\csig^c$ is Hermitian. 
The invariant solutions of eq. \eqref{eq:eqmotion} 
are then obtained for $a$ satisfying the self consistency equation:
\[
 a  = \frac{ -i\sqrt{\lambda}}{1 +i \sqrt{\lambda} Da } \; .
\]

This self consistency equation can only have real solutions for negative $\lambda$. 
This corresponds to an ill-defined tensor model in the original representation, as the action $S(\bar\bT,\bT)$ is not bounded from below and
the measure $d\nu(\bar\bT,\bT)$ is not integrable.

\subsubsection{The melonic vacuum}
A first solution, denoted $a_0$ and called the \emph{melonic vacuum}, is the sum of a power series in $\sqrt{\lambda}$ :
 \[
  a_0 = i\ \frac{1 - \sqrt{1+4D\lambda }}{2\sqrt{\lambda}D } = -i\sqrt{\lambda } \sum_{n\ge 0} \frac{1}{(n+1)!} \binom{2n}{n} (-\lambda D)^n \; .
\]
We will see below that this solution is the stable vacuum of the theory for $\lambda$ small enough.
For $\sqrt{\lambda}\in i\mathbb{R_+}$ and $-4D\lambda <1 $, $a_0$ is real, positive, and bounded by :
\[
 a_0  =  i\ \frac{1 - \sqrt{1+4D\lambda}}{ 2D \sqrt{\lambda} } =\frac{-2i\sqrt{\lambda} }{ 1 +  \sqrt{1+4D{\lambda}} }
  \le \frac{1}{\sqrt{D}} \;,
\]
as $ \frac{-2i \sqrt{\lambda} }{ 1 +  \sqrt{1+4D{\lambda}} }$  is increasing with $|\sqrt{\lambda}|$ and attains its maximum for $\sqrt{\lambda} = \frac{i}{2\sqrt{D}} $.

\subsubsection{The instanton}
The second solution, denoted $a_{\rm inst}$, is an instanton solution :
\[
  a_{\rm inst} = i\ \frac{1 + \sqrt{1+4D\lambda}}{2\sqrt{\lambda} D } \; .
\]
As $a_0  a_{\rm inst} = \frac{1}{D}$, it follows that for $\sqrt{\lambda}\in i\mathbb{R_+}$ and $-4D\lambda <1 $, $a_{\rm inst}$
is real, positive, and $a_{\rm inst} \ge \frac{1}{\sqrt{D}} $.

For $\sqrt{\lambda} = \frac{i}{2\sqrt{D}}$ the two solutions collapse: $a_0|_{\sqrt{\lambda} = \frac{i}{2\sqrt{D}} }=a_{\rm inst}|_{\sqrt{\lambda} = \frac{i}{2\sqrt{D}} }=\frac{1}{\sqrt{D}}$.

\subsection{Effective model}\label{subsec:translation}

The intermediate field model can now be rewritten in terms of fluctuation fields $\bM^c$ around the solution of the classical equations of motion,
\[
\csig^c =N^{\frac{D-1}{2}} (a \id + \bM^c)\ .
\]
In terms of the perturbation fields $\bM^c$, the quadratic terms write,
\begin{align}
 \frac{1}{2}\tr\left( N^{D-1} (a \id + \bM^c)^2\right) = \frac{N^{D-1}}2 \left(\,
 N a^2 + 2a \tr\ \bM^c + \tr\ \bM^{c\,2}\,
 \right)\ ,
\end{align}
whereas the inverse resolvent writes,
\begin{align}
\id^{\cD} + A(\sqrt{N^{D-1}} (a \id + \bM^c)) &= \id^{\cD} + i \sqrt{\frac{\lambda}{ N^{D-1}}}
\sum_{c=1}^D\  \id^{\cD \setminus \{c\} }
\otimes\sqrt{N^{D-1}} (a \id + \bM^c)\crcr
&= (1+iD\sqrt\lambda a) \left[\id^{\cD} - a
\sum_{c=1}^D\  \id^{\cD \setminus \{c\} }
\otimes  \bM^c\right]\ ,
\end{align}
and the action becomes :
\begin{align}
 S(\bM)&= \frac{1}{2} N^D D a^2 + N^D \log (1 +iD\sqrt\lambda a)
  + N^{D-1} a \sum_{c=1}^D \Tr_c\ \bM^c \crcr
  &+ \frac{1}{2}N^{D-1}\sum_{c=1}^D \Tr_{c}\ \bM^{c\,2} 
   + \Tr_{\cD} \left[ \log \left(\id^{\cD}  -a
\sum_{c=1}^D\  \id^{\cD \setminus \{c\} }
\otimes  \bM^c \right) \right] \; ,
\end{align}
A Taylor expansion of the logarithm shows that its first order cancels out the linear term of the action. 
\begin{equation}
 \Tr_{\cD} \left[ \log \left(\id^{\cD}  -a
\sum_{c=1}^D\  \id^{\cD \setminus \{c\} }
\otimes  \bM^c \right) \right]  = N^{D-1} a \sum_{c=1}^D \Tr_c\ \bM^c + O(\bM^2)\ .
\end{equation}
Therefore, defining $\log_2 (X) = \log (X) +1 -X$, the action writes as
\begin{align}\label{eq:effectiveaction}
 S(\bM)&= \frac{1}{2} N^D D a^2 + N^D \log (1 +iD\sqrt\lambda a) \crcr
  &+ \frac{1}{2}N^{D-1}\sum_{c=1}^D \Tr_{c}\ \bM^{c\,2} 
   + \Tr_{\cD} \log_2 \left(\id^{\cD}  -a
\sum_{c=1}^D\  \id^{\cD \setminus \{c\} }
\otimes  \bM^c \right) \; ,
\end{align}
which displays some interesting properties.
\begin{itemize}
 \item As expected, the effective action has no linear term and $\bM^c=0$ is therefore solution of the equations of motion. 
 In terms of Feynman expansion, this means that the Feynman maps generated by the fluctuation model cannot have leaves. 
 Consequently, plane trees, that were the dominant family for $\csig$,  do not appear in the Feynman expansion 
 of the fluctuation field.  Moreover, the addition of a leaf to a pre-existing map (and therefore the addition of a full tree), 
 which did not change the  exponent $\Omega$ for intermediate field maps, is impossible for the fluctuation field maps.
 \item According to \cite{BGR1202}, the free energy $\log Z$ of the original tensor model at large $N$, writes, in terms of $a_0$ as,
 \[
 \log Z \underset{N\to\infty}{\sim} -\frac{1}{2} N^D D a_0^2 - N^D \log (1 +iD\sqrt\lambda a_0) \ .
 \]
which, for the melonic vacuum $a=a_0$, corresponds precisely to the pre-factor in \eqref{eq:effectiveaction}. 
Note that this large $N$ free energy is the sum of the leading order in $1/N$ of vacuum maps, namely the \emph{melonic} trees, 
which do not participate to the map expansion of the fluctuation field $\bM$.
By translating the model to the melonic vacuum $a_0$, we summed the entire \emph{melonic} family of trees into a constant prefactor.
Therefore they no longer appear in the map expansion. A similar interpretation does not exist for the instanton $A_{\rm inst}$ 
as the prefactor in \eqref{eq:effectiveaction} no longer equals the leading (tree) order of the free energy.
 \item Quadratic terms arise from the logarithmic interaction term, for both the intermediate field $\csig$ and the fluctuation field $\bM$, which mix 
 fields of  different colours together. A careful study of the mass matrix is necessary to further characterise the new field theory. 
 A such inclusive study of the fluctuation field is performed in \cite{DG1504}.
\end{itemize}
\part{Constructive tensors}
\chapter{The constructive toolbox}

The perturbative expansions of Chapters \ref{chap:QTM} and \ref{chap:IF} were performed formally, without any regard for convergence and well-definedness.
A mathematically more correct study of tensor models 
and their graph expansions requires a new set of tools, that were inherited from the constructive study of
quantum field theory, and are therefore gathered under the name of \emph{constructive methods}.

\section{Borel sum}

\subsubsection{Definition}
Let $(a_k)_{k\ge 0}$ be a sequence, we define $A_n(z)$ as the partial sum
\[
 A_n(z) = \sum_{k=0}^n a_k z^k .
\]
We denote $A(z)$ the formal power series
\[
  A(z) = \sum_{k=0}^\infty a_k z^k ,
\]
even if the sum might not converge. The Borel transform of $A$ is the formal exponential series
\[
 \cB A (z) = \sum_{k=0}^\infty \frac1{k!} a_k z^k
\]
The formal power series $A(z)$ is called \emph{Borel summable} if the series $\cB A$ has non-zero convergence radius in $\setC$, 
converges on $\setR^+$  and its Borel sum $\hat A$  converges, with
\[
 \hat A(z) = \frac1z\int_0^\infty \cB A(t)e^{-\frac tz}dt \ .
\]
\subsubsection{Borel sum and perturbative expansion}

Borel sums are useful in constructive theory because many of the formal perturbative series encountered in tensor models and field theories 
are not summable,  but Borel summable, with the Borel sum being equal to the functional integral. 
The reason behind non-summability can be understood very easily by looking at the measure of the standard tensor model \eqref{eq:standardmodel}. 
For a positive coupling parameter $\lambda>0$, the interaction term of the measure is $\mu_\id$ integrable and the partition function $Z(\lambda)=\int d\nu$ as well as 
the moments are well defined as an absolutely convergent integral.  However, for $\lambda<0$, the integral diverges. The value $\lambda=0$ is on 
the boundary of the analyticity domain of $Z(\lambda)$, and any series expansion in powers of $\lambda$ is doomed to fail, as its convergence radius vanishes.

However, such series can be Borel summable, in which case the evaluated function can be reconstructed from the coefficients of the formal perturbative series : 
the perturbative expansion contains all the information about the function. Fortunately, by the Nevanlinna-Sokal \cite{Sokal1980} theorem, the
proof of the Borel summability only requires the satisfaction of two hypotheses. 

Denoting $D(a,b)$ the open complex disk of centre $a$ and radius $b$,
\begin{theorem}[Nevanlinna-Sokal]\label{thm:borel}
 Let $f : \setC\to\setC$ be a function. If
 \begin{itemize}
  \item $f(z)$ is analytic in a disc $D(R,R)$, with $R>0$
  \item $f$ admits a Taylor expansion at the origin, with its remainder obeying to the bound
  \[
   f(z)=\sum_{k=0}^{p-1} f_k z^k + R_p(z)\ ,\qquad |R_p(z)|\le K\ a^p\ p!\ |z|^p.
  \]
for some constants $K$ and $a$,
 \end{itemize}
then $\sum f_k z^k$ is Borel summable,
\[
 \cB f (t) =\sum_{k\ge0} \frac{1}{k!} f_k t^k 
\]
converges for $t\le a^{-1} $ and  admits an analytic continuation  on  the strip 
$ \cup_{x>0} D(x,a^{-1})$. Furthermore, the value of $f$ is given by the Borel sum :
\[
 f(z)=\frac1z\int_0^\infty \cB f(t)e^{-\frac tz}dt \qquad \forall\ z\in D(R,R)\ .
\]
\end{theorem}
Through misuse of language, a function $f$ verifying the  Nevanlinna-Sokal theorem is called \emph{Borel summable}.

\subsubsection{Uniform Borel summability}

When dealing with tensor models, one usually wants to vary the size $N$ of the tensor, and eventually set it to infinity. 
The above Borel summability theorem is therefore only good if summability is proven in a region of size independent of $N$, 
so it is valid for a non-zero radius even at large $N$. We therefore need to introduce the notion of uniform Borel summability, 
and an uniform version of the Nevanlinna-Sokal theorem.

\begin{theorem}\label{thm:unifborel}
 A function $f(\lambda, N)$ is Borel summable in $\lambda$ uniformly in $N$ if $f$ is analytic in $\lambda$ in a 
disk $D(R,R)$ with $R>0$ independent on $N$ and admits a Taylor expansion
\begin{align*}
 f(\lambda,N)=\sum_{k<r} A_k(N)\lambda^k \ +\ R_{r}(\lambda,N) \; , \;\;  |R_{r}(\lambda,N)|\leq r!  \; a^r  |\lambda|^r K(N) .
\end{align*}
for some $a$ independent of $N$. Then
\begin{itemize}
 \item $\sum f_k z^k$ is Borel summable.
$
 \cB f (t) =\sum_{k\ge0} \frac{1}{k!} f_k t^k 
$
converges for $t\le a^{-1} $ and admits an analytic continuation  on  the strip 
$ \cup_{x>0} D(x,a^{-1})$.
 \item $ \forall\ z\in D(R,R)\ ,\ \ $  $\ f(z)=\frac1z\int_0^\infty \cB f(t)e^{-\frac tz}dt $.
\end{itemize}

\end{theorem}

\section{Constructive expansions}

The forest and jungle expansions are two map expansions for statistical models and field theories that, applied to the intermediate field, allow for an 
absolutely convergent series expansion of the free energy and cumulants of tensor models. Both rely on the same interpolation formula,
Brydges-Kennedy-Abdesselam-Rivasseau forest formula.

\subsection{Forest formula}

The Brydges-Kennedy-Abdesselam-Rivasseau forest formula \cite{BK1987,AR1994} is a Taylor formula for functions of $n(n-1)/2$ variables which properties 
are well suited for the non perturbative study of tensor models.
Let $\cK_n$ be the complete graph with $n$ vertices. The set of edges $E(\cK_n)$ has $n(n-1)/2$ components.
Let $f : [0,1]^{n(n-1)/2}\rightarrow \setR$ be a smooth function, depending on edge variable $f\left((x_\e)_{\e\in E(\cK_n)}\right)$. 
The forests of $\cK_n$ are sub-graphs of $\cK_n$ with $n$ vertices which connected components are simply connected.
\begin{theorem}[Forest formula]\label{thm:bkar}
 \[
  f(1,\dots 1) = \sum_{\cF_n} \int_0^1 \dots\int_0^1 \left[
 \left( \prod_{\e\in E(\cF)}\frac{\partial}{\partial x_\e}\right) f
  \right]\left((w^\cF_{ij})_{i,j\in V(\cK_n)}\right)
  \prod_{\e\in E(\cF)}du_\e  \ ,
 \]
 where
 \begin{itemize}
  \item the sum runs over forests $\cF_n$ of $\cK_n$ including the empty one (with no edges). 
  To each edge $\e$ of $\cF$ is associated a variable $u_\e$ integrated from 0 to 1.
  \item The derivative is evaluated at the point $(w^\cF_{ij})_{i,j\in V(\cK_n)}$ defined as follow : 
  \begin{itemize}
  \item if i and j belong to the same connected component of $\cF$, then $\cP^\cF_{ij}$ denotes the unique path in $\cF$ joining the vertices i and j and
  \begin{equation}\label{eq:forestwdef}
 w^\cF_{ij} = \begin{cases}
           1 \qquad & i=j \\
           \min_{ \e\in \cP_{ij}  }\{u_\e\} \quad & i\neq j
          \end{cases} \; .
\end{equation}
  \item if i and j do not belong to the same connected component of $\cF$, $w^\cF_{ij} = 0$.
  \end{itemize}
 \end{itemize}

\end{theorem}
A useful result of \cite{AR1994} is that, for any values of $u_{ij}$, the matrix $w_{ij}$ is positive. 

\subsection{Forest expansion}\label{subsec:Forestexp}
%
%
The moment generating function of a model can be computed by expanding perturbatively at first, 
and then apply a replica trick and the forest formula instead of directly 
computing Gaussian integrals with the Feynman graphs. This new graph expansion, first introduced in \cite{Rivasseau0706}, sums over forests instead of general graphs. 
As forests proliferate at a more manageable rate with the number of vertices, this has been proven very useful to write convergent series expansions, 
and, together with the Hubbard–Stratonovich intermediate field transformation of section \ref{sec:IFformula}, forms the {\it loop vertex expansion}.

\begin{theorem}[Forest expansion]\label{thm:LVE1}
Let $\phi$ be a real vector field and $Z(J)=\langle e^{V(\phi,J)}\rangle_{\mu_\bC}$ the generating function of the moments, with $\bC^\top=\bC$ and 
$V(\phi,J)$ regroup the interaction and source terms. Then
\begin{align}\label{eq:lve1}
  Z(J)&=\sum_{n=0}^\infty \frac{1}{n!}
  \sum_{\cF_n} \int_0^1 \dots\int_0^1 \crcr
  &\int \left(\prod_{(ij)\in E(\cF_n)}\frac{\partial}{\partial\phi^i}\cdot\bC\frac{\partial}{\partial\phi^j}\right)
\prod_{i=1}^n V(\phi^i,J)\  d\mu_{\cF_n}(\{\phi^i \}) \
\prod_{\e\in E(\cF_n)}du_\e\ ,
\end{align}
where 
\begin{itemize}
  \item $\cF_n$ runs over forests with $n$ vertices. 
  To each edge $\e$ of $\cF_n$ is associated a variable $u_\e$ integrated from 0 to 1.
  \item To each vertex $i$ of the forest is associated an independent replica field $\phi^i$.
  \item The Gaussian measure over the replica $\phi$ fields is defined as
  \[
   \langle \phi^i_n \phi^j_m 
   \rangle_{\mu_{\cF_n}} = w^{\cF_n}_{ij} \bC_{nm}
  \]
  \begin{itemize}
  \item if i and j belong to the same connected component of $\cF_n$, then $\cP_{ij}$ denotes the unique path in $\cF_n$ joining the vertices i and j and
  \begin{equation}\label{eq:expwdef}
 w^{\cF_n}_{ij} = \begin{cases}
           1 \qquad & i=j \\
           \min_{ \e\in \cP_{ij}  }\{u_\e\} \quad & i\neq j
          \end{cases} \; .
\end{equation}
  \item if i and j do not belong to the same connected component of $\cF_n$, $w^{\cF_n}_{ij} = 0$.
  \end{itemize}
 \end{itemize}
\end{theorem}

\prf Using Corollary \ref{corollary:diffop}, the moment generating function writes: 
\begin{align}
 Z(J)=\left[ 
 e^{\frac12\frac{\partial}{\partial\phi}\cdot\bC\frac{\partial}{\partial\phi}}
e^{V(\phi,J) }  \right]_{\phi=0} \ .
\end{align}
A Taylor expansion gives
\[
  Z(J)=\sum_{n=0}^\infty \frac{1}{n!}\left[ 
 e^{\frac12\frac{\partial}{\partial\phi}\cdot\bC\frac{\partial}{\partial\phi}}
\left[V(\phi,J)\right]^n  \right]_{\phi=0} \ .
\]
We now use a replica trick. We rewrite $V(\phi,J)^n$ as a product of $n$ interaction terms with $n$ independent field variables:
\begin{equation}\label{eq:Forestreplica}
  Z(J)=\sum_{n=0}^\infty \frac{1}{n!}\left[ 
 e^{\frac12\sum_{i,j=1}^n\ \frac{\partial}{\partial\phi^i}\cdot\bC\frac{\partial}{\partial\phi^j}}
\prod_{i=1}^n V(\phi^i,J)  \right]_{\phi=0} \ ,
\end{equation}
the moment generating function now writes in terms of some new Gaussian measures over $n$ fields. The covariance of the new measure does not distinguish 
 the multiple replica field,
 \[
  \langle \phi^i_n\phi^j_m\rangle_{\tilde\mu_\bC} =  \langle \phi^i_n\phi^i_m\rangle_{\tilde\mu_\bC} = \bC_{nm} \ \forall i,j\in \{1\dots n\} \ ,
 \]
 therefore the expectation remains unchanged.
For each order $n$, we introduce a set of $n(n-1)/2$ interpolation parameters $x_n^{ii}$ and $x_n^{ij}=x_n^{ji}$ in the measure,
\[
  Z(J)=\sum_{n=0}^\infty \frac{1}{n!}\left[ 
 e^{\frac12\sum_{i,j=1}^n\ x_n^{ij} \frac{\partial}{\partial\phi^i}\cdot\bC\frac{\partial}{\partial\phi^j}}
\prod_{i=1}^n V(\phi^i,J)  \right]_{\substack{\phi^i=0\\ x_n^{ij}=1}} \ ,
\]
where the interpolation parameters are set to 1, such that the measure does not distinguish between replica fields. 
Each order of the moment generating function is now written as a function of $(n(n-1)/2$ edge variables,
associated with the edges of the complete graph $\cK_n$, evaluated at the point $x^{ij}=1,\ \forall (i,j)\in E(\cK_n)$. 
Furthermore, to each vertex $i$ of $\cK_n$ is associated an interaction term $V$ and a replica field $\phi^i$.

By applying the forest formula (Theorem \ref{thm:bkar}), we obtain :
\begin{align}
  Z(J)=\sum_{n=0}^\infty \frac{1}{n!}
  \sum_{\cF_n} \int_0^1 \dots\int_0^1 \Bigg[
 \bigg( \prod_{(ij)\in E(\cF_n)}\frac{\partial}{\partial x_n^{ij}}\bigg) 
  \crcr
 e^{\frac12\sum_{i,j=1}^n\ x_n^{ij} \frac{\partial}{\partial\phi^i}\cdot\bC\frac{\partial}{\partial\phi^j}}
\prod_{i=1}^n V(\phi^i,J)  \Bigg]_{\substack{\phi^i=0\\ x_n^{ij}=w^{\cF_n}_{ij}}} \
\prod_{\e\in E(\cF)}du_\e\ .
\end{align}
Taking into account
\[
 \frac{\partial}{\partial x_n^{ij}}\left[e^{\frac12\sum_{k,l=1}^n\ x_n^{kl} \frac{\partial}{\partial\phi^k}\cdot\bC\frac{\partial}{\partial\phi^l}}\right]
 =e^{\sum_{k,l=1}^n\ x_n^{kl} \frac{\partial}{\partial\phi^k}\cdot\bC\frac{\partial}{\partial\phi^l}}
 \left(\frac{\partial}{\partial\phi^i}\cdot\bC\frac{\partial}{\partial\phi^j}\right)
\]
leads to \eqref{eq:lve1}.
\qed

\subsubsection*{Trees and connected moments}
 
The connected components of forests being trees, using the result of Theorem \ref{thm:log}, 
the logarithm of the moment generating function can be computed as a sum over trees.

\begin{corollary}\label{cor:LVE2}
Let $\phi$ be a vector field. Using the same notations as in theorem \ref{eq:lve1}, the generating function of the cumulants writes
\begin{align}\label{eq:lve2}
  \log Z(J)&=\sum_{n=0}^\infty \frac{1}{n!}
  \sum_{\cT_n} \int_0^1 \dots\int_0^1 \crcr
  &\int \left(\prod_{(ij)\in E(\cT_n)}\frac{\partial}{\partial\phi^i}\cdot\bC\frac{\partial}{\partial\phi^j}\right)
\prod_{i=1}^n V(\phi^i,J)\  d\mu_{\cT_n}(\{\phi^i \}) \
\prod_{\e\in E(\cT_n)}du_\e\ ,
\end{align}
where $\cT_n$ runs over trees with $n$ vertices. 
\end{corollary}

 \subsubsection*{Intermediate field and loop vertex expansion}
 
 The loop vertex expansion is a non-perturbative expansion of quartic models which
 is obtained by applying the forest expansion to the intermediate field $\csig$ of a tensor model.
 The generating function of the moments \eqref{eq:ZLVE} writes:
 \[
  Z(\bJ,\bar \bJ)=\langle 
e^{-
\ \tr_{\cD} \big[ \log\left(\id^{\cD} + A(\csig)\right)   \big] 
+  \sum_{nm } \bar \bJ_{n} 
\left[\frac{1}{ \id^{\cD} + A(\csig) }\right]_{nm}\bJ_{m} }  \rangle_{\mu_\id} \ .
 \]
When applying the forest expansion \ref{thm:LVE1}, the differential operator associated with an edge writes,
\[
 \frac{\partial}{\partial\csig^i}\cdot\id\frac{\partial}{\partial\csig^j}\ =\ 
 \sum_{\cC\in\cQ}\ \sum_{a^\cC b^\cC}\frac{\partial}{\partial\csig_{a^\cC b^\cC}^{i\ \cC}}\frac{\partial}{\partial\csig_{b^\cC a^\cC}^{j\ \cC}}\ .
\]
In order to compute the forest expansion, the action of the derivative on the interaction and source terms must be computed.

 \subsection{Jungle expansion}\label{sec:jungleexp}
 
 Performing a non-perturbative graph expansion on a non invariant field theory may require a more involved version of the 
 forest expansion in order to avoid difficulty with processing of divergences.
 A two-level Fermionic jungle expansion has been developed in \cite{GR1312.7226}, specially designed to handle these renormalisable models.
 
 Consider a field theory $\phi$, with partition function
 \begin{equation}\label{eq:JungleZ}
  Z=\int e^{V(\phi)}\ d{\mu}(\phi)
 \end{equation}
 where $\mu$ is the standard Gaussian measure $d\mu = e^{\frac12\phi^2}d\phi$ and for which the interaction term can 
 be decomposed over some abstract scale parameter $j$ as
 \[
  V=\sum_{j=0}^{j_{\max}} V_j(\phi) \ .
 \]
The maximum scale $j_{\max}$ can be understood as a cut-off. We choose to decompose the interaction term instead of the covariance as the expansion 
is to be used on an intermediate field model. This is why we use the Gaussian measure with standard invariant covariance,
but an interaction term  that is somehow scale-related.

\begin{theorem}[Jungle expansion]\label{thm:jungle}
 \begin{align}\label{eq:jungle}
Z  = & \sum_{n=0}^\infty \frac{1}{n!}  \sum_{\cJ} \;\sum_{j_1=0}^{j_{\max }  } 
  \dots \sum_{j_n=0}^{j_{\max} }\crcr
  & \int_{w_\cJ}   \;  \int_{(\phi^a)} \int_{(\chi^{ \cB }_{j})}
\quad   \partial_\cJ   \Big[ \prod_{\cB} \prod_{a\in \cB}   \Bigl(    W_{j_a}   ( \phi^a )  
\chi^{ \cB }_{j_a}   \bar \chi^{\cB}_{j_a}  \Bigr)  \Big] \;d\mu_{ \cJ} (\phi,\chi) \; dw_\cJ
\ ,
\end{align}
where
\begin{itemize}

\item the sum over $\cJ$ runs over all two-level jungles with $n$ vertices labelled as $a \in \cV =\{1\dots n\}$,
hence over all ordered pairs $\cJ = (\cF_B, \cF_F)$ of two (each possibly empty) 
disjoint forests on $V$,
$\bar \cJ = \cF_B \cup \cF_F $ is still a forest on $V$. The forests $\cF_B$ and $\cF_F$ are the Bosonic and Fermionic components of $\cJ$.
The connected components $\cB$ of $\cF_B$ are called Bosonic blocks.

\item To each vertex $a\in \cV$ is associated a replica Bosonic field $\phi^a$. 
To each  Bosonic block $\cB \subset \cF_B$ and each scale index $j\in\cS=\{1\dots j_{\max}\}$ is associated
a Grassmannian field $\chi^\cB_j$ .
 
\item  $\int dw_\cJ$ means integration from 0 to 1 over parameters $w_\e$, one for each edge $\e \in \bar\cJ$, namely
$\int dw_\cJ  = \prod_{\e\in \bar \cJ}  \int_0^1 dw_\e  $.
There is no integration for the empty forest since by convention an empty product is 1. A generic integration point $w_\cJ$
is therefore made of $\vert \bar \cJ \vert$ parameters $w_\e \in [0,1]$, one for each $\e \in \bar \cJ$.

\item 
\begin{equation} \partial_\cJ  = \prod_{\genfrac{}{}{0pt}{}{\e_B \in \cF_B}{\e_B=(a,b)}} \Bigl(
\frac{\partial}{\partial \phi^a}\frac{\partial}{\partial \phi^b} \Bigr)
\prod_{\genfrac{}{}{0pt}{}{\e_F \in \cF_F}{\e_F=(d,e) } } \delta_{j_{d } j_{e } } \Big(
   \frac{\partial}{\partial \bar \chi^{\cB(d)}_{j_{d}  } }\frac{\partial}{\partial \chi^{\cB(e)}_{j_{e}  } }+ 
    \frac{\partial}{\partial \bar \chi^{ \cB( e) }_{j_{e} } } \frac{\partial}{\partial \chi^{\cB(d) }_{j_{d}  } }
   \Big) \; ,
\end{equation}
where $ \cB(d)$ denotes the Bosonic block to which the vertex $d$ belongs. 

\item Denoting $\id_\cS$ the identity matrix with indices in $\cS$,
the measure $d\mu_{\cJ}$ in Gaussian with covariance $ X (w_{\e_B})$ on Bosonic variables and $ Y (w_{\e_F}) \otimes \id_\cS  $  
on Fermionic variables, hence
\begin{equation}
\int F d\mu_{\cJ} = \biggl[e^{\frac{1}{2} \sum_{a,b=1}^n X_{ab}  \frac{\partial}{\partial \phi^a}\frac{\partial}{\partial \phi^b} 
   +  \sum_{\cB,\cB'} Y_{\cB\cB'}\sum_{\genfrac{}{}{0pt}{}{a\in \cB}{b\in \cB'}} \delta_{j_aj_b}
   \frac{\partial}{\partial \bar \chi_{j_a}^{\cB} } \frac{\partial}{\partial \chi_{j_b}^{\cB'} } }   F \biggr]_{\sigma = \bar\chi =\chi =0}\; .
\end{equation}

\item  $X_{ab} $  is the infimum of the $w_{\e_B}$ parameters for all the Bosonic edges $\e_B$
in the unique path $P^{\cF_B}_{a \to b}$ from $a$ to $b$ in $\cF_B$. The infimum is set to zero if such a path does not exists and 
to $1$ if $a=b$. 

\item  $Y_{\cB\cB'}$  is the infimum of the $w_{\e_F}$ parameters for all the Fermionic
edges $\e_F$ in any of the paths $P^{\cF_B \cup \cF_F}_{a\to b}$ from some vertex $a\in \cB$ to some vertex $b\in \cB'$. 
The infimum is set to $0$ if there are no such paths, and to $1$ if such paths exist but do not contain any Fermionic edges.

\end{itemize}
\end{theorem}

\subsubsection*{Pre-cooking with Fermions}
The partition function can be rewritten
\[
  Z=\int \prod_{j=0}^{j_{\max}} e^{V(\phi)}\ d{\mu}(\phi) \ .
\]
We define the normalized Grassmann Gaussian measure
\[
 d\mu(\bar\chi,\chi) =e^{-\bar\chi\chi}d\bar\chi d\chi \ .
\]
Then, for any number $a$,
\[
 a = \int e^{-\bar\chi a \chi}d\bar\chi d\chi =\int e^{-\bar\chi (a-1) \chi}d\mu(\bar\chi ,\chi) \ .
\]
Hence, denoting $W_j = e^{V_j} - 1$, we can rewrite the partition function \eqref{eq:JungleZ} .
 \begin{equation}\label{eq:jungleZ2}
  Z=\int 
  e^{-\sum_{j=0}^{j_{\max}} \bar\chi_j W_j(\phi)\chi_j
  }\
\left[ \prod_{j=0}^{j_{\max}} d\mu(\bar\chi_j,\chi_j)\right]\
  d{\mu}(\phi) \ .
 \end{equation}
 The previous expression \eqref{eq:jungleZ2} introduces independant Grassmannian fields $\chi$ for each scale $j$.
Defining the set of scales $ \cS = \{ 0\dots j_{\max} \} $, we rewrite the partition function as
 \begin{align}\label{eq:jungleZ3}
  &Z=\int 
  e^{-W
  }\ d\mu_\cS \ ,\crcr
 & d\mu_\cS =  \left[ \prod_{j=0}^{j_{\max}} d\mu(\bar\chi_j,\chi_j)\right]\
  d{\mu}(\phi) \ ,    
  \quad
 W= \sum_{j=0}^{j_{\max}} \bar\chi_j W_j(\phi)\chi_j
 \ .
 \end{align}
 We can now start the jungle expansion.
 
 \subsubsection*{Vertex replicas}
 We first expand the exponential in the partition function \eqref{eq:jungleZ3}
 \begin{align}
    &Z=\sum_{n=0}^{\infty}\frac1{n!}\int 
  (-W)^n\ d\mu_\cS \ .
 \end{align}
 Then introduce vertex-replicas for the $\phi$ field, using the same trick as for \eqref{eq:Forestreplica}.
 Defining the matrix ${\bf 1}_\cV$  with vertex indices $a,b \in \cV = \{ 1\dots n \} $, 
 with all coefficients equals to one, and the Gaussian measure $\nu_{\bf 1_\cV}$ as,
 \[
  \forall\, (a\, ,\, b)\in\cV^2 \; ,\qquad 
\braket{\phi_a\phi_b}_{\nu_{\bf 1_\cV}} = 1 =  ({\bf 1}_{\cV})_{ ab} \ ,
 \]
 we rewrite the partition function as,
  \begin{align}
    &Z=\sum_{n=0}^{\infty}\frac1{n!}\int 
  \prod_{a=1}^n(-W_a)\ d\mu_{\cS,\cV} \ , \crcr
   & d\mu_{\cS,\cV} =  \left[ \prod_{j=0}^{j_{\max}} d\mu(\bar\chi_j,\chi_j)\right]\
  d{\mu}_{{\bf 1}_\cV}(\{\phi_a\}) \ ,    
  \quad
 W_a= \sum_{j=0}^{j_{\max}} \bar\chi_j W_j(\phi_a)\chi_j
 \ .
 \end{align}
 Note that for the field $\phi$, we only introduced \emph{vertex} replicas, with a completely degenerated covariance $\bbone_\cV$. 
 For the Grassmann field $\chi$ however, we introduces \emph{scale} replicas, and not vertex replicas, and with a covariance,
 that we can encode with $\id_\cS$ as
 \[
   \left[ \prod_{j=0}^{j_{\max}} d\mu(\bar\chi_j,\chi_j)\right] = d\nu_{\id_\cS}(\{\bar\chi_j,\chi_j\})\ ,\qquad \braket{\bar\chi_i\chi_j}_{\nu_{\id_\cS}}=\delta_{ij}\ .
 \]

 From this point, the jungle expansion is obtained by applying two successive forest formulas. 
 \subsubsection*{Bosonic forest}
 As in section \ref{subsec:Forestexp}, the first forest formula is applied on the degenerate measure $\nu_{\bbone_V}$. We introduce $n$ interpolation parameters
 $n(n-1)/2$ interpolation parameters $x_n^{ab}=x_n^{ba}$, all set to one, 
 that can be seen as variables associated to the edges of the complete graph $\cK_n$. 
 \begin{equation}
Z = \sum_{n=0}^\infty \frac{1}{n!} 
\Bigl[ e^{\frac{1}{2} \sum_{a,b=1}^n x_{ab} 
  \frac{\partial}{\partial \phi^a}\frac{\partial}{\partial \phi^b} 
   +  \sum_{j=0}^{j_{\max}} \frac{\partial}{\partial \bar \chi_j  } \frac{\partial}{\partial \chi_j } } \; 
   \prod_{a=1}^n \Bigl( -  W^a \Bigr) 
   \Bigr]_{ \genfrac{}{}{0pt}{}{ \phi, \chi,  \bar\chi  =0}{x_{ab}=1 }} \;.
\end{equation}
 then, we apply the forest formula (Theorem \ref{thm:bkar}). Denoting $\cF_B$ a forest with $n$ vertices $i$ labelled from $1$ to $n$, the edges of $\cF_B$ are called
 $(a,b)\in \cF_B$ with $a<b$.
 \begin{align}
Z&    = \sum_{n=0}^\infty \frac{1}{n!} \sum_{\cF_{B}} \int_{w_{ab}=0}^1 
\; \;  \Bigg[  e^{\frac{1}{2} \sum_{a,b=1}^n X_{ab}  
    \frac{\partial}{\partial \phi^a}\frac{\partial}{\partial \phi^b} 
   +  \sum_{j=0}^{j_{\max}} \frac{\partial}{\partial \bar \chi_j  } \frac{\partial}{\partial \chi_j  } }
\crcr
&  \times 
   \prod_{(a,b) \in \cF_{B}} \Bigl(  
   \frac{\partial}{\partial \phi^{a}}\frac{\partial}{\partial \phi^{b}}  \Bigr) \;
 \prod_{a=1}^n \Bigl( - W_a   \Bigr) \Bigg]_{\phi ,\chi , \bar\chi =0 } \;\Bigl( \prod_{(a,b)\in \cF_B } dw_{ab} \Bigr) \ ,
\end{align}
where $X_{ab}$ is the infimum over the parameters $w_{ab}$ in the unique path
in the forest $\cF_B$ connecting $a$ to $b$. This infimum is set to 
$1$ if $a=b$ and to zero if $a$ and $b$ are not connected by the forest.
The connected components $\cB\subset\cF_B$ of the forest $\cF_B$ are called \emph{Bosonic blocks}. We denote $\cV\slash\cF_B$ the set of vertices 
$\cV$ where vertices belonging to a same bosonic block (connected component) $\cB$ are identified together. By construction $\cV\slash\cF_B$ is identified to
the set of Bosonic blocks of $\cF_B$.
 
 \subsubsection*{Fermionic forest}
 
 Using the same replica trick \eqref{eq:Forestreplica}, we introduce replica Grassmann fields $\chi_j^\cB$ for the 
 Bosonic blocks $\cB$ and interpolation parameters $y^{\cB\cB'}=y^{\cB'\cB}$, all set to one. Applying the forest formula on the $y$ parameters leads to a Forest over
 $\cV\slash\cF_B$ with a new set of Fermionic edges $\e_F$,
 \begin{align}
Z  &=  \sum_{n=0}^\infty \frac{1}{n!}  \sum_{\cF_{B}}\sum_{\cF_{F}} \;
 \;  \int  \biggl[e^{\frac{1}{2} \sum_{a,b} X_{ab}  \frac{\partial}{\partial \phi^a}\frac{\partial}{\partial \phi^b} 
   +  \sum_{j=0}^{j_{\max}}\sum_{\cB,\cB'} Y_{\cB\cB'}
   \frac{\partial}{\partial \bar \chi_{j}^{\cB} } \frac{\partial}{\partial \chi_{j}^{\cB'} } }  \crcr 
 &  \prod_{(\cB,\cB') \in \cF_F}
 \Big(
   \frac{\partial}{\partial \bar \chi^{\cB}_{j  } }\frac{\partial}{\partial \chi^{\cB'}_{j  } }+ 
    \frac{\partial}{\partial \bar \chi^{ \cB}_{j } } \frac{\partial}{\partial \chi^{\cB' }_{j  } }
   \Big)
\prod_{(a,b) \in \cF_{B}} \Bigl(  
   \frac{\partial}{\partial \phi^{a}}\frac{\partial}{\partial \phi^{b}}  \Bigr) \;    \prod_{a\in \cV}    \Big( -  W_{a}    \Big)
   \biggr]_{\phi,\chi =0} 
 \; dw_\cJ \ .
\end{align}
 The derivatives $\frac{\partial}{\partial \bar \chi^{ \cB}_{j } }$ associated with a Fermionic edge
 can be applied on the $W_a$ term of each vertex of the Bosonic block $\cB$, 
 \[
  \frac{\partial}{\partial \bar \chi^{ \cB}_{j } } \prod_{a\in \cB}    \Big( -  W_{a}    \Big) = 
  \sum_{a\in\cB} \delta_{j_a,j}\left[\frac{\partial(-W_{a})}{\partial \bar \chi^{ \cB}_{j_a } } \prod_{\substack{b\in\cB \\b\neq a}}    \Big( -  W_{b}    \Big) \right] \ ,
 \]
The sum over Fermionic forests $\cF_F$ over Bosonic blocks can therefore be refined as a sum over forests over the vertices $a\in\cV$, with Fermionic edges  connecting
vertices together $\e_F=(a,b),\ a,b\in \cV$. Finally, for scale indices that do not appear in a bosonic block, $j\not\in\{j_a, a\in\cB\}$, 
the integration over $\chi^{ \cB}_{j } $ is trivial, 
 leading to equation \eqref{eq:jungle}.
 
 \qed

\chapter{The constructive quartic model}\label{chap:borelQM}

This chapter will address the constructive study of standard invariant quartic tensor models using the loop vertex expansion. 

For rank $D$ tensors, the standard quartic model \eqref{eq:standardmodel} writes as 
\begin{align}
d\nu &=  e^{-\frac{\lambda}{2N^{D-1}}\sum_{\mathcal{C} \in \mathcal{Q}}  V_{\mathcal{C}}(\bar{\bf T},{ \bf T} ) }\ d\mu_\id(\bar\bT,\bT) \ .
\end{align}
In the intermediate field formalism, the partition function writes in terms of $|\cQ|$ matrix fields $\csig^\cC$ \eqref{eq:ZLVE},
\begin{align}\label{eq:intfieldZ}
Z(\bJ,\bar \bJ)=\int 
e^{
\ \tr_{\cD} \big[ \log\ R(\csig)   \big] 
+  \sum_{nm } \bar \bJ_{n} 
 R(\csig)_{nm}\bJ_{m} }  \; d\mu_\id(\csig) \ ,
\end{align}
where the resolvent operator $R$ is defined as
\begin{equation}
 R(\csig) = \frac{1}{ \id^{\cD} + i\ \sqrt{\frac{\lambda}{ N^{D-1}}}
 \sum_{\mathcal{C}\in \cal{Q}}\\  \id^{\cD \setminus \cC }
\otimes\csig^{\mathcal{C}}  
 } \ .
\end{equation}

\section{Loop vertex expansion}

The first step toward the constructive study of the model is to establish the loop vertex expansion \cite{Rivasseau0706} of the cumulants.

Using the forest expansion (Theorem \ref{thm:LVE1}) on the intermediate field model \eqref{eq:intfieldZ},
the loop vertex expansion allows to expand the cumulants in term of multicoloured plane trees.

\begin{definition}\label{def:planetree}
 A multicoloured plane tree is a simply connected multicoloured map.
\end{definition}

The key point is that the number of possible plane trees grows in a manageable manner with the number of vertices, 
leading to better summability properties if the amplitudes of these trees can be conveniently bounded.  

\begin{theorem}\label{th:LVE}
With the notations introduced in section \ref{subsec:Forestexp}, the cumulants of the measure $\mu$ are given by: 
\begin{align}\label{cumulants0}
&\kappa(\bT_{m_1}\bar{\bT}_{\bar{m}_1}...\bT_{m_k}\bar{\bT}_{\bar{m}_k})=
\sum_{\pi \bar\pi} \sum_{v\geq k}\frac1{v!}\ \frac{(-\lambda)^{v-1}}{N^{(D-1)(v-1)}}
\sum_{\mathcal{T}_{v,k}}
\crcr
&\times\int\dots\int_{u_{ij}=0}^1\int_{\csig} \sum_{n,m} 
   \left(\prod_{i=1}^v\prod_{p=1}^{\mathrm{cor}(i)} R(\csigma^i)_{n_{i,p}{\bar n}_{i,p}}\right)
\left(\prod_{l\in T_v} \ \delta_{\cD}^{l,{\mathcal{C}}(l)}\right) \crcr
&  \times \left(\prod_{d=1}^k\ \delta_{m_{\pi(d)} n_{i_d,q+1}}
  \delta_{\bar m_{\bar \pi(d)} \bar n_{i_d,q}}\right) \ d\mu_{T_v,u}(\csigma) \prod_{\e_{ij}\in T_v}du_{ij} \ ,
\end{align}
where  
\begin{itemize}
 \item $\pi$ and $\bar\pi$ are permutations over $k$ elements.
 \item $\cT_{v,k}$ are multicoloured plane trees 
with and $k$ ciliated vertices and $v-k$ non-ciliated vertices, $T_v$ is the combinatorial tree associated with $\cT_{v,k}$.
 \item ${\rm cor}(i)$ is the number of corners of a vertex $i$, each corner $p\in i$ bears a resolvent $R$ with inbound indices $n_{i,p}$ and outbound indices $\bar n_{i,p}$.
\end{itemize}
\end{theorem}

The plane trees from the loop vertex expansion are very similar to the multi-coloured maps of Chapter \ref{chap:IF}. 
They carry the same multi-stranded structure and index contraction patterns, and differs only by the addition of a \emph{resolvent} operator $R$ 
on each corners of the map, instead of straight index identifications.
The resolvents arise from derivative of the intermediate field interaction term.
While this complicates the amplitude computation, the resolvent operator, being of the form $(\id+ i\sqrt\lambda H)^{-1}$ with $H$ being Hermitian, 
stays bounded in a domain that includes positive $\lambda$. This property will allow for a bound of the tree amplitudes compatible with 
Borel summability.

If the forest expansion was used directly on the tensor model instead of the intermediate field, derivatives of the trace invariants being polynomials 
in $\bT$ and $\bar\bT$, the amplitude of the trees would not be bounded and the forest expansion would not succeed.

The expression \eqref{cumulants0} does not show the tensor invariant structure of the cumulants \eqref{eq:structurecumulants}, 
that was explicit in the perturbative formalism of section \ref{sec:tensmod}. As the resolvent operators are not unitary invariants, 
the indices of the source terms cannot be identified together and the invariant structure is hidden. 
The structure of the cumulant will be recovered later with a refined expansion, namely the mixed expansion, consisting in Taylor expanding over the LVE trees.
The structure can also be recovered using a trick \cite{Gurau1304} involving unitary operators and Weingarten functions, 
which allows to decouple the structure of the cumulant from the one of the LVE trees at the price of some extra twists and complications. 
This has been studied in detail in \cite{DGR1403}.

The remainder of this section will be devoted to the proof of Theorem \ref{th:LVE}.

\subsubsection*{Forest formula}\label{subsecFF}

To simplify notations we sometimes drop the superscript $\cC$ on the (multi) indices
of $\sigma^{\cC}$.

The idea of the loop vertex expansion is to apply the forest expansion (Theorem \ref{thm:LVE1}) to the intermediate field model. 
The process is to be adapted from section \ref{subsec:Forestexp} using a collection of Hermitian matrices 
$\csig=(\csig^\cC)_{\cC\in\cQ}$ instead of a real vector field. 
According to \eqref{eq:lve2}, the logarithm of $Z(\bJ,\bar \bJ)$ is:
\begin{align}
\lnz (\bJ,\bar \bJ)&=\sum_{v\geq 1}\frac1{v!} \sum_{T_v}\int_0^1 \left(\prod_{\e_{ij}\in T_v}du_{ij}\right)\int d\mu_{T_v,u}(\csigma)\crcr
&\times\prod_{\e_{ij}\in T_v}\left(
\sum_{\cC,ab }
\frac{\dr}{\dr\csigma_{ab}^{i\ {\mathcal{C}}}}\frac{\dr}{\dr\csigma_{ab}^{j\ {\mathcal{C}}}} 
\right)\crcr
&\times\prod_{i=1}^v \left(\tr_\cD \left[\log R(\csigma^i)\right]\ +\ \sum_{n\bar n}\bar \bJ_n R(\csigma^i)_{n\bar n}\ \bJ_{\bar n}\right),
\end{align}
where $T_v$ are combinatorial trees with $v$ vertices and the interpolated Gaussian measure $d\mu_{T_v, u}$ is degenerated over $\cC$:
\begin{align*}
 \int F(\sigma) \; d\mu_{T_v, u} = \left[e^{\frac12\sum_{i,j}w_{ij}\sum_{ \cC,ab }\left(
\frac{\dr}{\dr\csigma_{ab}^{i\ {\mathcal{C}} }}\frac{\dr}{\dr\csigma_{ab}^{j\ {\mathcal{C}}}} 
\right)}F(\csigma)\right]_{\csigma=0} \; ,
\end{align*}
and $w_{ij}$ is defined in equation \eqref{eq:expwdef}. Expanding the product over $i$ we get:
\begin{align}\label{eq:intermediar}
&\lnz (\bJ,\bar \bJ)=\sum_{v\geq 1}\frac1{v!} \sum_{T_v}\int_0^1 \left(\prod_{\e_{ij}\in T_v}du_{ij}\right)\int d\mu_{T_v,u}(\csigma)\crcr
& \;\; \times\left[\prod_{\e_{ij}\in T_v}\left(
\sum_{ \cC,ab }
\frac{\dr}{\dr\csigma_{ab}^{i\ {\mathcal{C}}}}\frac{\dr}{\dr\csigma_{ab}^{j\ {\mathcal{C}}}} 
\right)
\right]\sum_{k=1}^v\ \sum_{i_1<...< i_k} \crcr
& \;\; \times \prod_{d=1}^k\sum_{\bar nn}\,\bar \bJ_n\, R(\csigma^i)_{n\bar n}\, \bJ_{\bar n}
\prod_{i\neq i_1..i_k}\tr_\cD\ \mathrm{log} \ R(\csigma^i) \;.
\end{align}
The logarithm of $Z(J,\bar J)$ is then a sum over trees $T_{v,k,\{i_d\}}$ with $k$ ciliated vertices 
and $v-k$ regular vertices.
As in Chapter \ref{chap:IF}, the cilia are defined as pairs $\bJ\bar\bJ$ of source terms. 
The sum over ${\mathcal{C}}$ gives a sum over trees with coloured edges, each colouring corresponding 
to a set ${\mathcal{C}}\in\mathcal{Q}$.  

Before taking into account the action of the derivatives, to each ciliated vertex $i_d$ of the tree $T_{v,k,\{i_d\}}$ is associated a resolvent 
operator $R(\csigma^{i_d})=\left(\id^{\cD} + A(\csigma^{i_d}) \right)^{-1}$ and a cilium  $J\bar J$. 
To each non-ciliate vertex ($i\neq i_1..i_k$) is associated a $ \tr(\mathrm{log}\ R(\csigma^i) ) $ factor.

We now have to evaluate the action of the derivatives: 
\begin{align}\label{eq:derivative}
 & \frac{\partial}{\partial \csigma^{i \ \cC}_{a^{\cC} b^{\cC} }} \left[ R(\csigma^{i}) \right]_{n{\bar n}} = \crcr
 & = 
  \frac{\partial}{\partial \csigma^{i \ \cC}_{a^{\cC} b^{\cC} }} \left[ \sum_{q=0}^{\infty} \left( -i \sqrt{\frac{\lambda}{N^{D-1}}} \right)^q
   \left(    \sum_{\mathcal{C}}\id^{ \cD \setminus \cC }\otimes\csigma^{i \ \cC }   \right)^q \right]_{n{\bar n}}  \crcr
 & = \left( -i \sqrt{\frac{\lambda}{N^{D-1}}} \right) \sum_{a^{\cD\setminus \cC} b^{\cD\setminus \cC} }\sum_{q_1,q_2=0}^{\infty} \crcr
& \; \times \left( -i \sqrt{\frac{\lambda}{N^{D-1}}} \right)^{q_1}
   \left(    \sum_{\mathcal{C}}\id^{ \cD \setminus \cC }\otimes\csigma^{i \ \cC }  \right)^{q_1}_{n a} \crcr
& \; \times \left( -i \sqrt{\frac{\lambda}{N^{D-1}}} \right)^{q_2}
   \left(    \sum_{\mathcal{C}}\id^{ \cD \setminus \cC }\otimes \csigma^{i \ \cC }  \right)^{q_2}_{b {\bar n}}  
     \delta_{a^{\cD\setminus \cC} b^{\cD \setminus \cC} } \crcr
& = \left( -i \sqrt{\frac{\lambda}{N^{D-1}}} \right) \sum_{ a^{\cD\setminus \cC} b^{\cD\setminus \cC}  } 
   \left[ R(\csigma^{i}) \right]_{n a } \delta_{a^{\cD\setminus \cC} b^{\cD \setminus \cC} }  \left[ R(\csigma^{i}) \right]_{b {\bar n}},
   \end{align}
   \begin{align}
 &  \frac{\partial}{\partial \csigma^{i \ \cC}_{a^{\cC} b^{\cC} }} \left( - \tr\log[ R(\csigma^i)  ] \right) = \crcr
 & =  \frac{\partial}{\partial \csigma^{i \ \cC}_{a^{\cC} b^{\cC} }} \left[  \sum_{q=1}^{\infty}\frac{(-1)^q}{q} 
    \left( i  \sqrt{\frac{\lambda}{N^{D-1}}} \right)^q \tr
   \left(    \sum_{\mathcal{C}}\id^{ \cD \setminus \cC }\otimes \csigma^{i \ \cC }  \right)^q
 \right] \crcr
& =  \left( - i \sqrt{\frac{\lambda}{N^{D-1}}} \right) \sum_{q=0}^{\infty} \left( -  i\sqrt{\frac{\lambda}{N^{D-1}}} \right)^q \crcr
& \;\; \times \sum_{a^{\cD\setminus \cD} b^{\cD \setminus \cC}}  \delta_{a^{\cD \setminus \cD} b^{\cD \setminus \cC} }
 \left[ \left(    \sum_{\mathcal{C}}\id^{ \cD \setminus \cC }\otimes \csigma^{i \ \cC }  \right)^q \right]_{ba} \crcr
& = \left( - i\sqrt{\frac{\lambda}{N^{D-1}}} \right) \sum_{a^{\cD\setminus \cD} b^{\cD \setminus \cC}}  \delta_{a^{\cD \setminus \cD} b^{\cD \setminus \cC} }
     \left[ R(\csigma^{i}) \right]_{b a }   \; .
\end{align}

 Through its derivative operators, each edge of the tree adds a resolvent on both vertex it connects.
On each vertex, marked or not, acts at least one derivative operator, thus we obtain at least a resolvent per vertex. 
Each vertex is then the product of these resolvents (and a pair $\bJ\bar \bJ$ if ciliated).

The action of a $p+1$-th derivative acting on a $p$-valent vertex  is the sum of the $p$ positions on which one can add a resolvent into 
the partial trace of \eqref{eq:derivative}. There is a well defined cyclic ordering of the resolvents and half-edges at a vertex. 
The sum in \eqref{eq:intermediar} becomes thus a sum over plane trees, with well defined orderings
of the half edges at every vertex, with resolvents associated to the corners. Cilia act as regular half-edges.

We have thus expressed $\lnz$ as a sum over plane trees with coloured edges and marked vertices $\mathcal{T}_{v,\{i_d\},\{{\mathcal{C}}(l)\}}$. 
The contribution of a tree is a product of resolvents and $\bJ\bar \bJ$ terms with indices contracted in a certain pattern. 
To the edge $\e_{ij}$ connecting the vertices $i$ and $j$, incident at the corners $q$ and $q+1$ of the vertex $i$ and $p$ and $p+1$
of the vertex $j$, corresponds the contraction:
\begin{align*}
 \delta_{\cD}^{\e_{ij},{\mathcal{C}}(\e_{ij})}=  
 \left( \delta_{{\bar n}^{\cD \setminus \cC }_{i,q} n^{\cD  \setminus \cC }_{i,q+1} } \right) 
    \delta_{ {\bar n}^{\cC}_{i,q} n^{\cC}_{j,p+1} }      \delta_{ {\bar n}^{\cC}_{j,p} n^{\cC}_{i,q+1} }
 \left( \delta_{{\bar n}^{\cD \setminus \cC }_{j,p} n^{\cD \setminus \cC }_{j,p+1} } \right) \; .
\end{align*}

 Collecting everything we obtain: 
\begin{align}\label{eq:lnz1}
&\lnz (\bJ,\bar \bJ)=
\sum_{v\geq 1}\frac1{v!}\sum_{k=1}^v\ 
\sum_{\mathcal{T}_{v,\{i_d\},\{{\mathcal{C}}(\e)\}}}
\int_0^1 \left(\prod_{\e_{ij}\in T_v}du_{ij}\right)\int d\mu_{T_v,u}(\csigma) \sum_{n,m} \crcr
& \qquad \times \left(\prod_{i=1}^v\prod_{p=1}^{\mathrm{cor}(i)} R(\csigma^i)_{n_{i,p}{\bar n}_{i,p}}\right)
\left(\prod_{\e\in T_v}\frac{-\lambda}{N^{D-1}} \ \delta_{\cD}^{\e,{\mathcal{C}}(\e)}\right) \crcr
& \qquad \times \left(\prod_{d=1}^k\ \bar \bJ_{n_{i_d,q+1}}\bJ_{{\bar n}_{i_d,q}}\right) \; ,
\end{align}
where cor$(i)$, the number of corners of the vertex $i$, equals its degree for non-ciliated vertices, 
and its degree plus one for ciliated vertices. In the last line $q$ denotes the position of the $\bJ\bar \bJ$ cilium on the vertex 
$i_d$, and all the indices $n$ and ${\bar n}$ are summed.

\subsubsection*{Cumulants}\label{subsecCUMU}
 
 The cumulants are computed by evaluating the derivatives of eq. \eqref{eq:lnz1} with respect to $\bJ$ and $\bar \bJ$.
 As
 \begin{align}
 \frac{\partial^{(2k)}  }{\partial \bar{\bJ}_{m_1}
\partial \bJ_{\bar{m}_1}...\partial \bar{\bJ}_{m_k}\partial \bJ_{\bar{m}_k}} \ 
  \prod_{d=1}^k\ \bar \bJ_{n_{i_d,q+1}}\bJ_{\bar n_{i_d,q}} = 
  \sum_{\pi \bar\pi}  \prod_{d=1}^k\ \delta_{m_{\pi(d)} n_{i_d,q+1}}
  \delta_{\bar m_{\bar \pi(d)} \bar n_{i_d,q}} \ ,
 \end{align}
 where $\pi$ and $\bar \pi$ runs over permutations of $k$ elements,
 the cumulant of order $2k$ writes as
 \begin{align}\label{eq:cumulants1}
&\kappa(\bT_{n_1}\bar{\bT}_{\bar{n}_1}...\bT_{n_k}\bar{\bT}_{\bar{n}_k})=
\sum_{\pi \bar\pi} \sum_{v\geq k}\frac1{v!}\ 
\sum_{\mathcal{T}_{v,\{i_d\},\{{\mathcal{C}}(l)\}}}
\int\dots\int_{u_{ij}=0}^1\int_{\csig} \sum_{n,m} \crcr
& \qquad \times \left(\prod_{i=1}^v\prod_{p=1}^{\mathrm{res}(i)} R(\csigma^i)_{n_{i,p}m_{i,p}}\right)
\left(\prod_{\e\in T_v}\frac{-\lambda}{N^{D-1}} \ \delta_{\cD}^{\e,{\mathcal{C}}(\e)}\right) \crcr
& \qquad \times \left(\prod_{d=1}^k\ \delta_{m_{\pi(d)} n_{i_d,q+1}}
  \delta_{\bar m_{\bar \pi(d)} \bar n_{i_d,q}}\right) \ d\mu_{T_v,u}(\csigma) \prod_{\e_{ij}\in T_v}du_{ij} \ .
\end{align}
The sum runs over trees with $k$ cilia as other contributions are cancelled out either by the derivation or by the $\bJ=\bar\bJ=0$ prescription.
\qed

\section{Mixed Expansion} 

The loop vertex expansion of the cumulants can be refined to the \emph{mixed} expansion, an expansion over more general multicoloured maps which are seen as  
\emph{plane trees decorated by loop edges}. 
The loop edges are of the same nature as the tree edges: they have colours $\cC$ and represent identifications  of indices of the adjacent resolvents. 

The \emph{mixed expansion} of the cumulants is 
\begin{theorem}\label{th:mixed}
The cumulants of $\mu$ write:
\begin{align}\label{eq:mixedexp}
&\kappa(\bT_{n_1}\bar{\bT}_{\bar{n}_1}...\bT_{n_k}\bar{\bT}_{\bar{n}_k})=
\sum_{v\geq k}\frac1{v!}\ \frac{(-\lambda)^{v-1}}{N^{(D-1)(v-1) }} \sum_{\pi\bar\pi}    \sum_{\mathcal{T}} 
   \crcr
&\times
  \Bigg{[}  \sum_{q=0}^L \left(\frac{-\lambda}{N^{D-1}}\right)^q  \frac{1}{q!} \sum_{\cL, |\cL| = q}
  N^{F_{\rm int} (\cT,\cL ) } \crcr
& \qquad \times \left(\int_0^1  \prod_{l\in \cL } \frac{w_{i(l)j(l)}}2\prod_{\e_{ij}\in T_v}du_{ij}\right) \times 
\left(\prod_{d=1}^k\ \prod_{c=1}^D\delta_{m^c_{\pi(d)} \bar m^c_{\tau^c(\bar \pi(d))} }\right)
  \crcr
& \qquad + \left(\frac{-\lambda}{N^{D-1}}\right)^{L+1}  \frac{1}{L!}  \sum_{ \cL, |\cL| = L+1 } \crcr
& \qquad\times   \int_{t=0}^1  \; (1-t)^L  \int_{u_{ij}=0}^1 
   \left( \prod_{l\in \cL }   \frac{w_{i(l)j(l)}}2 \right) \crcr
& \qquad\quad \times
\int_{\csig}  \sum_{\bar n,n} \left(\prod_{i=1}^v\prod_{p=1}^{\mathrm{cor}(i)}  R( \sqrt{t}\csigma^i)_{n_{i,p} \bar n_{i,p}}  \right)
\crcr
&\qquad\quad \times \left(\prod_{l\in \cL }\delta_{\cD }^{l,{\mathcal{C}}(l)} \right)   \left(\prod_{\e\in T_v}\delta_{\cD}^{\e,{\mathcal{C}}(\e)}\right)
\left( \prod_{d=1}^k \delta_{m_{\pi(d)} n_{i_d,q+1}}
  \delta_{\bar m_{\bar \pi(d)} \bar n_{i_d,q}}\right)\crcr
  &\qquad\quad\times
  d\mu_{T_v,u}(\csigma)\left(\prod_{\e_{ij}\in T_v}du_{ij}\right)  dt
  \Bigg{]} 
\; .
\end{align}
\end{theorem}

%
%
%
%
%
%
\prf

Let us go back to \eqref{cumulants0},
 \begin{align}
&\kappa(\bT_{m_1}\bar{\bT}_{\bar{m}_1}...\bT_{m_k}\bar{\bT}_{\bar{m}_k})=
\sum_{\pi \bar\pi} \sum_{v\geq k}\frac1{v!}\ \frac{(-\lambda)^{v-1}}{N^{(D-1)(v-1)}}
\sum_{\mathcal{T}_{v,\{i_d\},\{{\mathcal{C}}(\e)\}}}
\int\dots\int_{u_{ij}=0}^1\crcr
&\times\int_{\csig} \sum_{n,m} 
   \left(\prod_{i=1}^v\prod_{p=1}^{\mathrm{res}(i)} R(\csigma^i)_{n_{i,p}{\bar n}_{i,p}}\right)
\left(\prod_{\e\in T_v} \ \delta_{\cD}^{\e,{\mathcal{C}}(\e)}\right) \crcr
&  \times \left(\prod_{d=1}^k\ \delta_{m_{\pi(d)} n_{i_d,q+1}}
  \delta_{\bar m_{\bar \pi(d)} \bar n_{i_d,q}}\right) \ d\mu_{T_v,u}(\csigma) \prod_{\e_{ij}\in T_v}du_{ij} \ .
\end{align}
and refine a term in this sum by Taylor expanding up to an order $L$ using the formula
\begin{align*}
f(\sqrt{\lambda})=& \sum_{q=0}^{L}\frac1{q!}\left[\frac{d^q}{dt^q}f(\sqrt{t\lambda})\right]_{t=0}\ \crcr
   & +\ \frac1{L !}\int_0^1 dt \; (1-t)^{L}\frac{d^{L+1} }{dt^{L+1} } \left( f(\sqrt{t\lambda}) \right) 
\end{align*}
on the contributions of the trees $\mathcal{T}_{v,\{i_d\},\{{\mathcal{C}}(\e)\}}$, using
\[ 
\frac{d}{dt} R(\sqrt{t} \csigma )_{ n {\bar n}}=
\frac1{2t} \left( \sum_{\mathcal{C}}\left( \csigma_{ab}^{i\ {\mathcal{C}}}\frac{\dr}{\dr\csigma_{ab}^{i\ {\mathcal{C}}}} 
\right)
\right) R( \sqrt{t} \sigma)_{n {\bar n}} \; ,
\]
where we dropped the superscript on the indices of $\csigma^{\cC}$. Integrating by parts 
\begin{align}
&\frac{d}{dt}\int \left(\prod_{i=1}^v\prod_{p=1}^{\mathrm{res}(i)}  R( \sqrt{t}\csigma^i)_{n_{i,p} {\bar n}_{i,p}}\right)d\mu_{T_v,u}(\csigma)\crcr
&=\frac{1}{2t}\int 
\sum_{i=1}^v\sum_{\mathcal{C}} 
\csigma_{ab}^{i\ {\mathcal{C}}}\frac{\dr}{\dr\csigma_{ab}^{i\ {\mathcal{C}}}} 
\ \prod_{i=1}^v\prod_{p=1}^{\mathrm{res}(i)} R( \sqrt{t}\csigma^i)_{n_{i,p} {\bar n}_{i,p}}   \ d\mu_{T_v,u}(\csigma)
\crcr
&=
\frac{1}{2t}\Bigg{[}e^{\sum_{i\leq j}w_{ij}\sum_{\mathcal{C}}
\frac{\dr}{\dr\csigma_{ab}^{i\ {\mathcal{C}}}}\frac{\dr}{\dr\csigma_{ab}^{j\ {\mathcal{C}}}}   } \
\sum_{i,\,{\mathcal{C}}}\left(
\csigma_{ab}^{i\ {\mathcal{C}}}\frac{\dr}{\dr\csigma_{ab}^{i\ {\mathcal{C}}}} 
\right)\left(\prod_{i,\,p}  R( \sqrt{t}\csigma^i)_{n_{i,p} {\bar n}_{i,p}} \right) \Bigg{]}_{\csigma=0}
\crcr
&=
\Bigg{[}e^{\sum_{i\leq j}w_{ij}\sum_{\mathcal{C}}
\frac{\dr}{\dr\csigma_{ab}^{i\ {\mathcal{C}}}}\frac{\dr}{\dr\csigma_{ab}^{j\ \cC}} 
}  
\  \sum_{i,j,\,{\mathcal{C}}}\frac{w_{ij}}{2t}\left(
\frac{\dr}{\dr\csigma_{ab}^{i\ {\mathcal{C}}}}\frac{\dr}{\dr\csigma_{ab}^{j\ {\mathcal{C}}}}
\right)\left(\prod_{i,\,p}   R( \sqrt{t}\csigma^i)_{n_{i,p} {\bar n}_{i,p}}    \right) \Bigg{]}_{\csigma=0} 
\; .
\end{align}
The sum over $i,j$ and $\cC$ is a sum over all the ways of adding a loop edge to the map 
$\mathcal{T}_{v,\{i_d\},\{{\mathcal{C}}(\e)\}}$. Evaluating the derivatives with respect to $\csigma^{\cC}$ and $\bar \csigma^{\cC}$ we see that the loop edge gives the 
same kind of coloured contraction $\delta^{l,{\mathcal{C}}}_{\cD}$ as a tree edge.
Furthermore, each loop edge brings a factor $\frac{-t\lambda}{N^{D-1}}$ (hence the $t$'s cancel), because the matrix $w_{ij}$ is symmetric
and the same loop edge $ij$ is generated by two terms: $\frac{\dr}{\dr\sigma_{ab}^{i\ {\mathcal{C}}}}\frac{\dr}{\dr\bar\sigma_{ab}^{j\ {\mathcal{C}}}}  $
and $\frac{\dr}{\dr\sigma_{ab}^{j\ {\mathcal{C}}}}\frac{\dr}{\dr\bar\sigma_{ab}^{i\ {\mathcal{C}}}} $.

Repeating this process $L$ times gives a sum of $\frac{(2v+k-3+2L)!}{(2v+k-3)!}$ 
terms labelled by trees $\mathcal{T}_{v,\{i_d\},\{{\mathcal{C}}(\e) \} }$ decorated with $L$ coloured,
labelled loop edges forming the set $\cL$,
\begin{align}
&\frac{d^q}{dt^q}
\int_0^1 \left(\prod_{\e_{ij}\in T_v}du_{ij}\right)\int d\mu_{T_v,u}(\csigma) 
\left(\prod_{i=1}^v\prod_{p=1}^{\mathrm{cor}(i)}  R( \sqrt{t}\csigma^i)_{n_{i,p} {\bar n}_{i,p}}    \right) 
\crcr
& \qquad \times \left(\prod_{\e\in T_v}\delta_{\cD}^{\e,{\mathcal{C}}(\e)}\right)
\left( \prod_{d=1}^k\ \delta_{m_{\pi(d)} n_{i_d,q+1}}
  \delta_{\bar m_{\bar \pi(d)} \bar n_{i_d,q}}\right)
\crcr
&=
\left(\frac{-\lambda}{N^{D-1}}\right)^q \sum_{ \cL, |\cL|=q }
\int_0^1 \left(\prod_{\e_{ij}\in T_v}du_{ij}\right)\int d\mu_{T_v,u}(\csigma) \left(\prod_{l\in \cL }\delta_{\cD }^{l,{\mathcal{C}}(l)} \frac{w_{i(l)j(l)}}2\right)
\crcr
&\qquad \times\left(\prod_{i=1}^v\prod_{p=1}^{\mathrm{cor}(i)}  R( \sqrt{t}\csigma^i)_{n_{i,p} \bar n_{i,p}}  \right)
\left(\prod_{\e\in T_v}\delta_{\cD}^{\e,{\mathcal{C}}(\e)}\right)
\left( \prod_{d=1}^k\ \delta_{m_{\pi(d)} n_{i_d,q+1}}
  \delta_{\bar m_{\bar \pi(d)} \bar n_{i_d,q}}\right).
\end{align}

Taking into account that $R(0) = {\id}^{\cD}$ proves the theorem because 
the first $L$ terms of the Taylor expansion up to order $L$ can be evaluated explicitly: one obtains a free sum for each of the 
internal faces of the map, and Theorem \ref{th:mixed} follows.

\qed

The mixed expansion allows to remove the resolvent operators in the explicit terms of the expansion, unveiling once again the tensor invariant structure of the
cumulants and their maps. The amplitude of the maps are once again computed as a power of N by face counting, as seen in  \eqref{eq:mixedexp}. 
Moreover, the identification of the indices of the source terms appears once again as in the perturbative expansion \eqref{eq:feynampcumulants}.

\section{Absolute convergence} 
In this section,  we will study the analyticity properties and the scaling with $N$ of the cumulants starting 
from the mixed expansion in \eqref{eq:mixedexp}.

We denote $C(\tau^{\cD})$ the number of connected components of the bipartite $D$-coloured graph associated to the $D$-uple of permutations $\tau^{\cD}$.

\begin{theorem}\label{th:conv}
 The series in (\ref{eq:mixedexp}) is absolutely convergent for $\lambda\in[0,\frac1{8|\mathcal{Q}|})$.
 In this domain the cumulants display a structure of tensor invariants,
 \begin{align}
 \label{eq:structurecumulantslve}
\kappa(\bT_{m_1}\bar{\bT}_{\bar{m}_1}...\bT_{m_k}\bar{\bT}_{\bar{m}_k})
=\sum_{\pi, \bar\pi}\sum_{\tau^{\cD} }\mathfrak{K}(\tau^{\cD})\prod_{d=1}^k  
\prod_{c=1}^D\delta_{m_{\pi(d)}^c \bar{m}_{\tau_c\bar\pi(d)}^c} \;,
\end{align}
 and obey the bound 
 \[
 | \mathfrak{K}(\tau^{\cD}) | \le N^{D - (D-1) k -  C(\tau^\cD)   }  K(\lambda) \; ,
 \]
 for some $K$ depending only on $\lambda$.
\end{theorem}
This theorem will be proved in the remainder of this section.
Note that the exponent of $N$ is exactly $-\Omega_{\min}$ for theorem \ref{thm:1/Nexp}, and we recover the polynomial bound of the perturbative expansion.

In order to establish the absolute convergence of the series in \eqref{eq:mixedexp}, we need to establish a bound on an individual term. 
The explicit terms (consisting in trees with up to $L$ loops) and the remainder (trees with $L+1$ loops) are bounded by very different methods,
explained in the next two subsections.

\subsubsection{Invariant structure}

The explicit terms of \eqref{eq:mixedexp} show the required invariant structure,
\[ \prod_{d=1}^k\ \prod_{c=1}^D \delta_{m^c_{\pi(d)} \bar m^c_{\tau^c(\bar \pi(d))}}.
\]
For a cumulant which index structure does not match the one of a tensor invariant ($D$-uple of permutation of its terms)
every term of the development vanish and the convergence is trivial. 
A proper bound on the remainder is sufficient to conclude.

\subsection{Bounds on the explicit terms}

The global scaling in $N$ of the (non-vanishing) term associated to the tree $\cT$ decorated with $q$ loop edges $\cL$ 
in eq. \eqref{eq:mixedexp} is
\begin{equation}\label{eq:globNscaling}
 \frac{ N^{F_{\rm int} (\cT,\cL)}}{N^{(D-1)(|\cL|+v-1) }} \;.
\end{equation}
We thus need to bound the number of internal  faces
of the  tree $\cT$ decorated by the loop edges $\cL$.

Recall that $C(\tau_{\cD }) $ denotes the number of connected components of the graph associated to the permutations $\tau_{\cD }$. 
As the term associated to $\cT$ does not vanish, the boundary graph of $(\cT,\cL)$ is the graph associated with $\tau^\cD$ ($\tau^cD=\tau_{(\cT,\cL)}^\cD$) and 
their numbers of connected components match, $C(\partial(\cT,\cL))=C(\tau^\cD)$.

By Lemma \ref{lem:facesbound}, the number of faces of a ciliated plane tree with loop edges $(\cT,\cL)$ is bounded by
\[
   F_{\rm int} (\cT,\cL)  \le \ 1\ -\ (D-1) k\ -\ C(\partial(\cT,\cL))+\ (D-1) v \  +\ \left\lfloor\frac{D}{2}\right\rfloor|\cL| \; .
\]
which according to \eqref{eq:globNscaling}, gives a global scaling in $N$ of 
\begin{equation}\label{eq:globscalingxplicit}
  \frac{ N^{F_{\rm int} (\cT,\cL)}}{N^{(D-1)(|\cL|+v-1) }}\leq N^{
  D - C(\partial(\cT,\cL))  -(D-1)k - \left\lceil\frac{D}{2}-1\right\rceil|\cL|
  }\ .
\end{equation}

\subsection{Bounds on the remainder}\label{subsec:CSbounds}

In order to establish a bound on the remainder
in equation \eqref{eq:mixedexp} we will use the technique introduced in \cite{MagRiv,Magnen:2009at}, and refined for tensors in \cite{DGR1403},
namely \emph{iterated Cauchy-Schwarz inequalities}.
For a pair of (multi-index) vectors $A$ and $B$, two linear operators  $R$ and $R'$, and ($\,\cdot\,$) the usual scalar product, the Cauchy-Schwarz inequality states, 
\begin{align}\label{eq:cauchy}
 | A\, \cdot\, (R\otimes R'\otimes \id^{\otimes p})\, B | \;   \leq  \;  \|R\|\,\|R'\|\,\sqrt{ A\cdot A}\,\sqrt{B\cdot B} \; .
\end{align}

\begin{lemma}\label{lem:boundrest}
We have the bound:
\begin{align}\label{eq:rest}
& \Bigg{|}\int \sum_{m,n} \left(\prod_{i=1}^v\prod_{p=1}^{\mathrm{res}(i)}  R( \sqrt{t}\csigma^i)_{n_{i,p} m_{i,p}}  \right)
 \left(\prod_{l\in \cL }\delta_{\cD }^{l,{\mathcal{C}}(l)} \right)  
\crcr
&\qquad \times \left(\prod_{\e\in T_v}\delta_{\cD}^{\e,{\mathcal{C}}(\e)}\right)
\left( \prod_{d=1}^k \delta_{m_{\pi(d)} n_{i_d,q+1}}
  \delta_{\bar m_{\bar \pi(d)} \bar n_{i_d,q}}\right)\  d\mu_{T_v,u}(\csigma)\Bigg{|}  \crcr
& \le  N^{D + (D-1)(v-1)+    |\cL| \frac{D}{2}}
 \; .
\end{align} 
\end{lemma}

\noindent{\bf Proof:} The remainder is a product of resolvents placed at the corners of the vertices of a tree 
decorated with loop edges and cilia.
On any tree, one can choose a corner to start, and then order the corners following the clockwise contour walk of the tree, indexing 
their resolvents from $R_1$ to $R_{2n}$ (or $R_{2n+1}$). 

\begin{figure}[h]
 \centering
  \includegraphics[scale=.75]{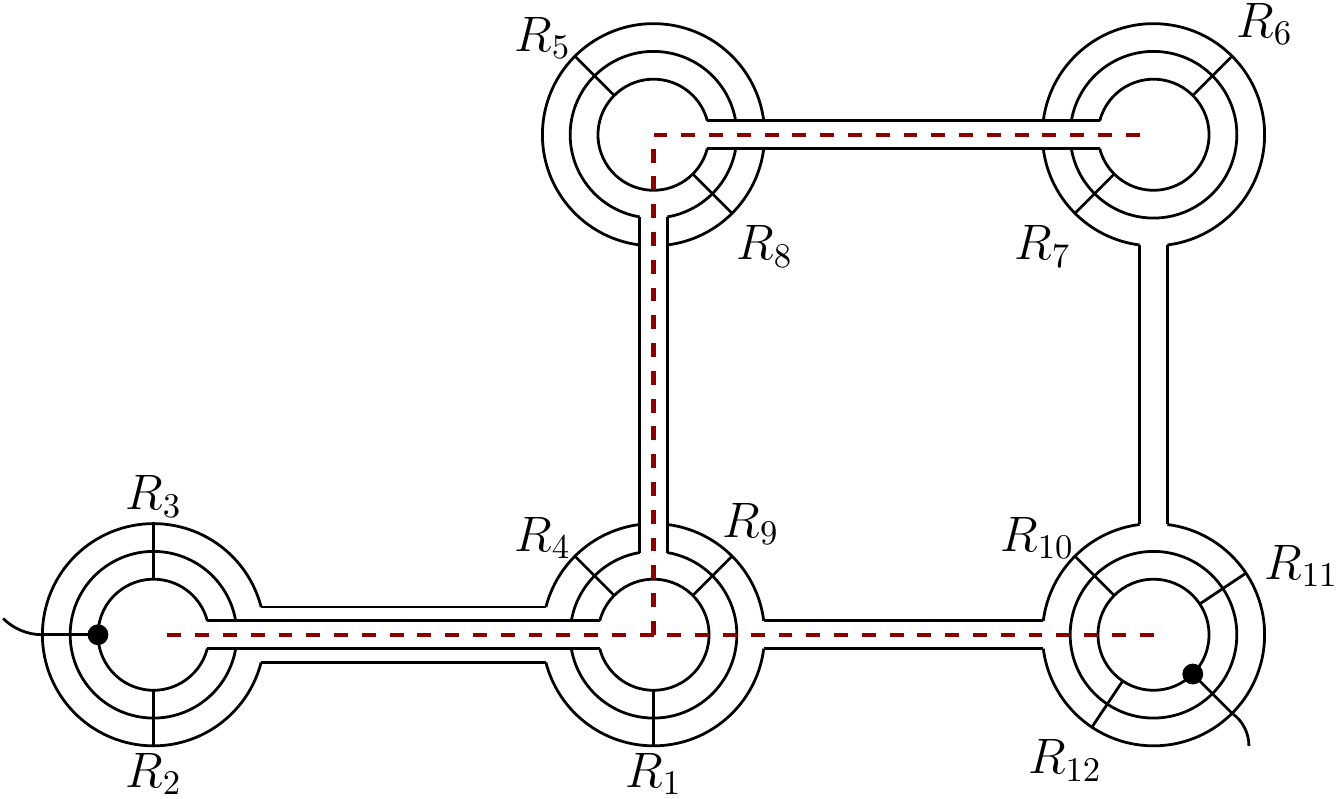}
\caption{A multicoloured map with spanning tree (dashed) and indexed resolvents.}
\label{graph2a}
  \end{figure}
  
\begin{figure}[h]
 \centering
  \includegraphics[scale=.75]{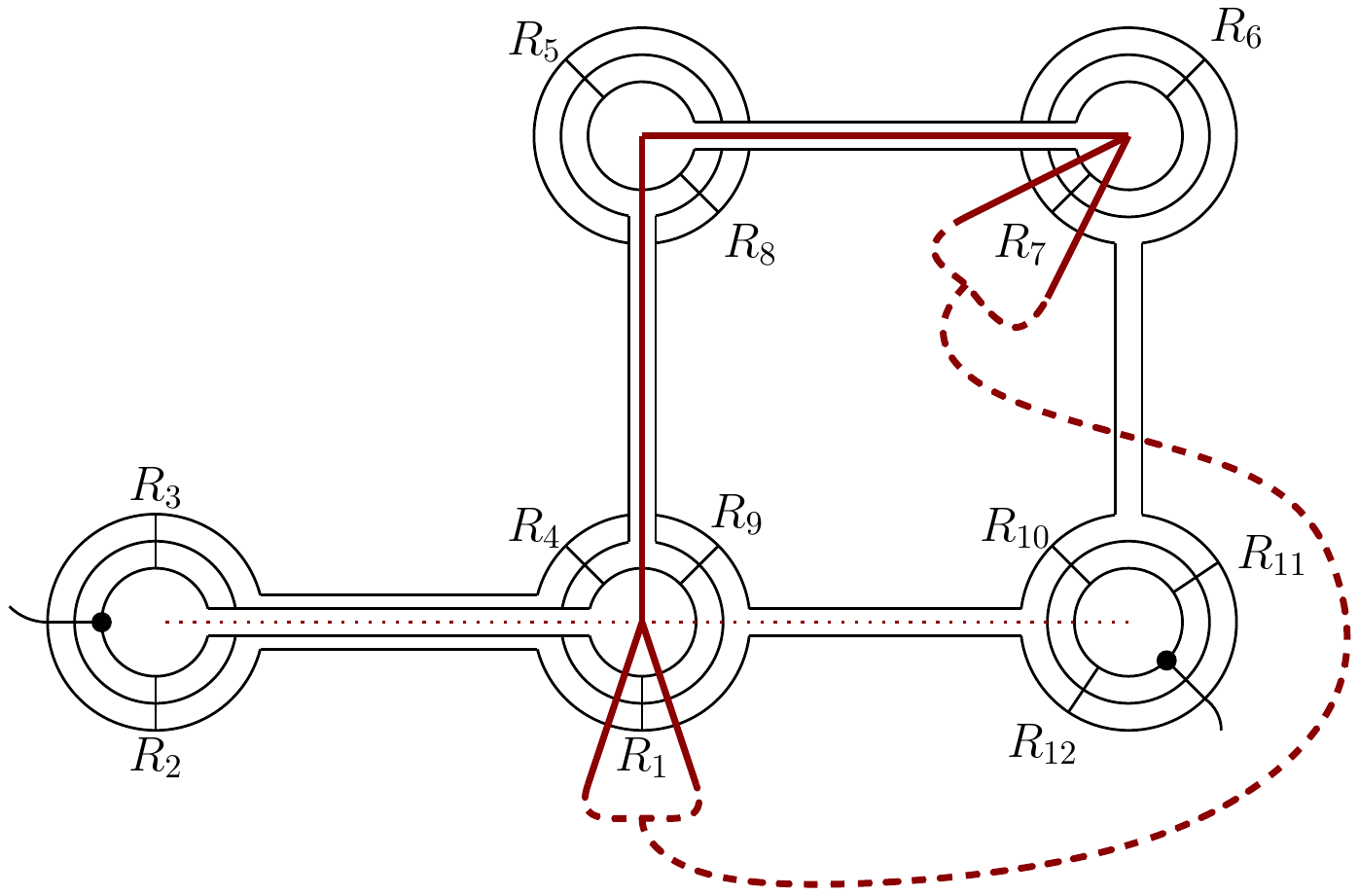}
\caption{Splitting the map in two parts in order to apply the Cauchy-Schwartz inequality. }
\label{graph2'}
  \end{figure}

\paragraph{Map splitting.}
Choosing $R_1$ and $R_{n+1}$ as the $R$ and $R'$ of formula \eqref{eq:cauchy}, the  
vector $A$ is made of all the resolvents from $R_2$ to $R_n$ and the contractions between them.
The vector $B$ is made of all the resolvents from $R_{n+2}$ to $R_{2n}$ (or $R_{2n+1}$) and the contractions between them.
The contractions of indices of the resolvents $R_2$ to $R_n$ with indices of the resolvents $R_{n+2}$ to $R_{2n}$ (or $R_{2n+1}$),
which can exist due to the loop edges, are encoded in the $\id^{\otimes p}$ operator.
If an index of $R_1$ is directly contracted with an index of $R_{n+1}$, the latter contributes with a Kronecker $\delta$ to the vector $A$ or $B$.
We represented such a splitting in Figure \ref{graph2'}.

We apply the formula \eqref{eq:cauchy} and, because the norm of the resolvent is bounded by $1$, $\|R(\csigma)\|\leq 1$, 
\begin{align}\label{eq:cauchy1}
 |\, {A}\,\cdot\, (R_1\otimes R_{n+1}\otimes \id^{\otimes p})\, B\, | \; \leq \; \sqrt{(A\cdot A)\ (B\cdot B)}\ .
\end{align}
The  splitting in $A$ and $B$ corresponds to cutting along the unique path in the tree going from corner $1$ to corner $n+1$ (the cut is represented in bold 
in Figure \ref{graph2'})  . 
Any occurring contraction strand between the part $A$ and the part $B$ (due to the loop edges) is also cut.

The scalar products $ {A\cdot A}$ and ${B\cdot B}$ are half maps merged with their mirror symmetric with respect to the splitting line.
They also have the structure of trees with loop edges and resolvents. However, in contrast 
with the original map, they can have several cilia on the same vertex. This happens whenever the resolvent 
$R_1$ or $R_{n+1}$ belongs to a ciliated vertex.

\begin{figure}[h]
\centering
\begin{tabular}{c}
\includegraphics[scale=.65]{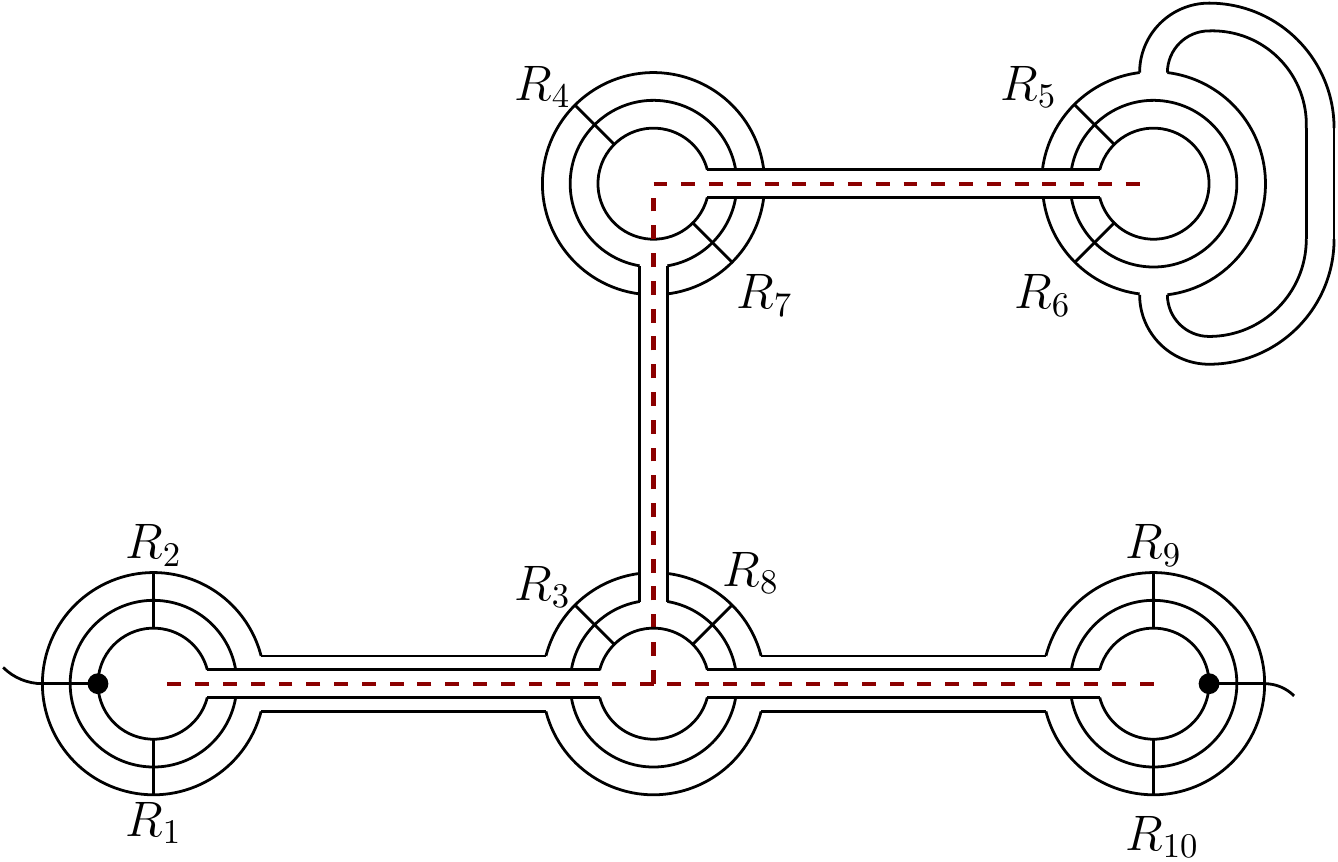} \\ \hspace{.5cm} \\ \includegraphics[scale=.65]{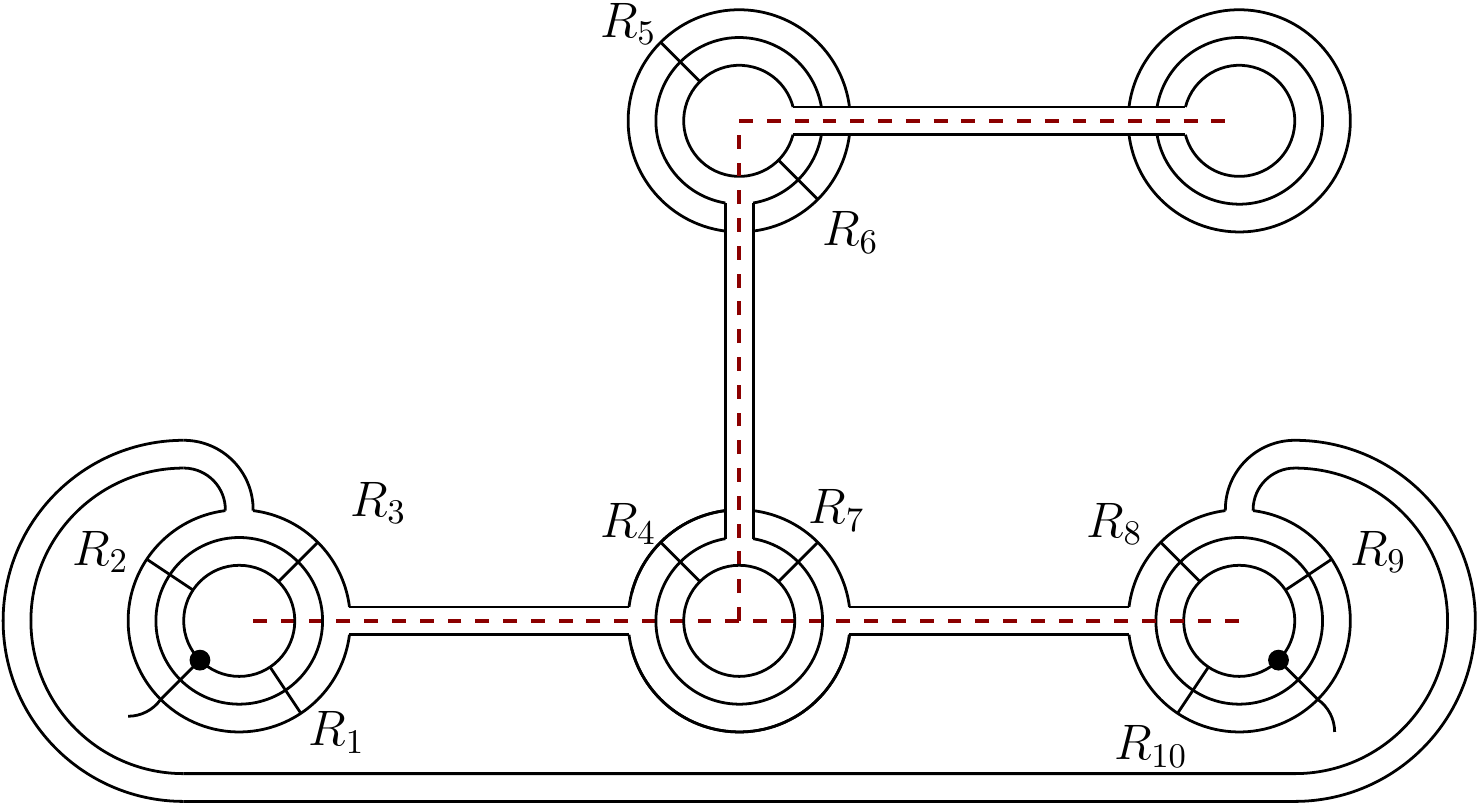} \end{tabular}
\caption{The scalar product graphs $\braket {A|A}$ (top) and $\braket{B|B}$ (bottom). The dashed lines represent their respective spanning trees, 
and their remaining resolvents have been indexed accordingly.}
\label{graph2ab}
\end{figure}

As the resolvents $R_1$ and $R_{n+1}$ have been taken out, the scars (i.e. the corners where $R_1$ and $R_{n+1}$ were connecting on $A$ and $B$)
are now direct identifications of indices. That is, in the maps merged with their mirror symmetric,
two resolvents have been set to the identity ${\id}^{\cD}$ operator. 
The corresponding corners will be represented as just $D$ parallel strands, with 
no resolvent, like in Figure \ref{graph2ab}.

Let us denote $v^A$, $k^A$, $\cT^{A}$, $\cL^{A}$ the number of vertices, cilia,
the tree and the set of loop edges corresponding to the map $\braket{A|A}$ and similarly for $B$. 
We have the following (in)equalities:
\begin{itemize}
 \item the number of vertices doubles
           \[ 2v = v^A + v^B  \; .\]
 \item the number of cilia doubles 
           \[ 2 k = k^A + k^B \; .\]
 \item the number of loop edges doubles
           \[ |\cL| = |\cL^A| + |\cL^B| \; . \] 
 \item the number of faces \emph{at least doubles}
          \begin{align*} 
          2F_{\rm int}(\cT, \cL)  \le F_{\rm int}(\cT^A, \cL^A) +  F_{\rm int}(\cT^B, \cL^B)   \; .
         \end{align*}
         This is because a face is either untouched by the splitting line, or it is cut in two pieces (and in both cases it leads to two faces in 
           the mirrored maps), or it is cut in at least four pieces in which case it leads to at least four faces in the mirrored maps.
\end{itemize}
 
Each mirrored map has an even number of resolvents and contributes to the bound by the square root of its amplitude. 
We can now iterate the process, each time eliminating two resolvents, until there are no resolvents left on any map. 

If we start with $2n$ resolvents we obtain $2^n$ final maps. If we start with $2n+1$ resolvents, the first iteration is asymmetrical as $A\cdot A$ has $2n-2$ 
resolvent and $B\cdot B$ has $2n$. The $B$ part thus requires an extra iteration of the process to cancel every resolvents.
The $2^n$ (resp. $2^{n-1}+2^{n}$) final maps we obtain are made solely 
of internal faces which represent traces of identity, hence  each  such face brings a factor $N$, and external faces that connect cilia together. 

We denote for, $q=1,\dots 2^n$ (or $2^{n-1}+2^{n}$) by $v(q)$, $k(q)$, $\cT^q$, $\cL^q$ the numbers of vertices, cilia, 
the tree and the set of loop edges of the final maps $q$.
Each cilium $a\in\{1\dots k(q)\}$ originates from a cilium $d(a)$ of the original map, and therefore corresponds to a source-term contraction operator
\[
 \delta^{[a,d(a)]}= \left(\delta_{m_{\pi(d(a))} n_{i_a,q+1}}
   \delta_{\bar m_{\bar \pi(d(a))} \bar n_{i_{a},q}}\right) \ ,
\]
which are contracted together by coloured external faces $f^c\in{}F_{\rm ext}^c(\cT^q, \cL^q)$
\[
 \sum_{c=1}^D\ \left(\sum_{f^c\in F_{\rm ext}^c(\cT^q, \cL^q)} \delta^{[f^c]}\right)\ \sum_{a\in\{1\dots k(q)\}} \delta^{[a, d(a)]}
 =\prod_{a=1}^{k(q)}\ 
  \prod_{c=1}^D \delta_{m^c_{\pi(d(a))} \bar m^c_{\tau^c_{[\cT^q,\cL^q]}(\bar \pi(d(a)))}} \ ,
\]
where the $D$-uple of permutation $\tau^c_{[\cT^q,\cL^q]}=\partial[\cT^q,\cL^q]$  is the boundary graph of the final map $q$, and has {\it a priori} 
no relation with the boundary of the original map $\partial[\cT,\cL]$, as the splitting and bounding process repeatedly severed the structure of the original map.
This is the reason why the bound \eqref{eq:rest} on the remainder is quite rough, and weaker than the bound established on the explicit terms.

The amplitude of a final map thus writes
\[
 A(q)= N^{F_{\rm int}(\cT^q, \cL^q)} \prod_{a=1}^{k(q)}\ 
  \prod_{c=1}^D \delta_{m^c_{\pi(d(a))} \bar m^c_{\tau^c_{[\cT^q,\cL^q]}(\bar \pi(d(a)))}}.
\]
The product of Kronecker $\delta$'s is bounded by 1 (which is as good as we can do, having no control whatsoever on its structure). 
The number of internal faces of a ciliated map $\cM$ is bounded by the number of internal faces of the same map with all cilia removed $\cM^{*}$, therefore
\[
 F_{\rm int}([\cT^q, \cL^q])\le F_{\rm int}\left([\cT^q, \cL^q]^*\right)\ .
\]
Indeed, removing a cilium can only increase the number of internal faces of a multicoloured map.

As before, adding a loop edge with colours $\cC$ to a final map can at most divide $|\cC|$ of its faces, hence
\[
 F_{\rm int}([\cT^q, \cL^q]^*)  \le  F_{\rm int}(\cT^{q\,*}) +  |\cL^q|\left\lfloor\frac{D}{2}\right\rfloor \;.
\]
where $\cT^{q\,*}$ is the tree corresponding to the final map $q$ with all cilia removed.

The number of internal faces of $\cT^q$ is at most $1 + (D-1) v^q $. If we start with $2n$ resolvents we get the bound
\begin{align*}
 & \Bigg{|}\int \sum_{m,n} \left(\prod_{i=1}^v\prod_{p=1}^{\mathrm{res}(i)}  R( \sqrt{t}\csigma^i)_{n_{i,p} m_{i,p}}  \right)
 \left(\prod_{l\in \cL }\delta_{\cD }^{l,{\mathcal{C}}(l)} \right)  
\crcr
&\qquad \times \left(\prod_{\e\in T_v}\delta_{\cD}^{\e,{\mathcal{C}}(\e)}\right)
\left( \prod_{d=1}^k \prod_{c=2}^D \delta_{n^c_{i_d,q+1}\, m_{i_{\tau_c(d)},q}^c}\right) d\mu_{T_v,u}(\csigma) \Bigg{|} \crcr
& \le N^{\frac{1}{2^n} \sum_{q=1}^{2^n}   F_{\rm int}(\cT^q, \cL^q) }
\le N^{\frac{1}{2^n} \sum_{q=1}^{2^n} \left( 1 + (D-1)v(q)      + |\cL^q|\left\lfloor\frac{D}{2}\right\rfloor   \right) } \crcr
& = N^{1 + (D-1)v+    |\cL| \left\lfloor\frac{D}{2}\right\rfloor}
\end{align*}
and the same holds if we start with $2n+1$ resolvents.
\begin{align*}
 \prod_{q=1}^{2^{n-1}} \left(N^{  F_{\rm int}(\cT^q, \cL^q) }\right)^{\frac{1}{2^{n-1}}}
 \prod_{q=2^{n-1}+1}^{2^{n-1}+2^{n}} \left(N^{  F_{\rm int}(\cT^q, \cL^q) }\right)^\frac{1}{2^{n}}
\le   N^{1 + (D-1)v+    |\cL| \left\lfloor\frac{D}{2}\right\rfloor} \ ,
\end{align*}
which conclude the proof of Lemma \ref{lem:boundrest}.

\qed
\subsection{Results}
We are now ready to prove Theorem \ref{th:conv}. Taking absolute values in \eqref{eq:mixedexp} and
using Lemmas \ref{lem:facesbound} and \ref{lem:boundrest} we find:
\begin{align*}
 &\kappa(\bT_{n_1}\bar{\bT}_{\bar{n}_1}...\bT_{n_k}\bar{\bT}_{\bar{n}_k})
 \le \sum_{\pi\bar\pi}  \sum_{v\geq k}\frac1{v!} \sum_{\mathcal{T}}  \crcr
& \qquad \Bigg{[}  \sum_{q=0}^L  \frac{   | \lambda|   ^{v-1+q}  }{q!} \sum_{\cL, |\cL| = q} N^{D - 2(D-1)k - {C}(\partial[\cT,\cL]) 
   - q\left\lceil\frac{D}{2}-1 \right\rceil } 
   \prod_{d=1}^k\ 
  \prod_{c=1}^D \delta_{m^c_{\pi(d)} \bar m^c_{\tau^c_{[\cT,\cL]}(\bar \pi(d))}}
   \crcr
& \qquad \; + \frac{  | \lambda|   ^{v+L}  }{L!} \sum_{\cL, |\cL| = L+1} N^{D - (D-1)k - (L+1) \left\lceil\frac{D}{2}-1 \right\rceil   } \Bigr] \; .
\end{align*}
\paragraph{For cumulants with invariant structure ($m^c_{\pi(d)}=\bar m^c_{\tau^c(\bar \pi(d))}$)}
only the maps with the right boundary contributes to the cumulants,
\[
 \partial[\cT,\cL] = \tau^\cD \ ,
\]
and as $ {C}(\partial[\cT,\cL]) \le k $, choosing $L \ge \frac{Dk}{D/2-1}-1$ ensures that 
\[
 D - (D-1)k - (L+1) \left\lceil\frac{D}{2}-1 \right\rceil  \le D - 2(D-1)k - {C}(\partial[\cT,\cL]) \; ,
\]
and the invariant cumulant obeys the bound
\begin{align}
 |\mathfrak{K}(\tau^{\cD})| & \le  N^{D - 2(D-1)k - \mathfrak{C}(\rho_{\cD}) } \crcr
&    \times  \sum_{v\geq k}\frac1{v!} \sum_{\mathcal{T}_{v}}  
   \sum_{q=0}^{L+1}  \frac{   | \lambda|  ^{v-1+q}  }{q!} \frac{(2v+k-3+2q)!}{(2v+k-3)!} \; ,
\end{align}
where the last combinatorial factor is the number of ways to add $q$ loop edges to a plane tree $\cT$.  
This bound is very broad as it does not account for the restriction on the structure of the boundary graph.
Taking into account that 
\begin{align*}
  & \sum_{\mathcal{T}_{v}}1 \crcr
  &=  |\mathcal{Q}|^{v-1}  
   \sum_{d_1 \dots d_v\ge 1 }^{\sum d_i=2(v-1)}\left(\frac{(v-2)!}{\prod_i (d_i-1)!} \prod_i (d_i-1)!\ \times \sum_{i_1,\dots i_k}^{i_k\neq i_{k'}} 
      d_{i_1}...d_{i_k}\right) \crcr
  &=|\cQ|^{v-1}\frac{v!(2v+k-3)!}{k!(v-k)!(v+k-1)!}\; , 
\end{align*}
with $|\mathcal{Q}|^{v-1}$ the number of edge colouring, $d_i$ the degree of the vertex $i$,
$\frac{(v-2)!}{\prod_i (d_i-1)!}$ the number of trees with fixed degrees and $\prod_i (d_i-1)!$ the number of associated plane
trees, and $\prod_{d=1}^k d_{i_d}$ the number of ways to put cilia on the vertices $i_d$, we obtain
\begin{align}\label{eq:boundseries}
 |\mathfrak{K}(\tau^{\cD})| & \le  N^{D - 2(D-1)k - {C}(\tau^{\cD}) } \crcr
&    \times   \sum_{q=0}^{L+1} \sum_{v\geq k} (| \lambda|\ |\mathcal{Q}| )^{v+q-1}
\frac{(2v+k-3+2q)!}{q!   (v-k)!(v+k-1)!} \; ,
\end{align}
and the sum over $v$ converges for $4|\mathcal{Q}|\, |\lambda|<1$.

\paragraph{For non-invariant cumulants}
the explicit development vanishes and the cumulant is equal to the remainder at every order, therefore,  
\[
 \kappa(\bT_{n_1}\bar{\bT}_{\bar{n}_1}...\bT_{n_k}\bar{\bT}_{\bar{n}_k}) = \lim_{L\to\infty}\left[{\rm Remainder}\right](L) =0\ ,
\]
which achieves the proof of Theorem \ref{th:conv}.

\qed

\section{Uniform Borel summability} 
We subsequently establish the uniform Borel summability of the cumulants at the origin.
\begin{theorem}\label{th:Borel}
The cumulants can be analytically continued for complex $\lambda=r e^{i\phi}$ with 
$r<\frac1{4|\mathcal{Q}|}\left(\mathrm{cos}\frac\phi2\right)^2$. In this domain they obey the bound:
\begin{align}\label{eq:Borelbound}
 | \mathfrak{K}(\tau^{\cD}) | \le N^{D - (D-1) k -  {C}(\tau^{\cD})   } 
 K\left( \frac{|\lambda|}  { \left(\cos \phi /2 \right)^2 } \right) \;,
\end{align}
and are Borel summable in $\lambda$ uniformly in $N$.
\end{theorem}
This theorem will be proved in the remainder of this section, by verifying the hypotheses of Theorem \ref{thm:unifborel}.

\subsubsection{Analyticity}\label{subsecANA}

To establish the convergence of the series in  \eqref{eq:mixedexp} in the domain in the complex plane $\lambda=r e^{i\phi}$, $\phi\in (-\pi,\pi)$
defined by $|\lambda|< \frac{1}{4|\cQ|} \left( \cos\frac{\phi}{2} \right)^2 $ it is enough to follow step by step the proof of 
Theorem \ref{th:conv} and to remark that the norm of the resolvent is bounded by
\[
  \|R(\csigma)\| \leq \frac{1}{\mathrm{cos}\frac\phi2} \; .
\]
The iterated Cauchy-Schwarz inequalities go through, and it is easy to see that the norm of each resolvent  
contributes to the power 1 to the amplitude of the map. The total number of resolvents of a map 
with $v$ vertices and $k$ marks is $2(v-1)+k$. Therefore each term of the overall bound in  \ref{eq:boundseries}
must be multiplied by $\left(\frac{1}{\mathrm{cos}\frac\phi2}\right)^{2(v-1)+k}$, which proves the convergence and 
\eqref{eq:Borelbound}.

The convergence domain 
 \begin{align*}
  |\lambda|<\frac1{4|\cQ|}\left(\mathrm{cos}\frac\phi2\right)^2 \; ,
 \end{align*}
contains a disk $\mathrm{Re}\frac1\lambda>\frac1R$. In this domain the cumulants  \eqref{cumulants0} 
are analytic as the resolvents themselves $R(\csigma^i)_{n_{i,p} m_{i,p}}$ are.

\subsubsection{Taylor expansion}\label{subsecTAYLOR}

The Taylor expansion in $\lambda$ of the cumulants up to order $r$ 
is obtained by using the mixed expansion 
in Theorem \ref{th:mixed}, but choosing the order $L$ up to which we develop the loop edges 
to depend on the number of vertices $v$ of the tree $L = \max (0, r - v)$. For $v\ge r+1$ we do not develop any loop edges.
Using the same bounds leading up to 
eq. \eqref{eq:boundseries}, and noting that the scaling with $N$ is always bounded by $N^D$, the remainder is bounded by 
\begin{align}
 |R_{r}(\lambda,N)| \le & N^{D}  \sum_{v\geq k} \Bigg{[}
 \left(|\mathcal{Q}|\,|\lambda|\right)^{v+q-1} \frac{(2v+k-3+2q)!}{q!   (v-k)!(v+k-1)!} \Bigg{]}_{q= \max(0,r+1-v)}\;,
\end{align}
hence up to irrelevant overall factors the remainder is bounded by 
\begin{align*}
   &  \sum_{v\ge r+1} \left(|\mathcal{Q}|\,|\lambda|\right)^{v -1} \frac{(2v+k-3)!}{(v-k)!(v+k-1)!} \crcr
 & \quad \quad      +\sum_{v=k}^{r+1}  \left(|\mathcal{Q}|\,|\lambda|\right)^{r} 
 \frac{\left[2v+k-3+2(r+1-v)\right]!}{(r+1-v)!   (v-k)!(v+k-1)!} \; .
\end{align*}
While the first term above is bounded by $|\lambda|^r$ times some constant for $\lambda$ small enough, 
the second one is bounded only as :
\[
 |\lambda|^r \frac{(2r+k-1)!}{(r-k)!} \le (2k-1)! 3^{2r+k-1} \;\;\; r! |\lambda|^r  \; .
\]

\subsection{The $1/N$ expansion} 

Furthermore, the mixed expansion in \eqref{eq:mixedexp} is the non perturbative $1/N$ expansion of the cumulants in the following sense
\begin{corollary}\label{th:1/N}
 The remainder in the mixed expansion is analytic 
 in the domain $r<\frac1{4|\mathcal{Q}|}\left(\mathrm{cos}\frac\phi2\right)^2$ and in this domain it admits the bound:
 \begin{align*}
  & 
  \sum_{v\geq k}\frac1{v!}\ \frac{(-\lambda)^{v-1}}{N^{(D-1)(v-1) }} \sum_{\pi\bar\pi}    \sum_{\mathcal{T}} 
%
   \crcr
&\qquad \times
   \left(\frac{-\lambda}{N^{D-1}}\right)^{L+1}  \frac{1}{L!}  \sum_{ \cL, |\cL| = L+1 }  \int_{t=0}^1  \; (1-t)^L 
   \crcr
& \qquad  \times   \int_{u_{ij}=0}^1   
   \left( \prod_{l\in \cL }   \frac12 w_{i(l)j(l)} \right) \left(\prod_{d=1}^k\ \prod_{c=1}^D\delta_{m^c_{\pi(d)} \bar m^c_{\tau^c(\bar \pi(d))} }\right)
   \crcr
& \qquad \times
\int_{\csig} \sum_{m,n} \left(\prod_{i=1}^v\prod_{p=1}^{\mathrm{res}(i)}  R( \sqrt{t}\csigma^i)_{n_{i,p} m_{i,p}}  \right)
\crcr
&\qquad \times \left(\prod_{l\in \cL }\delta_{\cD }^{l,{\mathcal{C}}(l)} \right)   \left(\prod_{\e\in T_v}\delta_{\cD}^{\e,{\mathcal{C}}(\e)}\right)
\left( \prod_{d=1}^k \prod_{c=2}^D \delta_{n^c_{i_d,q+1}\, m_{i_{\tau_c(d)},q}^c}\right) 
\crcr
&\qquad \times d\mu_{T_v,u}(\csigma) \left(\prod_{\e_{ij}\in T_v}du_{ij}\right) dt
\crcr
& \le N^{D - (D-1)k - (L+1) \left(\frac{D}{2}-1\right) }
\left( \frac{|\lambda|}  { \left(\cos \phi /2 \right)^2 } \right)^{L+k}
K'\left( \frac{|\lambda|}  { \left(\cos \phi /2 \right)^2 } \right) 
\; 
 \end{align*}
for some bounded function $K'$.
\end{corollary}


\chapter{The constructive $T^4_3$ field theory}
 \section{The Model}
\subsection{The bare model}
In this chapter we study the simplest renormalisable quartic tensor field theory in the constructive framework.
Our starting point is the rank $D=3$ tensor field theory,
\begin{align}\label{eq:CTFTZ}
d\nu &=  e^{-\frac12 \lambda\sum_{c =1}^3  V_{c}(\bar{\psit},{ \psit} ) }\ d\mu_\bC(\bar\psit,\psit) \ ,
\end{align}
where $\psit, \bar\psit$ are a Fourier transformed pair of conjugate $U(1)^3$ scalar fields, which can be considered as infinite-size 
complex tensors with indices in $\setZ^3$. The covariance $\bC$ is defined as,
\begin{equation}\label{eq:propdef}
 \bC_{\bar p p} = \frac{\delta_{\bar p p}}{ p_1^2+p_2^2+p_3^2 + 1} \ .
\end{equation}
 A quick study of the Feynman expansions reveals three divergent single-vertices graphs for each colour. 
 A logarithmically divergent 2-point graph $\cM$, a linearly divergent vacuum graph $\cV_1$ and a  logarithmically divergent vacuum graph $\cV_2$. 
 
 \begin{figure}[!h]
  \centering
  \includegraphics[width=0.55\textwidth]{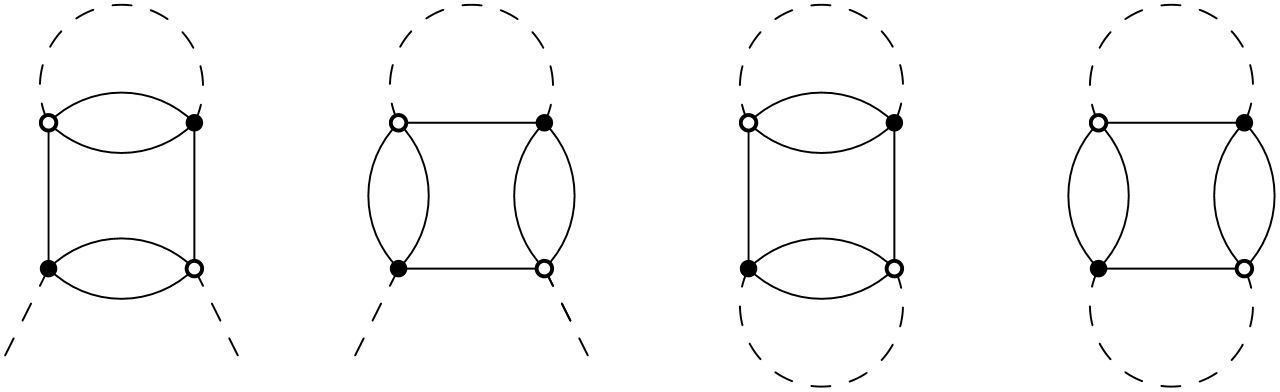}
  \caption{From left to right, the divergent self-loop $\cM$, the convergent self loop and the two vacuum connected graphs $\cV_1$ and $\cV_2$.}
  \label{divergences}
\end{figure}
 
  Renormalisation of the model \eqref{eq:CTFTZ} will thus be required.
To ensure that all quantities are well defined, we impose a cut-off $N$. We restrict the indices such as $n, \bar n$ to belong to $ \{-N\dots N\}^3$,
and therefore replace the Fourier-transform $U(1)^3$ field $\psit$ by a tensor field $\bT$ of size $2N+1$.

\subsection{Renormalisation}
\subsubsection{Mass renormalisation}
The bare amplitude for $\cM$ is the sum of three amplitudes with colour $c$, each of which is a function of the single incoming index $n_c$,
\begin{align}  
A(\cM)\, & = \sum_c   A(\cM_c),  \crcr
A(\cM_c) (n_c)\, & =  -\lambda  \sum_{p  \in  \{-N\dots N\}^3}  \frac{\delta (p_c -n_c) }{p^2 + 1} .
\end{align}
The sum over $p$ diverges logarithmically as $N \to \infty$. We define the mass counterterm as minus the value at $n_c =0$, namely
\begin{align} \label{eq:defdelta1}
\Delta m \ &=\, -\sum_c A(\cM_c) (0)\ =\, \lambda \sum_c  \delta m^c\ ,   \crcr
\delta m^c \  &=\, \sum_{p  \in  \{-N\dots N\}^3}  \frac{\delta(p_c) }{p^2 + 1} \ =\, 
\sum_{p  \in  \{-N\dots N\}^2}  \frac{1}{p^2 + 1}\ . 
\end{align}
Remark that $ \delta m^c $ is independent of $c$, so that in fact 
\begin{equation}  \Delta m = 3 \lambda  \sum_{p  \in  \{-N\dots N\}^2}  \frac{1}{p^2 + 1} .
\end{equation}
Note that the renormalised amplitude of $\cM$ at colour $c$ is a convergent sum, hence no longer requires the cut-off $N$:
\begin{align}  A^{ren}(\cM_c) (n_c)&=  A(\cM_c) (n_c)  + \delta m^c = - \lambda  \sum_{p   \in \{-N\dots N\}^3}   \frac{\delta (p_c -n_c)  - \delta (p_c) } 
{p^2+ 1} = \lambda  A(n_c)\\
A(n_c) &= \sum_{p   \in \{-N\dots N\}^2}   \frac{ n_c^2} {(n_c^2 + p^2+ 1)(p^2 +1)}  =  \underset{{N\to\infty}}O\left( \log (1+ \vert n_c \vert )\right) . \label{Abound}
\end{align}
The partition function with this mass counter term included is 
\begin{equation} 
Z_1 = \int   e^{-\frac{\lambda}{2}    \sum_c V_c (\bT, \bar \bT) + \lambda \sum_c \delta m^c  \bT\cdot_\cD \bar \bT  }    d\mu_{\bC} (\bT, \bar \bT).
\end{equation}
\subsubsection{Vacuum renormalisation}
\begin{figure}[!h]
  \centering
  \includegraphics{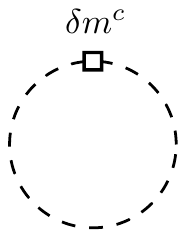}
  \caption{The divergent vacuum graph generated by the mass counter term.}
  \label{Fig:vacuumDM}
\end{figure}

We should similarly compute the vacuum counter-terms,  taking into account the presence of the $\Delta m$ counter term. 
The amplitude of $\cV_1$ is the sum over colours $c$ of the amplitudes of the graphs $\cV_1^c$. $\cV^1_1$ requires the counter-term
\begin{equation} 
\delta \cV_1^1 =  \frac{\lambda}{2} \sum_{n_1 , n_2, n_3, p_2, p_3 } \frac1{n_1^2+ n_2^2 +n_3^2 +1} \frac1{n_1^2 + p_2^2 +p_3^2 +1} .
\end{equation}
Similarly the amplitude of $\cV_2$ is the sum over colours of the amplitude of $\cV_2^c$. $\cV_2^1$ requires the counter term
\begin{equation} 
\delta \cV_2^1    =  \frac{\lambda}{2}\sum_{n_1 , n_2, n_3, p_1 } \frac1{n_1^2+n_2^2+n_3^2 +1} \frac1{p_1^2+n_2^2+n_3^2 +1} .
\end{equation}
Finally the mass counter-term $[\delta m^c\;  \bT \cdot\bar \bT] $ \eqref{eq:defdelta1} 
generates a divergent vacuum graph $\cV_{\Delta m}$ (Fig. \ref{Fig:vacuumDM}) which requires a different counter-term :
\begin{equation} 
\delta\cV_{\delta m}^1 = \delta m^c\;  \tr_\cD\bC = - \lambda  \sum_{n_1 , n_2, n_3, p_2, p_3 } \frac1{n_1^2+ n_2^2 +n_3^2 +1} \frac1{p_2^2 + p_3^2 +1}.
\end{equation}

The counter-terms for colours 2 and 3 are obtained by 
the same formulas with the appropriate colour permutation.
The renormalised partition function writes,
\begin{equation}  
Z = e^{ \sum_c (\delta \cV_1^c + \delta \cV_2^c + \delta\cV_{\delta m}^c) } 
\int   e^{-\frac{\lambda}{2} \sum_c V_c (\bT, \bar \bT) }e^{ \lambda \sum_c \delta m^c \sum_n \bT_n \bar \bT_n  }  d\mu_{C} (\bT, \bar \bT)\ .
\end{equation}

\subsection{Intermediate field representation}

As for the invariant quartic models, the constructive study of the $T^4_3$ field theory will be performed within the intermediate field formalism.
The intermediate field transformation is very similar to the one used for invariant models in section \ref{sec:IFformula}. 
\begin{equation}
 e^{-\frac{\lambda}{2} V_c(\bT, \bar \bT) }= \int e^{i\sqrt{\lambda} \sum_{n,  \bar n} \prod_{c'\neq c}\delta_{n_{c'} \bar n_{c'}} 
 \left( T_n\bar T_{\bar n}   \right)\csigma^c_{n_c\bar n_c}}    d\mu(\csigma^c).
\end{equation}
It will however require some adaptations to 
accommodate both the mass counter-term $\Delta m$ and the non-invariant covariance $\bC$.
%
%
\subsubsection*{Mass counterterm and Hubbard Stratonovich transformation}
The mass counter-term can be absorbed in a translation of the quartic interaction. Indeed remark that
\begin{align}   \label{reninter}
&\frac{1}{2}  V_c (\bT, \bar \bT)  -  \delta m^c \sum_n \bT_n \bar \bT_n  \ 
+\ \frac{1}{2}    \sum_{n, p} \delta_{n_c , p_c} \frac{  1 }{ (n^2 -n_c^2 )+ 1} \frac{  1 }{ (p^2 -p_c^2 )+ 1} 
 \crcr
&=\sum_{n,\bar n, p,\bar p}  \biggl[  \prod_{c'\neq c}\delta_{n_{c'} \bar n_{c'}}  \left( \bT_n\bar \bT_{\bar n} -   
 \frac{   \delta_{n_c \bar n_c}  }{ (n^2 -n_c^2 )+ 1}    \right)   \biggr]\crcr
&\times \delta_{n_c \bar p_c} \delta_{p_c \bar n_c} 
 \biggl[   \prod_{c'\neq c} \delta_{p_{c'} \bar p_{c'}}  \left(\ \bT_p\bar \bT_{\bar p} - 
 \frac{ \delta_{p_c \bar p_c} }{(p^2 -p_c^2)+ 1}   \right)  \biggr]  .
\end{align}
Therefore, defining
\begin{equation} 
\delta \cV_3^1    =  
\frac{\lambda}{2}\sum_{n_1 , n_2, n_3, p_2, p_3  \in  \{-N\dots N\}^5 } \frac1{n_2^2+n_3^2 +1} \frac1{p_2^2+p_3^2 +1} = \frac{\lambda}{2} N (\delta m^c)^2 ,
\end{equation}
and following the same definition for $\cV_3^1$ and $\cV_3^1$, $Z$ can be evaluated from \eqref{reninter} as
\begin{align}\label{eq:tfthubstr}
Z&=  e^{ \sum_c (\delta \cV_1^c + \delta \cV_2^c + \delta \cV_3^c + \delta\cV_{\delta m}^c) }\crcr
&\times\int   e^{ i\sqrt\lambda  \sum_c\sum_{n, \bar n} \left( \bT_n{\bar \bT}_{\bar n}
  \prod_{c'\neq c}\delta_{n_{c'} \bar n_{c'}}-  \delta m^c \delta_{n_c \bar n_c}   \right) \csigma^c_{n_c \bar n_c}  }   d\mu( \csigma)
  d\mu_{\bC}(\bT, \bar \bT) \ ,  
%
\end{align}
where we used the Hubbard Stratonovich  formula \eqref{eq:Hubbard}, introducing three Hermitian matrices $\csig^c$ with standard Gaussian measure $\mu$.  
\subsubsection*{Integration and resolvent}
Note that the previous equation can be re-written in order to isolate the contribution of the tensor entries
\begin{align}
Z&=  e^{ \sum_c (\delta \cV_1^c + \delta \cV_2^c + \delta \cV_3^c + \delta\cV_{\delta m}^c) }\crcr
&\times\int   e^{ i\sqrt\lambda  \sum_c\sum_{n, \bar n} \left( \bT_n{\bar \bT}_{\bar n}
  \prod_{c'\neq c}\delta_{n_{c'} \bar n_{c'}}   \right) \csigma^c_{n_c \bar n_c}  }   d\mu_{\bC}(\bT, \bar \bT)\
  e^{-i\sqrt\lambda\sum_c  \delta m^c \tr_c\csig^c}d\mu( \csigma)
  \ ,  
\end{align}
The integral over $\bT,\ \bar\bT$ now appears explicitly Gaussian, with effective covariance
\begin{equation}
 \bC_{\rm eff}(\csig) = \bC + i\sqrt{\lambda}\, \vec\csig\ , \quad {\rm with}\quad \vec\csig=\sum_{c=1}^3 \csig^c\otimes\id^{\cD\setminus\{c\}}
\end{equation}
After integration of $\bT,\ \bar\bT$, and recombining $\Det\,\bC_{\rm eff}$ with the $\Det\,\bC^{-1}$ in the measure $d\mu$, we obtain
\begin{align}
 Z=   Z'  \int    e^{-i\sqrt\lambda  \sum_c \delta m^c {\tr_c }\csigma^c   } e^{ 
  -{\rm \bf Tr} \log \left[\id-i\lambda    \bC \vec \csigma   \right]  } d\mu( \csigma)\ ,
\end{align}
where
\begin{equation} 
Z'  =  e^{ \delta \cV_1^c + \delta \cV_2^c + \delta \cV_3^c + \delta\cV_{\delta m}^c}\ .
\end{equation}
In the following developments, it will be usefull to rewrite the trace of the logarithm with a more symmetrical argument, which will be usefull when performing 
iterated Cauchy Schwarz bounds similar to those used in section \ref{subsec:CSbounds} .
Defining the resolvent,
\begin{equation} 
R (\csigma)  \equiv \frac{1}{  \id -i \sqrt\lambda \bC^{1/2}  \vec \csigma \bC^{1/2}  } ,
\end{equation}
and using the cyclicity of the trace, the partition function writes
\begin{align}
 Z=   Z'  \int    e^{-i\sqrt\lambda  \sum_c \delta m^c {\tr_c }\csigma^c   } e^{ 
  {\rm \bf Tr} \log R(\csig)  } d\mu( \csigma)\ ,
  \label{eq:tftif1}
\end{align}
We remark that the first term in the expansion in $\lambda$ of ${\rm \bf Tr} \log R(\csig) $ 
combines nicely with the mass counter-term
$-i\sqrt\lambda  \sum_c \delta m^c {\tr_c }\csigma^c  $, since
\begin{equation}
{\rm \bf Tr} \, \bC {\vec \csigma } - \sum_c \delta m^c {\tr_c }\csigma^c  = \sum_c  \sum_{n_c}A(n_c) \csigma^c_{n_c n_c}. \label{vecd}
\end{equation}
where $A(n_c)$ is the renormalised amplitude defined in \eqref{Abound}

Joining \eqref{eq:tftif1} and \eqref{vecd}
gives
\begin{equation}  \label{niceequat1}
Z = Z'   \int 
e^{ i \sqrt\lambda \sum_c  \sum_{n_c}A(n_c) \csigma^c_{n_c n_c} +{\rm \bf Tr} \log_2 \left[\id-i\lambda \bC^{1/2}  \vec \csigma \bC^{1/2}  \right]    }
d\mu( \csigma)\  ,
\end{equation}
where 
\[
\log_2 (1-x) = x+ \log (1-x) = O(x^2)\ .
\]

\subsubsection*{Vacuum counter-terms}

We will now study $Z'$ to check that it compensates indeed the divergent vacuum graphs of the $\csigma$ functional integral.
First remark that $ \delta \cV_1^c  + \delta \cV_3^c + \delta \cV_{\delta m}^c$ nicely recombine as
\[
 \sum_{n_c=-N}^N[A(n_c)]^2=2\left( \delta \cV_1^c  + \delta \cV_3^c + \delta \cV_{\delta m}^c\right) \ ,
\]
Therefore, we define the counter-term $\cD$ as,
\begin{equation}
\cD  = \sum_c \delta \cV_1^c  + \delta \cV_3^c + \delta \cV_{\delta m}^c  = \frac{\lambda}{2}  \sum_c \sum_{n_c}  A^2(n_c)
\end{equation}
Similarly we remark that the log-divergent counter term for $\cV_2$ can be written as a $\csigma$ integral,
\begin{equation}
\cE= \sum_c \delta \cV^c_2 = \frac{\lambda}{2}  \int  {\tr}_\cD  \left[( \bC {\vec \csigma })^2\right] d\mu( \csigma)
\ .
\end{equation}
Therefore
\begin{equation}  \label{niceequat2}
Z= \int  e^{- {\tr} \log_2 \left[\mathbb{I}-i\sqrt\lambda \bC^{1/2}  \vec \csigma \bC^{1/2}  \right] 
+  i \sqrt\lambda  \sum_c  \sum_{n_c}A(n_c) \csigma^c_{n_c n_c}  + \cD + \cE  }d\mu( \csigma) \ .
\end{equation}
A quick study of the first Feynman maps shows that $\cD$ and $\cE$ cancel exactly  the first order term in $\lambda$, so that
\begin{equation}  
\log Z  = O(\lambda^2)\ .
\end{equation}

We remark now that 
\begin{equation} 
 e^{ i \sqrt\lambda  \sum_c  \sum_{n_c}A(n_c) \csigma^c_{n_c n_c} + \cD } d\mu(\csigma)
\end{equation}
is exactly the Gaussian normalised measure for three fields $\csig^c$ translated by a diagonal matrix $M_{n_c\bar n_c}=A(n_c)\delta_{n_c\bar n_c}$.
Indeed, 
\begin{equation}
e^{ i \sqrt\lambda    \sum_{n_c}A(n_c) \csigma^c_{n_c n_c} + \frac{\lambda}{2}  \sum_{n_c}  A^2(n_c) } d\mu(\csigma^c) = 
\prod_{n_c} e^{- \frac{1}{2}  \tr\left[( \sigma ^c  - i \sqrt\lambda M)^2\right] }d\csig d\bar\csig.
\end{equation}

Let us define the diagonal operator $D (n, \bar n) =\delta_{n \bar n}  D( n) $ with eigenvalues 
\begin{equation}  D( n) \equiv  \sum_c \bC_{nn}  A(n_c) .
\end{equation}
This operator commutes with $\bC$ since they are both diagonal. It is bounded 
uniformly in $N $ since from 
\eqref{eq:propdef} and \eqref{Abound} we have 
\begin{equation} \label{Inbound}
\Vert  D( n) \Vert  \le   K\ ,
\end{equation}
for some finite constante $K$.
In fact $D$ is also compact as an infinite dimensional operator on $\ell^2 ({\mathbb Z}^3)$, hence  at $N = \infty$, and its square is trace class, since
\begin{equation} \label{squaretraceclass}
\sum_{n \in {\mathbb Z}^3} \sum_{c,c'}  \frac{\log (1+ \vert n_c \vert) \log (1+ \vert n_{c'} \vert)  }{(n^2 +1)^2}   = O(1)\ . 
\end{equation}

 
\begin{lemma} \label{lemmaresbounded} 
For $\lambda=|\lambda|e^{i\phi}$ in the small open cardioid domain $Card_\rho$
defined by $\vert \lambda \vert < \rho \cos [\phi/2]$ ,
the translated resolvent
\begin{equation}
R = [\id  -i\sqrt\lambda \bC^{1/2}  \vec \csigma \bC^{1/2} + \lambda D ]^{-1} 
\end{equation}
is well defined and uniformly bounded:
\begin{equation}\label{rescardbou}
\Vert R \Vert \le  \frac2{  \cos( \phi /2)}\ .
\end{equation}
\end{lemma}
\prf  In the cardioid domain  we have $\vert \phi \vert < \pi$ and for any self-adjoint operator $L$ 
we have 
\begin{equation} \Vert ( \id  -i\sqrt\lambda L )^{-1} \Vert \le \cos^{-1}( \phi /2).
\end{equation}
Taking  $\rho $ small enough so that $\rho\Vert  D( n) \Vert  < 1/2$, 
the Lemma follows from the power series expansion 
\begin{equation}
\Vert ( \id  -i\slambda L  +\lambda D )^{-1} \Vert \le \Vert J^{-1} \Vert  \sum_{q=0}^{\infty} \Vert \lambda D J^{-1} \Vert^q \le 2   \cos^{-1}( \phi /2)  ,
\end{equation}
with $J=  \id  -i \sqrt\lambda L $.
\qed

\begin{lemma} \label{lemmatrans} 
For $\lambda$ in the cardioid domain $Card_\rho$, the successive contour translations from $\csigma ^c_{n_c n_c}  $ to
$ \csigma ^c_{n_c n_c}  - i \sqrt\lambda A(n_c) $ do not cross any singularity of ${\trd} \log_2 \left[\id  -i\sqrt\lambda \bC^{1/2}  \vec \csigma \bC^{1/2}  \right]$.
\end{lemma}
\prf 
To prove that $ {\trd} \log_2 [\id  -i\slambda \bC^{1/2}  \vec \csigma \bC^{1/2}]$ is  analytic 
in the combined  translation band of imaginary width $\slambda A(n_c) $ 
for the $\csigma ^c_{n_c n_c}$ variables, one can write
\begin{equation}  \log_2 (1-x) =   - \int_0^1 \frac{tx^2}{1-tx} dt
\end{equation}
and then use the previous lemma to prove that, for $\lambda$ in the small open cardioid domain $Card_\rho$, 
the resolvent $R(t)= [\id  -it\slambda \bC^{1/2}  \vec \csigma \bC^{1/2} + t\slambda^2 D ]^{-1}$, is also well-defined for any $t \in [0,1]$ by a power series 
of analytic terms uniformly convergent  in the band of of imaginary width $\slambda A(n_c) $. Hence it is analytic in that band. \qed

Hence by Lemma \ref{lemmatrans}
\begin{equation}  \label{niceequat3}
Z = \int  e^{- {\trd} \log_2 \left[\id 
-i\slambda \bC^{1/2}  \vec \csigma \bC^{1/2} + \lambda D \right]  + \cE  }d\mu(\csigma)
= \int  e^{- V(\csigma)  } d\mu(\csigma)\ ,
\end{equation}
where the $\csigma$ interaction is now defined as
\begin{equation}
V(\csigma)  =   {\trd} \log_2 \left[\id -  U \right] - \cE ,  \quad U = i\slambda \bC^{1/2}  \vec \csigma \bC^{1/2}   - \lambda D .
\end{equation}

\section{The multi-scale loop vertex expansion}
\subsection{Scale slices in the intermediate field representation}\label{subsec:slicesIF}

The ``cubic" cut-off $[-N, N]^3$ of the previous section is not very well adapted to the multi-scale analysis of section \ref{sec:TFT}.
We will therefore use the scale slices of \eqref{eq:slices},
\begin{align}
S^1 &= \left\{ p\in\setZ^D,\ \,n^2 +1 \,\le\, M^{2} \right\}\crcr
 S^j &= \left\{ p\in\setZ^D,\ M^{2j-2}\,<\,n^2+1 \,\le\, M^{2j} \right\}\quad\forall j> 1 \ ,
\end{align}
and the corresponding cut-off of scale $j_{\max}$,
\[
 n^2+1 \leq M^{2j_{\max}}
\]
which no longer factorises over colours. This cut-off will later be applied on the covariance $\bC$, and thus on the interaction term of the intermediate field.
To properly define the intermediate field $\csig$ however,  requires the use of previous cut-off. This is why we only introduce the new cut-off now. It is required 
that the new cut-off is tighter that the cubic one, thus one must choose $N$ such that $M^{j_{\max}}\leq N$.

The action \eqref{niceequat3} has to be rewritten according to the new cut-off.
We define the slice characteristic functions and introduce an interpolation parameter $t_j$, 
which will help us compute the part of the interaction specific to a scale $j$.
\begin{align}
\bI_j =\bI_{S^j}\ ,\qquad
\bI_{\le j}(t_j) = \sum_{i=1}^{j-1} \bI_i + t_j \bI_j \ . 
\label{propmombound}
\end{align}
The interaction with cut-off $j_{\max}$ is $V_{\le j_{\max}}(1)$, where the interpolated 
%
interaction is defined as
\begin{align}
V_{\le j} (t_j) &=      {\tr}_\cD \log_2 \left[  \id - U_{\le j} (t_j) \right] -  \cE_{\le j}  (t_j) 
\ , \\
U_{\le j}(t_j)  &=   i \sqrt\lambda \bI_{\le j} (t_j)  \bC^{1/2} {\vec \csigma } \bC^{1/2} \bI_{\le j}(t_j) -  \lambda \bI_{\le j} (t_j)  D , \\
\cE_{\le j}   (t_j)  &=  \frac{\lambda}{2} \int  {\tr}_\cD  \left[( \bI^2_{\le j}   (t_j) \bC{\vec \csigma })^2  \right] d\mu(\csigma)
\ .
\end{align}

Remark that 
\begin{equation} \label{boundcoun}
0 \le   \cE_{\le j}  \le O(j)  , \quad  0 \le \cE_{\le j}  -   \cE_{\le j-1} \le O(1)  .
\end{equation}

We also define the interpolated resolvent
\begin{equation}
R_{\le j}(t_j)  = \frac{1}{  \id - U_{\le j} (t_j) }  .
\end{equation}
When the context is clear, we write simply $V_{\le j} $ for $V_{\le j} (t_j) $, $U_{\le j} $ for $U_{\le j} (t_j) $, $U'$ for $\frac{d}{dt_j}  U_{\le j}  $ and so on. 
We also write $\bC^{1/2}_{\le j}$ for $  \bI_{\le j}(t_j) \bC^{1/2} $,  $\bC^{1/2}_{j}$ for $  \bI_{j} \bC^{1/2} $,
$\bC_{j}$ for $\bI_{j} \bC $, $D_{\le j}$ for $ \bI_{\le j}(t_j) D$ and $D_{j}$ for $ \bI_{ j} D $. 
However beware that we shall write
$\bC_{\le j}$  for $ \bI^2_{\le j}(t_j) C$, as this is the natural expression which gives us the natural relations
\begin{equation}  [\bC^{1/2}_{\le j}  ]^2  = \bC_{\le j}, \quad  [ \bC^{1/2}_{j} ]^2 = \bC_j  .
\end{equation}

We now decompose the interaction over scales, defining the interaction term of scale $j$ as 
\begin{align} 
V_{j} &=  V_{\le j}(1)  - V_{\le j-1}(1) = \int_0^1  V'_{\le j}(t_j)\, dt_j \ , \\
{\rm with} \quad V'_{\le j}&=      {\tr}_\cD \;[
U'_{\le j}  (  \id  -  R_{\le j} )]  - \cE_{\le j}'  \ ,\\
U'_{\le j} &= i\sqrt\lambda    \bC^{1/2}_{j} \vec \csigma \bC^{1/2}_{\le j} +  i\sqrt\lambda  \bC^{1/2}_{\le j}\vec \csigma \bC^{1/2}_{j}  - \lambda D_j 
\ .
\end{align}
Now we use that $(  \id  -  R_{\le j} ) = - U_{\le j} R_{\le j} = - R_{\le j}U_{\le j}$ 
and the cyclicity of the trace, plus relations such as 
$ \bC^{1/2}_{\le j} \bC^{1/2}_{j} = t_j  \bC_j $  to write
\begin{align}  \label{keyequati}
\tr_\cD \; U'_{\le j}  (  \id  -  R_{\le j} )  =&
- i\sqrt\lambda \ {\tr}_\cD  [R_{\le j} U_{\le j}  \bC^{1/2}_{j} \vec \csigma \bC^{1/2}_{\le j}  \\
&+\ R_{\le j} \bC^{1/2}_{\le j}\vec \csigma \bC^{1/2}_{j} U_{\le j}  \
+\ i  \sqrt\lambda R_{\le j}U_{\le j} D_j ]  \\
=& \
\lambda\ {\tr}_\cD\  R_{\le j} [ 2 t_j \bC^{1/2}_{\le j}\vec \csigma   \bC_{j} \vec \csigma \bC^{1/2}_{\le j}   \\
  & +i \sqrt\lambda ( D_{\le j} \bC^{1/2}_{j} \vec \csigma \bC^{1/2}_{\le j}  + \bC^{1/2}_{\le j}\vec \csigma \bC^{1/2}_{j} D_{\le j} ) + U_{\le j} D_j   ] 
  \\
\tr_\cD \; \cE'_{\le j}  =& 2\lambda t_j\int   \tr_\cD ( \bC_{ j}  {\vec \csigma } \bC_{\le j}  {\vec \csigma })\ d\mu( \csigma) \ .
\end{align}

Remark that if we replace $R_{\le j}$ by $\id $ in the first term in \eqref{keyequati}  it would exactly 
cancel the $\cE'_{\le j} $ term. This is nothing but again the exact cancellation
of the last vacuum graph in Figure \ref{divergences} with its counter term.

We now have 
\begin{equation}
Z(j_{max})= \int \prod_{j =0}^{j_{\max}}     e^{ - V_j } \; d\mu(\csigma) , \label{factoredintera}
 \end{equation}
and we can apply the jungle expansion from section \ref{sec:jungleexp}.

\subsection{Jungle expansion}

With the intermediate field interaction decomposed over scale slices, we can now perform the two-level jungle expansion of \ref{sec:jungleexp}.
According to the theorem \ref{thm:jungle}, the logarithm of the partition function writes

\begin{align} \label{eq:logZjungle}  
\log Z=&  \sum_{n=1}^\infty \frac{1}{n!}  \sum_{\cJ \;{\rm connected}} \;\sum_{\{j_a\} \in \cS^{n} } \\
 &\times\; \int_{w}\ \int_{\csig}\ \int_\chi
\  \partial_\cJ   \Big[ \prod_{\cB} \prod_{a\in \cB}   \Bigl(    W_{j_a}   (  \vec \sigma^a )  
 \chi^{ \cB }_{j_a} \bar \chi^{\cB}_{j_a} \Bigr)    \Big] \;dw_\cJ  \; d\mu_{ \cJ}   , 
\end{align}
where
\begin{itemize}
\item the sum over $\cJ$ runs over all connected two-level jungles, the first level of which is a 3-coloured forest $\cF_B$ called the Bosonic forest,
The second level (the Fermionic forest) $\cF_F$ is uncoloured. And
$\bar \cJ = \cF_B \cup \cF_F $ is connected (hence, a tree).
Bosonic edges $\e_B \in \cF_B$ have a well-defined colour $c(\e) \in \{1,2,3\}$ 
and Fermionic edges $\e_F \in \cF_F$ are uncoloured.
 
\item  $\int dw_\cJ$ means integration from 0 to 1 over parameters $w_\e$, one for each edge $\e \in \bar\cJ$, namely
$\int dw_\cJ  = \prod_{\e\in \bar \cJ}  \int_0^1 dw_\e  $.
There is no integration for the empty forest since by convention an empty product is 1. A generic integration point $w_\cJ$
is therefore made of $\vert \bar \cJ \vert$ parameters $w_\e \in [0,1]$, one for each $\e \in \bar \cJ$.

\item 
\begin{equation} \partial_\cJ  = \prod_{\genfrac{}{}{0pt}{}{\e_B \in \cF_B}{\e_B=(a,b)}} \Bigl(
\frac{\partial}{\partial (\csigma^{c(\e_B)})^a}\frac{\partial}{\partial (\csigma^{c(\e_B)})^b} \Bigr)
\prod_{\genfrac{}{}{0pt}{}{\e_F \in \cF_F}{\e_F=(d,e) } } \delta_{j_{d } j_{e } } \Big(
   \frac{\partial}{\partial \bar \chi^{\cB(d)}_{j_{d}  } }\frac{\partial}{\partial \chi^{\cB(e)}_{j_{e}  } }+ 
    \frac{\partial}{\partial \bar \chi^{ \cB( e) }_{j_{e} } } \frac{\partial}{\partial \chi^{\cB(d) }_{j_{d}  } }
   \Big) \; ,
\end{equation}
where $ \cB(d)$ denotes the Bosonic block to which the vertex $d$ belongs. 

\item The measure $d\mu_{\cJ}$ has covariance $ X (w_{\e_B}) \otimes \bbone_\cS $ on Bosonic variables and $ Y (w_{\e_F}) \otimes \id_\cS  $  
on Fermionic variables, hence
\begin{equation}
\int F d\mu_{\cJ}  = \biggl[e^{\frac{1}{2} \sum_{a,b=1}^n X_{ab}  \sum_{c } \frac{\partial}{\partial (\csigma^c)^a}\frac{\partial}{\partial (\csigma^c)^b} 
   +  \sum_{\cB,\cB'} Y_{\cB\cB'}\sum_{\genfrac{}{}{0pt}{}{a\in \cB}{b\in \cB'} } \delta_{j_aj_b}
   \frac{\partial}{\partial \bar \chi_{j_a}^{\cB} } \frac{\partial}{\partial \chi_{j_b}^{\cB'} } }   F \biggr]_{\csigma  =\chi =0}\; .
\end{equation}

\item  $X_{ab} (w_{\e_B} )$  is the infimum of the $w_{\e_B}$ parameters for all the Bosonic edges $\e_B$
in the unique path $P^{\cF_B}_{a \to b}$ from $a$ to $b$ in $\cF_B$. The infimum is set to zero if such a path does not exists and 
to $1$ if $a=b$. 

\item  $Y_{\cB\cB'}(w_{\e_F})$  is the infimum of the $w_{\e_F}$ parameters for all the Fermionic
edges $\e_F$ in any of the paths $P^{\cF_B \cup \cF_F}_{a\to b}$ from some vertex $a\in \cB$ to some vertex $b\in \cB'$. 
The infimum is set to $0$ if there are no such paths, and to $1$ if such paths exist but do not contain any Fermionic edges.

\end{itemize}

The previous expression \eqref{eq:logZjungle} factorises over the Fermionic and Bosonic part. And as the Bosonic blocks are completely independents, 
it also factorises over blocks as,
\begin{align}  \label{eq:logZfactorised}
&\log Z= \sum_{n=1}^\infty \frac{1}{n!}  \sum_{\cJ \;{\rm connected}} \;\sum_{\{j_a\} \in \cS^{n} }  \int_{w}\\
&\left[ \int_{\chi}
   \prod_{\genfrac{}{}{0pt}{}{\e_F \in \cF_F}{\e_F=(d,e) } } \delta_{j_{d } j_{e } } \Big(
   \frac{\partial}{\partial \bar \chi^{\cB(d)}_{j_{d}  } }\frac{\partial}{\partial \chi^{\cB(e)}_{j_{e}  } }+ 
    \frac{\partial}{\partial \bar \chi^{ \cB( e) }_{j_{e} } } \frac{\partial}{\partial \chi^{\cB(d) }_{j_{d}  } }
   \Big)
    \left(\prod_{a\in \cV}    
 \chi^{ \cB(a) }_{j_a} \bar \chi^{\cB(a)}_{j_a} \right)\
 d\mu_{Y\otimes\id_\cS}
\right]
\\ &\times \prod_\cB \left[ \int_{\csig}
 \prod_{\genfrac{}{}{0pt}{}{\e_B \in \cF_B\cap\cB}{\e_B=(a,b)}} \Bigl(
\frac{\partial}{\partial (\csigma^{c(\e_B)})^a}\frac{\partial}{\partial (\csigma^{c(\e_B)})^b} \Bigr)
\prod_{a\in \cB}   \Bigl(    W_{j_a}   (  \vec \csigma^a )  
  \Bigr) \
d\mu_{X \otimes \bbone_\cS}
\right]  \  dw_F dw_B, 
\end{align}

\section{Convergence and analyticity}

The main result of the present chapter is the absolute convergence of the multi-scale loop vertex expansion ,

\begin{theorem} \label{T43theorem} Fix  $\rho >0$ small enough.
The series \eqref{eq:logZjungle} is absolutely and uniformly in $j_{\max}$ convergent for $\lambda$ in the small open cardioid domain ${\rm Card}_\rho$
defined by $\vert \lambda \vert < \rho \cos [({\rm Arg} \; \lambda )/2]$.
Its ultraviolet limit $\log Z (g) = \lim_{j_{max}  \to \infty}  \log Z(g, j_{max})$ is therefore well-defined and
analytic in that cardioid domain. 
\end{theorem}

The rest of the section is devoted to the proof of this Theorem.

\subsection{Grassmann Integral}

We define the symmetric matrix ${\bY}_{ab} = Y_{\cB(a)\cB(b)}\delta_{j_a j_b}$. 
Using the shorthand notation $\partial_X=\frac{\partial}{\partial X}$, the Grassmann integral in \eqref{eq:logZfactorised} writes,
\begin{align}
 \left[
 e^{
 \sum_{a,b\in\cV} \bY_{ab}
    \partial_{ \bar \chi_{j_a}^{\cB(a)} } \partial_{ \chi_{j_b}^{\cB(b)} } 
    }
   \prod_{\genfrac{}{}{0pt}{}{\e_F \in \cF_F}{\e_F=(d,e) } } \delta_{j_{d } j_{e } } \Big(
   \partial_{ \bar \chi^{\cB(d)}_{j_{d}  } }\partial_{ \chi^{\cB(e)}_{j_{e}  } }+ 
    \partial_{ \bar \chi^{ \cB( e) }_{j_{e} } } \partial_{ \chi^{\cB(d) }_{j_{d}  } }
   \Big)
    \prod_{a\in \cV}    
 \chi^{ \cB(a) }_{j_a} \bar \chi^{\cB(a)}_{j_a} 
 \right]_{\chi=0}
 \ .
\end{align}
Note that after expanding the exponential, the above expression can be written in terms of a polynomial $P$ of degree $2n$, and a power series $Q$ as,
\[
 Q(\{\partial_{\bar\chi_j^\cB},\partial_{\chi_j^\cB}\})
 P(\left\{\chi_j^\cB,\bar\chi_j^\cB\right\}) \ ,
\]
where $P$ and $Q$ can be written in the general form $P(\{\chi,\bar\chi\}) = \sum_{k} P_k \chi_1\dots\chi_k\bar\chi_k\dots\chi_1$, and
therefore,
\[
 Q(\partial_{\bar\chi_k},\partial_{\chi_k})
 P(\chi_l,\bar\chi_l) = \sum_{k,l} \delta_{kl} Q_k P_l 
 = P(\partial_{\chi_k},\partial_{\bar\chi_k})
 Q(\chi_l,\bar\chi_l) \ ,
\] 
which is a polynomial as $P_l=0$ for $l>n$, hence the sum is finite. The Grassmannian Gaussian integral thus writes,
\begin{align}
 \left[\left(
 \prod_{a\in \cV}    
 \partial_{\chi^{ \cB(a) }_{j_a}}\partial_{ \bar \chi^{\cB(a)}_{j_a}} \right)
 e^{
 \sum_{a,b} 
    {  \chi_{j_a}^{\cB(a)} }{\bY}_{ab} {\bar \chi_{j_b}^{\cB(b)} } 
    }
   \prod_{\genfrac{}{}{0pt}{}{\e_F \in \cF_F}{\e_F=(d,e) } } \delta_{j_{d } j_{e } } \Big(
   {  \chi^{\cB(d)}_{j_{d}  } }{ \bar\chi^{\cB(e)}_{j_{e}  } }+ 
    {  \chi^{ \cB( e) }_{j_{e} } } {\bar \chi^{\cB(d) }_{j_{d}  } }
   \Big)
 \right]_{\chi=0}
 \ .
\end{align}
As for Grassmannian integrals, $\int F d\chi d\bar\chi = \frac{\partial F}{\partial\chi\partial\bar\chi}$, the above expression can be written as a integral,
\begin{align}
 \int 
 e^{-
 \sum_{a,b} 
    { \bar \chi_{j_a}^{\cB(a)} }{\bY}_{ab} { \chi_{j_b}^{\cB(b)} } 
    }
   \prod_{\genfrac{}{}{0pt}{}{\e_F \in \cF_F}{\e_F=(d,e) } } \delta_{j_{d } j_{e } } \Big(
   { \bar \chi^{\cB(d)}_{j_{d}  } }{ \chi^{\cB(e)}_{j_{e}  } }+ 
    { \bar \chi^{ \cB( e) }_{j_{e} } } { \chi^{\cB(d) }_{j_{d}  } }
   \Big)
   \left(
 \prod_{a\in \cV}    
 d \chi^{ \cB(a) }_{j_a} d \bar \chi^{\cB(a)}_{j_a} \right)
 \ .
\end{align}
Note that if two vertices $a$ and $b$ belonging to the same block $\cB$ have the same scale $j_a=j=b$, 
the above integral vanishes, as the same variable $\chi^\cB_{j_a}$ is integrated twice.

We denote $k=|E_F|$ the number of Fermionic edges, and for any matrix $\bM$,
\[
 \bM^{b_1\dots b_k}_{a_1\dots a_k} =\int e^{-\sum_{ab}\bar\phi_a \bM_{ab} \phi_b} \left(\prod_{i=1}^k \phi_{a_i}\bar\phi_{b_i}\right)\ \prod_a d\bar\phi_a d\phi_a
\]
which, up to a minus sign, are equal to the minors of $\bM$. 
The Grassmann integral thus writes,
\begin{align}
 \Bigl( \prod_{\genfrac{}{}{0pt}{}{\e_F \in \cF_F}{\e_F=(d,e) } } \delta_{j_{d } j_{e} } \Bigr)
 \Bigl( {\bf Y }^{ e_1 \dots  e_k}_{ d_1 \dots  d_k}  + 
 {\bf Y }^{ d_1 \dots  e_k}_{ e_1 \dots  d_k}+\dots + {\bf Y }_{ e_1 \dots  e_k}^{ d_1 \dots  d_k}   \Bigr) \; ,
\end{align} 
where the sum runs over the $2^k$ ways to exchange an $d_i$ and a $e_i$.

According to the forest formula, $\bY$ is positive, with its highest coefficient being $1$, on the diagonal. This  means that the $Y$ 
minors are all bounded by 1 \cite{AR1994,GR1312.7226}, namely
for any $a_1,\dots a_k$ and $b_1,\dots b_k$,
 \begin{equation}
   \Big{|}  {\bf Y }^{ a_1 \dots  a_k}_{ b_1 \dots  b_k} \Big{|}\le 1 \; .
 \end{equation}
and the Grassmanian integral is bounded by
\begin{equation}\label{eq:Grassbound}
 2^{|E_F|}
 \Bigl( \prod_{\cB} \prod_{\genfrac{}{}{0pt}{}{a,b\in \cB}{a\neq b}} (1-\delta_{j_aj_b}) \Bigr)
 \Bigl( \prod_{\genfrac{}{}{0pt}{}{\e_F \in \cF_F}{\e_F=(d,e) } } \delta_{j_{d } j_{e} } \Bigr)
 \; ,
\end{equation}
where the second term expresses the $j_a\not=j_b$ constraint inside a Bosonic block $\cB$.

\subsection{Bosonic Integrals}

The Bosonic integral in \eqref{eq:logZfactorised} is factorised over the Bosonic blocks,
In which the Bosonic forest restricts to a three-coloured Bosonic tree 
$\cT_{\cB}= \cF_B\cap \cB$, 
and the Bosonic Gaussian measure restricts to $d \nu_\cB$ defined by
\begin{equation}
\int F_\cB \ d \mu_\cB   = \biggl[ e^{\frac{1}{2} \sum_{a,b\in \cB} X_{ab} 
\sum_{c =1}^3 \frac{\partial}{\partial (\csigma^c)^a}\frac{\partial}{\partial (\csigma^c)^b}} F_\cB \biggr]_{\csigma =0}.
\end{equation}
The Bosonic integrand $F_\cB =  \prod_{\e \in \cT_{\cB} , \e_B=(a,b)} \Bigl(
\frac{\partial}{\partial (\csigma^{c(\e_B)})^a}\frac{\partial}{\partial (\csigma^{c(\e_B)})^b} \Bigr) \prod_{a\in \cB}  
\Bigl(    W_{j_a}   (  \vec \csigma^a ) \Bigr)
$ can be written in shorter notations as 
\begin{equation}   F_\cB =  \prod_{a\in \cB}    \bigl[ \prod_{s \in S^a_\cB} \partial_{\sigma(s)} W_{j_a}   \bigr]
\end{equation}
where $S^a_\cB$ runs over the set of all edges in $\cT_{\cB}$ which end at vertex $a$, hence $\vert S^a_\cB \vert = d_a( \cT_{\cB} ) $, the
degree or coordination of the tree $\cT_{\cB}$ at vertex $a$. To each element $s$ is therefore associated a well-defined colour and 
well-defined matrix elements
(which have to be summed later after identifications are made through the edges of $\cT_{\cB}$).

When $\cB$ has more than one vertex,
since $\cT_{\cB}$ is a tree, each vertex $a \in \cB$ is touched by at least one derivative and we can replace 
$W_{j_a} =e^{- V_{j_a}} -1$ by $ e^{- V_{j_a}}$ (the derivative of 1 giving 0) and write
\begin{equation} \label{manyder}  F_\cB =    \prod_{a\in \cB}    \bigl[ \prod_{s \in S^a_\cB} \partial_{\sigma(s)}  e^{- V_{j_a}}  \bigr].
\end{equation}
We can evaluate the derivatives in \eqref{manyder} through the Fa\`a di Bruno formula:
\begin{equation}
\prod_{s\in S} \partial_{\sigma(s)}   f\bigl( g( \csigma ) \bigr) =   \sum_{\pi } f^{(|\pi|)}\bigl( g( \csigma ) \bigr) \prod_{b\in \pi} 
\left(\bigl[ \prod_{s\in b} \partial_{\sigma(s)} \bigr] g (\csigma)\right) \; ,
\end{equation}
where $\pi$ runs over the partitions of the set $S$ and $b$ runs through the blocks of the partition $\pi$. In our case $f$, the exponential
function, is its own derivative, hence the formula simplifies to
\begin{align}\label{eq:partitio}
F_\cB&=   \prod_{a\in \cB}   e^{- V_{j_a}}   \biggl[  \sum_{\pi^a} \prod_{b^a\in \pi^a} \;  
\bigl[\prod_{s\in b^a} \partial_{\sigma(s)}\bigr]  (-V_{j_a})  \biggr]\; ,
\end{align}
where $\pi^a$ runs over partitions of $S^a_\cB$ into blocks $b^a$.

The Bosonic integral in a block can therefore be written as:
\begin{align}\label{eq:bosogauss1}
& \int   F_\cB \,d \mu_\cB =  \int  \prod_{a\in \cB}  e^{-V_{j_a} (\csigma_a) }  A_{\cT_\cB} (\csigma)
 \;d \mu_\cB\ ,
\end{align} 
where the amplitude  $A_{\cT_\cB}( \csigma)$ of the tree $\cT_\cB$ gather the derivatives. As for invariant models (Chapter \ref{chap:borelQM}), 
the action of the derivatives will express the amplitude of a tree as a sum over plane trees (definition \ref{def:planetree}). 

\subsubsection*{Tree derivatives}
As in \eqref{eq:derivative} for the invariant models, tree derivatives develop a vertex into cycles of operators and edge-contractions. 
Because of the more complicated structure of the interaction term, the exact expressions are unfortunately lengthy. 
There main characteristics and implications, which are sufficient for future treatment and bounding, will be summarised thereafter.

We recall that, with the notations of Section \ref{subsec:slicesIF}
\begin{align} \label{interaj4c}
V_{j} &= \lambda \int_0^1     \biggl(
\tr_\cD\ R_{\le j}  \biggl[ 2 t_j \bC^{1/2}_{\le j}\vec \csigma   \bC_{j} \vec \csigma \bC^{1/2}_{\le j}   
  + U_{\le j} D_j 
\crcr
&+i \sqrt\lambda   ( D_{\le j}\bC^{1/2}_{j} \vec \csigma \bC^{1/2}_{\le j}  + \bC^{1/2}_{\le j}\vec \csigma \bC^{1/2}_{j} D_{\le j}) \biggr] 
 -2 t_j \int d\nu(\vec \csigma)   \bC_{ j}  {\vec \csigma } C_{\le j}  {\vec \csigma } \biggr) \ dt_j
\\
&= \cD_j  + 2\lambda \int_0^1 t_j   \biggl[ 
{\rm \bf Tr} R_{\le j}     \bC^{1/2}_{\le j}\vec \csigma   \bC_{j} \vec \csigma \bC^{1/2}_{\le j}   
- \int d\nu(\vec \csigma)   \bC_{ j}  {\vec \csigma } \bC_{\le j}  {\vec \csigma }    \biggr] \ dt_j\ ,
\end{align}
where $\cD_j$ gathers all terms with a $D$ factor:
\begin{align}\label{defDj}
\cD_j &=  \lambda \int_0^1    {\tr_\cD} \, R_{\le j}  \bigl[ i \sqrt\lambda ( D_{\le j}  \bC^{1/2}_{j} \vec \csigma \bC^{1/2}_{\le j}  
+ \bC^{1/2}_{\le j}\vec \csigma \bC^{1/2}_{j}  D_{\le j} )  + U_{\le j} D_j ]\ dt_j \\
&= i\lambda^{3/2} \int_0^1   {\tr_\cD} \, R_{\le j} \bigl[ D_{\le j}  \bC^{1/2}_{j} \vec \csigma \bC^{1/2}_{\le j} 
+\bC^{1/2}_{\le j}\vec \csigma \bC^{1/2}_{j}  D_{\le j}    \crcr
&\qquad\qquad\qquad\qquad +\bC^{1/2}_{\le j}  \vec \csigma \bC^{1/2}_{\le j}D_j   +i\sqrt\lambda D_{\le j}   D_j \bigr] \  dt_j\ .
\end{align}
The last term in \eqref{interaj4c} is the $\cE'_{\le j} $  constant term, which does not depend on $\vec \csigma$. 
Hence remembering that $\partial \csigma_s$  really stand for a derivative with well defined colour and matrix elements
$\partial \csigma^{c_s}_{n^{c_s},\bar n^{c_s}}$, we get
following the order of the terms in \eqref{interaj4c}
\begin{align}\label{eq:developcycles}
 & \partial_{\sigma(1)} ( -V_j ) =  i\sqrt\lambda^{3/2}\int_0^1  \biggl( {\tr_\cD} \
  i\lambda R_{\le j}\bC^{1/2}_{\le j}\Delta^1 \bC^{1/2}_{\le j}R_{\le j} \\
 & \times \bigl[ t_j^2 D_{ j}  \bC^{1/2}_{j} \vec \csigma \bC^{1/2}_{\le j}  +
t^2_j \bC^{1/2}_{\le j}\vec \csigma \bC^{1/2}_{j}  D_{ j}  +  t_j \bC^{1/2}_{\le j}  \vec \csigma \bC^{1/2}_{j}D_j   +i\sqrt\lambda t_j^2 D_{ j}   D_j \bigr] \\
& +{\tr_\cD}\ R_{\le j} \bigl[ t_j^2 D_{j}  \bC^{1/2}_{j} \Delta^1  \bC^{1/2}_{\le j} 
+t^2_j \bC^{1/2}_{\le j}\Delta^1  \bC^{1/2}_{j}  D_{ j}  + t_j \bC^{1/2}_{\le j}  \Delta^1  \bC^{1/2}_{ j}D_j   \bigr]
\biggr)\ dt_j \\
&+ 2\lambda \int_0^1 t_j \biggl(  \ i\sqrt\lambda {\tr_\cD}\ R_{\le j}\bC^{1/2}_{\le j}\Delta^1
\bC^{1/2}_{\le j}R_{\le j}\left[   \bC^{1/2}_{\le j}\vec \csigma   \bC_{j} \vec \csigma \bC^{1/2}_{\le j}\right]\\ 
& + {\tr_\cD}\ R_{\le j}  \left[   \bC^{1/2}_{\le j}\Delta^1   \bC_{j} \vec \csigma \bC^{1/2}_{\le j}
+  \bC^{1/2}_{\le j}\vec \csigma   \bC_{j} \Delta^1 \bC^{1/2}_{\le j}  \right]\biggr)\ dt_j\ .
\end{align}
The formula for several successive derivations is similar and straightforward although longer:
\begin{align} \label{developcycles2}
  &\prod_{s=1}^k\partial_{ \sigma(s)} (-V_j ) = \sum_{\tau} i\lambda^{3/2}\int_0^1  \biggl( {\tr_\cD} \
  \bigl[ \prod_{s=1}^{k} i\sqrt\lambda R_{\le j} \bC^{1/2}_{\le j} \Delta^{\tau (s)} \bC^{1/2}_{\le j}] R_{\le j}\\
&\times \bigl[ t_j^2 D_{ j}  \bC^{1/2}_{j} \vec \csigma \bC^{1/2}_{\le j}  +t^2_j \bC^{1/2}_{\le j}\vec \csigma \bC^{1/2}_{j}  D_{ j} 
+ t_j \bC^{1/2}_{\le j}  \vec \csigma \bC^{1/2}_{ j}D_j   +i\sqrt\lambda t_j D_{ j}   D_j \bigr] \\
& +{\tr_\cD}\bigl[ \prod_{s=2}^{k} i\sqrt\lambda R_{\le j} \bC^{1/2}_{\le j} \Delta^{\tau (s)} \bC^{1/2}_{\le j}] R_{\le j}\\
&\times \bigl[t_j^2 D_{ j}  \bC^{1/2}_{j} \Delta^{\tau (1)}  \bC^{1/2}_{\le j}  + t^2_j \bC^{1/2}_{\le j}\Delta^{\tau (1)}
\bC^{1/2}_{j}  D_{ j}  +  t_j \bC^{1/2}_{\le j}  \Delta^{\tau (1)}  \bC^{1/2}_{ j}D_j   \bigr]\biggr)  dt_j
\\
&+ 2\lambda \int_0^1 t_j  \biggl(  {\tr_\cD}\bigl[ \prod_{s=1}^{k} i\sqrt\lambda R_{\le j} \bC^{1/2}_{\le j} \Delta^{\tau (s)}
\bC^{1/2}_{\le j}]R_{\le j}\left[   \bC^{1/2}_{\le j}\vec \csigma   \bC_{j} \vec \csigma \bC^{1/2}_{\le j}\right]\\
&+{\tr_\cD}\bigl[ \prod_{s=2}^{k} i\sqrt\lambda R_{\le j} \bC^{1/2}_{\le j} \Delta^{\tau (s)} \bC^{1/2}_{\le j}] R_{\le j} 
\left[   \bC^{1/2}_{\le j}\Delta^{\tau (1)}   \bC_{j} \vec \csigma \bC^{1/2}_{\le j} +  \bC^{1/2}_{\le j}\vec \csigma   C_{j} 
\Delta^{\tau (1)} \bC^{1/2}_{\le j} \right] \\
&+{\tr_\cD}\bigl[ \prod_{s=3}^{k} i\sqrt\lambda R_{\le j} \bC^{1/2}_{\le j} \Delta^{\tau (s)} \bC^{1/2}_{\le j}] R_{\le j} \\
&\times\left[   \bC^{1/2}_{\le j}\Delta^{\tau (1)}   \bC_{j} \Delta^{\tau (2)}  \bC^{1/2}_{\le j} +  \bC^{1/2}_{\le j} \Delta^{\tau (2)}  
\bC_{j} \Delta^{\tau (1)} \bC^{1/2}_{\le j} \right]\biggr)\ dt_j\  . 
 \end{align}
 We used $\bC^{1/2}_{ j}D_{\le j} =t_j \bC^{1/2}_{\le j}D_{ j} = t_j^2 \bC^{1/2}_{ j}D_j $. In \eqref{developcycles2} the sum over $\tau $ runs over the
 permutations of $[1,k]$ and $\Delta^s$, defined as
 \[
 (\Delta^s)_{m \bar m} = \frac{\partial  \vec \csigma_{m\bar m}  }{\partial \csigma^{c(s)}_{n^{c(s)},\bar n^{c(s)}}} 
 =\delta_{m^{c(s)}n^{c(s)}}\delta_{\bar m^{c(s)}\bar n^{c(s)}}\prod_{c\neq c(s)} \delta_{m^c\bar m^c} 
 \ .
 \]

 \paragraph{Outlines :}
 These formulas express the derivatives of the trace as a sum over all cycles with exactly $k$ derivatives, zero or one operator $D$, and 
\emph{up to two remaining numerator $\csigma$ fields} (one at most if the cycle contains an operator $D$). Rewriting $D_j$ as
\begin{align}
D_j = \bC^{1/2}_{ j} A_j \bC^{1/2}_{ j}\, ,\qquad (A_j)_{n\bar n} = {\bI}_j(n)A(n) \delta_{n\bar n}\ ,
\end{align}
the cycles have $2k$ to $2k+4$ numerator half-propagators $\bC^{1/2}$, 
and two of them form a $\bC_j$, hence each cycles bears a single propagator of scale $j$. 

\subsubsection{Map decomposition}
The Bosonic integral in a block $\cB$ can therefore be written as a sum over maps $\cM(T)$ corresponding to the combinatorial tree $T$ , 
\begin{align}\label{eq:bosogauss}
& \int   F_\cB\, d \mu_\cB = \sum_{\cM(T)} \int  \prod_{a\in \cB}  e^{-V_{j_a} (\csigma_a) }  A_G (\csigma)
 \;d \mu_\cB\ ,
\end{align} 
These maps $\cM$ are still forests, but not necessarily connected (because of the Fa'a di Bruno partitioning \eqref{eq:partitio}), 
with an effective loop vertex for each $b^a \in \pi^a, a \in \cB$,  
each of them expressed as a trace of a product of three-stranded operators 
by \eqref{eq:developcycles}, with $k = \vert b^a\vert$.
Each such effective vertex of $G$ bears \emph{at most two} $\csigma$ insertions plus 
exactly $\vert b^a \vert$ $\Delta$ insertions, which are contracted together via the coloured edges of the tree $\cT_\cB$. 

\begin{figure}[!h]
\begin{center}
{\includegraphics[width=4.5cm]{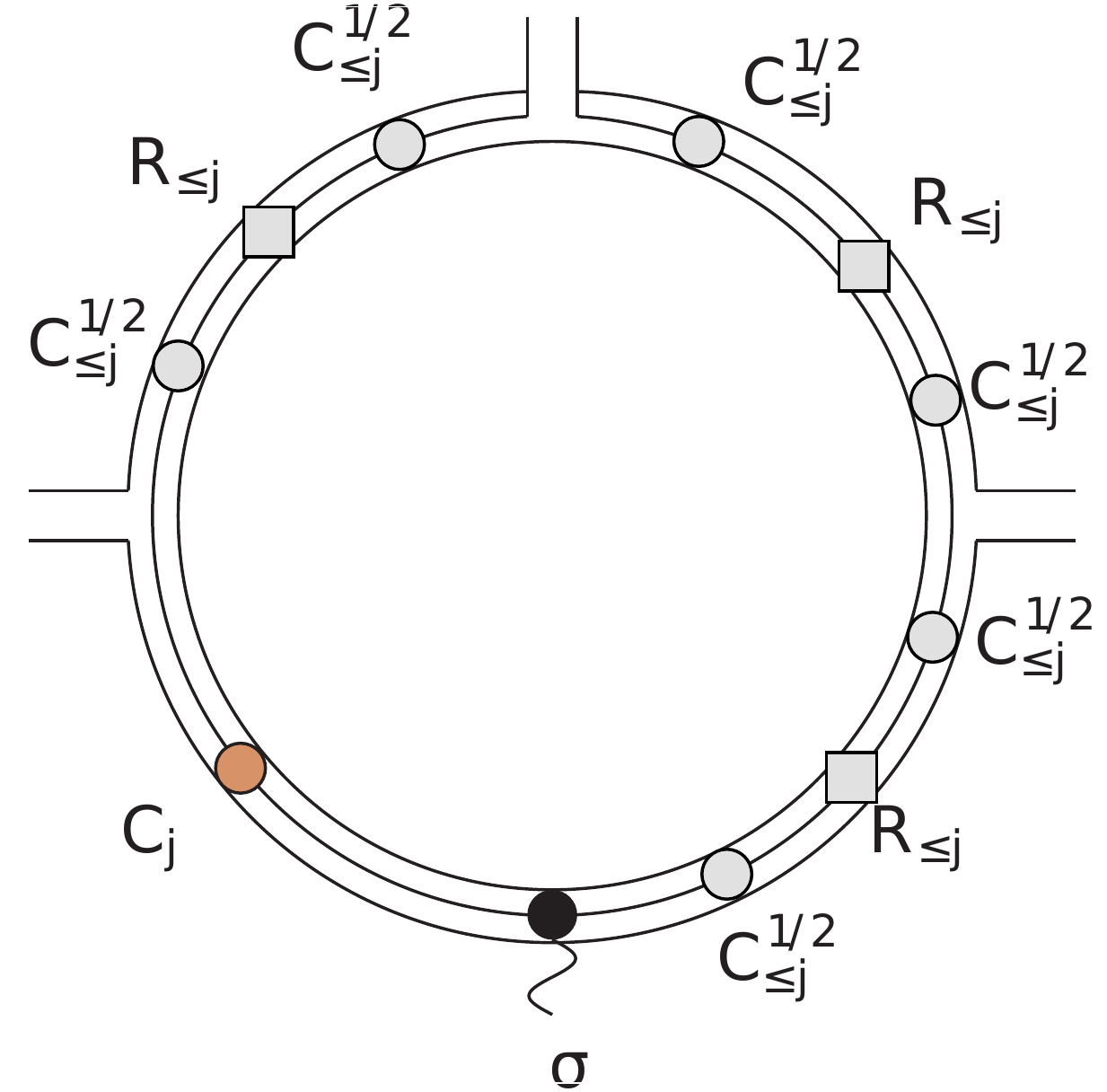}}
\end{center}
\caption{A detailed three-stranded vertex with its cycle of operators. Open strands corresponds to $\Delta$ operators, and to half edges of the tree $\cT_\cB$.}
\label{fig:t43loopvertex}
\end{figure}

We define the corners of the map $\cM$ as pairs of two consecutive insertions of either $\Delta$,  $\csigma$ or  $A_j$ operators.  
Therefore, each corner of the vertices bears a $\bC^{1/2}_{(\le) j_a}R_{\le j_a} \bC^{1/2}_{(\le) j_a}$ operator 
(the $\bC^{1/2}$'s being either $\bC^{1/2}_j$ or $\bC^{1/2}_{\le j}$), except one distinguished corner which bears a $\bC_{j_a}$ operator and no resolvent
(see Fig. \ref{fig:t43loopvertex}).

Note that to each initial $W_{j_a}$ may correspond several  effective loop vertices $V_{b^a}$,
depending of the partitioning of $S^a_\cB$ in \eqref{eq:partitio}. Therefore although
at fixed $|\cB|$ the number of (coloured) edges for any $\cM$ in the sum \eqref{eq:bosogauss} is exactly $|E(\cM)|=|\cB|-1$,
the number of connected components $C(\cM)$ 
is not fixed but simply bounded by $|\cB|-1$ (each edge can belong to a single connected component). Similarly
the number $|V(\cM)|= C(\cM) + |E(\cM)|$ of effective loop vertices of $G$ is not fixed, and simply obeys the bounds
\begin{equation}   \vert \cB \vert  \le |V(\cM)| \le 2(|\cB|-1) . \label{foreboun}
\end{equation}

From now on we shall simply call ``vertices" the effective loop vertices of $G$, as we shall no longer meet the initial $W_{j_a}$ vertices.

We then apply a Cauchy-Schwarz inequality with respect to the positive measure $d\nu_\cB$ to separate the map amplitudes from the 
exponential factor,
\begin{align}  \label{CS} 
\vert  \int F_\cB\; d \mu_\cB   \vert \le 
 \sum_\cM \Bigl( \int   \prod_a  e^{2 \vert V_{j_a} (\csigma_a) \vert } d \mu_\cB  \Bigr)^{1/2} 
\Bigl( \int   \vert  A_\cM (\csigma)  \vert^2  d \mu_\cB \Bigr)^{1/2}\ .
\end{align}
Both parts will be bounded respectively in sections \ref{subsec:Expbound} and \ref{subsec:T43graphbound}.

\subsubsection*{Single vertex block}
When the block $\cB$ is reduced to a single vertex $a$, we have a simpler contribution for which
an important cancellation occurs due to the presence of the logarithmically divergent counter term in $V_j$.
More precisely (writing simply $j$ for $j_a$)
\begin{align}\label{singbloc}
&\int   F_\cB\ d \mu_\cB  =  \int   \bigl[ e^{- V_{j}(\vec \sigma) } - 1\bigr]d \mu ( \csigma) = \int_{t=0}^1  
\int_{\csig}    V_{j}(\vec \csigma)  e^{- tV_{j}(\vec \csigma) }d \mu ( \csigma) dt\\
&= \int_{t=0}^1
\int\cD_je^{- tV_{j}(\vec \csigma) } \ d \mu (\csigma)dt   
\ +\ 2\lambda
\int_{t=0}^1  \int_{t_j=0}^1   
\int_{\csig} \\
&\times t_j  
\biggl[ 
{\tr_\cD} R_{\le j}     \bC^{1/2}_{\le j}\vec \csigma   \bC_{j} \vec \csigma \bC^{1/2}_{\le j}
- \int_{\csig'}   \bC_{ j}  {\vec \csigma' } \bC_{\le j}  {\vec \csigma' } d\mu( \csigma')    \biggr]
e^{- tV_{j}(\vec \csigma) }\ d \mu ( \csigma) dt_j dt
\ ,
\end{align}
and, using $R_{\le j} = 1 +  i\sqrt \lambda R_{\le j}  (i \sqrt\lambda D_{\le j} + \bC^{1/2}_{\le j}  \vec \csigma \bC^{1/2}_{\le j} ) $,
\begin{align}
&\int  t_j  \biggl[ 
{\tr_\cD} R_{\le j}     \bC^{1/2}_{\le j}\vec \csigma   \bC_{j} \vec \csigma \bC^{1/2}_{\le j}    
- \int    \bC_{ j}  {\vec \csigma' } \bC_{\le j}  {\vec \csigma' }  d\mu( \csigma')  \biggr]
e^{- tV_{j}(\vec \csigma) }d \mu ( \csigma)\\
&= \int  t_j  \biggl[ 
 i\sqrt\lambda{\tr_\cD} R_{\le j}  \bC^{1/2}_{\le j}  \vec \csigma   \bC_{\le j}\vec \csigma   \bC_{j} \vec \csigma \bC^{1/2}_{\le j} 
-\lambda {\tr_\cD} R_{\le j}  D_{\le j}   \bC^{1/2}_{\le j}\vec \csigma   \bC_{j} \vec \csigma \bC^{1/2}_{\le j} \\
&\qquad\qquad +\ 
{\tr_\cD}  \bC^{1/2}_{\le j}\vec \csigma   \bC_{j} \vec \csigma \bC^{1/2}_{\le j}     
- \int   \bC_{ j}  {\vec \csigma' } \bC_{\le j}  {\vec \csigma' }  d\mu( \csigma')   \biggr]
e^{- tV_{j}(\vec \csigma) }d \mu ( \csigma)\\
&= \int  t_j  \biggl[ 
 i\sqrt\lambda{\tr_\cD} R_{\le j}  \bC^{1/2}_{\le j}  \vec \csigma   \bC_{\le j}\vec \csigma   \bC_{j} \vec \csigma \bC^{1/2}_{\le j} 
-\lambda {\tr_\cD} R_{\le j}  D_{\le j}   \bC^{1/2}_{\le j}\vec \csigma   \bC_{j} \vec \csigma \bC^{1/2}_{\le j} \\
&\qquad\qquad +\ 
t{\tr_\cD}  \bC^{1/2}_{\le j}\vec \csigma   \bC_{j} \vec \Delta \bC^{1/2}_{\le j} \cdot  \left( \partial \vec \csigma ( -V_j )\right)   \biggr]
e^{- tV_{j}(\vec \csigma) }d \mu (\csigma)\ ,
\end{align}
where in the last line we used integration by parts with respect to one $\sigma$ to explicit the cancellation in the last term.
In the last term the dot means a scalar product between the $\Delta$ and insertions of both the trace and the vertex derivative.
As expected this formula shows that the vacuum expectation value of the graph made of a single vertex has been successfully canceled by the counter term.
The contribution of a single vertex corresponds therefore again to perturbatively convergent graphs
with either at least two vertices, or one vertex and an operator $D$, multiplied by the exponential of the interaction, and can be treated therefore exactly as the ones with two or more vertices.

\subsection{Bound on the exponential}\label{subsec:Expbound}

\begin{lemma}\label{boundv}
For $g$ in the cardioid domain $\cC  ard_\rho$
we have
\begin{align}\label{boundlemmanopert}
\vert V_j (\csigma) \vert \le \rho \;  O(1)   \bigl[ 1 +  {\tr_\cD}  \bigl(  \bC_{\le j}  \vec \csigma \bC_j   \vec \csigma \bigr)  \bigr]. 
\end{align}
\end{lemma}
\prf  
Starting from \eqref{interaj4c}-\eqref{defDj} let us write $V_j = \cV_j + \cD_j$. Using the bound \eqref{boundcoun} for the $\cE'_j$ term, we get
\begin{align} 
\vert \cV_j (\csigma) \vert \le   \vert \lambda \vert  \biggl( O(1) + 2 \int_0^1  t_j 
\vert {\tr}_\cD  \, R_{\le j}  \bC^{1/2}_{\le j}\vec \csigma   \bC_{j} \vec \csigma \bC^{1/2}_{\le j}    \vert  dt_j  \biggr) .
\label{noper0}
\end{align}
For $A$ positive\footnote{We usually simply say positive for ``non-negative", i. e. each eigenvalue is strictly 
positive or zero.} Hermitian and $B$ bounded we have $\vert  \Tr A B \vert \le \Vert B \Vert \Tr A $. 
Indeed if $B$ is diagonalizable with eigenvalues $\mu_i$, computing the trace in a 
diagonalizing basis we have $\vert \sum_i   A_{ii} \mu_i \vert \le \max_i \vert \mu_i \vert \sum_i   A_{ii}  $;
if $B$ is not diagonalizable we can use a limit argument. 
Hence using \eqref{rescardbou}
\begin{align}
{\tr_\cD} \vert  R_{\le j} \bC_{\le j} \vec  \csigma    \bC_j \vec \csigma \vert 
\ \le&\ 2 \cos^{-1} (\phi /2)  {\trd}  \bigl(  \bC_{\le j}^{1/2} \vec \csigma \bC_j   \vec \csigma \bC_{\le j}^{1/2}  \bigr) \crcr
&= 2\cos^{-1} (\phi /2)
{\trd}  \bigl(  \bC_{\le j}  \vec \csigma \bC_j   \vec \csigma \bigr) . \label{noper1} 
\end{align}
We conclude that $\cV_j$ obeys the bound \eqref{boundlemmanopert}, since in the cardioid $\vert \lambda\vert \cos^{-1} (\phi /2) \le \rho$. 
\qed

It remains to check it for the $\cD_j$ term. Returning to \eqref{defDj}
\begin{align} \vert \cD_j \vert  &\le  \vert \lambda \vert \int_0^1  
 \vert {\trd} \, \sqrt{\vert \lambda \vert} R_{\le j} \bigl( D_{\le j}  \bC^{1/2}_{j} \vec \csigma \bC^{1/2}_{\le j} 
 +  \bC^{1/2}_{\le j}\vec \csigma \bC^{1/2}_{j} D_{\le j} +  \bC^{1/2}_{\le j}  \vec \csigma \bC^{1/2}_{\le j}D_j  \bigr)  \vert dt_j  \crcr
&+  \vert \lambda \vert  \vert {\trd}\  R_{\le j}  D_{\le j}   D_j \vert . \label{Djboundb}
\end{align}
We use the Hilbert-Schmidt bound $ \vert {\tr} AB \vert \le {\tr} AA^\star + {\tr} BB^\star$. Remember \eqref{squaretraceclass}: $D$
is Hermitian positive and square trace class 
and so are also $D_j $ and $D_{\le j}$. Hence 
\begin{equation}  
\vert {\trd}  R_{\le j}  D_{\le j}   D_j \vert  \le {\trd}  R^\star_{\le j} 
R_{\le j}  D_{\le j}^2   +  {\trd}    D_j^2 \le O(1) [1+\cos^{-2} (\phi /2)] .
\end{equation}
Similarly 
\begin{align}
\vert {\trd} \, \sqrt{\vert \lambda \vert}   R_{\le j} D_{\le j}  \bC^{1/2}_{j} \vec \csigma \bC^{1/2}_{\le j} \vert &\le 
\vert \lambda \vert  {\trd}\, R^\star_{\le j}  R_{\le j} D^2_{\le j} +   {\trd} \,  \bC_{\le j} \vec \csigma \bC_{j}  \vec \csigma\ , \\
\vert {\trd} \,\sqrt{\vert \lambda \vert} D_{\le j} R_{\le j}  \bC^{1/2}_{\le j} \vec \csigma \bC^{1/2}_{j}  \vert &\le 
\vert \lambda \vert {\trd}\,  D^2_{\le j}  R^\star_{\le j}  R_{\le j} +   {\trd}\   \bC_{\le j} \vec \csigma \bC_{j}  \vec \csigma\ .
\end{align}
Finally for the last term $\vert {\trd} \,  \sqrt{\vert \lambda \vert}  R_{\le j} \bC^{1/2}_{\le j} \vec \csigma \bC^{1/2}_{\le j} D_{j} \vert$, 
we remark that $ \bC^{1/2}_{\le j} D_{j} = t_j \bC^{1/2}_{j} D_{j}$. Then
\begin{equation}
t_j \vert {\trd} \, \sqrt{\vert \lambda \vert} D_{j}   R_{\le j} \bC^{1/2}_{\le j} \vec \csigma \bC^{1/2}_{j} \vert \le
\vert \lambda \vert {\trd}\, R^\star_{\le j}  R_{\le j} D^2_{j} +   {\trd} \,  \bC_{\le j} \vec \csigma \bC_{j}  \vec \csigma .
\end{equation}
Using again the inequality $\vert  \tr A B \vert \le \Vert B \Vert \tr A $ for $A$ positive and $B$ bounded, we can get rid of the resolvents:
\begin{equation}
\vert \lambda \vert  {\trd}\  R^\star_{\le j}  R_{\le j} D^2_{\le j} \le O(1)\vert \lambda \vert \cos^{-2} (\phi /2) , \quad  \vert \lambda \vert  
{\trd} R^\star_{\le j}  R_{\le j} D^2_{j} \le O(1)\vert \lambda \vert \cos^{-2} (\phi /2).
\end{equation}
Hence we can conclude that the three first terms in \eqref{Djboundb} obey the bound \eqref{boundlemmanopert}  since in the cardioid
$\vert \lambda \vert \cos^{-1} (\phi /2) \le \rho$. 
\qed

We can now bound the first factor in the Cauchy-Schwarz inequality \eqref{CS}.
\begin{theorem}[Bosonic Integration]\label{BosonicIntegration}
For $\rho $ small enough and for any value of the $w$ interpolating parameters
\begin{align}  \Bigl{(} \int \prod_{a \in \cB}   e^{ 2 \vert V_{j_a} (\sigma_a) \vert } d\mu_\cB   \Bigr{)}^{1/2}  &\le
e^{ O(1) \rho  \vert \cB \vert }.
\end{align}
\end{theorem}
\prf  As $\prod_{a \in \cB} e^{c \rho}=e^{ c \rho  \vert \cB \vert }$, applying Lemma \ref{boundv} we get
\begin{equation}
 \int   \prod_{a \in \cB}   e^{ 2 \vert V_{j_a} (\sigma^a)\vert  }d\mu_\cB  \le e^{ c \rho  \vert \cB \vert }
\int   \; e^{ \; < \csigma , \bQ \csigma > }d\mu_\cB
\end{equation}
where $\bQ$ is a symmetric positive matrix in the vector space  $\bV$ of collections of matrices $((\csig^c)^a)_{a\in\cB,c\in\cD}$ .
$\bV$ has dimension $N_\bV = 3 \vert \cB \vert N^2$. 
The $N_\bV$ by $N_\bV$ matrix is block diagonal, with,
\begin{align} 
< \csigma , \bQ \csigma >  &= \sum_{a \in \cB}  < \csigma^a , Q^a \csigma^a >, \\  < \csigma^a , Q^a \csigma^a > &=
2\rho  \int_0^1  {\trd}  \bigl(  \bC_{\le j_a}  \vec \csigma^a C_{j_a}   \vec \csigma^a \bigr)dt_{j_a}\ .
\end{align}
Hence $\bQ = \sum_{a \in \cB} \bQ^a $,
where $\bQ^a $ is the $N_\bV$ by $N_\bV$ matrix with all elements zero except the $3 N^2 $ by $3  N^2$
which have both vertex indices equal to $a$. These non zero elements form the
$3 N^2 $ by $3  N^2$ positive symmetric matrix $Q^a$ with matrix elements
\begin{align}
Q^a_{c,m,n;\; c' m' n'} &= Q^{a,1}_{c,m,n;\; c' m' n'} + Q^{a,2}_{c,m,n;\; c' m' n'}\\
Q^{a,1}_{c,m,n;\; c' m' n'}&=
 2\rho \delta_{c,c'} \int_0^1 
 \delta_{m,m'} \delta_{n,n'}  [ t_{j_a}   \cQ_{j_a,j_a,1} (m,n)  + \sum_{k=0}^{j_a -1} \cQ_{j_a,k,1} (m,n) ]  dt_{j_a}     \\
 Q^{a,2}_{c,m,n;\; c' m' n'}&= 
 2\rho (1-  \delta_{c,c'})  \int_0^1  \delta_{m,n}\delta_{m',n'}   [ t_{j_a}  \cQ_{j_a,j_a, 2} (m,m') + \sum_{k=0}^{j_a -1}  \cQ_{j_a,k, 2} (m,m')  ]  dt_{j_a}\ ,
 \end{align} 
where the $\cQ$ factors are defined respectively as the colour-diagonal and colour off-diagonal part of a bubble with 
two propagators of slices $j $ and $k$:
\begin{align}
\cQ_{j,k,1} (m,n) &=   \sum_{m_2, m_3}  (\bC_{k})_{(m, m_2, m_3)(m, m_2, m_3)  } (\bC_{ j})_{(n, m_2, m_3)(n, m_2, m_3)} ,
\label{boundd1}\\
\cQ_{j,k, 2} (m,m') &=  \sum_{m_3}  (\bC_{k})_{(m, m', m_3)(m, m', m_3) } (\bC_{ j})_{(m, m', m_3)(m, m', m_3)} \label{boundd2}.
\end{align}
The big matrix $\bQ$ has elements
$\bQ_{a,c,m,n;\; a',c' m' n'} =  \delta_{a,a'}  Q^a_{c,m,n;\; c' m' n'} $. 
Using the bounds \eqref{propmombound} it is easy to check that
\begin{align}
\cQ_{j,k,1} (m,n) &\le O(1)  M^{-2j}e^{- M^{-j} \vert n \vert }e^{- M^{-k} \vert m \vert }, \\
\cQ_{j,k , 2} (m,m') &\le O(1)  M^{-2j-k} e^{- M^{-k} (\vert m \vert + \vert m' \vert )} .
\end{align}

\begin{lemma}
The following bounds hold uniformly in $j_{\max}$ and $N$
\begin{align}
{\rm \bf Tr}\; Q^a &\le  O(1)  \rho ,\label{bonnetrace}\\
\Vert Q^{a} \Vert  &\le O(1) \rho  j_a  M^{-2 j_a} \label{bonnenorm}.
\end{align}
\end{lemma}
\prf  The first bound is easy. Since we compute a trace, only $Q^{a,1}$ contributes and the bound follows from \eqref{boundd1}
which implies that $\sum_{m, n} \cQ_{j,1} (m,n) \le O(1) $.
Since $Q^{a,1}$ is diagonal both in component and colour space, from \eqref{boundd1}
we deduce that $\sup_{m, n} \cQ_{j,1} (m,n) \le O(1) j M^{-2j}  $, hence
\begin{equation}
\Vert Q^{a,1} \Vert  \le O(1)  \rho  j_a  M^{-2 j_a} \label{bonnenorm1}.
\end{equation}
Finally to bound  $\Vert Q^{a,2} \Vert $
we use first a triangular inequality to sum over the 6 pairs of colours $c, c'$ and over $k$
\begin{equation}
\Vert Q^{a,2} \Vert  \le 12\rho \sum_{k=0}^{j} \Vert E_{j_a,k,2}\Vert    
\end{equation}
where $ E_{j_a,k,2} $ is the (component space) matrix with matrix elements 
\begin{equation} E_{j_a,k,2} (m,n ; m',n')=\delta_{m,n}\delta_{m',n'}  \cQ_{j_a,k,2} (m,m') .
\end{equation}
The operator norm of $E_{j_a,k,2} $ is bounded by its Hilbert Schmidt norm
\begin{equation}  \Vert E_{j_a,k,2}\Vert_2 = [ \sum_{m,m'} \cQ^2_{j_a,k,2} (m,m')]^{1/2}  \le   O(1) M^{-2j_a - k} [M^{2k}]^{1/2} = O(1) M^{-2j_a} .
\end{equation}
It follows that
\begin{equation}
\Vert Q^{a,2} \Vert   \le O(1) \rho  j_a  M^{-2 j_a}, \label{bonnenorm2}
\end{equation}
and gathering \eqref{bonnenorm1} and \eqref {bonnenorm2} proves \eqref{bonnenorm}.
\qed

The covariance $\bX$ of the Gaussian measure $d\mu_\cB$ is also a symmetric matrix on the big space $\bV$, but which is the tensor product 
of the identity in colour and component space times the matrix $X_{ab} (w_{\e_B} )$ in the vertex space. 
Defining $\bA \equiv \bX \bQ$, we have
\begin{lemma}
The following bounds hold uniformly in $j_{\max}$ and $N$
\begin{align}
\tr \; \bA &\le  O(1)  \rho  \, \vert \cB \vert  ,\label{bonnegtrace}
\\
\Vert \bA \Vert  &\le O(1) \rho  \label{bonnegnorm}.
\end{align}
\end{lemma}
\prf Since $\bQ = \sum_{a \in \cB} \bQ^a $ we find that 
\begin{equation}
\Tr \; \bA = \sum_{a \in \cB} \Tr \bX \bQ^a = \sum_{a \in \cB}  X_{aa} (w_{\e_B} ) \Tr  Q^a = \sum_{a \in \cB} {\rm \bf Tr}\; Q^a \le  O(1)  \rho  \, \vert \cB \vert .
\end{equation}
where in the last inequality we used \eqref{bonnetrace}. Furthermore by the triangular inequality and \eqref{bonnenorm}
\begin{align}
\Vert \bA \Vert  \le   \sum_{a \in \cB} \Vert \bX \bQ^a \Vert = \sum_{a \in \cB}  X_{aa} (w_{\e_B} ) \Vert  Q^a \Vert = \sum_{a \in \cB}  \Vert  Q^a \Vert 
\le  \sum_{j =0}^{\infty} 
O(1)  \rho  jM^{-2j} \le O(1).
\end{align}
where we used the fundamental fact that all vertices $a\in\cB$ have \emph{different scales} $j_a$.
\qed

We can now complete the proof of Theorem \ref{BosonicIntegration}. 
By \eqref{bonnegnorm} for $\rho$ small enough the series $\sum_{n=1}^\infty (\Tr \bA^n)/n $ converges and we have
\begin{align} \int   e^{ \; < \sigma , \bQ \sigma > }\; d\mu_\cB &= [\det (1  - \bA  )]^{-1/2}  = e^{-(1/2)\Tr \log  (1  - \bA  )  } = 
e^{(1/2) \sum_{n=1}^\infty (\Tr \bA^n)/n  } 
\\
&\le e^{(1/2) \Tr \bA  (\sum_{n=1}^\infty \Vert \bA \Vert ^{n-1})  } 
\le e^{ O(1)  \rho  \vert \cB \vert}. 
\label{goodtra}
\end{align}
\qed

\subsection{Map Bounds} \label{subsec:T43graphbound}

We still have to bound the second factor in \eqref{CS}, namely
$\Bigl( \int   \vert  A_\cM (\sigma)  \vert^2\,  d \mu_\cB \Bigr)^{1/2}$.
We recall that at fixed $|\cB|$, the maps $\cM$ are forests with $E(G)= |\cB|-1$ coloured edges joining 
$|V(\cM)| = C(\cM) + |E(\cM)|$ (effective) vertices, each of which has a weight given by 
\eqref{eq:developcycles} and \eqref{developcycles2}. The number of connected components $C(\cM)$ is bounded by $ |\cB|-1$, hence
\eqref{foreboun} holds.

This squared amplitude can be represented as the square root of an ordinary amplitude but for a map $\cM'= \cM \cup \cM^*$ 
which is the (disjoint) union of the map $\cM$ and its conjugate map $\cM^*$ with mirror-symmetric structure 
and on which each operator has been replaced by its Hermitian conjugate. This overall map $\cM'$  
has thus twice as many vertices, edges, resolvents,  $\vec\csigma^a$ insertions and connected components than the initial map $\cM$.

\subsubsection*{Deletion of the remaining $\csig$}
To evaluate the amplitude $A_{\cM'}= \int \vert  A_\cM (\csigma)\, d \mu_\cB  \vert^2$, we first delete every $\vec\csigma^a$ insertion using 
repeatedly integration by parts
\begin{align}
\int \sigma^c_{n^c} F(\vec \csigma) d\mu (\csigma)=\int  \frac{\partial}{\partial \csigma^c_{n^c}}F(\vec \csigma) d\mu (\csigma) .
\end{align}
The derivatives $\frac{\partial}{\partial \csigma^c_{n^c}}$ will act on any resolvent $R_{ j_a}$ or remaining $\vec \csigma^a$ insertion of $\cM'$,
creating a new contraction edge.
When it acts on a resolvent, it creates a new corner, bearing a product of operators $\bC^{1/2}_{\le j_a} R_{\le j_a} \bC^{1/2}_{\le j_a}$ 
(or $\bC^{1/2}_{\le j_a}R_{\le j_a}^{*} \bC^{1/2}_{\le j_a}$ if the resolvent is on a mirror vertex).

Remark that at the end of this process we have a sum over new maps $\mathfrak M$ with no longer any $\vec\csigma^a$ insertion,
\[
 A(\cM') = \sum_{\kM(\cM)} A(\kM)\ ,
\]
where $\kM(\cM)$ runs over all possible maps obtainbed by replacing $\csig$ insertions by coloured edges, 
but the
number of edges, resolvents and connected components at the end of this contraction process typically has changed. However we have a bound
on the number of new edges generated by the contraction process. Since each vertex of $\cM$ contains at most two $\vec\csigma^a$ insertions,
$\cM'$ contains at most $4 |V(\cM)|$ insertions, hence using \eqref{foreboun} at most $8(|\cB|-1)$ insertions to contract. Each such contraction creates 
at most one new edge. 
Therefore each map $\mathfrak M$ contains  the initial $2(|\cB|-1)$ coloured edges of $\cM'$ decorated with up to at most $8(|\cB|-1)$ additional new edges.

\subsubsection{Scale decomposition}
Until now, the amplitude $A(\mathfrak M)$ contains $\bC_{\le j}^{1/2}=\sum_{j'< j}\bC^{1/2}_{j'}+t_j \bC^{1/2}_j$ operators. We now develop the product
of all such $\bC^{1/2}_{\le j}$ factors as a sum over scale assignments. 
It means that each former $\bC^{1/2}_{\le j}$ is replaced by a fixed scale $\bC^{1/2}_{j'}$ operator (the $t_j$ factor being bounded by 1)
with  scale attribution $j'\leq j$. The amplitude
at fixed scale attribution $\mu$ is noted $A(\mathfrak M_\mu)$ and we shall now bound each such amplitude. The sum over $\mu$ will be standard to bound
after the key estimate \eqref{decayattri} is established. Similarly the sums over $\cM$ and over $\mathfrak M$ only generate
a finite power of $\vert \cB \vert !$, hence will be no problem using the huge
decay factors of  \eqref{decayattri}.

\begin{theorem}[Map bound]\label{thmgraphbound}
The amplitude of a map $\mathfrak M$ with scale attribution $\mu$ is bounded by
\begin{align}
|A(\mathfrak M_\mu)| &\leq [O(1) \rho]^{|E(\mathfrak M)|}
M^{-\frac12\sum_{v\in \cB} j_{a(v)}}.
\end{align}
\end{theorem}
 
\prf
We work at fixed value of each $\csigma$, and denote $\mathfrak C$ the connected components of a map $\mathfrak M$,
thus $A(\mathfrak M_\mu)=\prod_{\mathfrak C}A(\mathfrak C_\mu)$.
The amplitude $A(\mathfrak C_\mu)$ of a connected component $\mathfrak C$ can be bounded by iterated Cauchy-Schwarz inequalities, as for invariant models in section
\ref{subsec:CSbounds}.
 A spanning tree is chosen for each connected component $\mathfrak C$, resolvents $R_{\le j}$ are ordered along the clockwise 
 contour walk of the tree and used  as $R$ and $R'$ in formula \eqref{eq:cauchy}.
For a connected component with $2n$ resolvents, this processs gives a geometric mean over $2^n$ final maps ${\mathfrak m}$ bearing no $R_{\le j}$ at all, 
times a product of norms of resolvents, which are all bounded by $2\cos^{-1}(\phi/2)$ for $\lambda=|\lambda|e^{i\phi}$ in the cardioid. 
Since these graphs no longer have
any dependence on $\csigma$, the normalised measure $\int d\mu ( \csigma)$ simply evaluates to 1, and we are left with a bound:
\begin{align}
|A({\mathfrak C_\mu})| &\leq  \prod_{i=1}^{2n}\|R_i\|\ \left( \prod_{\mathfrak m} A({\mathfrak m}) \right)^{\frac1{2^n}}
 \leq 4^n \left[\cos\frac\phi2\right]^{-2n} \left( \prod_{\mathfrak m} A({\mathfrak m}) \right)^{\frac1{2^n}}.
\end{align}
The Cauchy-Schwarz process keeps track of a number of items. Indeed, at every iteration, 
each vertex and edge of the bounded map gives respectively two vertices and two edges of the next-stage maps.
The $\bC^{1/2}_j$ and $A_j$ operators follow the same rule, and each $\bC_j^{1/2}$ operator of the original map will generate $2^n$ identical
$\bC_j^{1/2}$ operators in the final maps, which repartition is {\it a priori} unknown. 
Finally, each vertex of the original map bearing at least one resolvent, each vertex of the final map has been cut at least once and is thus mirror-symmetric.
 
In a final map $\mathfrak{m}$, each corner bears either a $\bC_j$ operator or a product of two identical $\bC_{j'}^{1/2}$ operators (thus one full $\bC_{j'}$), 
vertices may also bear $A_j$ insertions, and the strands represent contractions of their indices. All operators left are diagonal, and bounded as 
\begin{align}
(\bC_j^{1/2})_{n, \bar n}&= \delta_{n,\bar n}  \sqrt{\frac{1}{n^2+1}} {\bf I}_{j}(n)\le \frac{1}{M^{j-1}}\delta_{n,\bar n}
\ \prod_{c=1}^3\ {\bf I}_{n_c^2 \le M^{2j}}, \\
A_j(n, \bar n)&= {\bf I}_j(n)\delta_{n\bar n} \sum_c A(n_c)  \le \delta_{n\bar n}\sum_c O(1) {\rm log} M^j =j O(1)\delta_{n\bar n}
\ .
\end{align}
 Then, for a final map $\mathfrak{m}$ with $2n$ corners that we index by $\eta$, each bearing a $\bC^\eta$ operator, 
 and denoting $a(\eta)$ the vertex of the original graph that bore the operator,
\begin{align}
A(\mathfrak m) &\leq |\lambda|^{E(\mathfrak m)}\sum_{\{\vec n\}} \prod_\eta \bC^{\eta}(n^{\eta} \bar n^{\eta} )\delta_{\bar n^\eta n^{\eta}}\ 
\prod_{{\rm strands}\ s} \delta_{n_{c_s}^s \bar n_{c_s}^s}\prod_{{\rm operators}\ A}O(1)j_{a(A)} \\
&\leq|\lambda|^{E(\mathfrak m)}\sum_{\{\vec n\}}\  \prod_\eta \delta_{n^\eta,\bar n^{\eta}}
\frac{1}{M^{2j_\eta-2}}\left(\prod_c {\bf I}_{|n_c^{ \eta}| \le M^{j_\eta}}\right)\  
\prod_{ s} \delta_{n_{i_s}^s \bar n_{i_s}^s} \prod_{A}O(1)j_{a(A)}\\
&= [M^2 |\lambda|]^{E(\mathfrak m)}\ M^{-2\sum_\eta j_\eta}\ \prod_{{\rm faces}\ f}\sum_{ n_f}\prod_{\eta, \eta'\in f}\left( {\bf 1}_{n_f \le M^{j_{\eta}}}\right)
\prod_{A}O(1)j_{a(A)} ,
\end{align}
where $j_\eta \in \{0...j_{a(\eta)}\}$ is the scale assignment of the corresponding $\bC$ operator, $j_{a(A)}$ the scale of the vertex 
bearing the operator $A_{j}$, $c_s$ is the colour of a strand $s$ and $f$ are the coloured faces. 
In the bound, the $\bC$ operators being removed, the 
faces are closed cycles of $\delta$ operators multiplied by scale factors and cut-offs. Hence only one index $n_f$ remains for each coloured face $f$. 
Thus the amplitude of a final map $\mathfrak m$ is bounded by,
\begin{align}
A(\mathfrak m) &\leq  [M^2 |\lambda|]^{E(\mathfrak m)} M^{-2\sum_\eta j_\eta}\ \prod_{f} M^{j_{\min}(f)} \prod_{A}O(1)j_{a(A)}\\
&\leq  [O(1) |\lambda|]^{E(\mathfrak g)} M^{\sum_f j_{\min}(f)-2\sum_\eta j_\eta} \prod_{A}O(1)j_{a(A)}.
\end{align}
where $j_{\min}(f)=\min_{\eta\in f} j_\eta$.
Corners of final maps that were generated by distinguished corners (without resolvents) of the original map will be denoted $\eta^* \in H^*$, 
as opposed to regular corners $\eta$. Those corners bear $\bC_{j_{a(\eta^*)}}$ operators that we want to keep track of. 
Other corners bear $C$ operators of scale $j_\eta\leq j_{a(\eta)}$.
The amplitude is thus bounded by
\begin{align}
|A({\mathfrak C_\mu})| &\leq\left[\frac12 \cos\frac\phi2\right]^{-2n} \left( \prod_{\mathfrak m} 
[O(1) |\lambda|]^{E(\mathfrak m)} \prod_{A}O(1)j_{a(A)}\right)^{\frac1{2^n}}
M^{\frac1{2^n}\sum_{\mathfrak m}\left[\sum_f j_{\min}(f)-2 j_\eta \right]}\\
&\leq   [O(1) \rho]^{E(\mathfrak C)}\left(\prod_{v}j_{a(v)}\right)
M^{-\frac12\sum_v j_{a(v)}}\ M^{\frac1{2^n}\sum_{\mathfrak m}\left[\sum_f j_{\min}(f)-2\sum_{\eta} j_\eta +\frac{1}{2}\sum_{H^*}j_{\eta*}\right]}\\
&\leq   [O(1) \rho]^{E(\mathfrak C)}
M^{-\frac14\sum_v j_{a(v)}}\ M^{\frac1{2^n}\sum_{\mathfrak m}\left[\sum_f j_{\min}(f)-2\sum_{\eta} j_\eta +\frac{1}{2}\sum_{H^*}j_{\eta*}\right]},
\end{align}
where we use the conservation of the number of distinguished corners during the Cauchy-Schwarz process 
$\sum_{\mathfrak m}\left[\sum_{H^*}j_{a(\eta*)}\right]=
2^n \sum_v j_{a(v)}$, along with the fact that for any map, $2n< 2E(\mathfrak C)$.
We also used the conservation of $A_j$ operators, and the fact that there is at most one $A_j$ per original vertex.

As in section \ref{subsec:CSbounds}, for a connected component with an odd number $2n+1$ of resolvents, we first proceed to a slightly asymmetric
Cauchy-Schwarz splitting of the map, choosing $R^1$ and $R^{n+1}$ as $R$ and $R'$. Both scalar product graphs will then have an even number
of resolvents and the previous results stand.

\begin{lemma} 
For any connected components $\mathfrak C_\mu$ with final maps $\mathfrak m$,
 \begin{align}
\sum_{\mathfrak m}\left[\sum_f j_{\min}(f)-2\sum_{\eta} j_\eta +\frac{1}{2}\sum_{H^*}j_{\eta*}\right]\leq0.
\end{align}
\end{lemma}

\prf
A final map consists of the gluing of two mirror symmetric maps along a path whose ends are undistinguished corners $\eta\not\in H^*$. 
Thus a final map bears at least two undistinguished corners. Therefore,
\begin{align}
\sum_f 1-2\sum_{\eta} 1 +\frac{1}{2}\sum_{H^*}1 = F - 2C + \frac12|H^*| \leq F-\frac32C -1.
\end{align}
For any tree, the relationship between the number $C$ of corners $\eta$, the number $F$ of faces $f$ and the number of $A_j$ insertions $|A|$ is $F-C+|A|=3$. 
This can be proved starting with a single isolated vertex and adding extra vertices, edges and $A_j$'s one by one. 
Each new vertex and edge comes with two new faces and two new corners, each $A_j$ with one corner, and the isolated vertex had three faces and no corner.

Any loop edge adds two corners and may increase or decrease the number of faces by one. 
Thus, for a tree $\mathcal T$ decorated with $L$ loop edges $\e\in\mathcal L$, 
\begin{equation} (F-\frac32C-1)_{\mathcal T+\mathcal L} \leq (F-\frac32C-1)_{\mathcal T}-2L=3-V-\frac{3}{2}|A|-2L. \end{equation} 
For any map with at least $3$ vertices, or with at least one loop edge, or with $A_j$ insertions ($|A|$ is always even for a final map), this is lower than $0$. 
A pathological final map cannot be a single vertex without edges, because final maps have at least two corners. 
A final map composed of two vertices, no loops and no $A_j$ can only arise from a map which had two consecutive corners bearing resolvents, 
separated by an edge of the chosen tree, before the last iteration of the Cauchy-Schwarz process.

\begin{figure}[!h]
  \begin{center}
  {\includegraphics[width=0.4\textwidth]{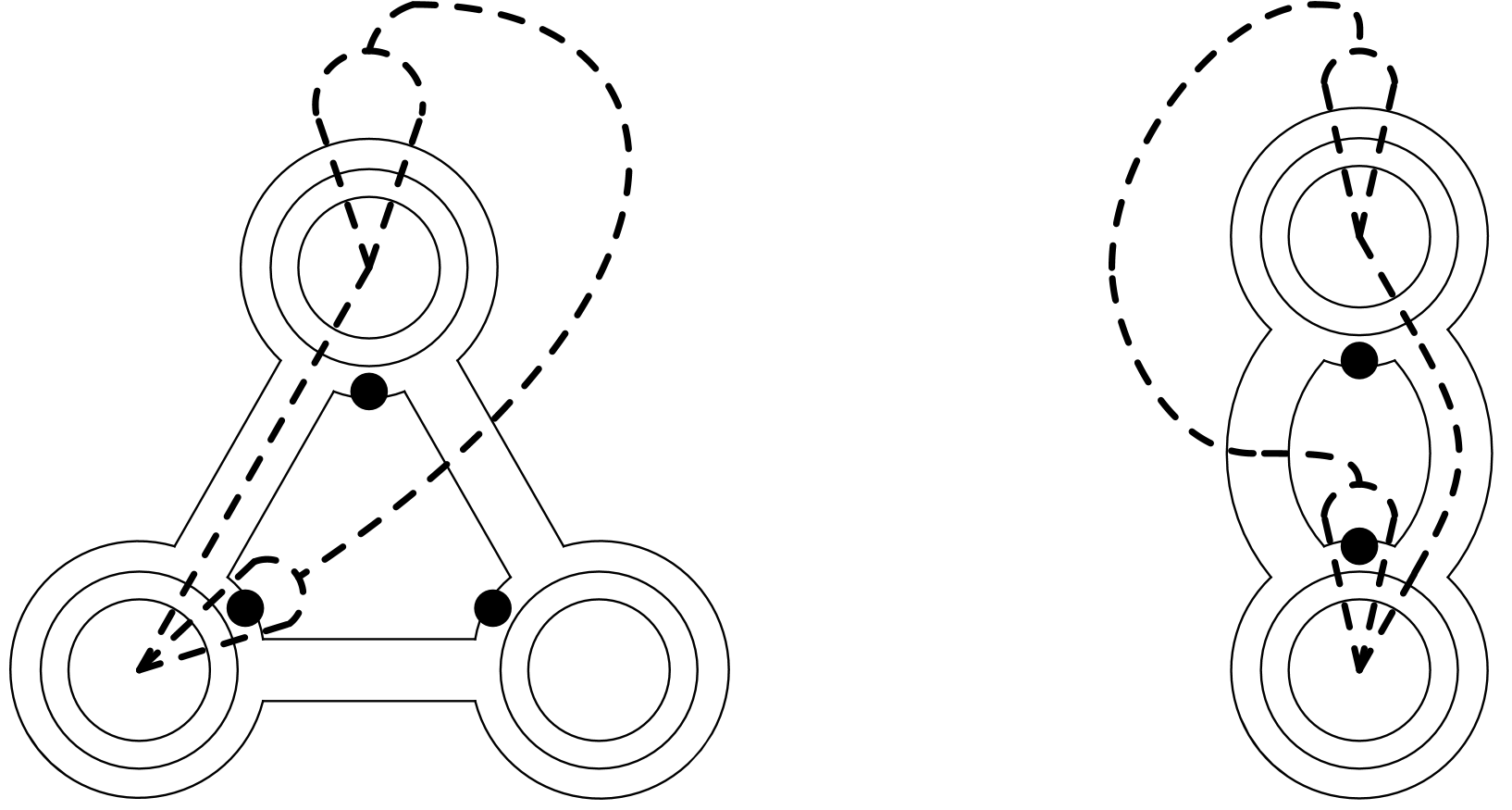}}
   \end{center}
  \caption{The maps $S_3$ (left) and $S_2$ (right) with dashed lines representing the Cauchy-Schwarz splitting used to avoid pathological maps.
  Dotted corners bears resolvents $R_{\le j}$. 
  When the dashed line crosses an un-dotted corner, propagator $C_j$ must be rewritten as 
  $C^{1/2}{\mathbf 1}C^{1/2}$ and the identity matrix $\mathbf{1}$ is used instead of a resolvent.}
  \label{S3S2}
\end{figure}

If a mirror-symmetric graph has only two resolvents, then those resolvents are mirror symmetric and therefore are on 
each side of the symmetry axis, which is a path between two ``cleaned'' corners (bearing no resolvent), therefore there is at least one corner between them. 
Therefore, a map with only two remaining resolvents, which are on consecutive corners, cannot arise from the bounding process. 
There are only two families of original maps with less than four resolvents, two being separated only by an edge. 
We will call them $S_2$ and $S_3$, and deal with them with an adapted Cauchy-Schwarz bound that avoid pathological final graphs (Fig. \ref{S3S2}).

Therefore, for any final map, the $j$s brought by corners (2 for undistinguished ones, and $3/2$ for distinguished ones)
is large enough to cancel the number of $j_{\min}$ brought by the faces. However, each $j_{\min}$ must be canceled individually by a higher $j$.

First, we consider a distinguished corner of scale $j_{a(\eta^*)}$. Such a corner is generated by a corner without resolvent and thus 
cannot be used in a Cauchy-Schwarz bound. Thus, each vertex being mirror symmetric, they carry an even number of distinguished corners.
If a vertex only bears distinguished corners, it is then made of $2k$ replicas of the same corner (and thus brings $3kj_{a}$), and has degree $2k$. 
A vertex of degree $2k$ can belong to at most $2+2k$ faces. For $k>1$, the $3kj_a$ are enough to cancel the $j_{\min}$ of all faces the vertex belongs to. 
For k=1 (vertex of degree 2), if the two edges are of different colours, the vertex belongs to only three faces, that are cancelled out by the $3j_a$. 
If the two edges are of the same colour $c$, the vertex can belong to two distinct faces of colour $c$. 
If any of those faces also goes through a vertex of degree two bearing two undistinguished
corners (and thus bringing $4j_\eta$, enough to cancel all the faces running through it),
or a vertex of degree $\geq 2$, its $j_{min}$ will be cancelled out by this vertex.
If both those faces run only through vertices of degree two bearing only distinguished corners, 
then the final map must be a closed cycle of vertices of degree two bearing only distinguished corners, which is impossible.
Therefore, $\frac32\sum_{H^*}j_{\eta^*}$ is enough to cancel out every potential faces with $j_{min}=j_{a(\eta^*)}$.

For vertices bearing undistinguished corners, the situation is actually better. Each vertex of degree $>1$ brings enough $j_\eta$ to cancel each faces it belongs to. 
Only the leaf without $A_j$ has one more face than $j_\eta$s. However, one face running through a leaf will also run through its only neighbouring vertex, 
which is of degree two or more (recall that the two-leaves-graph is excluded), and will be cancelled out by this vertex. If the neighbour is of degree two, 
then it has only $3$ faces running through it. If it is of degree $>2$, then it has more than enough $j_\eta$.  

Therefore on any final graph, all the $j_{\min}$ can be cancelled individually by a $j_\eta$, hence we have
\begin{align}
\sum_{\mathfrak m}\left[\sum_f j_{\min}(f)-2\sum_{\eta} j_\eta +\frac{1}{2}\sum_{H^*}j_{\eta*}\right]\leq0,
\end{align}
and thus,
\begin{align}
|A(\mathfrak M_\mu)| =\prod_{\mathfrak C_\mu} |A(\mathfrak C_\mu)|&\leq [O(1)\rho]^{E(\mathfrak M)}
M^{-\frac14\sum_{v\in \kM} j_{a(v)}}.  \label{decayattri}
\end{align}
Theorem \ref{thmgraphbound} follows as $\sum_{v\in \kM} j_{a(v)} = 2\sum_{v\in \cB} j_{a(v)}$.
\qed

\subsection{Overall bound}

Collecting the free energy
\eqref{eq:logZfactorised} with the hard-core constraint \eqref{eq:Grassbound} for the Grassmannian integral, the map decomposition and Cauchy Schwarz bound
of \eqref{CS}, the exponential bound of
theorem \ref{BosonicIntegration} and the power-counting of theorem \ref{thmgraphbound},
\begin{align}  \label{eq:logZbound1}
&|\log Z|\le \sum_{n=1}^\infty \frac{1}{n!}  \sum_{\cJ \;{\rm connected}} \;\sum_{\{j_a\} \in \cS^{n} }  \\
&\left[ 
2^{|E_F|}
 \Bigl( \prod_{\cB} \prod_{\genfrac{}{}{0pt}{}{a,b\in \cB}{a\neq b}} (1-\delta_{j_aj_b}) \Bigr)
 \Bigl( \prod_{\genfrac{}{}{0pt}{}{\e_F \in \cF_F}{\e_F=(d,e) } } \delta_{j_{d } j_{e} } \Bigr)
\right]
\\
&\times
\prod_\cB \left[
\sum_{\kM(T_\cB)}
e^{ O(1) \rho  \vert \cB \vert }
 [O(1) \rho]^{|E(\mathfrak M)|}
M^{-\frac12\sum_{v\in \cB} j_{a(v)}}
\right]  \  , 
\end{align}
where, for $\cM(T)$ the (possibly disconnected) maps corresponding to a tree $T$ after edges-derivations and the corresponding Fa\`a di Bruno partitioning of 
\eqref{eq:partitio}. $\kM(T)$ are then the different maps corresponding to $T$  after merging the $\cM$ with their mirror symmetric and the addition of additional
edges corresponding to numerator $\csig$ fields. A bound on the number of those maps, along with the number of connected jungles, is necessary to achieve the proof.

\subsubsection{Combinatorial bounds}

The sum over connected jungles can be reorganised as,
\begin{enumerate}
 \item We choose a partition $\cP$ of $\cV=\{1\dots n\}$ into $|\cP|$ Bosonic blocks $\cB(i),\ i\in\{1\dots|\cP|\}$.
 \item We choose a Fermionic forest $\cF_F$ :  We choose a set $E_F$  of Fermionic edges forming a tree over Bosonic blocks, with coordination $D_\cB$ at each block
 $\cB\in\cP$. Then, we choose a particular vertex to hook each incident Fermionic edges inside a block.
 Each half-edges incident to a block $\cB$ has $|\cB|$ possible vertices to hook at.
 \item We choose a Bosonic tree $\cT_\cB$ in each block $\cB$.
\end{enumerate}
And the sum writes,
\[
 \sum_\cJ A(\cF_F)\prod_{\cB A(\cT_\cB)} = \sum_\cP \sum_{\cF_F}A(\cF_F) \left(\prod_{i=1}^{|\cP|} \sum_{\cT_{\cB(i)}} A(\cT_{\cB(i)}) \right)\ .
\]

The number of trees $\cT_\cB$ with $|\cB|$ labelled vertices of fixed degree $d_i,\ i\in\{1\dots |\cB|\}$ is
\[
 \frac{(|\cB|-2)!}{\prod_i (d_i-1)!}\ .
\]

From there, the maps $\kM$ corresponding to a tree $\cT_\cB$ are obtaiend by,
\begin{itemize}
 \item Applying the derivatives:
For a $d_i$-valent vertex, the number of (Fa\`a di Bruno) partitioning of the set of incident half edges into $m_1$ 1-valent effective vertices, 
$m_2$ 2-valent and so on is,
\[
 \frac{d_i!}{\prod_{k\ge 1} m_k ! (k!)^{m_k}}\ .
\]
The number of cyclic orderings of $m_k$ half edges is \[
(m_k-1)!\] 
\item Merging the graph with its mirror symmetric and developing extra edges to absorb numerator $\csig$. 
The number of possible ways to add $n$ additional edges to a map with $2(|\cB|-1)$ edges
is,
\[
 \frac{\left(4|\cB|-4+2n\right)!}{(4|\cB|-4)!}\ .
\]

The number of additional edges being bounded by $8(|\cB|-1)$, the number of possible maps $\kM$ for a given Bosonic block $\cB$ is bounded by,
\[
\frac{1}{|\cB| !}
\sum_{d_1 \dots d_{|\cB|}\ge 1}^{\sum d_i=2(|\cB|-1) }
(|\cB|-2)!\prod_{i=1}^{|\cB|}
d_i
\left(\sum_{m_1\dots \ge0}^{\sum_k k m_k = d_i}
\prod_{k\ge 1}
 \frac{1}{ m_k  (k!)^{m_k}}\right) \times\
 \frac{\left(10|\cB|-10\right)!}{(4|\cB|-4)!}
 \ .
 \]
 \end{itemize}

 Using the bound
 \[
  \sum_{m_1\dots \ge0}^{\sum_k k m_k = d_i}
\prod_{k\ge 1}
 \frac{d_i}{ m_k  (k!)^{m_k}} \le O(1)^{d_i}\ ,
 \]
 we can write,
 \[
  \sum_{\kM} 1 \le \frac{\left(10|\cB|\right)!O(1)^{2(|\cB|-1)}}{|\cB| (|\cB|-1)}
\sum_{d_1 \dots d_{|\cB|}\ge 1}^{\sum d_i=2(|\cB|-1) } 1
  \le \left(10|\cB|\right)!\;(O(1)|\cB|)^{2|\cB|}
 \ .
 \]
 
 Finally, the number of Fermionic forest is given by an adapted Cayley's formula. The Cayley's Formula \cite{Cayley1889} states that the number of trees over $v$ labelled vertices
 is $v^{v-2}$. Accounting for the fact that each Fermionic half-edge incident to $\cB$ has $|\cB|$ possible vertices to hook on, the formula becomes,
 \[
  \sum_{\cF_F} 1 = \left(\sum_{\cB\in\cP} |\cB|\right)^{|\cP|-2} \prod_{\cB\in\cP} |\cB|\ .
 \]
\prf
Pr\"ufer's code is a bijection between trees over $v$ labelled vertices and Pr\"ufer sequences of $v-2$ elements of $\{1\dots v\}$ \cite{Prufer1918}.
The conversion algorithm from a tree to a sequence is :
{\emph Take the leaf with the smallest label, delete it from the tree and write down the value of its only neighbour.
Repeat this process  until only one vertex remains.}

In order to encode our Fermionic forests, the above code can be modified as follow :
Let $\cP$ be a partition of $\cV$ in $|\cP|$ Bosonic blocks $\cB(i)$ labelled from $1$ to $|\cP|$, $\cF_F$ be a Fermionic forest such that, if each Bosonic block 
is replaced by a single effective vertex, $\cF_F$ becomes a tree $T_F$.  The vertices of $T_F$ are the Bosonic blocks labelled by $i\in\{1\dots |\cP|\}$.
 Take the leaf of $T_F$ with the smallest label $i$, delete it from the tree and, denoting $(a,b)\in E(\cF_F)$ the only Fermionic 
 edge incident to the block $\cB(i)$, with $a\in\cV\setminus\cB(i)$ and $b\in\cB(i)$.
 \begin{itemize}
  \item  write down in $A$ the index of the vertex $a\in\cV\setminus\cB(i)$ to
 which the only Fermionic edge incident to $\cB(i)$ was hooked.
 \item $B_i=b$ with $b\in\cB(i)$ the index of the vertex to
 which the only Fermionic edge incident to $\cB(i)$ was hooked.
 \end{itemize}
Repeat this process  until only one vertex remains. Then $B_{|\cP|}$ is the index of the vertex of $\cB(|\cP|)$ to which was hooked the last edge to be deleted.

This algorithm wrote two sequences :
\begin{enumerate}
 \item $A$ is a generalised Pr\"ufer sequence of of $|\cP|-2$ elements of $\cV=\bigcup_{i=1}^{|\cP|} \cB(i)$. The number of such sequences is given by,
 \[
  \left(\sum_{\cB\in\cP} |\cB|\right)^{|\cP|-2} = n^{|\cP|-2}\ .
 \]
\item $B$ chooses a particular vertex $B^i\in \cB(i)$ for each $\cB(i)\in\cP$. The number of such sequences is given by,
\[
 \prod_{\cB\in\cP} |\cB|\ .
\]
\end{enumerate}
 Pr\"ufer's proof of bijectivity stands with minor adjustments. 
 \qed

\subsubsection{Amplitude bound}
The power of $M$ arising from map amplitudes in \eqref{eq:logZbound1} can be further bounded as
the hard core constraint $ \prod_{a,b\in \cB} (1-\delta_{j_aj_b})$ inside each block imposes that the slices indices of its vertices are different, and,
\begin{align}
\sum_{v\in \cB} j_{a(v)}\ &\ge\  j_{\min} + (j_{\min}+1) \cdots +(j_{\min}+|\cB|-1)\\
 &\qquad =
 |\cB| j_{\min} + \frac{|\cB|^2-|\cB|}{2} \ \ge \  \frac{|\cB|^2}{2}  \ ,
\end{align}
and therefore,
\[
 \sum_{v\in \cB} j_{a(v)} = 2\left(\frac12\sum_{v\in \cB} j_{a(v)}\right)\ge \frac{|\cB|^2}{4} + \frac12\sum_{v\in \cB} j_{a(v)} \ .
\]
The sum over scales can be bounded as, inside a block $\cB$,
\begin{align}
 &\sum_{\{j_a\}\in\cS^{|\cB|}}\left[\prod_{a,b\in \cB} (1-\delta_{j_aj_b})\right] M^{-\frac12\sum_{v\in \cB} j_{a(v)}} \le
 M^{-\frac{|\cB|^2}8}\sum_{\{j_a\}\in\cS^{|\cB|}} M^{-\frac14\sum_{v\in \cB} j_{a(v)}}\\
\qquad\qquad &\le M^{-\frac{|\cB|^2}8}   \left(\sum_{j\ge 1} M^{-\frac14 j}\right)^{|\cB|} \le O(1)^{|\cB|} M^{-\frac{|\cB|^2}8} \ ,
\end{align}
as the geometric series converges for any $M>1$.

\subsubsection*{Final bound}

Gathering all previous bounds, and  dropping the Fermionic edges $\delta$'s,
\begin{align}  \label{eq:logZbound2}
|\log Z|&\le \sum_{n=1}^\infty \frac{1}{n!}   \sum_{B_1\dots \ge0}^{\sum_k k B_k = n} \frac{n!}{\prod_{k\ge 1} B_k ! (k!)^{B_k}} \\
&\times\ n^{\sum B_k-2}\ 2^{\sum B_k-1}\ \prod_{k\ge 1}\left[
 O(1)  \rho^{8k}  M^{-\frac{k^2}{16}} \right]^{B_k}
 \ ,
\end{align}
where we used,
\[
 |\cB| \left(10|\cB|\right)!\;(O(1)|\cB|)^{2|\cB|}
e^{ O(1) \rho  \vert \cB \vert }
 [O(1) \rho]^{|E|}
 O(1)^{|\cB|}  M^{-\frac{|\cB|^2}8} \le O(1)  \rho^{8|\cB|}  M^{-\frac{|\cB|^2}{16}}
 \ .
\]
Using $n^{\sum B_k-2}\le e^n (\sum B_k)!$, we write,
\begin{align}  \label{eq:logZbound3}
|\log Z|&\le \sum_{n=1}^\infty   \sum_{B_1\dots \ge0}^{\sum_k k B_k = n} \frac{(\sum B_k)!}{\prod_{k} B_k ! }  \ \prod_{k\ge 1}\left[\frac{e^k}{k!}
 O(1)  \rho^{8k}  M^{-\frac{k^2}{16}} \right]^{B_k}
 \ ,
\end{align}
Choosing $M$ high enough ensures that $\frac{e^k}{k!}
 O(1) M^{-\frac{k^2}{32}}\le 1$, and the series above is bounded by
 \begin{equation}\label{eq:logZbound4}
  |\log Z|\le \sum_{B=0}^\infty \left[ \sum_{k= 1}^\infty  \rho^{8k}  M^{-\frac{k^2}{32}} \right]^{B}
 \ ,
 \end{equation}
where the sum over $k$ converges and is bounded,
\[
  \sum_{k= 1}^\infty  \rho^{8k}  M^{-\frac{k^2}{32}} \le  \sum_{k= 1}^\infty  \rho^{8k}  M^{-\frac{k}{32}} = \frac{1}{1-\rho^8M^{1/32}}< 1,
\]
which ensures the convergence of the sum over $B$ in \eqref{eq:logZbound4}. 
Therefore, the multiscale loop vertex expansion of the free energy \eqref{eq:logZjungle}  is absolutely convergent, which achieves the proof of theorem \ref{T43theorem}.

\part{Enhanced models}
\chapter{Enhanced tensor models}\label{chap:enh}
 \section{Limitations of the standard quartic model}

The standard invariant quartic tensor model \eqref{eq:standardmodel} at rank $D\geq 3$ is defined with a single 
coupling parameter $\lambda$ and a set of interactions $\cQ$ as,
\begin{equation} \label{eq:standmodintro}
d\mu =  e^{-\frac{\lambda }{2}N^{\alpha}\sum_{\mathcal{C} \in \mathcal{Q}}  V_{\mathcal{C}}(\bar{\bf T},{ \bf T} ) }\ d\mu_\id(\bar\bT,\bT)\ .
\end{equation}
The scaling in $N$, namely the power of $N$ factor in the interaction term, has been chosen carefully as $\alpha=1-D$, to ensure that, 
\begin{itemize}
 \item the cumulants obey a polynomial bound in $1/N$, of type $\kK(\tau^\cD)\le K(\lambda) N^{-\Omega_{\min}(\tau^\cD)  }$, 
 with $\Omega_{\min}>0$ (Theorem \ref{th:conv}), 
 and so does the rescaled free energy $N^{-D} \log Z$,
 \item the free energy and cumulants admit a $1/N$ expansion, and there are infinitely many Feynman graphs with the same exponent in $N$ (section \ref{subsec:1/Nexp}).
\end{itemize}
The first condition ensure that the free energy and observables do not diverge at large $N$. Along with the existence of the $1/N$ expansion, 
they require $\alpha \leq 1-D$, such that for graphs of larger size,
the higher number of faces is compensated by the $N^{\alpha}$ factor brought by each interaction bubbles. 
Choosing exactly $\alpha = 1-D$ further ensures that, in the intermediate field representation, adding a
mono-coloured leaf to a map does not change the power of $N$, therefore arbitrarily large maps can be built with the same power of $N$.

For the free energy $\log Z$ and cumulants with trivial structure $\kK(\id_k),\ k\geq1$, the leading order in $1/N$ is therefore the family of the plane trees
made solely of mono-coloured edges. Such edges correspond to the {\emph melonic} interaction terms, with $|\cC|=1$, and in the $D+1$-coloured graph representation, 
such trees correspond to {\emph melonic} graphs \cite{GuRy1302}.

Graphs with non-melonic interaction bubbles (with $|\cC|\geq2$) do not, however, participate in the leading order in $1/N$ of the perturbative expansion of the
standard quartic model. At large $N$, non-melonic interaction are therefore suppressed and do not contribute to the behaviour of the standard model. 

In order to make relevant the addition of non-melonic bubbles in the interaction term of the invariant model, we need to move away from the standard quartic model
and fine tune the scaling in $N$ of the different interactions independently.

\section{Models with maximally enhanced interactions}\label{sec:max}

The non-melonic interactions do not participate in the leading order in $1/N$ of the standard quartic model because the power of $N$ in the associated pre-factor 
(namely $\alpha$ in \eqref{eq:standmodintro}) is to low. Therefore, a non-melonic bubble in a Feynamn graph 
(or a multi-coloured edge in the corresponding intermediate field map) comes with a power of $N$ factor to small for the graph to be leading order. 

In order to build a well defined tensor model where the non-melonic interactions survive at large $N$, we need to find a family 
$(\alpha_\cC)_{\cC\in\cQ}$ of scalings in $N$ such that, for the {\it enhanced} tensor model,
\begin{equation} \label{eq:enhancedmod1}
d\mu =  e^{-\frac{\lambda}{2}\sum_{\mathcal{C} \in \mathcal{Q}}  N^{\alpha_\cC} V_{\mathcal{C}}(\bar{\bf T},{ \bf T} ) }\ d\mu_\id(\bar\bT,\bT)\ .
\end{equation}
\begin{itemize}
 \item the rescaled free energy $N^{-D} \log Z$ and the cumulants still obey a polynomial bound in $1/N$, 
 of type $\kK(\tau^\cD)\le K(\lambda) N^{-\Omega_{\min}(\tau^\cD)  }$,  with $\Omega_{\min}>0$,
 \item the free energy and cumulants admit a $1/N$ expansion, and there are graphs with arbitrary many bubbles of each type $\cC\in\cQ$ in each non-empty order in $1/N$.
 \end{itemize}
The first condition, being the same as for standard models, will give us a bound on the scaling in $N$ of each interaction. The last condition is verified if, for any 
interaction of colours $\cC$, the free energy of the $\cC$-model,
\[
 d\mu_{\cC} =  e^{-\frac{\lambda }{2} N^{\alpha_\cC}  V_{\mathcal{C}}(\bar{\bf T},{ \bf T} ) }\ d\mu_\id(\bar\bT,\bT)\ ,
\]
is properly bounded and admits infinitely many graphs in the leading order in $1/N$. 
If such a $\alpha_\cC$ exists, we say that the interaction $V_{\cC}$ can be {\it maximally enhanced}. maximally refers to the fact that if $\alpha_\cC$ is further 
increased, the boundedness of the cumulants and free energy is lost.

The remainder of this chapter shall be written in the intermediate field representation, we remind that with our conventions, for any $\cC\in\cQ$,
$|\cC|\leq\left\lfloor D/2\right\rfloor$ and if $|\cC|=D/2$, then $1\in\cC$. 
\subsubsection{Enhancing the interactions}
In the intermediate field representation, the partition function of the
$\cC$-model writes,
\begin{align}\label{eq:Cmodel}
Z_\cC=\int 
e^{-
\ \tr_{\cD} \big[ \log\left(\id^{\cD} + i \sqrt{N^{\alpha_\cC}\lambda}\: \id^{\cD \setminus \cC }
\otimes\csig^{\mathcal{C}}\right)   \big] }
  \; d\mu_\id(\csig^\cC) \ .
\end{align}

Let $\cM$ be a Feynman map contributing to the order $N^{-\Omega(\cM)}$ in the free energy $\log Z_\cC$, and $\cT\subset\cM$ a spanning tree. 
By construction, every edges of $\cM$ are of colours $\cC$.
The deletion of an edge $\ell\in\cM\setminus\cT$  can change the number of faces of $\cM$ by at most $|\cC|$, therefore,
\[
|F_{\rm int}(\cM\setminus\ell)|\geq |F_{\rm int}(\cM)| - |\cC|\ ,
 \]
 and, during the deletion of the loop edge $\ell$, the exponent of $1/N$ varies as,
 \begin{equation}\label{eq:MvT}
  \Omega(\cM\setminus\ell)\leq \Omega(\cM) + |\cC| + \alpha_\cC \ .
\end{equation}
Iterating $|E|-|V|+1$ times until only $\cT$ remains,
\[
  \Omega(\cT)\leq \Omega(\cM) + (|E|-|V|+1) (|\cC| + \alpha_\cC) \ .
\]

The deletion of a leaf of $\cT$ (deletion of a monovalent vertex and the only incident edge) decrease the number of faces by $D-|\cC|$,
iterating up to a single remaining vertex shows that
the number of faces of any tree is 
\[
|F_{\rm int} (\cT)|= D+ (|V(\cT)|-1)(D-|\cC|)\ .
\]
and its exponent of $1/N$ is,
\begin{equation}\label{eq:OmT}
  \Omega(\cT)= -  D - (|V(\cT)|-1)(\alpha_\cC +D-|\cC|) \ .
\end{equation}

Gathering the results of \eqref{eq:MvT} and \eqref{eq:OmT},
\begin{itemize}
 \item 
 For $\alpha_\cC > |\cC|-D$, the exponent of $1/N$ for trees, $\Omega(\cT)$, is not bounded from bellow, 
 therefore the free energy $N^{D}\log Z_\cC$ is not bounded.
 \item
 For $\alpha_\cC \leq |\cC|-D$, $\Omega(\cT)>-D$ for any plane tree. Moreover, for any map $\cM$ and  spanning tree $\cT$, 
 $\Omega(\cM)\geq\Omega(\cT)\geq -D$. The perturbative expansion of the free energy is properly bounded in $N$, and the free energy admits a $1/N$ expansion.
 \item
For  $\alpha_\cC = |\cC|-D$, $\Omega(\cT)=-D$ for any tree $\cT$, the whole family of tree contributes to the leading order in $1/N$.
\end{itemize}
Therefore, at any rank $D$, any quartic invariant $V_\cC$ can be maximally enhanced, by choosing the scaling in $N$.
\[
 \alpha_\cC = |\cC|-D \ .
\]
The maximally enhanced invariant quartic tensor model with a set of interactions $\cQ$ and coupling parameter $\lambda$ writes, 
\begin{equation} \label{eq:enhancedmod}
d\mu =  e^{-\frac{\lambda}{2}\sum_{\mathcal{C} \in \mathcal{Q}}  N^{|\cC|-D} V_{\mathcal{C}}(\bar{\bf T},{ \bf T} ) }\ d\mu_\id(\bar\bT,\bT)\ .
\end{equation}
\section{Feynman expansion}
In the intermediate field representation, the moment generating function of the
maximally enhanced quartic model writes,
\begin{align}
Z(\bJ,\bar \bJ)=\int 
e^{-
\ \tr_{\cD} \big[ \log\left(\id^{\cD} + A(\csig)\right)   \big] 
+  \sum_{nm } \bar \bJ_{n} 
\left[\frac{1}{ \id^{\cD} + A(\csig) }\right]_{nm}\bJ_{m} }  \; d\mu_\id(\csig) \ ,
\end{align}
with
\begin{align}
 A(\csig)=\sum_{\mathcal{C}\in \cal{Q}}\ i\ \sqrt{N^{|\cC|-D}\lambda}\  \id^{\cD \setminus \cC }
\otimes\csig^{\mathcal{C}} \ .
\end{align}
The Feynman expansion of the intermediate fields is similar as for standard invariant model (Section \ref{subsec:IFFeyn}) and
the amplitude of a map writes,
\[
 A(\cM) =  \left(-\lambda\right)^{|E(\cM)|} N^{-\Omega(\cM)}\ ,
\]
where the exponent of $1/N$ is,
\begin{equation}\label{eq:enhexpon}
 \Omega(\cM) = \sum_{\cC\in\cQ}(D-|\cC|)|E_\cC(\cM)|-|F_{\rm int}(\cM)|\ .
\end{equation}
where $E_\cC(\cM)$ is the set of edges of colours $\cC$.

\subsection{The $1/N$ expansion}
The choice of maximally enhanced observable ensure the existence of the $1/N$ expansion, 
as the exponent of $1/N$ of the Feynman maps is  bounded from bellow. 
\begin{theorem}\label{thm:enh1/Nexp}
 For any $D$-uple  $\tau^\cD$ of permutations over $k$ elements, we define the minimal exponent,
 \[
  \Omega_{\min}(\tau^\cD) = -D + \left\lceil D/2\right\rceil k +  C(\tau^\cD) \ .
 \]
where $C(\tau^\cD)$ is the number of connected components of the $D$-coloured graph associated with $\tau^\cD$. Then,
with the exponent $\Omega$ defined in \eqref{eq:enhexpon},
 \begin{itemize}
 \item For $k>0$,
 for any multicoloured map $\cM$ with $\partial\cM=\tau^\cD$,  $\Omega(\cM)\geq\Omega_{\min}(\tau_\cD)\geq 0$.
 \item For a vacuum map $\cM$, $\Omega(\cM)\geq-D = \Omega_{\min}(\emptyset)$.
 \end{itemize}
\end{theorem}
\prf
The proof is quite similar to the proof of the first part of Theorem \ref{thm:1/Nexp} and its structure
follows precisely the proof of Lemma \ref{lem:facesbound}. However, instead of tracking the number of internal faces $F_{\rm int}$, 
we must track the whole exponent $-\Omega(\cM)$ in order to account for the different $\alpha_\cC$'s.
Let $\cM$ be a multicoloured map, we introduce the map $\cM^*$ with (mono-coloured) external edges as for Lemma \ref{lem:facesbound}. 
The external edges are not associated with $N^{1-D}$ factors and the exponent of $\cM^*$ follows,
\[
\Omega (\cM^*) = \Omega (\cM)- Dk(\partial\cM)\ .
\]
We choose a spanning tree $\cT$ and define $\cT^*$ as for  Lemma \ref{lem:facesbound}. 
Deleting an edge $\e$ of $\cM^*\setminus\cT^*$ can change the number of
internal faces by at most $|\cC(\e)|$ while removing a factor $N^{|\cC|-D}$. As $|\cC|\le \left\lfloor D/2\right\rfloor$, we have
$
\Omega (\cM^*\setminus\e) \le \Omega (\cM^*)
$.
Iterating until all edges in $\cM^*\setminus\cT^*$ are deleted, we obtain,
\[
\Omega (\cT^*) \le \Omega (\cM^*)\ .
\]
 It is now sufficient to prove that for any $\cT^*$,
$ C(\partial\cT^*)-\Omega(\cT^*)\le D+\lfloor D/2\rfloor k$.
Let us denote $\ell$ a leaf of $\cT^*$ and let us denote $\hat\cT^*$ the tree obtained from $\cT^*$ after deleting $\ell$ as follows.

{\noindent \it Deleting a non-ciliated leaf} connected to the rest of map by an edge of colour $\cC$ decreases the number of face by $D-|\cC|$, removes 
a factor $N^{|\cC|-D}$, and does not affect the boundary graph, therefore, 
\[
 C(\partial\hat\cT^*)-\Omega(\hat\cT^*) = C(\partial\cT^*)-\Omega(\cT^*)\ .
\]
{\noindent \it Deleting a non-ciliated leaf} with cilium $h$, connected to the rest of map by an edge of colour $\cC$,
follows the two-steps process of Section \ref{subsec:1/Nexp}. The discussion about the changes in the 
boundary graphs and the number of internal faces does not change but we must account for the factor $N^{|\cC|-D}$.
The four combined cases are then treated as follows
\begin{itemize}
 \item no external face is looped at $h$. Then:
     \[ 
      -\Omega (\cT^*)+C(\partial\cT^*) \le \left[-\Omega (\hat \cT^*) + |\cC| +(|\cC|-D)\right] + \left[ C(\partial \hat \cT^*) + 1\right]
      \;. 
     \]
 \item all the external faces are looped at $h$. Then:
      \[
        -\Omega (\cT^*)+C(\partial\cT^*) \le \left[-\Omega (\hat \cT^*) + (D-|\cC|) + (|\cC|-D)\right] + \left[ C(\partial \hat \cT^*) + 1 \right]
         \;. 
      \]
\item At least one external face is  looped at $h$, but not all. Then:
       \[
        -\Omega (\cT^*)+C(\partial\cT^*) \le \left[ F_{\rm int} (\hat \cT^*)  + D 
        + (|\cC|-D)\right] + \left[C(\partial \hat \cT^*)\right]  \;.
      \]
\end{itemize}
In all cases, by deleting a ciliated leaf:
 \begin{align}
  C(\partial\cT^*)-\Omega(\cT^*)  \leq
  C(\partial \hat \cT^*)-\Omega (\hat \cT^*)+ \lfloor D/2\rfloor \; .
 \end{align}
where we used $1\le|\cC|\le\lfloor D/2\rfloor$.
Iterating up to the last vertex, which, ciliated or not, has $D$ internal faces, we get,
\[
  C(\partial\cT^*)-\Omega(\cT^*)\le D+ \lfloor D/2\rfloor k\ ,
\]
and we conclude.

\qed
\subsection{Leading order maps}

In this section we focus on vacuum maps, and characterise the leading order in the $1/N$ expansion of $\log Z$. 
The leading order of the 2-point cumulant $\kK(\id_2^\cD)$ is similar, with the only difference being the addition of a single cilium to the maps, 
which turns $D$ internal faces into external ones. The leading order maps for higher order cumulants are 
more complicated as they must respect a non-trivial boundary graph.  

A vacuum connected map $\cM$ is leading order if $\Omega(\cM)=-D$. The single vertex map is leading order. 
Let $\cM$ be a map. Let $\ell\in\cM$ be a leaf (a mono-valent vertex). The deletion of $\ell$ does not change the exponent of a map, 
\begin{equation}\label{eq:deletleaf}
\Omega(\cM\setminus\ell)=\Omega(\cM)\ ,
\end{equation}
therefore, plane trees are leading order.

Let $\cT\subset\cM$ be a spanning tree. Let $\e\in\cM\setminus\cT$ be an edge, then
\begin{equation}\label{eq:deletloop}
 \Omega(\cM\setminus\e)\le\Omega(\cM)+ D-2|\cC(\e)|\ .
\end{equation}
Therefore, 
\begin{itemize}
\item the deletion of a loop-edge cannot increase the exponent $\Omega$.
\item if $|\cC(\e)|\neq D/2$, the deletion of the loop-edge $\e$ decreases the exponent $\Omega$. 
\end{itemize}
\subsubsection{Odd rank models}
The second proposition shows that for tensor models of odd rank $D$, the leading order maps are the trees.
\begin{proposition}
 If $D\not\in2\setN$, then a connected vacuum map $\cM$ is leading order if and only if $\cM$ is a plane tree.
\end{proposition}
This is always true, as for odd $D$, $\max_{\cC\in\cQ}|\cC|<D/2$.

\subsection{Leading order for even rank models}
In this subsection we characterise the leading order maps for maximally enhanced models of even rank $D\in2\setN$ (with $D\ge4$). 
We assume that the set of interactions $\cQ$ is chosen such that there is some $\cC\in\cQ$ with $|\cC|=D/2$.
\begin{definition}
 Let $\cM$ be a $\cD$-multicoloured map with $D\in2\setN$. A maximally coloured edge is an edge $\e\in\cM$ with $|\cC(\e)|=D/2$. 
 Edges with $|\cC(\e)|<D/2$ are called non-maximally coloured.
 We define the set of maximally coloured interactions as,
 \[
  \cQ_{\max} = \{ \cC\in\cQ , \ |\cC|=D/2\} \ .
 \]
\end{definition}
The term \emph{maximally coloured} is quite natural as $D/2$ is the maximum amount of colours that an edge can bear.

\begin{lemma} \label{corsub}
If a connected map $\cM$ is leading order, all its connected sub-maps $\cM' \subset \cM$ must also be leading order.
\end{lemma}

\prf 
Suppose $\cM' \subset \cM$
is vacuum and \emph{not} leading, $\Omega(\cM')>\Omega(\cM)=-D$. We can pick a spanning tree $\cT \subset \cM$ whose restriction $\cT'$ to $\cM'$ is 
a spanning tree of $\cM'$. Then $ \Omega(\cM) = \Omega(\cT) = \Omega(\cT') < \Omega(\cM')$.
\begin{itemize}
\item Let $\tilde{\cM} = \cM' \cup (\cT\setminus\cT')$ be the map obtained from $\cM$ by deleting one by one the edges
which are neither in $\cT$ nor in $\cM'$. According to \eqref{eq:deletloop}, $\Omega(\cM)\geq \Omega(\tilde{\cM})$. 

\item Deleting the vertices of $\cT\setminus\cT'$ from $\tilde{\cM}$ can be done by deleting in sequence the leaves of the tree, which, according to 
\eqref{eq:deletloop} does not change the value of $\Omega$.  Therefore, $\Omega(\cM) \geq \Omega(\tilde{\cM}) = \Omega(\cM')$.
\end{itemize} 
\qed

Using the above lemma, we can now  characterize the leading order maps. 
\begin{proposition}
For any leading order connected vacuum map,
\begin{enumerate}
\item\label{TreeCase} Any non-maximally coloured edge is a bridge. Therefore any connected sub-map made of non-maximally coloured edges is a tree.
\item\label{PlanarCase} For a given $\cC\in\cQ_{\max}$, any connected sub-map formed by maximally coloured edges of colours $\cC$ is planar.
\end{enumerate}
\end{proposition}
\prf 
\begin{itemize}
\item {\it Item \ref{TreeCase}}\;\;
It is a direct consequence of \eqref{eq:deletloop} : if $\cM$ is leading order, it cannot contain a non-maximally coloured loop edge. 
The tree characterisation is a trivial consequence of the previous statement, but can be recovered another way. 
Such sub-maps correspond to a sector deprived of maximally coloured interaction, and therefore follow the same behaviour as odd rank enhanced models. 
%
%
 \item {\it Item \ref{PlanarCase}}\;\;
Let $\cC$ be a maximal set of colours, we denote $\cM_{\cC}$ the connected map. As an ordinary map (that is, forgetting the edge colours),
we denote the number of faces, edges and vertices $F_{\cC}, E_{\cC}$ and $V_{\cC}$. Notice that for $c,\in\cC$, its faces of colour $c$ 
coincide with those of the ordinary map, which implies $|F^{c} |=|F_{\cC}|$. Moreover, each vertex contributes to one face for each colour $c'\not\in\cC$. 
Therefore $|F^{c'}  |= |V_{\cC}|$. Consequently, the exponent $\Omega$ evaluated on such a connected submap reduces to
\begin{equation*}
\Omega(\cM_{\cC}) = \frac D2\,|E_{\cC}| - \sum_{c=1}^D |F^c| = -\frac D2 (F_{\cC} - E_{\cC} + V_{\cC}) =  D (g_{\cC} - 1),
\end{equation*}
where the genus $g_{\cC}\geq 0$ of $\cM_{\cC}$ seen as an ordinary map has been introduced, in a similar way as for matrices in section \ref{sec:MatrixModels}.
Therefore, minimising $\Omega(\cM_{\cC})$ is equivalent to the map being planar, 
\[
 \Omega(\cM_{\cC})=-D\ \iff\ g_{\cC}=0 \ .
\]
 \end{itemize}
\qed

The second item could be expected. For $\cC\in\cQ_{\max}$,
we can re-organise the indices and write $\bT_n=\bT_{n^\cC n^{\cD\setminus\cC}}$ as a size $N^{D/2}$ matrix. The $\cC$-model \eqref{eq:Cmodel} now writes 
exactly as the quartic matrix model \eqref{eq:matrixmodel} of size $N^{D/2}$. Denoting $\bT^*_{ n^{\cD\setminus\cC}n^\cC}=\bar\bT_{n^\cC n^{\cD\setminus\cC}}$,
\[
 Z_\cC = \int\left(\prod_{a,b}\frac{d\bT_{ab}d\bT^*_{ba}}{2i\pi}\right) 
 e^{ -\tr_\cC(\bT\bT^*) - \frac{\lambda}{2N^{D/2}} \tr_\cC\ (\bT\bT^*)^2
 }
 \ ,
\]
which generates Feynman maps on amplitude $\lambda^{|B|}(N^{\frac D2})^{2-2g}$.

We also want to analyse how the various non-maximally coloured and maximally coloured connected sub-maps are connected to one another in a leading order map.
%
%
%

Consider a vacuum leading order map $\cM$ in the full quartic model. We know that non-maximally coloured edges are bridges and that
the maximally coloured components of given colours $\cC$ are planar. 
It remains to characterize the way the maximally coloured planar components can be attached to one another.

\begin{definition}\label{def:enhtreecarac}
Let $\cM$ be a vacuum, connected map.
\begin{itemize}
\item For $\cC\in\cQ_{\max}$, remove all edges but those of colour type $\cC$ as well as the isolated vertices this creates, 
and denote the connected components thus obtained $\cM_{\cC}^{(1)},\dots,$ $ \cM_{\cC}^{(R_\cC)}$. 
\item Pick up a spanning tree $\cT_{\cC}^{(\rho_\cC)}$ for each $\cC\in\cQ_{\max}$ and $\rho_\cC = 1,\dots, R_\cC$. Let $\cT_\cM\subset\cM$ be the sub-map which contains all those spanning trees as well as all monocoloured edges.
\end{itemize}
\end{definition}

\begin{proposition} \label{prop:FullQuartic}
Let $\cM$ be a vacuum, connected, leading order map of the enhanced model. Then $\cT_\cM$ is a tree.
\end{proposition}

\prf 
$\cT_\cM$ obviously contains all the vertices of $\cM$ and is a connected map. Let us assume that $\cM$ is leading order while
$\cT_\cM$ has a cycle. This cycle cannot contain a non-maximally coloured edge. It is therefore assumed to be made of maximally coloured edges only.
Furthermore, those maximally coloured edges cannot be all of the same colours $\cC$. Indeed, if that were the case, then they would all belong to
a single connected component $\cM_{\cC}^{(\rho_\cC)}$ 
(by definition of the latter and because the cycle is connected), but this is impossible as only a spanning tree of $\cM_{\cC}^{(\rho_\cC)}$ is part of $\cT_\cM$.

The cycle, say $\kC$, thus has at least two maximally coloured edges of two different colours. As a vacuum map itself, it is easy to 
check that its degree takes a non leading order value, $\Omega(\kC)\geq -2$. According to Lemma \ref{corsub}, it cannot be a sub-map of a leading order map,
which is a contradiction.

\qed

Proposition \ref{prop:FullQuartic} puts explicit restrictions on the gluing of the planar components in the large $N$ limit.
In particular, $\cM_{\cC}^{(\rho_\cC)}$ and $\cM_{\cC'}^{(\rho_{\cC'})}$ can at most share one vertex. If not, there would be a path 
in $\cT_{\cC}^{(\rho_\cC)}$ and another path in $\cT_{\cC'}^{(\rho_{\cC'})}$ joining the same two vertices, thus creating a cycle in $\cT_\cM$.

Furthermore, there cannot be ``closed chains'' of planar components of different colours. Therefore,if $\cM_{\cC}^{(\rho_\cC)}$ and $\cM_{\cC'}^{(\rho_{\cC'})}$
are not incident to one another, then they can be connected through either non-maximally coloured bridges, or other connected sub-maps $\cM_{\cC''}^{(\rho_{\cC''})}$
whose removal disconnects $\cM$.

\begin{figure} 
\centering
\includegraphics[scale=.35]{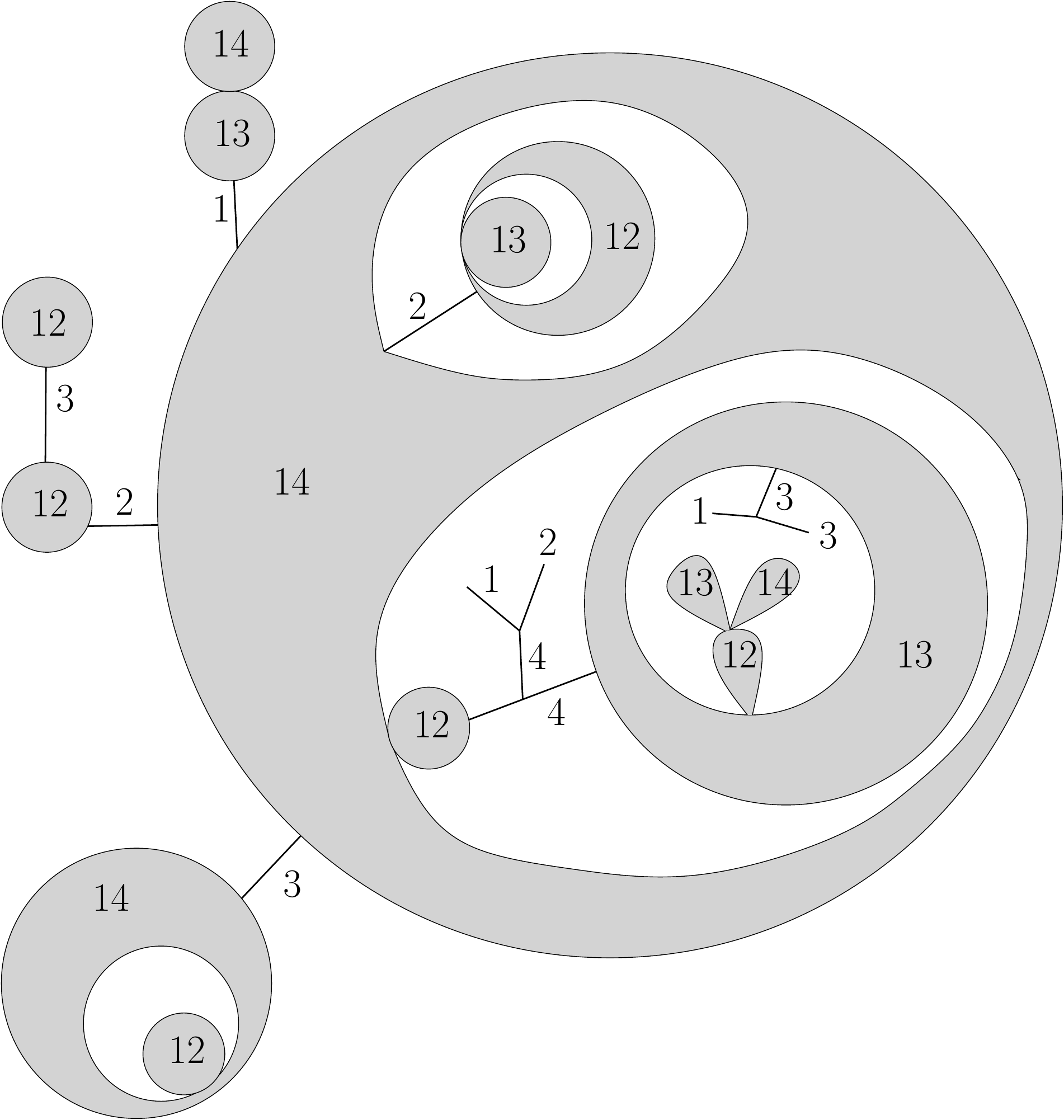}
\caption{\label{fig:multitreedisk} This represents the structure of the leading order maps in the rank $D=4$ enhanced model. 
The grey areas are connected components of given colour set $\cC$. A bi-coloured connected component can be attached to another 
one on a single vertex, without forming cycles of such components.}
\end{figure}

Figure \ref{fig:multitreedisk} therefore shows the structure of the leading order maps. They are planar, and made of trees of non-maximally coloured edges
which connect maximally coloured planar objects. The latter can touch one another at a single vertex at most and do not form closed chains, 
thus displaying a ``cactus'' structure.

\section{Further discussions}

At even rank, the enhanced quartic model generates a whole new family of leading order graphs, that mixes 
the tree behaviour of standard tensor models with the planar behaviour of matrix model.  
The natural continuation, which falls beyond the scope of this thesis, would be to solve the model in the large $N$ limit, by enumerating the leading order
maps contributing to the free energy and the cumulants (see Section \ref{sec:K2} for the second cumulant of the standard quartic model).
This can usually be done by the introduction of a wider class of interactions and the use of Schwinger-Dyson equations \cite{BonGurRyaTan1404,DarGurRiv1307}. 
Unfortunately, the presence of multiple matrix-like maximally coloured interactions, 
with varying index structure, complicates those equations greatly. 

In \cite{BonDelRiv1502}, a simpler, restricted model is introduced at rank $D=4$ with a single maximally coloured interaction ($|\cQ_{\max}|=1$). 
For this restricted case, the model is solved using a generalised class of interaction called trees of necklaces, and the Schwinger-Dyson equations 
reduce to the same equations as in matrix models with multi-trace interactions \cite{AGBC1992,DDSW1990}. It results that free energy admits a critical 
point $\lambda_c$, where it behaves as $(\lambda-\lambda_c)^{2-\gamma}$. Using different coupling for mono-coloured and bi-coloured interactions, 
the different sectors can be tuned to criticality independently,  while the contribution of other sectors is washed away.
This leads to a phase of pure $2D$ gravity ($\gamma=-1/2$)\cite{DFGZJ1993} and another phase of branched polymers ($\gamma=1/2$)\cite{BiaBur1996,GuRy1302}. 
Both sectors can also be tune to criticality simultaneously, 
leading to the proliferation of baby universes ($\gamma=1/3$)\cite{JaiMat1992} as branching competes with the proliferation of planar components.

\chapter{Enhanced tensor field theory}

The usual tensorial group field theories over $U(1)^D$  display the same behaviour as standard invariant tensor models, with their divergent sub-graphs 
- or rather their divergent sub-maps, in the intermediate field representation - having a structure of plane trees. 
A similar {\it enhancement} to the one of invariant models is however possible, with perspective, once again, of an extended family of divergent graphs. 
Such an enhancement must however diverge from the framework of {\it tensorial} field theory, as the tensor invariance of the interaction is not respected.
This chapter, based on  unpublished work in collaboration with Vincent Lahoche, 
introduces the simplest maximally enhanced field theory of even rank, namely the enhanced $T^4_4$.

\section{The Group Field Theory}
%
%
\subsubsection*{$\bf U(1)^4$ field theories}
Our starting point is rank 4 tensor field theory with covariance,
\[
 \bC_{\bar p p} = \frac{\delta_{\bar p p}}{(\bar p\cdot p)^\eta + m^{2\eta}}
\]
where the exponent $\eta$ has been introduce in order to fine-tune the renormalisation properties of the theory.

Denoting $\mathbf{\Psi}_{p},\  p=(p^1,p^2,p^3,p^4),$ an infinite size tensor, the generating functional and the kinetic part of the action are, 
\begin{align}\label{genfuncfourier2}
\mathcal{Z}[\mathbf{J},\bar{\bf J}]&=\int \prod_{p \in \mathbb{Z}^4} d{\bf\Psi}_{ p}\; d\bar{\bf\Psi}_{ p}\
e^{-\: S_{int}[\psit,\bar{\psit}]\: -\: S_{kin}[\psit,\bar{\psit}]\: +\: \sum_{ p}\:  \bar{\bf J}_{ p} \psit_{ p}\:
+\: \bar{\psit}_{ p}{\bf J}_{ p}} \ ,\\
S_{kin}[\psit,\bar{\psit}]&=\sum_{{ p} \in \mathbb{Z}^4}\left(({ p}^{\:2})^{\eta}+m^{2\eta}\right)\bar{\psit}_{ p} \psit_{p} 
\ . 
\end{align}

\begin{figure}
\centering
\includegraphics{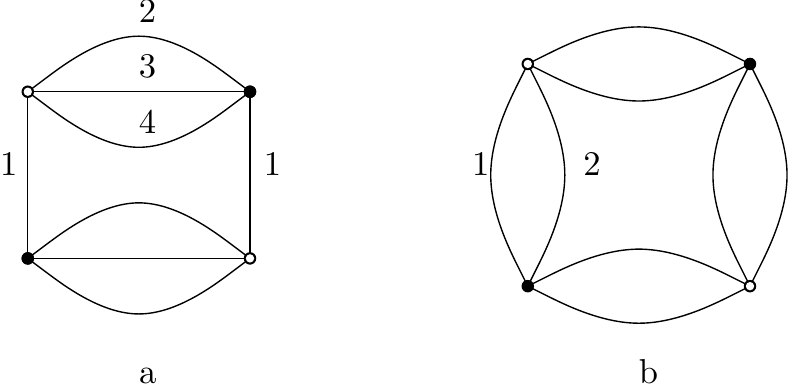} 
\caption{The two families of quartic invariants. Figure a represents $V_1$, a \emph{melonic} bubble ($|\cC|=1$). Figure b shows $V_{12}$, a necklace ( $|\cC|=2$).}\label{fig:fig1}
\end{figure}

The interaction terms for tensorial field theories are the quartic tensor invariants. For $D=4$, they are divided 
into two families :
the melonic invariants $V_c$ with $c\in\{1\dots4\}$ and the necklaces $V_{1c}$ with $c\in\{2\dots4\}$,
\begin{align}
 {\rm V}'_{c}(\psit,\bar\psit) &= i \sum_{ p p' q q'}
 \psit_{ p}\bar\psit_{ q}\ \psit_{ p'}\bar\psit_{ q'}\ \delta_{p^c q'^c}\delta_{q^c p'^c} \prod_{c'\neq c}\delta_{p^{c'} q^{c'}}\delta_{p'^{c'} q'^{c'}}
 \\
 {\rm V}'_{1c}(\psit,\bar\psit) &= i \sum_{ p p' q q'}
 \psit_{ p}\bar\psit_{ q}\ \psit_{ p'}\bar\psit_{ q'}\ \prod_{c'\neq1,c}\delta_{p^{c'} q^{c'}}\delta_{p'^{c'} q'^{c'}}
 \prod_{c'=1,c}\delta_{p^{c'} q'^{c'}}\delta_{q^{c'} p'^{c'}} 
\end{align}
which are both represented in Figure \ref{fig:fig1}.

\subsubsection*{Enhanced interactions}
In Chapter \ref{chap:enh}, tensor models with leading order non-melonic contribution have been introduced using a proper rescaling of the coupling constants associated to non-melonic interactions, by a specific power of the tensor size $N$.
In a group field theory framework, the simplest way to implement a similar enhancement would be with derivative couplings.
The non-local structure of the quartic tensorial interactions offers eight independent possible derivatives.
In Fourrier space, these derivations become multiplications by the corresponding moment.
In order to preserve symmetry, we make a democratic choice, summing derivatives for each colour. The enhanced interaction term for colours $\cC=\{1,i\}$ writes,
\begin{align}
 {\rm V}'_{1i}(\psit,\bar\psit) &= i \sum_{ p p' q q'} \left(\sum_{i=1}^4 |p^i| + |p'^i|\right)
 \psit_{ p}\bar\psit_{ q}\ \psit_{ p'}\bar\psit_{ q'}\ \prod_{c\neq1,i}\delta_{p^c q^c}\delta_{p'^c q'^c}
 \prod_{c=1,i}\delta_{p^c q'^c}\delta_{q^c p'^c} \\ 
 &=\sum_{k=1}^8 {\rm B}^{\prime k}_{1i}(\psit,\bar\psit)   \ ,\label{quarticenhanced}
\end{align}
where for $k\in\{1..4\}$, ${\rm B}^{\prime k}_{1i}$ is the interaction term corresponding to the momentum $p(k)=p^k$, and for $k\in\{5..8\}$, ${\rm B}^{\prime k}_{1i}$ is the term corresponding to $p(k)=p^{\prime k-4}$.
\begin{equation}
 {\rm B}^{\prime k}_{1i} (\psit,\bar\psit)= i \sum_{ p  p' q q'}\ |p(k)|
 \psit_{p}\bar\psit_{ q}\ \psit_{ p'}\bar\psit_{q'}\ \prod_{c\not\in\cC}\delta_{p^c q^c}\delta_{p'^c q'^c}
 \prod_{c\in\cC}\delta_{p^c q'^c}\delta_{q^c p'^c} \; .
\end{equation}
We use the absolute value of the momentum in order to avoid some unfortunate cancellations in the further developments. 
One can then build a new interaction term $S_{int}$:
\begin{equation}
S_{int}[\psit,\bar{\psit}]=\lambda_1\sum_{i=1}^4{\rm V}_{i}[\psit,\bar{\psit}]+\lambda_2\sum_{i=2}^4\sum_{k=1}^8{\rm B}^{\prime k}_{1i}[\psit,\bar{\psit}],\label{interaction}
\end{equation}
where ${\rm V}_{i}$ is the standard quartic melonic trace invariant associated with the colour $i$ \eqref{quarticbubbles}, and ${\rm B}^{\prime k}_{1i}$ is the enhanced necklace bubble associated with colours $1,i$ and the momentum $p(k)$.

The enhanced bubbles, with derivative couplings, are no longer trace invariants, and thus the interactions \eqref{interaction} do not define a standard \emph{tensorial} group field theory.

\section{Power counting}
\subsection{Multi-scale analysis}
\subsubsection*{Scale slicing}

Once again, we introduce scale slices similar to \eqref{eq:slices},
\begin{align}\label{eq:enhslices}
S^1 &= \left\{ p\in\setZ^D,\ \,m^{2\eta}+(p\cdot p)^\eta \,\le\, M^{2\eta} \right\}\crcr
 S^j &= \left\{ p\in\setZ^D,\ M^{2\eta (j-1)}\,<\,m^{2\eta}+(p\cdot p)^\eta  \,\le\, M^{2\eta j} \right\}\quad\forall j> 1
\end{align}
and the covariance of scale $j$,
\begin{align}
 \bC_j = \bI_{j} \bC \ ,\qquad\qquad (\bI_j)_{\bar n n} =\bI_{S^j}(n)\ \delta_{\bar nn}\ .
\end{align}
The regularisation procedure consists in choosing a $j_{\max}$ and replacing the covariance by $ \bC_{\le j_{\max}} = \sum_{j=1}^{j_{\max}} \bC_j$.

\subsubsection*{Perturbative Expansion}

According to \eqref{eq:feynmanamplitude}, the regularised amplitude of a Feynman graph writes,
\begin{align}\label{eq:TFTfeynmanamplitude}
 A^{\rm reg}(\cG)&=\prod_{\cC\in\cQ}\prod_{\cB^\cC\subset\cG} \left(-\lambda_\cC
 \sum_{n_\cB, \bar n_\cB, m_\cB, \bar m_\cB}   \delta_{\bar n_\cB^{\cD\setminus \cC} n_\cB^{\cD\setminus \cC}  }  \ 
   \delta_{n_\cB^{\cC}\bar m_\cB^{\cC}} \delta_{   \bar n_\cB^{\cC}  m_\cB^{\cC}}\ 
     \delta_{\bar m_\cB^{\cD\setminus \cC}  m_\cB^{\cD\setminus \cC}  }  \   \;
 \right)\crcr
 &\times \prod_{\ell\in E^0} (\bC_{\le j_{\max}})_{n_\ell\bar n_\ell}\ \times\ \prod_{\cB\in B_N}p_\cB(k_\cB)
 \ , 
\end{align}
where $B_N$ is the set of necklace bubbles of $\cG$, and the $p_\cB(k_\cB)$ are the corresponding momenta arising from derivative couplings.
As usual, the amplitude decomposes in
\[
 A^{\rm reg}(\cG) =  \lambda_1^{|B_M|}\lambda_2^{|B_N|} \cA^{\rm reg}_{\cG}\ . 
\]

\subsubsection*{Scale decomposition}
Using the scale decomposition  of the regularized propagator, one can write
\begin{align}
 \cA^{\rm reg}_{\cG}
 &=\sum_\mu {\cA}_{G}^\mu
 \ .
\end{align}
where $\mu=\{j_\ell,\ \ell\in E^0\}\in\{1\dots j_{\max}\}^{|E^0|}$
and ${\cA}_{G}^\mu$ is the corresponding amplitude,
\begin{align} \label{amplitude3}
 {\cA}_{G}^\mu
 &=\sum_{\{n\}}\ \,\prod_{\cB\subset\cG}
    \delta^{\cB}  \   \;
 \prod_{\ell\in E^0} \left[(\bC_{ j_\ell(\mu)})_{n_\ell\bar n_\ell}\right]  \times\ \prod_{\cB\in B_N}p_\cB(k_\cB)
\ .
\end{align}

We can now established the power counting theorem.
\begin{theorem}\label{thm:enhpowcount}
The amplitude $\mathcal{A}_{\mathcal{G}\mu}$ for the scale attribution $\mu$ admit the following uniform bound:
\begin{equation}
|\mathcal{A}_{\mathcal{G}\mu}|\leq M^{2\eta|E^0|}\prod_{i}\prod_{k=1}^{\kappa(i)}M^{\omega(\mathcal{G}_i^k)},
\end{equation}
where $K$ is a constant depending on the graph $\mathcal{G}$ and, denoting $E^0 ,\ F_{\rm int}  ,\ F_{\rm ext}$ the sets of internal colour 0 edges, internal and external faces of a graph, the degree of divergence $\omega$ is defined by:
\begin{equation}
\omega(\mathcal{G}_i^k) = -2\eta |E^0(\mathcal{G}_i^k)|+|F_{\rm int}(\cG_i^k)|+\sum_{f\in F_{\rm int}(\cG_i^k)}\alpha(f) \ ,\label{eq:enhdivdegree}
\end{equation}
whith $\alpha(f)$ the number of occurrences of derivative couplings $p(k)$ along the face $f$.
\end{theorem}

\prf
The proof follow closely the one of theorem \ref{thm:powercounting}. Note that with the scale slices of \eqref{eq:enhslices},
 the amplitude is bounded by, 
\begin{align}
 {\cA}_{\cG}^\mu&\leq 
 \sum_{\{n\}}\ \,\prod_{\cB\subset\cG}
    \delta^{\cB}  \   \;
 \prod_{\ell\in E^0} \left[
 M^{(2-2j_\ell)\eta} \prod_{c\in \cD} \delta_{n^c_\ell\bar n^c_\ell}\ \bI_{|n^c_\ell|\leq M^{j_\ell}}
 \right]\ \prod_{\cB\in B_N}p_\cB(k_\cB)
 \ .
\end{align}
For a momentum $p_B$ corresponding to a coloured edge belonging to a face $f$, we can write,
\[
 |p_\cB|\le M^{j_{\min}(f)} \ ,
\]
therefore, as each $p_\cB$ belong to one and only one face,
\[
 \prod_{\cB\in B_N}p_\cB(k_\cB) \leq \prod_f  M^{\alpha(f)\ j_{\min}(f)}\ ,
\]
and we can conclude, using the same trick as for Theorem \ref{thm:powercounting}.

\qed

\subsection{Divergent maps}

\subsubsection*{Intermediate field representation}

Once again, it is convenient to study the structure of the divergent graphs in the intermediate field representation. The Hubbard Stratonovich field formulation
will not be used, as the purely graphical correspondence of Section \ref{sec:graphIFrep} is sufficient for our needs.
Note that, in the intermediate field representation, $E^0(\cG)$ is the set of \emph{corners} of the associated map, $B_M$ the set of \emph{mono-coloured edges} and $B_N$ the set of \emph{bi-coloured edges}. A graph and its associated map will be identified and therefore share the same notation.

\subsubsection*{Divergence degree}
In a high sub-map $\cG_i^k$, each bi-coloured edge $\e\in B_N(\cG_i^k)$ bears a momentum $p(\e)$ on one of its strands $s$. If the strand $s$ belongs to an internal 
face $f\in F_{\rm int}$, then the momentum insertion contributes to $\alpha(f)$ and adds one to the divergent degree $\omega(\cG_i^k)$. But if the strand $s$ 
belongs to an external face, the momentum insertion does not contribute to the divergent degree of the high sub-map $\cG_i^k$. 
Therefore, the set $B_N(\cG_i^k)$ must be divided between the set $B_N^{\rm int}(\cG_i^k)$ of bi-coloured edges bearing a momentum in an internal 
face of $\cG_i^k$ and the set $B^{\rm ext}_N(\cG_i^k)$  of bi-coloured edges bearing a momentum in an external face, that does not contribute to $\omega(\cG_i^k)$. 

We write $P$ the number of cilia, such that $2P$ is the number of external edges in the ordinary representation, 
and $B$ the set intermediate field-edges, or bubbles in the original representation. Using
\begin{align}
 \sum_{f\in F_{\rm int}(\cG_i^k)}\alpha(f) = | B^{int}_N(\cG_i^k)|\qquad {\rm and}\qquad |E^0(\cG_i^k)|=2| B(\cG_i^k)| - p(\cG_i^k)\ ,
\end{align}
the divergent degree \eqref{eq:enhdivdegree} can be rewritten
\begin{align}
\omega
&=-4\eta (|B_M|+ |B^{ext}_N|)\ +\ (1-4\eta)|B^{\rm int}_N|\ +\ |F_{\rm int}| +2\eta P \ . \label{eqdivdegreewithbubbles}
\end{align}
Counting faces and vertices immediately gives us the following results,
\begin{itemize}
 \item Let $\cG$ be a map and $\e$ an edge that is not a bridge. Denote $\cG_{\backslash \e}$ the map obtained by removing the edge $\e$. Then
 \begin{enumerate}
  \item  if $\e\in B_M$, $\omega(\cG) \leq \omega(\cG_{\backslash \e})-4\eta +1$
 \item if $\e\in B_N^{\rm ext}$,  $\omega(\cG) \leq \omega(\cG_{\backslash \e})-4\eta +2$
 \item if $\e\in B_N^{\rm int}$, $\omega(\cG) \leq \omega(\cG_{\backslash \e})-4\eta +3$
 \end{enumerate}
 \item Let $\cG$ be a map. $\cG'$ a map obtained by adding a non ciliated leaf to $\cG$ i.e. a non ciliated vertex connected 
 to a corner of $\cG$ by one edge $\e$, either mono- or bi-coloured. Then,
 \begin{enumerate}
 \item if $\e\in B_M\ {\rm or}\ B_N^{\rm int}$, $\omega(\cG) = \omega(\cG')+4\eta -3$,
  \item if $\e\in B_N^{\rm ext}$, $\omega(\cG) = \omega(\cG')+4\eta -2$.
 \end{enumerate}
 \item Let $\cG$ be a map. $\cG'$ a map obtained by adding a ciliated leaf to $\cG$. Then $\omega(\cG) \geq \omega(\cG')+2\eta -1$.
 \end{itemize}
 
 This allows us to establish the following propositions.
 \begin{proposition}\label{prop:treedegree}
 \begin{itemize}
  \item For a vacuum tree $\cT_0$ (with $|\cV(\cT)|+1$ vertices), $\omega(\cT_0)= 4+(3-4\eta)|B|$  .
  \item For a 2-point tree (with one cilium) $\cT_2$, $\omega(\cT_2)= 2\eta+(3-4\eta)|B|-|B_N^{\rm ext}|$ .
  \item For a tree $\cT_{2P}$ with $P$ cilia, $\omega(\cT_{2P}) \leq 2\eta + (3-4\eta)(|B|+1-P)+ (1-2\eta)(P-1)-|B_N^{\rm ext}|$. 
 \end{itemize}
 \end{proposition}
 \begin{proposition}\label{prop:loopdegree}
  Let $\cG$ be a map and $\cT\in\cG$ a spanning tree. Then, 
  \begin{equation}
   \omega(\cG)\leq \omega(\cT) +(3-4\eta)\left(|B(\cG)|-|B(\cT)|\right) .
  \end{equation}
 \end{proposition}

 Therefore, 
  \begin{itemize}
   \item For $\eta> 3/4$, the divergent degree of a map proliferates with the number of edges. Divergent $2P$-point high sub-maps occur for any $P\in\setN$
   and the theory is not renormalisable.
   \item For $\eta < 3/4$, only graphs with a finite number of bubbles, or maps with a finite number of edges, can diverge.
   Thus only a finite number of maps diverges,and  the model is super-renormalisable.
   \item For $\eta=3/4$, only the vacuum and up to 8-point maps can have positive degree and diverge.
   However, the divergent degree of trees does not depend on the number of edges, thus infinitely many maps diverge, 
   the theory is just-renormalisable. Under certain conditions, a map with loop edges can diverge. 
  \end{itemize}

  \subsubsection*{Leading order graphs for $\eta=3/4$.}
  
  In the just-renormalisable case $\eta=3/4$, the divergent degree becomes
  \begin{equation}
   \omega=-3 (|B_M|+ |B^{\rm ext}_N|)\ -\ 2 |B^{int}_N|\ +\ |F_{\rm int}| +\frac32 P 
   \ . \label{eqdivdegree3/4}
  \end{equation}
  According to Propositions \ref{prop:treedegree} and \ref{prop:loopdegree}, for $\eta = 3/4$, The degree of any $2P$-point map $\cG_{2P}$ is bounded by
  \begin{equation}\omega(\cG_0) \leq 4\ ,\qquad \omega(\cG_{2P})\leq 2 - \frac{P}{2}\ {\rm for}\ P\geq 1\ .
  \end{equation}
Therefore only vacuum, and up to 8-point maps can diverge. The characterisation of the leading order of divergent maps, which is the set of maps with highest degree, 
for $2P\leq4$, follows the one of enhanced invariant models. 

\begin{proposition}\label{propcaractLO1}
 For $\cG$ a leading order map,
 \begin{enumerate}
 \item $|B_N^{\rm ext}(\cG)|=0$.
  \item Any mono-coloured edge is a bridge.
  \item For all $i\neq1$, any connected sub-map made of bi-coloured edges of colour $1i$ is planar. 
  \item Let us choose a sub-map $\cT_\cG$ following the construction of Definition \ref{def:enhtreecarac}, then $\cT_\cG$ is a tree.
 \end{enumerate}
 \end{proposition}
%

\subsubsection{Counter-terms}

Finally, let us discuss the structure of the counter-terms required to renormalise the theory:
\begin{itemize}
 \item a vacuum term to absorb the (up to) quartic divergences of the free energy,
 \item a quadratic, tensor invariant, mass counter-term, for divergences in $p^{3/2}$,
 \item a non invariant mass counter-term of type $\sum_p p\psit_p\bar\psit_p$ to absorb $p^{1/2}$ divergences that arise whenever an external face goes through 
  the ``enhanced" strand of a bi-coloured bridge,
 \item a set of 4 invariant quartic melonic counter-terms,  and 3 necklace invariant counter-terms to absorb linear divergences,
 \item a disconnected quadratic term of type $(\psit\cdot_\cD\bar\psit)^2$, to absorb logarithmic divergences,
 \item a full set of non-invariant quartic melonic and necklace invariants, with ''derivative`` couplings $p$, to absorb logarithmic divergences.
 \item invariant necklaces, melons and necklaces with melonic insertions with 6 and 8 vertices, for respectively square-root and logarithmic divergences.
 \end{itemize}


\begin{thebibliography}{99}

   \bibitem{AR1994}
A.~Abdesselam and V.~Rivasseau,  
Trees, forests and jungles: a botanical garden for cluster expansions,
 Lect. Notes Phys. 446 (1995) 7-36,
   arXiv:hep-th/9409094
   
   \bibitem{AGBC1992}
   L. Alvarez-Gaume, J.L.F. Barbon and C. Crnkovic,
    A Proposal for Strings at $D>1$,
    Nucl.Phys. B 394 (1993) 383-422,
    arXiv:hep-th/9208026.
    
    \bibitem{ABJ1991}
    J. Ambjørn, B. Durhuus and T. Jonsson,
    Three-dimensional simplicial quantum gravity and generalized matrix models,
Mod. Phys. Lett. A6 (1991) 1133-1146.
%
   
   \bibitem{BeBeOr1411}
   D. Benedetti, J. Ben Geloun and D. Oriti,
    Functional Renormalisation Group Approach for Tensorial Group Field Theory: a Rank-3 Model,
    JHEP 1503 (2015) 084,
  arXiv:1411.3180 [hep-th].
   
   
   \bibitem{BengeRiv1111}
   J.~Ben Geloun and V.~Rivasseau,
   A Renormalizable 4-Dimensional Tensor Field Theory,
   Commun. Math phys, 2013, 318, pp.69-109,
   arXiv:1111.4997 [hep-th].
   
   
   \bibitem{BiaBur1996}
P. Bialas and Z. Burda,   
Phase transition in fluctuating branched geometry,
Phys.Lett. B 384 (1996) 75-80,   
arXiv:hep-lat/9605020.


\bibitem{BGRR1105}
V. Bonzom, R. Gurau, A. Riello and V. Rivasseau
 Critical behavior of colored tensor models in the large N limit,
 Nucl. Phys. B 853 (2011) 174-195,
arXiv:1105.3122 [hep-th].
   
 \bibitem{BGR1202}
V.~Bonzom, R.~Gurau, and V.~Rivasseau, 
Random tensor models in the large N
  limit: Uncoloring the colored tensor models,
 Phys. Rev. D 85, 084037 (2012), 
 arXiv:1202.3637 [hep-th].
 
 \bibitem{BonGurRyaTan1404}
   V.~Bonzom, R.~Gurau, J.~P.~Ryan and A.~Tanasa,
  The double scaling limit of random tensor models,
  JHEP  1409, 051 (2014),
  arXiv:1404.7517 [hep-th].
 
 \bibitem{BonDelRiv1502}
 V.~Bonzom, T.~Delepouve and V.~Rivasseau,
  Enhancing non-melonic triangulations: A tensor model mixing melonic and planar maps,
  Nuclear Physics B895 (2015) 161-191,
  arXiv:1502.01365 [math-ph].
  
\bibitem{BoLiRi1508}
V. Bonzom, L. Lionni and V. Rivasseau,
 Colored triangulations of arbitrary dimensions are stuffed Walsh maps,
 arXiv:1508.03805 [math.CO].
 
  \bibitem{Bou92}
   D. Boulatov,
 A Model of Three-Dimensional Lattice Gravity,
Mod.Phys.Lett. A7 (1992) 1629-1646,
ArXiv:hep-th/9202074.

\bibitem{BIPZ1978}
E. Br\'ezin, C. Itzykson, G. Parisi, and J. B. Zuber,
Planar diagrams,
Comm. Math. Phys. 59 (1978) 35-51.

   \bibitem{BK1987}
   D. Brydges and T. Kennedy,
Mayer expansions and the Hamilton-Jacobi equation,
Journal ofStatistical Physics, 48, (1987) 19-49.

\bibitem{CarOriRiv1207}
S. Carrozza, D. Oriti and V. Rivasseau,
 Renormalization of Tensorial Group Field Theories: Abelian U(1) Models in Four Dimensions,
 Commun. Math. Phys. 327(2014) 603-641,
  arXiv:1207.6734 [hep-th].
  
  \bibitem{Carrozza2014}
  S. Carrozza,
  {\it Tensorial Methods and Renormalization in Group Field Theories},
      Springer International Publishing (2014).
      
      \bibitem{CarLah1612}
 S. Carrozza and V. Lahoche,
  Asymptotic safety in three-dimensional SU(2) Group Field Theory: evidences in the local potential approximation,
  arXiv:1612.02452 [hep-th].
      
  
  \bibitem{Cayley1889}
  A. Cayley, 
  A theorem on trees, 
  Quart. J. Math 23 (1889) 376-378.
  
  
  \bibitem{DarGurRiv1307}
  S.~Dartois, R.~Gurau and V.~Rivasseau,
  Double Scaling in Tensor Models with a Quartic Interaction,
  JHEP  1309 (2013) 088,
  arXiv:1307.5281 [hep-th].
  
  \bibitem{DDSW1990}
  S. R. Das, A. Dhar, A. M. Sengupta and S. R. Wadia,
  New Critical Behavior in d = 0 Large-N Matrix Models,
  Mod. Phys. Lett. A 05 (1990) 1041.
  
   \bibitem{DGR1403}
  T.~Delepouve, R.~Gurau and V.~Rivasseau,
  Universality and Borel summability of arbitrary quartic tensor models, 
  Ann. Inst. H. Poincaré Probab. Statist. 52 (2016), no. 2, 821-848,
  arXiv:1403.0170 [hep-th].

 \bibitem{DG1504}
T.~Delepouve and R.~Gurau,  
Phase Transition in Tensor Models,
 JHEP 06 (2015) 178,
   arXiv:1202.3637 [hep-th].
   
   \bibitem{DR1412}
     T.~Delepouve and V.~Rivasseau,
   Constructive Tensor Field Theory: The T43 Model,
   Commun. Math. Phys. 345 (2016) 477,
   arXiv:1412.5091 [math-ph].
   
   
   \bibitem{DFGZJ1993}
   P. Di Francesco, P. Ginsparg and J. Zinn-Justin,
   2D Gravity and Random Matrices
   Phys. Rept. 254 (1995) 1-133,
   arXiv:hep-th/9306153.
   
   \bibitem{Freidel2005}
   L. Freidel,
   Group Field Theory: An Overview,
    Int. J. Theor. Phys. 44 (2005) 1769-1783,
    arXiv:hep-th/0505016.
    
    \bibitem{Gross1992}
    M. Gross,
    Tensor models and simplicial quantum gravity in $>2-D$,
    Nuc. Phys. B Proc. Suppl. 25 (1992) 144–149.
   
\bibitem{Gurau0907}
R.~Gurau,
 Colored Group Field Theory,
 Commun. Math. Phys. 304 (2011) 69-93,
 arXiv:0907.2582 [hep-th].
 
 \bibitem{Gurau0911}
 R.~Gurau,
  Topological Graph Polynomials in Colored Group Field Theory,
  Annales Henri Poincare 11 (2010) 565-584,
  arXiv:0911.1945 [hep-th].
  
   
\bibitem{Gurau1006}
R.~Gurau,
 Lost in Translation: Topological Singularities in Group Field Theory,
  Class. Quant. Grav. 27 (2010) 235023,
  arXiv:1006.0714 [hep-th].
  
\bibitem{Gurau1011}
R.~Gurau,
The 1/N expansion of colored tensor models,
Ann. H. Poincar\'e 12 (2011) 829-847,
 arXiv:1011.2726 [gr-qc]

\bibitem{Gurau1102}
R.~Gurau,
The Complete 1/N Expansion of Colored Tensor Models in Arbitrary Dimension,
Ann. Henri Poincar\'e 13 (2012) 399,
arXiv:1102.5759 [gr-qc].
   
   \bibitem{GuRy1109}
   R.~Gurau and J.~P.~Ryan,
    Colored Tensor Models - a Review,
   SIGMA 8 (2012), 020,
   arXiv:1109.4812 [hep-th]

\bibitem{Gurau1111}
R.~Gurau, Universality for random tensors, 
Ann. Inst. H. Poincar\'e Probab. Statist. 50 (2014), no. 4, 1474-1525,
arXiv:1111.0519 [math.PR].
  
\bibitem{GuRi1101}
R. Gurau and V. Rivasseau,
 The 1/N expansion of colored tensor models in arbitrary dimension,
 Europhys. Lett. 95 (2011) 50004,
 arXiv:1101.4182 [gr-qc].
  
   \bibitem{GuRy1302}
  R.~Gurau and J.~P.~Ryan,
  Melons are branched polymers,
  Ann. Henri Poincaré 15 (2014) 2085-2131,
  arXiv:1302.4386 [math-ph].
  
    \bibitem{Gurau1304}
  R.~Gurau,
  The 1/N Expansion of Tensor Models Beyond Perturbation Theory,
  Commun.Math.Phys. 330 (2014) 973-1019,
  arXiv:1304.2666 [math-ph] .
   
   \bibitem{GR1312.7226}
  R.~Gurau and V.~Rivasseau,
  The Multiscale Loop Vertex Expansion,
   Ann. Henri Poincare 16 (2015) 1869-1897,
  arXiv:1312.7226 [math-ph].
  
  \bibitem{GurauHDR}
  R.~Gurau,
  {\it Random Tensors},
  Oxford University Press (2016).
  
  \bibitem{Gurau1611}
   R.~Gurau,
  1/N expansion of a SYK-like tensor model,
 arXiv:1611.04032 [hep-th].
  
  \bibitem{Hubbard}
  J. Hubbard,
  Calculation of Partition Functions,
Phys. Rev. Lett. 3  (1959) 77 .
  
  \bibitem{JaiMat1992}
  S. Jain and S. D. Mathur,
   World-Sheet Geometry and Baby Universes in 2-D Quantum Gravity,
   Phys.Lett. B 286 (1992) 239-246,
   arXiv:hep-th/9204017.
   
   \bibitem{Kit2015}
   A. Kitaev, 
   A simple model of quantum holography,
KITP strings seminar and Entanglement
2015 program (Feb. 12, April 7, and May 27, 2015)
.
  
  \bibitem{LeGall2007}
  J. F. Le Gall,
  The topological structure of scaling limits of large planar maps,
  Invent. math. 169 (2007) 621-670,
  arXiv:math/0607567 [math.PR].
  
  \bibitem{LeGall1105}
  J. F. Le Gall,
   Uniqueness and universality of the Brownian map, 
  Ann. Probab. 41 (2013) 2880-2960,
   arXiv:1105.4842 [math.PR].    
  
   \bibitem{MagRiv}
 J.~Magnen and V.~Rivasseau,
 Constructive $\phi^4$ field theory without tears,
Annales Henri Poincare 9 (2008) 403-424, 
arXiv:0706.2457 [math-ph].
  
  \bibitem{Magnen:2009at} 
  J.~Magnen, K.~Noui, V.~Rivasseau and M.~Smerlak,
  Scaling behaviour of three-dimensional group field theory,
  Class.\ Quant.\ Grav.\  26 (2009) 185012,
  arXiv:0906.5477 [hep-th].
  
   
   \bibitem{MatSal1954}
   P.T. Matthews and A. Salam,
  The Green's functions of quantized fields,
    Nuovo Cim. 12 (1954) 563-565.
    
    \bibitem{Muh1962}
    B. M\"uhlschlegel,
     Asymptotic Expansion of the Bardeen-Cooper-Schrieffer Partition Function by Means of the Functional Method,
J. Math. Phys. 3 (1962) 522-530. 

\bibitem{NguDarEyn1409}
V. A. Nguyen, S. Dartois and B. Eynard,
An analysis of the intermediate field theory of $T^4$ tensor model,
JHEP 01 (2015) 13,
arXiv:1409.5751 [math-ph].


\bibitem{Oriti2001}
D. Oriti,
Spacetime geometry from algebra: spin foam models for non-perturbative quantum gravity,
Rept. Prog. Phys. 64 (2001) 1703-1756,
arXiv:gr-qc/0106091.


  \bibitem{Oriti0912}
  D. Oriti,
   The group field theory approach to quantum gravity: some recent results,
   The Planck Scale: Proceedings of the XXV Max Born Symposium (2009),
    arXiv:0912.2441 [hep-th].
    
     \bibitem{Oriti1110}
  D. Oriti,
   The microscopic dynamics of quantum space as a group field theory,
   in 
{\it Foundations of space and time},
 G. Ellis, J. Murugan (eds.), Cambridge University Press (2011),
   arXiv:1110.5606 [hep-th].
   
  \bibitem{Oriti1211}
  D. Oriti,
  The quantum geometry of tensorial group field theories,
   proceedings of The XXIX International Colloquium on Group-Theoretical Methods in Physics, Chern Institute of Mathematics, Tianjin (2012),
   arXiv:1211.5714 [hep-th].
   
   
   \bibitem{Perez2003}
   A. Perez,
  Spin foam models for quantum gravity,
  Class .Quant. Grav. 20 (2003) R43, 
  arXiv:gr-qc/0301113.
  
  \bibitem{PFKR99}
R. De Pietri, L. Freidel, K. Krasnov and C. Rovelli,  
  Barrett-Crane model from a Boulatov-Ooguri field theory over a homogeneous space,
  Nucl.Phys. B574 (2000) 785-806,
  arXiv:hep-th/9907154.
  
\bibitem{PolRos1601}
J. Polchinski and V. Rosenhaus,
 The Spectrum in the Sachdev-Ye-Kitaev Model,
JHEP 04 (2016) 1,
arXiv:1601.06768 [hep-th].
 
  
  \bibitem{Prufer1918}
 H. Pr\"ufer, 
 Neuer Beweis eines Satzes über Permutationen, 
 Arch. Math. Phys. 27 (1918) 742–744.
 
  
  \bibitem{Rivasseau1991}
  V. Rivasseau,
  {\it From Perturbative to Constructive Renormalization},
  Princeton University Press (1991).

  \bibitem{Rivasseau0706}
  V. Rivasseau,
   Constructive Matrix Theory,
  JHEP 0709 (2007) 008,
   arXiv:0706.1224 [hep-th].

\bibitem{SacYe1992}
S. Sachdev and J. Ye,
 Gapless Spin-Fluid Ground State in a Random Quantum Heisenberg Magnet,
 Phys. Rev. Lett. 70 (1993) 3339,
 arXiv:cond-mat/9212030.
 

 \bibitem{Sasakura1991}
N. Sasakura, 
Tensor model for gravity and orientability of manifold,
Mod. Phys. Lett. A6 (1991) 2613-2624.
   
   \bibitem{Sokal1980}
A.~D.~Sokal,  
An improvement on Watson's theorem on Borel summability,
 J. Math. Phys. 21 (1980) 261-263,
   arXiv:1202.3637 [hep-th].
   
   \bibitem{Stratonovich}
  R. L. Stratonovich, 
  Soviet Physics Doklady 2 (1957) 416.
  
  \bibitem{tHooft1974}
  G. 't Hooft,
  A Planar Diagram Theory for Strong Interactions,
  Nucl.Phys. B 72 (1974) 461.
  
  \bibitem{Wishart1928}
  J. Wishart,
  The generalised product moment distribution in samples from a normal multivariate population. Biometrika 20 (1928) 32-52. 
  
 \bibitem{Witten1610}
 E. Witten,
 An SYK-Like Model Without Disorder,
arXiv:1610.09758 [hep-th].
  
  
\end{thebibliography}
\end{document}